%% file: main.tex
\documentclass[acmsmall,screen,nonacm]{acmart}

\usepackage{subcaption}
\usepackage{geometry}
\usepackage{adjustbox}
\usepackage{tikz-cd}
\usepackage{tikz}
\usetikzlibrary{cd,arrows}
\usepackage{mathpartir}
\usepackage{subcaption}
\usepackage{amsfonts}
\usepackage[renew-matrix,renew-dots]{nicematrix}
\usepackage{float}
\usepackage{enumitem}
\usepackage{xcolor}
\usepackage{booktabs}
\usepackage{rotating}
\usepackage{multirow}
\usepackage{comment}
\usepackage{braket}
\usepackage{enumitem}
\usepackage{xparse}
\usepackage{amsthm}
\usepackage{mathtools}
\usepackage{xifthen}
\usepackage{makecell}

\colorlet{tapeBg}{red!30}
\colorlet{tapeBorder}{red!60}

\DeclareMathSymbol{\mathinvertedexclamationmark}{\mathclose}{operators}{'074}
\DeclareMathSymbol{\mathexclamationmark}{\mathclose}{operators}{'041}

\makeatletter
\newcommand{\raisedmathinvertedexclamationmark}{\mathclose{\mathpalette\raised@mathinvertedexclamationmark\relax}}
\newcommand{\raised@mathinvertedexclamationmark}[2]{\raisebox{\depth}{$\m@th#1\mathinvertedexclamationmark$}}
\begingroup\lccode`~=`! \lowercase{\endgroup
  \def~}{\@ifnextchar`{\raisedmathinvertedexclamationmark\@gobble}{\mathexclamationmark}}
\mathcode`!="8000
\makeatother

\usepackage{etoolbox}

\makeatletter
\patchcmd{\endalign}{\restorealignstate@}{\global\let\df@label\@empty\restorealignstate@}{}{}
\makeatother

\makeatletter
\newcommand{\mylabel}[2]{\def\@currentlabel{#2}\label{#1}}
\makeatother

\makeatletter
\newcommand\newtag[2]{#1\def\@currentlabel{#1}\label{#2}}
\makeatother

\newcommand{\op}[1]{#1^{\dagger}}

\newcommand{\CNum}{\mathbb{C}}

\newcommand{\multiset}[1]{\left\{\!\left| \, #1 \, \right|\! \right\}} 

\newcommand{\lbbd}{\left \llbracket}
\newcommand{\rbbd}{\right \rrbracket}
\newcommand{\dsem}[1]{\lbbd #1 \rbbd}

\newcommand{\CBdsem}[1]{\dsem{#1}}
\newcommand{\TCBdsem}[1]{\dsem{#1}^{\sharp}}

\makeatletter
\newsavebox{\@brx}
\newcommand{\Myllangle}[1][]{\savebox{\@brx}{\(\m@th{#1\langle}\)}%
  \mathopen{\copy\@brx\kern-0.5\wd\@brx\usebox{\@brx}}}
\newcommand{\Myrrangle}[1][]{\savebox{\@brx}{\(\m@th{#1\rangle}\)}%
  \mathclose{\copy\@brx\kern-0.5\wd\@brx\usebox{\@brx}}}
\makeatother

\newcommand{\dsemRel}[1]{\Myllangle #1\Myrrangle_\interpretation}

\newcommand{\id}[1]{{id}_{#1}}

\newcommand{\perG}{\odot} \newcommand{\unoG}{I} 

\newcommand{\piu}{\oplus} \newcommand{\per}{\otimes} 

  \newcommand{\perT}{\per_{\CatTape}}

\newcommand{\kron}{\circledast} \newcommand{\perM}{\per_{\Cat{Mat}}}

\NewDocumentCommand\Piu{oom}{\bigoplus_{\IfNoValueTF{#1}{}{#1}}^{\IfNoValueTF{#2}{}{#2}}{#3}} \NewDocumentCommand\Per{oom}{\bigotimes_{\IfNoValueTF{#1}{}{#1}}^{\IfNoValueTF{#2}{}{#2}}{#3}} 

\NewDocumentCommand\PiuPar{oom}{\Piu[#1][#2] {(#3)}} \NewDocumentCommand\PerPar{omm}{\Per[#1][#2] {(#3)}}

\NewDocumentCommand\PiuL{oom}{\bigoplus\limits_{\IfNoValueTF{#1}{}{#1}}^{\IfNoValueTF{#2}{}{#2}}{#3}} \NewDocumentCommand\PerL{oom}{\bigotimes\limits_{\IfNoValueTF{#1}{}{#1}}^{\IfNoValueTF{#2}{}{#2}}{#3}} 

\NewDocumentCommand\PiuParL{oom}{\PiuL[#1][#2] {(#3)}} \NewDocumentCommand\PerParL{omm}{\PerL[#1][#2] {(#3)}} 

\newcolumntype{C}[1]{>{\centering\let\newline\\\arraybackslash\hspace{0pt}}p{#1}}
\newcolumntype{L}[1]{>{\raggedright\let\newline\\\arraybackslash\hspace{0pt}}p{#1}}
\newcolumntype{R}[1]{>{\raggedleft\let\newline\\\arraybackslash\hspace{0pt}}p{#1}}

\newcommand{\sort}{\mathcal{S}}
\newcommand{\sign}{\Sigma}
\renewcommand{\terms}[1]{\mathcal{T}_{#1}} 

\newcommand{\gen}{s}

\newcommand{\Cat}[1]{\mathbf{#1}} \newcommand{\CatString}{\Cat{C}_\sign} \newcommand{\CatTape}{\Cat{T}_\sign}   
  \newcommand{\fssbRig}{\mathbf{ssR}_\sign^\flat}
\newcommand{\fssbRigM}{\mathbf{ssR}_{\sign_M}^\flat}
\newcommand{\sCat}[1]{{\Cat{#1}}} \newcommand{\sCatT}[1]{\overline{\Cat{#1}}}

\newcommand{\CMon}{\Cat{CMon}} \newcommand{\CAT}{\Cat{Cat}} \newcommand{\CMonCat}{\CMon\CAT} \newcommand{\fbCat}{\Cat{FBC}} \newcommand{\FBC}{\Cat{FBC}} \newcommand{\fCMon}[1]{#1^+} \newcommand{\Mat}[1]{\Cat{Mat}(#1)} \newcommand{\SMC}{\Cat{SMC}}
\newcommand{\ZX}{\Cat{ZX}}
\newcommand{\TZX}{\Cat{T_{ZX}}}

\newcommand{\Rel}{\Cat{Rel}}
\newcommand{\sRel}{\Rel}
\newcommand{\sRelT}{\sCatT{\Rel}}

 \newcommand{\CB}{\Cat{CB}_{\sign}} \newcommand{\CatTapeCB}{\Cat{T}_{\CB}}

\newcommand{\monomial}{\Cat{Mnm}}

\newcommand{\GFunct}{\beta}

\newcommand{\FF}{\mathcal F}
\newcommand{\GG}{\mathcal G}
\newcommand{\Gg}{\overline{\GG}}

\newcommand{\assoc}[3]{\alpha_{#1,#2,#3}} \newcommand{\lunit}[1]{\lambda_{#1}} \newcommand{\runit}[1]{\rho_{#1}} \newcommand{\Iassoc}[3]{\alpha^{-}_{#1,#2,#3}} \newcommand{\Ilunit}[1]{\lambda^{-}_{#1}} \newcommand{\Irunit}[1]{\rho^{-}_{#1}} \NewDocumentCommand\symm{gg}{\sigma\IfNoValueTF{#1}{}{_{#1,#2}}}

\newcommand{\assocp}[3]{\alpha^\piu_{#1,#2,#3}} \newcommand{\lunitp}[1]{\lambda^\piu_{#1}} \newcommand{\runitp}[1]{\rho^\piu_{#1}} \newcommand{\zero}{\ensuremath{0}} \NewDocumentCommand\symmp{gg}{\sigma^\piu\IfNoValueTF{#1}{}{_{#1,#2}}} 

\newcommand{\Iassocp}[3]{\alpha^{-\piu}_{#1,#2,#3}}

\newcommand{\assoct}[3]{\alpha^\per_{#1,#2,#3}}  \newcommand{\runitt}[1]{\rho^\per_{#1}} \newcommand{\uno}{1} \NewDocumentCommand\symmt{gg}{\sigma^\per\IfNoValueTF{#1}{}{_{#1,#2}}} 

   \newcommand{\Isymmt}[2]{\sigma^{-\per}_{#1,#2}}

\newcommand{\dl}[3]{\delta^l_{#1,#2,#3}} \newcommand{\dr}[3]{\delta^r_{#1,#2,#3}} 

\newcommand{\annl}[1]{\lambda^\bullet_{#1}} \newcommand{\annr}[1]{\rho^\bullet_{#1}} 

\newcommand{\Idl}[3]{\delta^{-l}_{#1,#2,#3}} \newcommand{\Idr}[3]{\delta^{-r}_{#1,#2,#3}} 

\newcommand{\Iannl}[1]{\lambda^{-\bullet}_{#1}} \newcommand{\Iannr}[1]{\rho^{-\bullet}_{#1}}

\NewDocumentCommand\Dl{gg}{\Delta^l\IfNoValueTF{#1}{}{_{#1,#2}}}
\NewDocumentCommand\Dr{gg}{\Delta^r\IfNoValueTF{#1}{}{_{#1,#2}}}
\NewDocumentCommand\IDl{gg}{\Delta^{-l}\IfNoValueTF{#1}{}{_{#1,#2}}}
\NewDocumentCommand\IDr{gg}{\Delta^{-r}\IfNoValueTF{#1}{}{_{#1,#2}}}
\NewDocumentCommand\D{gg}{\Delta\IfNoValueTF{#1}{}{_{#1,#2}}}
\NewDocumentCommand\ID{gg}{\Delta^{-}\IfNoValueTF{#1}{}{_{#1,#2}}}

\newcommand{\copier}[1]{\blacktriangleleft_{#1}} \newcommand{\discharger}[1]{\mbox{!}_{#1}} 

\newcommand{\proj}[1]{\pi_{#1}} 

\newcommand{\cocopier}[1]{\blacktriangleright_{#1}} \newcommand{\codischarger}[1]{\mbox{!`}_{#1}} 

\newcommand{\inj}[1]{\mu_{#1}} \newcommand{\pairing}[1]{\langle #1 \rangle}
\newcommand{\copairing}[1]{[ #1 ]}
\newcommand{\diagg}[2]{\diag{#1}^{#2}}
\newcommand{\codiagg}[2]{\codiag{#1}^{#2}}

\newcommand{\matr}[1]{M(#1)}

\newcommand{\diag}[1]{\lhd_{#1}}
\newcommand{\bang}[1]{\rotatebox{90}{$\multimap$}_{#1}}

\newcommand{\codiag}[1]{\rhd_{#1}}
\newcommand{\cobang}[1]{\rotatebox{90}{$\multimapinv$}_{#1}}

\newcommand{\precongR}[1]{\,{\leq}_{#1}\,}

\newcommand{\precongRA}[1]{\,{\lesssim}_{#1}\,}
\newcommand{\wiskR}[1]{\hat{#1}}
\newcommand{\basicR}{\mathbb{I}}
\newcommand{\wiskbasicR}{\wiskR{\basicR}}

\newcommand{\WprecongBA}{\precongRA{\wiskR{\basicR}}}

\newcommand{\precongB}{\precongR{{\basicR}}}

\newcommand{\interpretation}{\mathcal{I}}

\newcommand{\LW}[2]{L_{#1}({#2})} \newcommand{\RW}[2]{R_{#1}({#2})}   \newcommand\defeq{\stackrel{\text{\tiny def}}{=}}

\newcommand{\axeq}[1]{\stackrel{({#1})}{=}}
\newcommand{\axsubeq}[1]{\stackrel{({#1})}{\leq}}

  \newcommand{\PPiuL}[5]{\PiuL[#1=1][#2]{\PiuL[#3=1][#4]{#5}}}

\newcommand{\tape}[1]{\overline{\underline{\; #1 \vphantom{d_A} \;}}} \newcommand{\tapeFunct}[1]{\underline{\smash{\overline{\; #1 \vphantom{t} \;}}}}

\newcommand{\tapesymm}[2]{\tape{\sigma_{#1,#2}}} \renewcommand{\t}{\mathfrak{t}}
\newcommand{\s}{\mathfrak{s}}
\newcommand{\m}{\mathfrak{m}}

\newcommand{\zerotape}{\mathfrak{o}}
\newcommand{\ar}{\mathit{ar}}
\newcommand{\coar}{\mathit{coar}}

\theoremstyle{plain}
\newtheorem{theorem}{Theorem}[section]
\newtheorem{lemma}[theorem]{Lemma}
\newtheorem{corollary}[theorem]{Corollary}

\theoremstyle{definition}
\newtheorem{definition}[theorem]{Definition}
\newtheorem{proposition}[theorem]{Proposition}
\newtheorem{remark}[theorem]{Remark}
\newtheorem{example}[theorem]{Example}
\newtheorem*{notation}{Notation}

\newcommand{\PreCatTapeI}[1]{\Cat{T}_{\sign,#1}} \newcommand{\posetification}[1]{#1^\sim}
\newcommand{\CatTapeI}[1]{\posetification{\PreCatTapeI{#1}} } 
\newcommand{\CatTapeIGamma}[1]{\posetification{\Cat{T}_{\Gamma,#1}}}
\newcommand{\downsetification}[1]{{#1}^\downarrow}
\newcommand{\downward}[1]{#1\downarrow}

\newcommand{\CR}{\mathsf{CR}}
\newcommand{\CRS}{\mathsf{CR}_{\sign}}
\newcommand{\encoding}[1]{\mathcal{E}(#1)}
\newcommand{\minorExpression}{\leq_{\CR}}

\newcommand{\minorinTCB}{\leq_{T}}
\newcommand{\minorinCB}{\leq_{\CB}}

\usepackage{tikz}
\usetikzlibrary{cd,arrows,backgrounds,shapes.geometric,shapes.misc,matrix,arrows.meta,decorations.markings}
\pgfdeclarelayer{background}
\pgfdeclarelayer{edgelayer}
\pgfdeclarelayer{nodelayer}
\pgfsetlayers{background,edgelayer,nodelayer,main}

\tikzset{baseline=-0.5ex,>=stealth'}
\tikzset{every picture/.append style={scale=0.3}}
\tikzstyle{none}=[inner sep=0pt]
\tikzstyle{label}=[fill=none, draw=none, shape=circle, scale=0.7]
\tikzstyle{dots}=[fill=none, draw=none, shape=circle, scale=0.5]
\tikzstyle{black}=[circle, draw=black, fill=black, inner sep=0pt, minimum size=2.5pt]
\tikzstyle{white}=[circle, draw=black, fill=white, inner sep=0pt, minimum size=2.5pt]
\tikzstyle{reg}=[draw, fill=white, rounded rectangle, rounded rectangle left arc=none, minimum height=0.8em, minimum width=1em, node font={\tiny}]
\tikzstyle{coreg}=[draw, fill=white, rounded rectangle, rounded rectangle right arc=none, minimum height=0.8em, minimum width=1em, node font={\tiny}]
\tikzstyle{box}=[draw, fill=white, rectangle, minimum height=1em, minimum width=1em, node font={\tiny}]
\tikzstyle{triang}=[draw, fill=white, isosceles triangle, isosceles triangle apex angle=60, minimum size=0.5em, node font={\tiny}]

\tikzstyle{tapeFill} = [fill=tapeBg]
\tikzstyle{tapeNoFill} = [draw=tapeBorder, line width=0.5pt]
\tikzstyle{tape} = [fill=tapeBg, draw=tapeBorder, line width=0.5pt]

\tikzcdset{arrow style=tikz}

\newcommand{\sr}{\rule[-0.45cm]{0pt}{0.9cm}}

\newcommand{\comonoid}[1]{
    \begin{tikzpicture}
        \begin{pgfonlayer}{nodelayer}
            \node [style=none] (81) at (-0.75, 0) {};
            \node [style=black] (107) at (0, 0) {};
            \node [style=none] (108) at (0.5, -0.375) {};
            \node [style=none] (109) at (0.5, 0.375) {};
            \node [style=none] (110) at (0.75, -0.375) {};
            \node [style=none] (111) at (0.75, 0.375) {};
            \ifthenelse{\isempty{#1}}{}{
            \node [style=label] (112) at (1.075, -0.375) {$#1$};
            \node [style=label] (113) at (1.075, 0.375) {$#1$};
            \node [style=label] (114) at (-1.075, 0) {$#1$};
            }
        \end{pgfonlayer}
        \begin{pgfonlayer}{edgelayer}
            \draw [bend right] (109.center) to (107);
            \draw [bend right] (107) to (108.center);
            \draw (81.center) to (107);
            \draw (111.center) to (109.center);
            \draw (108.center) to (110.center);
        \end{pgfonlayer}
    \end{tikzpicture}           
 }

 \newcommand{\counit}[1]{
    \begin{tikzpicture}
        \begin{pgfonlayer}{nodelayer}
            \node [style=none] (81) at (-0.75, 0) {};
            \node [style=black] (107) at (0, 0) {};
            \ifthenelse{\isempty{#1}}{}{
            \node [style=label] (114) at (-1.075, 0) {$#1$};
            }
        \end{pgfonlayer}
        \begin{pgfonlayer}{edgelayer}
            \draw (81.center) to (107);
        \end{pgfonlayer}
    \end{tikzpicture}      
}

 \newcommand{\monoid}[1]{
    \begin{tikzpicture}
        \begin{pgfonlayer}{nodelayer}
            \node [style=none] (81) at (0.75, 0) {};
            \node [style=black] (107) at (0, 0) {};
            \node [style=none] (108) at (-0.5, -0.375) {};
            \node [style=none] (109) at (-0.5, 0.375) {};
            \node [style=none] (110) at (-0.75, -0.375) {};
            \node [style=none] (111) at (-0.75, 0.375) {};
            \ifthenelse{\isempty{#1}}{}{
            \node [style=label] (112) at (-1.075, -0.375) {$#1$};
            \node [style=label] (113) at (-1.075, 0.375) {$#1$};
            \node [style=label] (114) at (1.075, 0) {$#1$};
            }
        \end{pgfonlayer}
        \begin{pgfonlayer}{edgelayer}
            \draw [bend left] (109.center) to (107);
            \draw [bend left] (107) to (108.center);
            \draw (81.center) to (107);
            \draw (111.center) to (109.center);
            \draw (108.center) to (110.center);
        \end{pgfonlayer}
    \end{tikzpicture}                     
 }

\newcommand{\unit}[1]{
    \begin{tikzpicture}
        \begin{pgfonlayer}{nodelayer}
            \node [style=none] (81) at (-0.325, 0) {};
            \node [style=black] (107) at (-1.075, 0) {};
            \ifthenelse{\isempty{#1}}{}{
            \node [style=label] (114) at (0, 0) {$#1$};
            }
        \end{pgfonlayer}
        \begin{pgfonlayer}{edgelayer}
            \draw (81.center) to (107);
        \end{pgfonlayer}
    \end{tikzpicture}         
}

\newcommand{\boxCirc}[1]{
    \begin{tikzpicture}
        \begin{pgfonlayer}{nodelayer}
            \node [style=none] (60) at (-1, 0) {};
            \node [style=none] (61) at (1, 0) {};
            \node [style=box] (62) at (0, 0) {$#1$};
        \end{pgfonlayer}
        \begin{pgfonlayer}{edgelayer}
            \draw (60.center) to (62);
            \draw (62) to (61.center);
        \end{pgfonlayer}
    \end{tikzpicture}    
}

\newcommand{\wire}[1]{
    \begin{tikzpicture}
        \begin{pgfonlayer}{nodelayer}
            \node [style=label] (8) at (1.5, 0) {$#1$};
            \node [style=label] (11) at (-1.5, 0) {$#1$};
        \end{pgfonlayer}
        \begin{pgfonlayer}{edgelayer}
            \draw (11) to (8);
        \end{pgfonlayer}
    \end{tikzpicture}    
}

\newcommand{\Tcomonoid}[1]{
    \ifthenelse{\equal{#1}{\uno}}{
        \begin{tikzpicture}
            \begin{pgfonlayer}{nodelayer}
                \node [style=none] (1) at (1, 1) {};
                \node [style=none] (2) at (1, -1) {};
                \node [style=none] (6) at (-1.5, 0.5) {};
                \node [style=none] (7) at (-1.5, -0.5) {};
                \node [style=none] (8) at (1, -1.5) {};
                \node [style=none] (9) at (1, 1.5) {};
                \node [style=none] (10) at (1, 0.5) {};
                \node [style=none] (11) at (0.5, 0) {};
                \node [style=none] (12) at (1, -0.5) {};
                \node [style=none] (13) at (-0.5, -0.5) {};
                \node [style=none] (14) at (-0.5, 0.5) {};
                \node [style=none] (15) at (-1.5, 0) {};
            \end{pgfonlayer}
            \begin{pgfonlayer}{edgelayer}
                \path [style=tape] (9.center)
                     to [bend right] (14.center)
                     to (6.center)
                     to (7.center)
                     to (13.center)
                     to [bend right] (8.center)
                     to (12.center)
                     to [bend left=45] (11.center)
                     to [bend left=45] (10.center)
                     to cycle;
            \end{pgfonlayer}
        \end{tikzpicture}        
    }
    {
        \begin{tikzpicture}
            \begin{pgfonlayer}{nodelayer}
                \node [style=none] (0) at (0, 0) {};
                \node [style=none] (1) at (1, 1) {};
                \node [style=none] (2) at (1, -1) {};
                \ifthenelse{\isempty{#1}}{}{
                    \node [style=label] (3) at (-2, 0) {$#1$};
                    \node [style=label] (4) at (1.5, 1) {$#1$};
                    \node [style=label] (5) at (1.5, -1) {$#1$};
                }
                \node [style=none] (6) at (-1.5, 0.5) {};
                \node [style=none] (7) at (-1.5, -0.5) {};
                \node [style=none] (8) at (1, -1.5) {};
                \node [style=none] (9) at (1, 1.5) {};
                \node [style=none] (10) at (1, 0.5) {};
                \node [style=none] (11) at (0.5, 0) {};
                \node [style=none] (12) at (1, -0.5) {};
                \node [style=none] (13) at (-0.5, -0.5) {};
                \node [style=none] (14) at (-0.5, 0.5) {};
                \node [style=none] (15) at (-1.5, 0) {};
            \end{pgfonlayer}
            \begin{pgfonlayer}{edgelayer}
                \path [style=tape] (9.center)
                     to [bend right] (14.center)
                     to (6.center)
                     to (7.center)
                     to (13.center)
                     to [bend right] (8.center)
                     to (12.center)
                     to [bend left=45] (11.center)
                     to [bend left=45] (10.center)
                     to cycle;
                \draw [bend right=45] (0.center) to (2.center);
                \draw [bend left=45] (0.center) to (1.center);
                \draw (0.center) to (15.center);
            \end{pgfonlayer}
        \end{tikzpicture}           
    }             
}

\newcommand{\TLcomonoid}[2]{
    \begin{tikzpicture}
        \begin{pgfonlayer}{nodelayer}
            \node [style=none] (0) at (-0.5, 0.25) {};
            \node [style=none] (1) at (0.5, 1.15) {};
            \node [style=none] (2) at (0.5, -0.65) {};
            \node [style=label] (4) at (1.575, 1.4) {$#2$};
            \node [style=label] (5) at (1.575, 0.65) {$#1$};
            \node [style=none] (6) at (-1.75, 0.65) {};
            \node [style=none] (7) at (-1.75, -0.65) {};
            \node [style=none] (8) at (1, -1.55) {};
            \node [style=none] (9) at (1, 1.55) {};
            \node [style=none] (10) at (1, 0.25) {};
            \node [style=none] (12) at (1, -0.25) {};
            \node [style=none] (13) at (-0.75, -0.65) {};
            \node [style=none] (14) at (-0.75, 0.65) {};
            \node [style=none] (15) at (-1.75, 0.25) {};
            \node [style=none] (16) at (-0.5, -0.25) {};
            \node [style=none] (17) at (0.5, 0.65) {};
            \node [style=none] (18) at (0.5, -1.15) {};
            \node [style=none] (19) at (-1.75, -0.25) {};
            \node [style=label] (20) at (1.575, -0.65) {$#2$};
            \node [style=label] (21) at (1.575, -1.4) {$#1$};
            \node [style=label] (22) at (-2.325, 0.5) {$#2$};
            \node [style=label] (23) at (-2.325, -0.25) {$#1$};
            \node [style=none] (24) at (0.25, 0.25) {};
            \node [style=none] (25) at (0.25, -0.25) {};
            \node [style=none] (26) at (-0.025, 0) {};
            \node [style=none] (27) at (1, 1.15) {};
            \node [style=none] (28) at (1, -0.65) {};
            \node [style=none] (29) at (1, 0.65) {};
            \node [style=none] (30) at (1, -1.15) {};
            \node [style=none] (31) at (0.475, 1.55) {};
            \node [style=none] (32) at (0.575, -1.55) {};
        \end{pgfonlayer}
        \begin{pgfonlayer}{edgelayer}
            \draw [style=tape] (6.center)
                 to (7.center)
                 to (13.center)
                 to [bend right] (32.center)
                 to (8.center)
                 to (12.center)
                 to (25.center)
                 to [bend left=45] (26.center)
                 to [bend left=45] (24.center)
                 to (10.center)
                 to (9.center)
                 to (31.center)
                 to [bend right] (14.center)
                 to cycle;
            \draw [bend right=45] (0.center) to (2.center);
            \draw [bend left=45] (0.center) to (1.center);
            \draw (0.center) to (15.center);
            \draw [bend right=45] (16.center) to (18.center);
            \draw [bend left=45] (16.center) to (17.center);
            \draw (16.center) to (19.center);
            \draw (27.center) to (1.center);
            \draw (17.center) to (29.center);
            \draw (28.center) to (2.center);
            \draw (18.center) to (30.center);
        \end{pgfonlayer}
    \end{tikzpicture}                    
}

\newcommand{\Tcounit}[1]{
    \begin{tikzpicture}
        \begin{pgfonlayer}{nodelayer}
            \node [style=none] (0) at (1.25, 0) {};
            \ifthenelse{\isempty{#1}}{}{
                \node [style=label] (3) at (-0.75, 0) {$#1$};
            }
            \node [style=none] (6) at (-0.25, 0.5) {};
            \node [style=none] (7) at (-0.25, -0.5) {};
            \node [style=none] (13) at (0.75, -0.5) {};
            \node [style=none] (14) at (0.75, 0.5) {};
            \node [style=none] (15) at (-0.25, 0) {};
        \end{pgfonlayer}
        \begin{pgfonlayer}{edgelayer}
            \path [style=tape] (13.center)
                 to [bend right=45] (0.center)
                 to [bend right=45] (14.center)
                 to (6.center)
                 to (7.center)
                 to cycle;
            \draw (0.center) to (15.center);
        \end{pgfonlayer}
    \end{tikzpicture}      
}

\newcommand{\Tmonoid}[1]{
    \ifthenelse{\equal{#1}{\uno}}{
        \begin{tikzpicture}
            \begin{pgfonlayer}{nodelayer}
                \node [style=none] (1) at (-1, 1) {};
                \node [style=none] (2) at (-1, -1) {};
                \node [style=none] (6) at (1.5, 0.5) {};
                \node [style=none] (7) at (1.5, -0.5) {};
                \node [style=none] (8) at (-1, -1.5) {};
                \node [style=none] (9) at (-1, 1.5) {};
                \node [style=none] (10) at (-1, 0.5) {};
                \node [style=none] (11) at (-0.5, 0) {};
                \node [style=none] (12) at (-1, -0.5) {};
                \node [style=none] (13) at (0.5, -0.5) {};
                \node [style=none] (14) at (0.5, 0.5) {};
                \node [style=none] (15) at (1.5, 0) {};
            \end{pgfonlayer}
            \begin{pgfonlayer}{edgelayer}
                \path [style=tape] (9.center)
                     to [bend left] (14.center)
                     to (6.center)
                     to (7.center)
                     to (13.center)
                     to [bend left] (8.center)
                     to (12.center)
                     to [bend right=45] (11.center)
                     to [bend right=45] (10.center)
                     to cycle;
            \end{pgfonlayer}
        \end{tikzpicture}         
    }
    {
        \begin{tikzpicture}
            \begin{pgfonlayer}{nodelayer}
                \node [style=none] (0) at (0, 0) {};
                \node [style=none] (1) at (-1, 1) {};
                \node [style=none] (2) at (-1, -1) {};
                \ifthenelse{\isempty{#1}}{}{
                    \node [style=label] (3) at (2, 0) {$#1$};
                    \node [style=label] (4) at (-1.5, 1) {$#1$};
                    \node [style=label] (5) at (-1.5, -1) {$#1$};
                }
                \node [style=none] (6) at (1.5, 0.5) {};
                \node [style=none] (7) at (1.5, -0.5) {};
                \node [style=none] (8) at (-1, -1.5) {};
                \node [style=none] (9) at (-1, 1.5) {};
                \node [style=none] (10) at (-1, 0.5) {};
                \node [style=none] (11) at (-0.5, 0) {};
                \node [style=none] (12) at (-1, -0.5) {};
                \node [style=none] (13) at (0.5, -0.5) {};
                \node [style=none] (14) at (0.5, 0.5) {};
                \node [style=none] (15) at (1.5, 0) {};
            \end{pgfonlayer}
            \begin{pgfonlayer}{edgelayer}
                \path [style=tape] (9.center)
                     to [bend left] (14.center)
                     to (6.center)
                     to (7.center)
                     to (13.center)
                     to [bend left] (8.center)
                     to (12.center)
                     to [bend right=45] (11.center)
                     to [bend right=45] (10.center)
                     to cycle;
                \draw [bend left=45] (0.center) to (2.center);
                \draw [bend right=45] (0.center) to (1.center);
                \draw (0.center) to (15.center);
            \end{pgfonlayer}
        \end{tikzpicture}          
    }  
}

\newcommand{\Tunit}[1]{
    \begin{tikzpicture}
        \begin{pgfonlayer}{nodelayer}
            \node [style=none] (0) at (-1.25, 0) {};
            \ifthenelse{\isempty{#1}}{}{
                \node [style=label] (3) at (0.75, 0) {$#1$};
            }
            \node [style=none] (6) at (0.25, 0.5) {};
            \node [style=none] (7) at (0.25, -0.5) {};
            \node [style=none] (13) at (-0.75, -0.5) {};
            \node [style=none] (14) at (-0.75, 0.5) {};
            \node [style=none] (15) at (0.25, 0) {};
        \end{pgfonlayer}
        \begin{pgfonlayer}{edgelayer}
            \path [style=tape] (13.center)
                 to [bend left=45] (0.center)
                 to [bend left=45] (14.center)
                 to (6.center)
                 to (7.center)
                 to cycle;
            \draw (0.center) to (15.center);
        \end{pgfonlayer}
    \end{tikzpicture}      
}

\newcommand{\Tsymmp}[2]{
    \begin{tikzpicture}
        \begin{pgfonlayer}{nodelayer}
            \node [style=label] (3) at (-1.5, 1) {$#1$};
            \node [style=none] (6) at (-1, -0.5) {};
            \node [style=none] (7) at (-1, -1.5) {};
            \node [style=none] (15) at (-1, -1) {};
            \node [style=none] (17) at (-1, 1.5) {};
            \node [style=none] (18) at (-1, 0.5) {};
            \node [style=none] (21) at (-1, 1) {};
            \node [style=none] (22) at (2.5, 1) {};
            \node [style=none] (25) at (2.5, 0.5) {};
            \node [style=none] (26) at (2.5, 1.5) {};
            \node [style=none] (28) at (2.5, -1) {};
            \node [style=none] (31) at (2.5, -1.5) {};
            \node [style=none] (32) at (2.5, -0.5) {};
            \node [style=label] (34) at (3, -1) {$#1$};
            \node [style=label] (35) at (3, 1) {$#2$};
            \node [style=label] (36) at (-1.5, -1) {$#2$};
        \end{pgfonlayer}
        \begin{pgfonlayer}{edgelayer}
            \path [style=tapeFill] (26.center)
                 to [in=0, out=-180, looseness=0.75] (6.center)
                 to (7.center)
                 to [in=-180, out=0, looseness=0.75] (25.center)
                 to cycle;
            \path [style=tapeFill] (18.center)
                 to (17.center)
                 to [in=180, out=0, looseness=0.75] (32.center)
                 to (31.center)
                 to [in=0, out=-180, looseness=0.75] cycle;
            \path [style=tapeNoFill] (26.center)
                 to [in=0, out=-180, looseness=0.75] (6.center)
                 to (7.center)
                 to [in=-180, out=0, looseness=0.75] (25.center)
                 to cycle;
            \path [style=tapeNoFill] (18.center)
                 to (17.center)
                 to [in=180, out=0, looseness=0.75] (32.center)
                 to (31.center)
                 to [in=0, out=-180, looseness=0.75] cycle;
            \draw [in=-180, out=0, looseness=0.75] (15.center) to (22.center);
            \draw [in=180, out=0, looseness=0.75] (21.center) to (28.center);
        \end{pgfonlayer}
    \end{tikzpicture}    
}

\newcommand{\Twire}[1]{
    \ifthenelse{\equal{#1}{\uno}}{
        \begin{tikzpicture}
            \begin{pgfonlayer}{nodelayer}
                \node [style=none] (0) at (-1.5, 0.5) {};
                \node [style=none] (3) at (-1.5, -0.5) {};
                \node [style=none] (4) at (1.5, -0.5) {};
                \node [style=none] (5) at (1.5, 0.5) {};
                \node [style=none] (6) at (-1.5, 0) {};
                \node [style=none] (7) at (1.5, 0) {};
            \end{pgfonlayer}
            \begin{pgfonlayer}{edgelayer}
                \path [style=tape] (5.center)
                     to (4.center)
                     to (3.center)
                     to (0.center)
                     to cycle;
            \end{pgfonlayer}
        \end{tikzpicture}        
    }
    {
        \begin{tikzpicture}
            \begin{pgfonlayer}{nodelayer}
                \node [style=none] (0) at (-1.5, 0.5) {};
                \node [style=none] (3) at (-1.5, -0.5) {};
                \node [style=none] (4) at (1.5, -0.5) {};
                \node [style=none] (5) at (1.5, 0.5) {};
                \node [style=none] (6) at (-1.5, 0) {};
                \node [style=none] (7) at (1.5, 0) {};
                \node [style=label] (8) at (-2, 0) {$#1$};
                \node [style=label] (9) at (2, 0) {$#1$};
            \end{pgfonlayer}
            \begin{pgfonlayer}{edgelayer}
                \path [style=tape] (5.center)
                        to (4.center)
                        to (3.center)
                        to (0.center)
                        to cycle;
                \draw (6.center) to (7.center);
            \end{pgfonlayer}
        \end{tikzpicture}        
    }   
}

\newcommand{\TLwire}[2]{
    \begin{tikzpicture}
        \begin{pgfonlayer}{nodelayer}
            \node [style=label] (3) at (-2, -0.25) {$#1$};
            \node [style=none] (17) at (-1.5, 0.75) {};
            \node [style=none] (18) at (-1.5, -0.75) {};
            \node [style=none] (21) at (-1.5, -0.25) {};
            \node [style=none] (40) at (1.5, 0.75) {};
            \node [style=none] (41) at (1.5, -0.75) {};
            \node [style=none] (42) at (1.5, -0.25) {};
            \node [style=none] (46) at (0, -0.25) {};
            \node [style=none] (47) at (0, -0.75) {};
            \node [style=none] (48) at (0, 0.75) {};
            \node [style=label] (49) at (2, -0.25) {$#1$};
            \node [style=label] (50) at (-2, 0.5) {$#2$};
            \node [style=none] (51) at (-1.5, 0.25) {};
            \node [style=none] (52) at (1.5, 0.25) {};
            \node [style=none] (53) at (0, 0.25) {};
            \node [style=label] (54) at (2, 0.5) {$#2$};
        \end{pgfonlayer}
        \begin{pgfonlayer}{edgelayer}
            \path [style=tape] (41.center)
                 to [in=0, out=180, looseness=0.75] (47.center)
                 to [in=0, out=-180, looseness=0.75] (18.center)
                 to (17.center)
                 to [in=180, out=0, looseness=0.75] (48.center)
                 to [in=180, out=0, looseness=0.75] (40.center)
                 to cycle;
            \draw (42.center) to (46.center);
            \draw (21.center) to (46.center);
            \draw (52.center) to (53.center);
            \draw (51.center) to (53.center);
        \end{pgfonlayer}
    \end{tikzpicture}    
}

\newcommand{\TRwire}[2]{
    \TLwire{#2}{#1}
}

\newcommand{\Tcirc}[3]{
    \begin{tikzpicture}
        \begin{pgfonlayer}{nodelayer}
            \node [style=none] (6) at (-1.5, 0.75) {};
            \node [style=none] (7) at (-1.5, -0.75) {};
            \node [style=none] (15) at (-1.5, 0) {};
            \node [style=none] (22) at (1.5, 0) {};
            \node [style=none] (25) at (1.5, -0.75) {};
            \node [style=none] (26) at (1.5, 0.75) {};
            \node [style=label] (35) at (2, 0) {$#3$};
            \node [style=label] (36) at (-2, 0) {$#2$};
            \node [style=none] (37) at (-0.5, 0.5) {};
            \node [style=none] (38) at (-0.5, -0.5) {};
            \node [style=none] (39) at (-0.5, 0) {};
            \node [style=none] (40) at (0.5, 0) {};
            \node [style=none] (41) at (0.5, -0.5) {};
            \node [style=none] (42) at (0.5, 0.5) {};
            \node [style=none] (43) at (0, 0) {$#1$};
        \end{pgfonlayer}
        \begin{pgfonlayer}{edgelayer}
            \path [style=tape] (26.center)
                 to (6.center)
                 to (7.center)
                 to (25.center)
                 to cycle;
            \draw [fill=white] (42.center)
                 to (37.center)
                 to (38.center)
                 to (41.center)
                 to cycle;
            \draw (15.center) to (39.center);
            \draw (40.center) to (22.center);
        \end{pgfonlayer}
    \end{tikzpicture}      
}

\newcommand{\monGen}[3]{
    \begin{tikzpicture}
        \begin{pgfonlayer}{nodelayer}
            \node [style=none] (6) at (-2, 1) {};
            \node [style=none] (7) at (-2, -1) {};
            \node [style=none] (22) at (2, 0) {};
            \node [style=none] (25) at (2, -1) {};
            \node [style=none] (26) at (2, 1) {};
            \node [style=label] (35) at (3, 0) {$#3$};
            \node [style=none] (37) at (-1, 0.75) {};
            \node [style=none] (38) at (-1, -0.75) {};
            \node [style=none] (40) at (1, 0) {};
            \node [style=none] (41) at (1, -0.75) {};
            \node [style=none] (42) at (1, 0.75) {};
            \node [style=none] (43) at (0, 0) {$#1$};
            \node [style=none] (44) at (-2, 0) {};
            \node [style=none] (45) at (-1, 0) {};
            \node [style=label] (46) at (-3, 0) {$#2$};
        \end{pgfonlayer}
        \begin{pgfonlayer}{edgelayer}
            \path [style=tape] (26.center)
                 to (6.center)
                 to (7.center)
                 to (25.center)
                 to cycle;
            \draw [fill=white] (42.center)
                 to (37.center)
                 to (38.center)
                 to (41.center)
                 to cycle;
            \draw (40.center) to (22.center);
            \draw (44.center) to (45.center);
        \end{pgfonlayer}
    \end{tikzpicture}       
}

\newcommand{\monGenDue}[4]{
    \begin{tikzpicture}
        \begin{pgfonlayer}{nodelayer}
            \node [style=none] (6) at (-2, 1) {};
            \node [style=none] (7) at (-2, -1) {};
            \node [style=none] (15) at (-2, -0.5) {};
            \node [style=none] (22) at (2, 0) {};
            \node [style=none] (25) at (2, -1) {};
            \node [style=none] (26) at (2, 1) {};
            \node [style=label] (35) at (3, 0) {$#4$};
            \node [style=label] (36) at (-3, -0.5) {$#3$};
            \node [style=none] (37) at (-1, 0.75) {};
            \node [style=none] (38) at (-1, -0.75) {};
            \node [style=none] (39) at (-1, -0.5) {};
            \node [style=none] (40) at (1, 0) {};
            \node [style=none] (41) at (1, -0.75) {};
            \node [style=none] (42) at (1, 0.75) {};
            \node [style=none] (43) at (0, 0) {$#1$};
            \node [style=none] (44) at (-2, 0.5) {};
            \node [style=none] (45) at (-1, 0.5) {};
            \node [style=label] (46) at (-3, 0.5) {$#2$};
        \end{pgfonlayer}
        \begin{pgfonlayer}{edgelayer}
            \path [style=tape] (26.center)
                 to (6.center)
                 to (7.center)
                 to (25.center)
                 to cycle;
            \draw [fill=white] (42.center)
                 to (37.center)
                 to (38.center)
                 to (41.center)
                 to cycle;
            \draw (15.center) to (39.center);
            \draw (40.center) to (22.center);
            \draw (44.center) to (45.center);
        \end{pgfonlayer}
    \end{tikzpicture}       
}

\newcommand{\Tdl}[4]{
    \begin{tikzpicture}
        \begin{pgfonlayer}{nodelayer}
            \node [style=label] (3) at (-3, 1.275) {$#1$};
            \node [style=none] (21) at (-2.25, 1.2) {};
            \node [style=none] (28) at (2.25, -0.725) {};
            \node [style=label] (34) at (3, -0.65) {$#1$};
            \node [style=label] (35) at (-3, 0.675) {$#4$};
            \node [style=none] (36) at (-2.25, 0.7) {};
            \node [style=none] (37) at (2.25, -1.175) {};
            \node [style=label] (38) at (3, -1.25) {$#4$};
            \node [style=none] (40) at (-2.25, 0.45) {};
            \node [style=none] (41) at (2.25, -1.45) {};
            \node [style=none] (44) at (-2.25, 1.45) {};
            \node [style=none] (45) at (2.25, -0.45) {};
            \node [style=label] (58) at (3, 2.8) {$#1$};
            \node [style=none] (59) at (2.25, 2.725) {};
            \node [style=none] (60) at (-2.25, 2.725) {};
            \node [style=label] (61) at (-3, 2.8) {$#1$};
            \node [style=label] (62) at (3, 2.2) {$#3$};
            \node [style=none] (63) at (2.25, 2.275) {};
            \node [style=none] (64) at (-2.25, 2.275) {};
            \node [style=label] (65) at (-3, 2.2) {$#3$};
            \node [style=none] (66) at (2.25, 2) {};
            \node [style=none] (67) at (-2.25, 2) {};
            \node [style=none] (68) at (2.25, 3) {};
            \node [style=none] (69) at (-2.25, 3) {};
            \node [style=label] (70) at (3, -2.2) {$#2$};
            \node [style=none] (71) at (2.25, -2.275) {};
            \node [style=none] (72) at (-2.25, -2.275) {};
            \node [style=label] (73) at (-3, -2.2) {$#2$};
            \node [style=label] (74) at (3, -2.775) {$#4$};
            \node [style=none] (75) at (2.25, -2.725) {};
            \node [style=none] (76) at (-2.25, -2.725) {};
            \node [style=label] (77) at (-3, -2.775) {$#4$};
            \node [style=none] (78) at (2.25, -3) {};
            \node [style=none] (79) at (-2.25, -3) {};
            \node [style=none] (80) at (2.25, -2) {};
            \node [style=none] (81) at (-2.25, -2) {};
            \node [style=label] (82) at (3, 1.275) {$#2$};
            \node [style=none] (83) at (2.25, 1.2) {};
            \node [style=none] (84) at (-2.25, -0.725) {};
            \node [style=label] (85) at (-3, -0.65) {$#2$};
            \node [style=label] (86) at (3, 0.675) {$#3$};
            \node [style=none] (87) at (2.25, 0.7) {};
            \node [style=none] (88) at (-2.25, -1.175) {};
            \node [style=label] (89) at (-3, -1.25) {$#3$};
            \node [style=none] (90) at (2.25, 0.45) {};
            \node [style=none] (91) at (-2.25, -1.45) {};
            \node [style=none] (92) at (2.25, 1.45) {};
            \node [style=none] (93) at (-2.25, -0.45) {};
        \end{pgfonlayer}
        \begin{pgfonlayer}{edgelayer}
            \draw [style=tapeFill] (41.center)
                 to [in=0, out=180, looseness=0.75] (40.center)
                 to (44.center)
                 to [in=180, out=0, looseness=0.75] (45.center)
                 to cycle;
            \draw [style=tapeFill] (91.center)
                 to [in=180, out=0, looseness=0.75] (90.center)
                 to (92.center)
                 to [in=0, out=180, looseness=0.75] (93.center)
                 to cycle;
            \draw [style=tapeNoFill] (41.center)
                 to [in=0, out=180, looseness=0.75] (40.center)
                 to (44.center)
                 to [in=180, out=0, looseness=0.75] (45.center)
                 to cycle;
            \draw [style=tapeNoFill] (91.center)
                 to [in=180, out=0, looseness=0.75] (90.center)
                 to (92.center)
                 to [in=0, out=180, looseness=0.75] (93.center)
                 to cycle;
            \draw [style=tape] (69.center)
                 to (67.center)
                 to (66.center)
                 to (68.center)
                 to cycle;
            \draw [style=tape] (81.center)
                 to (79.center)
                 to (78.center)
                 to (80.center)
                 to cycle;
            \draw [in=180, out=0, looseness=0.75] (21.center) to (28.center);
            \draw [in=180, out=0, looseness=0.75] (36.center) to (37.center);
            \draw [in=0, out=180, looseness=0.75] (59.center) to (60.center);
            \draw [in=0, out=180, looseness=0.75] (63.center) to (64.center);
            \draw [in=0, out=180, looseness=0.75] (71.center) to (72.center);
            \draw [in=0, out=180, looseness=0.75] (75.center) to (76.center);
            \draw [in=0, out=180, looseness=0.75] (83.center) to (84.center);
            \draw [in=0, out=180, looseness=0.75] (87.center) to (88.center);
        \end{pgfonlayer}
    \end{tikzpicture}                  
}

\newcommand{\Tsymmt}[4]{
    \begin{tikzpicture}
        \begin{pgfonlayer}{nodelayer}
            \node [style=label] (3) at (-3, 1.275) {$#1$};
            \node [style=none] (21) at (-2.25, 1.2) {};
            \node [style=none] (28) at (2.25, -0.725) {};
            \node [style=label] (34) at (4.25, -0.65) {$#4$};
            \node [style=label] (35) at (-3, 0.675) {$#4$};
            \node [style=none] (36) at (-2.25, 0.7) {};
            \node [style=none] (37) at (2.25, -1.175) {};
            \node [style=label] (38) at (4.25, -1.25) {$#1$};
            \node [style=none] (40) at (-2.25, 0.45) {};
            \node [style=none] (41) at (2.25, -1.45) {};
            \node [style=none] (44) at (-2.25, 1.45) {};
            \node [style=none] (45) at (2.25, -0.45) {};
            \node [style=label] (58) at (4.25, 2.8) {$#3$};
            \node [style=none] (59) at (2.25, 2.725) {};
            \node [style=none] (60) at (-2.25, 2.725) {};
            \node [style=label] (61) at (-3, 2.8) {$#1$};
            \node [style=label] (62) at (4.25, 2.2) {$#1$};
            \node [style=none] (63) at (2.25, 2.275) {};
            \node [style=none] (64) at (-2.25, 2.275) {};
            \node [style=label] (65) at (-3, 2.2) {$#3$};
            \node [style=none] (66) at (2.25, 2) {};
            \node [style=none] (67) at (-2.25, 2) {};
            \node [style=none] (68) at (2.25, 3) {};
            \node [style=none] (69) at (-2.25, 3) {};
            \node [style=label] (70) at (4.25, -2.2) {$#4$};
            \node [style=none] (71) at (2.25, -2.275) {};
            \node [style=none] (72) at (-2.25, -2.275) {};
            \node [style=label] (73) at (-3, -2.2) {$#2$};
            \node [style=label] (74) at (4.25, -2.775) {$#2$};
            \node [style=none] (75) at (2.25, -2.725) {};
            \node [style=none] (76) at (-2.25, -2.725) {};
            \node [style=label] (77) at (-3, -2.775) {$#4$};
            \node [style=none] (78) at (2.25, -3) {};
            \node [style=none] (79) at (-2.25, -3) {};
            \node [style=none] (80) at (2.25, -2) {};
            \node [style=none] (81) at (-2.25, -2) {};
            \node [style=label] (82) at (4.25, 1.275) {$#3$};
            \node [style=none] (83) at (2.25, 1.2) {};
            \node [style=none] (84) at (-2.25, -0.725) {};
            \node [style=label] (85) at (-3, -0.65) {$#2$};
            \node [style=label] (86) at (4.25, 0.675) {$#2$};
            \node [style=none] (87) at (2.25, 0.7) {};
            \node [style=none] (88) at (-2.25, -1.175) {};
            \node [style=label] (89) at (-3, -1.25) {$#3$};
            \node [style=none] (90) at (2.25, 0.45) {};
            \node [style=none] (91) at (-2.25, -1.45) {};
            \node [style=none] (92) at (2.25, 1.45) {};
            \node [style=none] (93) at (-2.25, -0.45) {};
            \node [style=none] (94) at (3.5, -0.725) {};
            \node [style=none] (95) at (3.5, -1.175) {};
            \node [style=none] (96) at (3.5, -1.45) {};
            \node [style=none] (97) at (3.5, -0.45) {};
            \node [style=none] (98) at (3.5, 2.725) {};
            \node [style=none] (99) at (3.5, 2.275) {};
            \node [style=none] (100) at (3.5, 2) {};
            \node [style=none] (101) at (3.5, 3) {};
            \node [style=none] (102) at (3.5, -2.275) {};
            \node [style=none] (103) at (3.5, -2.725) {};
            \node [style=none] (104) at (3.5, -3) {};
            \node [style=none] (105) at (3.5, -2) {};
            \node [style=none] (106) at (3.5, 1.2) {};
            \node [style=none] (107) at (3.5, 0.7) {};
            \node [style=none] (108) at (3.5, 0.45) {};
            \node [style=none] (109) at (3.5, 1.45) {};
        \end{pgfonlayer}
        \begin{pgfonlayer}{edgelayer}
            \draw [style=tapeFill] (96.center)
                 to (41.center)
                 to [in=0, out=180, looseness=0.75] (40.center)
                 to (44.center)
                 to [in=180, out=0, looseness=0.75] (45.center)
                 to (97.center)
                 to cycle;
            \draw [style=tapeFill] (90.center)
                 to (108.center)
                 to (109.center)
                 to (92.center)
                 to [in=0, out=180, looseness=0.75] (93.center)
                 to (91.center)
                 to [in=180, out=0, looseness=0.75] cycle;
            \draw [style=tapeNoFill] (96.center)
                 to (41.center)
                 to [in=0, out=180, looseness=0.75] (40.center)
                 to (44.center)
                 to [in=180, out=0, looseness=0.75] (45.center)
                 to (97.center)
                 to cycle;
            \draw [style=tapeNoFill] (90.center)
                 to (108.center)
                 to (109.center)
                 to (92.center)
                 to [in=0, out=180, looseness=0.75] (93.center)
                 to (91.center)
                 to [in=180, out=0, looseness=0.75] cycle;
            \draw [style=tape] (100.center)
                 to (66.center)
                 to (67.center)
                 to (69.center)
                 to (68.center)
                 to (101.center)
                 to cycle;
            \draw [style=tape] (80.center)
                 to (81.center)
                 to (79.center)
                 to (78.center)
                 to (104.center)
                 to (105.center)
                 to cycle;
            \draw [in=180, out=0, looseness=0.75] (21.center) to (28.center);
            \draw [in=180, out=0, looseness=0.75] (36.center) to (37.center);
            \draw [in=0, out=180, looseness=0.75] (59.center) to (60.center);
            \draw [in=0, out=180, looseness=0.75] (63.center) to (64.center);
            \draw [in=0, out=180, looseness=0.75] (71.center) to (72.center);
            \draw [in=0, out=180, looseness=0.75] (75.center) to (76.center);
            \draw [in=0, out=180, looseness=0.75] (83.center) to (84.center);
            \draw [in=0, out=180, looseness=0.75] (87.center) to (88.center);
            \draw [in=180, out=0] (59.center) to (99.center);
            \draw [in=0, out=-180] (98.center) to (63.center);
            \draw [in=180, out=0] (83.center) to (107.center);
            \draw [in=0, out=-180] (106.center) to (87.center);
            \draw [in=-180, out=0] (37.center) to (94.center);
            \draw [in=0, out=180] (95.center) to (28.center);
            \draw [in=-180, out=0] (71.center) to (103.center);
            \draw [in=0, out=-180] (102.center) to (75.center);
        \end{pgfonlayer}
    \end{tikzpicture}       
}

\newcommand{\Cgen}[3]{
    \begin{tikzpicture}
        \begin{pgfonlayer}{nodelayer}
            
            \ifthenelse{\isempty{#2}}{
                \node [style=none] (15) at (-1.25, 0) {};
                \node [style=none] (22) at (1.25, 0) {};
            }{
                \node [style=none] (15) at (-1.5, 0) {};
                \node [style=none] (22) at (1.5, 0) {};
                \node [style=label] (35) at (2, 0) {$#3$};
                \node [style=label] (36) at (-2, 0) {$#2$};
            }     
            \node [style=none] (37) at (-0.5, 0.5) {};
            \node [style=none] (38) at (-0.5, -0.5) {};
            \node [style=none] (39) at (-0.5, 0) {};
            \node [style=none] (40) at (0.5, 0) {};
            \node [style=none] (41) at (0.5, -0.5) {};
            \node [style=none] (42) at (0.5, 0.5) {};
            \node [style=none] (43) at (0, 0) {$#1$};
        \end{pgfonlayer}
        \begin{pgfonlayer}{edgelayer}
            \draw [fill=white] (42.center)
                 to (37.center)
                 to (38.center)
                 to (41.center)
                 to cycle;
            \draw (15.center) to (39.center);
            \draw (40.center) to (22.center);
        \end{pgfonlayer}
    \end{tikzpicture}
}

\newcommand{\Csymm}[2]{
    \begin{tikzpicture}
        \begin{pgfonlayer}{nodelayer}
            \node [style=label] (8) at (1.75, 0.5) {$#2$};
            \node [style=label] (11) at (-1.75, 0.5) {$#1$};
            \node [style=label] (12) at (-1.75, -0.5) {$#2$};
            \node [style=label] (13) at (1.75, -0.5) {$#1$};
        \end{pgfonlayer}
        \begin{pgfonlayer}{edgelayer}
            \draw [in=-180, out=0, looseness=1.25] (11) to (13);
            \draw [in=-180, out=0, looseness=1.25] (12) to (8);
        \end{pgfonlayer}
    \end{tikzpicture}             
}

\newcommand{\Creg}[2]{
    \begin{tikzpicture}
        \begin{pgfonlayer}{nodelayer}
            \node [style=reg] (0) at (0, 0) {$#1$};
            \node [style=none] (1) at (1.5, 0) {};
            \node [style=none] (2) at (-1.5, 0) {};
            \ifthenelse{\equal{#2}{\uno}}{}{
            \node [style=label] (3) at (-2.5, 0) {$#2$};
            \node [style=label] (4) at (2.5, 0) {$#2$};
            }
        \end{pgfonlayer}
        \begin{pgfonlayer}{edgelayer}
            \draw (2.center) to (0);
            \draw (0) to (1.center);
        \end{pgfonlayer}
    \end{tikzpicture}       
}

\newcommand{\Ccoreg}[2]{
    \begin{tikzpicture}
        \begin{pgfonlayer}{nodelayer}
            \node [style=coreg] (0) at (0, 0) {$#1$};
            \node [style=none] (1) at (1.5, 0) {};
            \node [style=none] (2) at (-1.5, 0) {};
            \ifthenelse{\equal{#2}{\uno}}{}{
            \node [style=label] (3) at (-2.5, 0) {$#2$};
            \node [style=label] (4) at (2.5, 0) {$#2$};
            }
        \end{pgfonlayer}
        \begin{pgfonlayer}{edgelayer}
            \draw (2.center) to (0);
            \draw (0) to (1.center);
        \end{pgfonlayer}
    \end{tikzpicture}
}

\newcommand{\Cunit}[1]{
    \begin{tikzpicture}
        \begin{pgfonlayer}{nodelayer}
            \ifthenelse{\equal{#1}{\uno}}{}{
            \node [style=label] (4) at (1, 0) {$#1$};
            }
            \node [style=none] (11) at (0, 0) {};
            \node [style=black] (12) at (-1, 0) {};
        \end{pgfonlayer}
        \begin{pgfonlayer}{edgelayer}
            \draw (12) to (11.center);
        \end{pgfonlayer}
    \end{tikzpicture}    
}

\newcommand{\Ccounit}[1]{
    \begin{tikzpicture}
        \begin{pgfonlayer}{nodelayer}
            \ifthenelse{\equal{#1}{\uno}}{}{
            \node [style=label] (4) at (-1, 0) {$#1$};
            }
            \node [style=none] (11) at (0, 0) {};
            \node [style=black] (12) at (1, 0) {};
        \end{pgfonlayer}
        \begin{pgfonlayer}{edgelayer}
            \draw (12) to (11.center);
        \end{pgfonlayer}
    \end{tikzpicture}    
}

\newcommand{\Ctotrel}[1]{
    \Ccounit{#1} \;\; \Cunit{#1}
}

\newcommand{\Tcodiagdiag}[1]{
    \ifthenelse{\equal{#1}{\uno}}{
        \begin{tikzpicture}
            \begin{pgfonlayer}{nodelayer}
                \node [style=none] (1) at (-0.75, 1.25) {};
                \node [style=none] (8) at (-0.75, -1.75) {};
                \node [style=none] (9) at (-0.75, 1.75) {};
                \node [style=none] (10) at (-0.75, 0.75) {};
                \node [style=none] (11) at (-1.5, 0) {};
                \node [style=none] (12) at (-0.75, -0.75) {};
                \node [style=none] (13) at (-2.5, -0.5) {};
                \node [style=none] (14) at (-2.5, 0.5) {};
                \node [style=none] (35) at (-3.75, -0.5) {};
                \node [style=none] (36) at (-3.75, 0.5) {};
                \node [style=none] (81) at (0.75, 1.25) {};
                \node [style=none] (83) at (0.75, -1.75) {};
                \node [style=none] (84) at (0.75, 1.75) {};
                \node [style=none] (85) at (0.75, 0.75) {};
                \node [style=none] (86) at (1.5, 0) {};
                \node [style=none] (87) at (0.75, -0.75) {};
                \node [style=none] (88) at (2.5, -0.5) {};
                \node [style=none] (89) at (2.5, 0.5) {};
                \node [style=none] (91) at (3.75, -0.5) {};
                \node [style=none] (92) at (3.75, 0.5) {};
            \end{pgfonlayer}
            \begin{pgfonlayer}{edgelayer}
                \draw [style=tape] (83.center)
                     to (8.center)
                     to [bend left] (13.center)
                     to (35.center)
                     to (36.center)
                     to (14.center)
                     to [bend left] (9.center)
                     to (84.center)
                     to [bend left] (89.center)
                     to (92.center)
                     to (91.center)
                     to (88.center)
                     to [bend left] cycle;
                \path [style=tapeNoFill, fill=white] (87.center)
                     to (12.center)
                     to [bend left=45] (11.center)
                     to [bend left=45] (10.center)
                     to (85.center)
                     to [bend left=45] (86.center)
                     to [bend left=45] cycle;
            \end{pgfonlayer}
        \end{tikzpicture}                 
    }
    {
        \begin{tikzpicture}
            \begin{pgfonlayer}{nodelayer}
                \node [style=none] (0) at (-2, 0) {};
                \node [style=none] (1) at (-0.75, 1.25) {};
                \node [style=none] (2) at (-0.75, -1.25) {};
                \node [style=none] (8) at (-0.75, -1.75) {};
                \node [style=none] (9) at (-0.75, 1.75) {};
                \node [style=none] (10) at (-0.75, 0.75) {};
                \node [style=none] (11) at (-1.5, 0) {};
                \node [style=none] (12) at (-0.75, -0.75) {};
                \node [style=none] (13) at (-2.5, -0.5) {};
                \node [style=none] (14) at (-2.5, 0.5) {};
                \node [style=none] (24) at (-3.75, 0) {};
                \node [style=none] (35) at (-3.75, -0.5) {};
                \node [style=none] (36) at (-3.75, 0.5) {};
                \node [style=none] (50) at (-0.75, 1.25) {};
                \node [style=none] (80) at (2, 0) {};
                \node [style=none] (81) at (0.75, 1.25) {};
                \node [style=none] (82) at (0.75, -1.25) {};
                \node [style=none] (83) at (0.75, -1.75) {};
                \node [style=none] (84) at (0.75, 1.75) {};
                \node [style=none] (85) at (0.75, 0.75) {};
                \node [style=none] (86) at (1.5, 0) {};
                \node [style=none] (87) at (0.75, -0.75) {};
                \node [style=none] (88) at (2.5, -0.5) {};
                \node [style=none] (89) at (2.5, 0.5) {};
                \node [style=none] (90) at (3.75, 0) {};
                \node [style=none] (91) at (3.75, -0.5) {};
                \node [style=none] (92) at (3.75, 0.5) {};
                \node [style=none] (93) at (0.75, 1.25) {};
                \node [style=label] (94) at (-4.25, 0) {$#1$};
                \node [style=label] (95) at (4.25, 0) {$#1$};
            \end{pgfonlayer}
            \begin{pgfonlayer}{edgelayer}
                \path [style=tape] (83.center)
                     to (8.center)
                     to [bend left] (13.center)
                     to (35.center)
                     to (36.center)
                     to (14.center)
                     to [bend left] (9.center)
                     to (84.center)
                     to [bend left] (89.center)
                     to (92.center)
                     to (91.center)
                     to (88.center)
                     to [bend left] cycle;
                \path [style=tapeNoFill, fill=white] (87.center)
                     to (12.center)
                     to [bend left=45] (11.center)
                     to [bend left=45] (10.center)
                     to (85.center)
                     to [bend left=45] (86.center)
                     to [bend left=45] cycle;
                \draw [bend right=45] (0.center) to (2.center);
                \draw [bend left=45] (0.center) to (1.center);
                \draw (24.center) to (0.center);
                \draw [bend left=45] (80.center) to (82.center);
                \draw [bend right=45] (80.center) to (81.center);
                \draw (90.center) to (80.center);
                \draw (50.center) to (93.center);
                \draw (82.center) to (2.center);
            \end{pgfonlayer}
        \end{tikzpicture}                   
    }     
}

\newcommand{\Tconv}[4]{
    \begin{tikzpicture}
        \begin{pgfonlayer}{nodelayer}
            \node [style=none] (8) at (-0.75, -2) {};
            \node [style=none] (9) at (-0.75, 2) {};
            \node [style=none] (10) at (-0.75, 1) {};
            \node [style=none] (11) at (-1.75, 0) {};
            \node [style=none] (12) at (-0.75, -1) {};
            \node [style=none] (13) at (-2.75, -0.5) {};
            \node [style=none] (14) at (-2.75, 0.5) {};
            \node [style=none] (24) at (-4, 0) {};
            \node [style=none] (35) at (-4, -0.5) {};
            \node [style=none] (36) at (-4, 0.5) {};
            \node [style=none] (83) at (0.75, -2) {};
            \node [style=none] (84) at (0.75, 2) {};
            \node [style=none] (85) at (0.75, 1) {};
            \node [style=none] (86) at (1.75, 0) {};
            \node [style=none] (87) at (0.75, -1) {};
            \node [style=none] (88) at (2.75, -0.5) {};
            \node [style=none] (89) at (2.75, 0.5) {};
            \node [style=none] (90) at (4, 0) {};
            \node [style=none] (91) at (4, -0.5) {};
            \node [style=none] (92) at (4, 0.5) {};
            \node [style=label] (94) at (-3.25, 1) {$#3$};
            \node [style=label] (95) at (3.25, 1) {$#4$};
            \node [style=none] (96) at (-0.75, 0.5) {};
            \node [style=none] (97) at (0.75, 0.5) {};
            \node [style=none] (98) at (0.75, 2.5) {};
            \node [style=none] (99) at (-0.75, 2.5) {};
            \node [style=none] (100) at (-0.75, -0.5) {};
            \node [style=none] (101) at (0.75, -0.5) {};
            \node [style=none] (102) at (-0.75, -2.5) {};
            \node [style=none] (103) at (0.75, -2.5) {};
            \node [style=none] (104) at (0, 1.5) {$#1$};
            \node [style=none] (105) at (0, -1.5) {$#2$};
            \node [style=dots] (106) at (3.25, 0.025) {$\vdots$};
            \node [style=dots] (107) at (-3.25, 0.025) {$\vdots$};
        \end{pgfonlayer}
        \begin{pgfonlayer}{edgelayer}
            \path[style=tape] (102.center)
                 to (8.center)
                 to [bend left] (13.center)
                 to (35.center)
                 to (36.center)
                 to (14.center)
                 to [bend left] (9.center)
                 to (99.center)
                 to (98.center)
                 to (84.center)
                 to [bend left] (89.center)
                 to (92.center)
                 to (91.center)
                 to (88.center)
                 to [bend left] (83.center)
                 to (103.center)
                 to cycle;
            \draw [style=tapeNoFill, fill=white] (96.center)
                 to (97.center)
                 to (85.center)
                 to [bend left=45] (86.center)
                 to [bend left=45] (87.center)
                 to (101.center)
                 to (100.center)
                 to (12.center)
                 to [bend left=45] (11.center)
                 to [bend left=45] (10.center)
                 to cycle;
        \end{pgfonlayer}
    \end{tikzpicture}    
}

\newcommand{\Tbox}[3]{
    \begin{tikzpicture}
        \begin{pgfonlayer}{nodelayer}
            \node [style=none] (35) at (-2, -0.5) {};
            \node [style=none] (36) at (-2, 0.5) {};
            \node [style=dots] (90) at (1.5, 0.025) {$\vdots$};
            \node [style=none] (91) at (2, -0.5) {};
            \node [style=none] (92) at (2, 0.5) {};
            \node [style=label] (94) at (-1.5, 1) {$#2$};
            \node [style=label] (95) at (1.5, 1) {$#3$};
            \node [style=none] (96) at (-0.75, -1) {};
            \node [style=none] (97) at (0.75, -1) {};
            \node [style=none] (98) at (0.75, 1) {};
            \node [style=none] (99) at (-0.75, 1) {};
            \node [style=none] (104) at (0, 0) {$#1$};
            \node [style=dots] (105) at (-1.5, 0.025) {$\vdots$};
            \node [style=none] (106) at (-0.75, -0.5) {};
            \node [style=none] (107) at (0.75, -0.5) {};
            \node [style=none] (108) at (0.75, 0.5) {};
            \node [style=none] (109) at (-0.75, 0.5) {};
        \end{pgfonlayer}
        \begin{pgfonlayer}{edgelayer}
            \draw [style=tape] (97.center)
                 to (96.center)
                 to (106.center)
                 to (35.center)
                 to (36.center)
                 to (109.center)
                 to (99.center)
                 to (98.center)
                 to (108.center)
                 to (92.center)
                 to (91.center)
                 to (107.center)
                 to cycle;
        \end{pgfonlayer}
    \end{tikzpicture}      
}

\newcommand{\Ttconv}[3]{
    \begin{tikzpicture}
        \begin{pgfonlayer}{nodelayer}
            \node [style=none] (8) at (-0.75, -2) {};
            \node [style=none] (9) at (-0.75, 2) {};
            \node [style=none] (10) at (-0.75, 1) {};
            \node [style=none] (11) at (-1.75, 0) {};
            \node [style=none] (12) at (-0.75, -1) {};
            \node [style=none] (13) at (-2.75, -0.5) {};
            \node [style=none] (14) at (-2.75, 0.5) {};
            \node [style=none] (35) at (-6.75, -0.5) {};
            \node [style=none] (36) at (-6.75, 0.5) {};
            \node [style=none] (83) at (-0.25, -2) {};
            \node [style=none] (84) at (-0.25, 2) {};
            \node [style=none] (85) at (-0.25, 1) {};
            \node [style=none] (86) at (0.75, 0) {};
            \node [style=none] (87) at (-0.25, -1) {};
            \node [style=none] (88) at (1.75, -0.5) {};
            \node [style=none] (89) at (1.75, 0.5) {};
            \node [style=none] (91) at (3, -0.5) {};
            \node [style=none] (92) at (3, 0.5) {};
            \node [style=label] (94) at (-6.25, 1) {$#2$};
            \node [style=label] (95) at (2.25, 1) {$#3$};
            \node [style=none] (100) at (-5.5, 1) {};
            \node [style=none] (101) at (-4, 1) {};
            \node [style=none] (102) at (-5.5, -1) {};
            \node [style=none] (103) at (-4, -1) {};
            \node [style=none] (105) at (-4.75, 0) {$#1$};
            \node [style=dots] (106) at (-6.25, 0.025) {$\vdots$};
            \node [style=dots] (107) at (2.25, 0.025) {$\vdots$};
            \node [style=label] (108) at (-3.25, 1) {$#3$};
            \node [style=dots] (109) at (-3.25, 0.025) {$\vdots$};
            \node [style=none] (110) at (-4, -0.5) {};
            \node [style=none] (111) at (-4, 0.5) {};
            \node [style=none] (112) at (-5.5, -0.5) {};
            \node [style=none] (113) at (-5.5, 0.5) {};
        \end{pgfonlayer}
        \begin{pgfonlayer}{edgelayer}
            \draw [style=tape] (84.center)
                 to [bend left] (89.center)
                 to (92.center)
                 to (91.center)
                 to (88.center)
                 to [bend left] (83.center)
                 to (8.center)
                 to [bend left] (13.center)
                 to (110.center)
                 to (103.center)
                 to (102.center)
                 to (112.center)
                 to (35.center)
                 to (36.center)
                 to (113.center)
                 to (100.center)
                 to (101.center)
                 to (111.center)
                 to (14.center)
                 to [bend left] (9.center)
                 to cycle;
            \draw [style=tapeNoFill, fill=white] (87.center)
                 to (12.center)
                 to [bend left=45] (11.center)
                 to [bend left=45] (10.center)
                 to (85.center)
                 to [bend left=45] (86.center)
                 to [bend left=45] cycle;
        \end{pgfonlayer}
    \end{tikzpicture}       
}

\newcommand{\Czero}[1]{
    \begin{tikzpicture}
        \begin{pgfonlayer}{nodelayer}
            \ifthenelse{\equal{#1}{\uno}}{}{
            \node [style=label] (4) at (1, 0) {$#1$};
            }
            \node [style=none] (11) at (0, 0) {};
            \node [style=white] (12) at (-1, 0) {};
        \end{pgfonlayer}
        \begin{pgfonlayer}{edgelayer}
            \draw (12) to (11.center);
        \end{pgfonlayer}
    \end{tikzpicture}    
}

\newcommand{\Cuno}[1]{
    \begin{tikzpicture}
        \begin{pgfonlayer}{nodelayer}
            \ifthenelse{\equal{#1}{\uno}}{}{
            \node [style=label] (4) at (-1, 0) {$#1$};
            }
            \node [style=none] (11) at (0, 0) {};
            \node [style=none] (12) at (1, 0) {};
            \node [style=none] (13) at (1, 0.25) {};
            \node [style=none] (14) at (1, -0.25) {};
        \end{pgfonlayer}
        \begin{pgfonlayer}{edgelayer}
            \draw (12.center) to (11.center);
            \draw (14.center) to (13.center);
        \end{pgfonlayer}
    \end{tikzpicture}    
}

\newcommand{\TPolyCounit}[1]{
    \begin{tikzpicture}
        \begin{pgfonlayer}{nodelayer}
            \node [style=label] (3) at (0, 1) {$#1$};
            \node [style=none] (23) at (-0.5, 0.5) {};
            \node [style=none] (24) at (-0.5, -0.5) {};
            \node [style=none] (25) at (0.75, -0.5) {};
            \node [style=none] (26) at (0.75, 0.5) {};
            \node [style=dots] (27) at (0, 0.025) {$\vdots$};
        \end{pgfonlayer}
        \begin{pgfonlayer}{edgelayer}
            \path [style=tape] (24.center)
                 to (25.center)
                 to [bend right=90, looseness=1.75] (26.center)
                 to (23.center)
                 to cycle;
        \end{pgfonlayer}
    \end{tikzpicture}    
}

\newcommand{\TPolyUnit}[1]{
    \begin{tikzpicture}
        \begin{pgfonlayer}{nodelayer}
            \node [style=label] (3) at (0.25, 1) {$#1$};
            \node [style=none] (23) at (0.75, 0.5) {};
            \node [style=none] (24) at (0.75, -0.5) {};
            \node [style=none] (25) at (-0.5, -0.5) {};
            \node [style=none] (26) at (-0.5, 0.5) {};
            \node [style=dots] (27) at (0.25, 0.025) {$\vdots$};
        \end{pgfonlayer}
        \begin{pgfonlayer}{edgelayer}
            \path [style=tape] (24.center)
                 to (25.center)
                 to [bend left=90, looseness=1.75] (26.center)
                 to (23.center)
                 to cycle;
        \end{pgfonlayer}
    \end{tikzpicture}        
}

\newcommand{\TPolycounitunit}[2]{
    \TPolyCounit{#1} \; \TPolyUnit{#2}
}

\newcommand{\TPolyWire}[1]{
    \begin{tikzpicture}
        \begin{pgfonlayer}{nodelayer}
            \node [style=none] (0) at (-1.5, 0.5) {};
            \node [style=none] (3) at (-1.5, -0.5) {};
            \node [style=none] (4) at (1.5, -0.5) {};
            \node [style=none] (5) at (1.5, 0.5) {};
            \node [style=dots] (8) at (-1, 0.025) {$\vdots$};
            \node [style=dots] (9) at (1, 0.025) {$\vdots$};
            \node [style=label] (10) at (1, 1) {$#1$};
            \node [style=label] (11) at (-1, 1) {$#1$};
        \end{pgfonlayer}
        \begin{pgfonlayer}{edgelayer}
            \path [style=tape] (5.center)
                 to (4.center)
                 to (3.center)
                 to (0.center)
                 to cycle;
        \end{pgfonlayer}
    \end{tikzpicture}          
}

\newcommand{\TPolyDiag}[1]{
    \begin{tikzpicture}
        \begin{pgfonlayer}{nodelayer}
            \node [style=none] (8) at (0.75, -1.75) {};
            \node [style=none] (9) at (0.75, 1.75) {};
            \node [style=none] (10) at (0.75, 0.75) {};
            \node [style=none] (11) at (0, 0) {};
            \node [style=none] (12) at (0.75, -0.75) {};
            \node [style=none] (13) at (-1, -0.5) {};
            \node [style=none] (14) at (-1, 0.5) {};
            \node [style=none] (35) at (-2.25, -0.5) {};
            \node [style=none] (36) at (-2.25, 0.5) {};
            \node [style=label] (93) at (-1.5, 1) {$#1$};
            \node [style=dots] (96) at (-1.5, 0.025) {$\vdots$};
        \end{pgfonlayer}
        \begin{pgfonlayer}{edgelayer}
            \path[style=tape] (12.center)
                 to (8.center)
                 to [bend left] (13.center)
                 to (35.center)
                 to (36.center)
                 to (14.center)
                 to [bend left] (9.center)
                 to (10.center)
                 to [bend right=45] (11.center)
                 to [bend right=45] cycle;
        \end{pgfonlayer}
    \end{tikzpicture}    
}

\newcommand{\TPolyCodiag}[1]{
    \begin{tikzpicture}
        \begin{pgfonlayer}{nodelayer}
            \node [style=none] (8) at (-0.75, -1.75) {};
            \node [style=none] (9) at (-0.75, 1.75) {};
            \node [style=none] (10) at (-0.75, 0.75) {};
            \node [style=none] (11) at (0, 0) {};
            \node [style=none] (12) at (-0.75, -0.75) {};
            \node [style=none] (13) at (1, -0.5) {};
            \node [style=none] (14) at (1, 0.5) {};
            \node [style=none] (35) at (2.25, -0.5) {};
            \node [style=none] (36) at (2.25, 0.5) {};
            \node [style=label] (93) at (1.5, 1) {$#1$};
            \node [style=dots] (96) at (1.5, 0.025) {$\vdots$};
        \end{pgfonlayer}
        \begin{pgfonlayer}{edgelayer}
            \path[style=tape] (12.center)
                 to (8.center)
                 to [bend right] (13.center)
                 to (35.center)
                 to (36.center)
                 to (14.center)
                 to [bend right] (9.center)
                 to (10.center)
                 to [bend left=45] (11.center)
                 to [bend left=45] cycle;
        \end{pgfonlayer}
    \end{tikzpicture}    
}

\newcommand{\TLcounit}[2]{
    \begin{tikzpicture}
        \begin{pgfonlayer}{nodelayer}
            \node [style=label] (5) at (-1.75, -0.25) {$#1$};
            \node [style=none] (43) at (0.25, 0.65) {};
            \node [style=none] (45) at (0.25, -0.65) {};
            \node [style=none] (46) at (-1, 0.65) {};
            \node [style=none] (48) at (-1, -0.65) {};
            \node [style=label] (52) at (-1.75, 0.5) {$#2$};
            \node [style=none] (53) at (1, 0) {};
            \node [style=none] (54) at (-1, -0.25) {};
            \node [style=none] (55) at (-1, 0.25) {};
            \node [style=none] (56) at (0.925, 0.25) {};
            \node [style=none] (57) at (0.925, -0.25) {};
        \end{pgfonlayer}
        \begin{pgfonlayer}{edgelayer}
            \draw [style=tape] (53.center)
                 to [bend right=45] (43.center)
                 to (46.center)
                 to (48.center)
                 to (45.center)
                 to [bend right=45] cycle;
            \draw (55.center) to (56.center);
            \draw (57.center) to (54.center);
        \end{pgfonlayer}
    \end{tikzpicture}            
}

\newcommand{\TPolySymmp}[2]{
    \begin{tikzpicture}
        \begin{pgfonlayer}{nodelayer}
            \node [style=label] (3) at (-1.5, 2.25) {$#1$};
            \node [style=none] (6) at (-2, -0.75) {};
            \node [style=none] (7) at (-2, -1.75) {};
            \node [style=none] (17) at (-2, 1.75) {};
            \node [style=none] (18) at (-2, 0.75) {};
            \node [style=none] (25) at (2, 0.75) {};
            \node [style=none] (26) at (2, 1.75) {};
            \node [style=none] (31) at (2, -1.75) {};
            \node [style=none] (32) at (2, -0.75) {};
            \node [style=label] (34) at (1.5, -0.1) {$#1$};
            \node [style=label] (35) at (1.5, 2.25) {$#2$};
            \node [style=label] (36) at (-1.5, -0.1) {$#2$};
            \node [style=dots] (37) at (-1.5, 1.125) {$\vdots$};
            \node [style=dots] (38) at (1.5, 1.125) {$\vdots$};
            \node [style=dots] (39) at (1.5, -1.1) {$\vdots$};
            \node [style=dots] (40) at (-1.5, -1.1) {$\vdots$};
        \end{pgfonlayer}
        \begin{pgfonlayer}{edgelayer}
            \draw [style=tapeFill] (26.center)
                 to [in=0, out=-180, looseness=0.75] (6.center)
                 to (7.center)
                 to [in=-180, out=0, looseness=0.75] (25.center)
                 to cycle;
            \draw [style=tapeFill] (18.center)
                 to (17.center)
                 to [in=180, out=0, looseness=0.75] (32.center)
                 to (31.center)
                 to [in=0, out=-180, looseness=0.75] cycle;
            \draw [style=tapeNoFill] (26.center)
                 to [in=0, out=-180, looseness=0.75] (6.center)
                 to (7.center)
                 to [in=-180, out=0, looseness=0.75] (25.center)
                 to cycle;
            \draw [style=tapeNoFill] (18.center)
                 to (17.center)
                 to [in=180, out=0, looseness=0.75] (32.center)
                 to (31.center)
                 to [in=0, out=-180, looseness=0.75] cycle;
        \end{pgfonlayer}
    \end{tikzpicture}        
}

\newcommand{\EntryUxU}[2]{
    \begin{tikzpicture}
        \begin{pgfonlayer}{nodelayer}
            \node [style=none] (46) at (-0.5, 0.25) {};
            \node [style=none] (47) at (-0.5, 1.25) {};
            \node [style=none] (48) at (-0.5, 0.75) {};
            \node [style=none] (49) at (0.5, 0.75) {};
            \node [style=none] (50) at (0.5, 1.25) {};
            \node [style=none] (51) at (0.5, 0.25) {};
            \node [style=none, scale=0.9] (52) at (0, 0.75) {$#1$};
            \node [style=none] (53) at (-2.5, 0.75) {};
            \node [style=none] (57) at (2.5, 0.75) {};
            \node [style=none] (81) at (-0.5, -1.25) {};
            \node [style=none] (82) at (-0.5, -0.25) {};
            \node [style=none] (83) at (-0.5, -0.75) {};
            \node [style=none] (84) at (0.5, -0.75) {};
            \node [style=none] (85) at (0.5, -0.25) {};
            \node [style=none] (86) at (0.5, -1.25) {};
            \node [style=none, scale=0.9] (87) at (0, -0.75) {$#2$};
            \node [style=none] (88) at (-2.5, -0.75) {};
            \node [style=none] (91) at (2.5, -0.75) {};
        \end{pgfonlayer}
        \begin{pgfonlayer}{edgelayer}
            \draw [fill=white] (51.center)
                 to (46.center)
                 to (47.center)
                 to (50.center)
                 to cycle;
            \draw (53.center) to (48.center);
            \draw (57.center) to (49.center);
            \draw [fill=white] (86.center) to (81.center);
            \draw [fill=white] (81.center) to (82.center);
            \draw [fill=white] (82.center) to (85.center);
            \draw [fill=white] (85.center) to (86.center);
            \draw (88.center) to (83.center);
            \draw (91.center) to (84.center);
        \end{pgfonlayer}
    \end{tikzpicture}    
}

\newcommand{\EntryUxD}[3]{
    \begin{tikzpicture}
        \begin{pgfonlayer}{nodelayer}
            \node [style=none] (46) at (-0.5, 0.25) {};
            \node [style=none] (47) at (-0.5, 1.25) {};
            \node [style=none] (48) at (-0.5, 0.75) {};
            \node [style=none] (49) at (2.5, 0.75) {};
            \node [style=none] (50) at (0.5, 1.25) {};
            \node [style=none] (51) at (0.5, 0.25) {};
            \node [style=none, scale=0.9] (52) at (0, 0.75) {$#1$};
            \node [style=none] (53) at (-2.5, 0.75) {};
            \node [style=none] (57) at (0.5, 0.75) {};
            \node [style=none] (81) at (-1.25, -1.25) {};
            \node [style=none] (82) at (-1.25, -0.25) {};
            \node [style=none] (83) at (-1.25, -0.75) {};
            \node [style=none] (84) at (-0.25, -0.75) {};
            \node [style=none] (85) at (-0.25, -0.25) {};
            \node [style=none] (86) at (-0.25, -1.25) {};
            \node [style=none, scale=0.9] (87) at (-0.75, -0.75) {$#2$};
            \node [style=none] (88) at (-2.5, -0.75) {};
            \node [style=none] (89) at (0.25, -1.25) {};
            \node [style=none] (90) at (0.25, -0.25) {};
            \node [style=none] (91) at (0.25, -0.75) {};
            \node [style=none] (92) at (1.25, -0.75) {};
            \node [style=none] (93) at (1.25, -0.25) {};
            \node [style=none] (94) at (1.25, -1.25) {};
            \node [style=none, scale=0.9] (95) at (0.75, -0.75) {$#3$};
            \node [style=none] (103) at (2.5, -0.75) {};
        \end{pgfonlayer}
        \begin{pgfonlayer}{edgelayer}
            \draw [fill=white] (51.center)
                 to (46.center)
                 to (47.center)
                 to (50.center)
                 to cycle;
            \draw (53.center) to (48.center);
            \draw (57.center) to (49.center);
            \draw [fill=white] (86.center) to (81.center);
            \draw [fill=white] (81.center) to (82.center);
            \draw [fill=white] (82.center) to (85.center);
            \draw [fill=white] (85.center) to (86.center);
            \draw (88.center) to (83.center);
            \draw [fill=white] (94.center)
                 to (89.center)
                 to (90.center)
                 to (93.center)
                 to cycle;
            \draw (91.center) to (84.center);
            \draw (92.center) to (103.center);
        \end{pgfonlayer}
    \end{tikzpicture}    
}

\newcommand{\EntryUxT}[4]{
    \begin{tikzpicture}
        \begin{pgfonlayer}{nodelayer}
            \node [style=none] (46) at (-0.5, 0.25) {};
            \node [style=none] (47) at (-0.5, 1.25) {};
            \node [style=none] (48) at (-0.5, 0.75) {};
            \node [style=none] (49) at (2.5, 0.75) {};
            \node [style=none] (50) at (0.5, 1.25) {};
            \node [style=none] (51) at (0.5, 0.25) {};
            \node [style=none, scale=0.9] (52) at (0, 0.75) {$#1$};
            \node [style=none] (53) at (-2.5, 0.75) {};
            \node [style=none] (57) at (0.5, 0.75) {};
            \node [style=none] (81) at (-2, -1.25) {};
            \node [style=none] (82) at (-2, -0.25) {};
            \node [style=none] (83) at (-2, -0.75) {};
            \node [style=none] (84) at (-1, -0.75) {};
            \node [style=none] (85) at (-1, -0.25) {};
            \node [style=none] (86) at (-1, -1.25) {};
            \node [style=none, scale=0.9] (87) at (-1.5, -0.75) {$#2$};
            \node [style=none] (88) at (-2.5, -0.75) {};
            \node [style=none] (89) at (-0.5, -1.25) {};
            \node [style=none] (90) at (-0.5, -0.25) {};
            \node [style=none] (91) at (-0.5, -0.75) {};
            \node [style=none] (92) at (0.5, -0.75) {};
            \node [style=none] (93) at (0.5, -0.25) {};
            \node [style=none] (94) at (0.5, -1.25) {};
            \node [style=none, scale=0.9] (95) at (0, -0.75) {$#3$};
            \node [style=none] (103) at (1, -0.75) {};
            \node [style=none] (104) at (1, -1.25) {};
            \node [style=none] (105) at (1, -0.25) {};
            \node [style=none] (107) at (2, -0.75) {};
            \node [style=none] (108) at (2, -0.25) {};
            \node [style=none] (109) at (2, -1.25) {};
            \node [style=none, scale=0.9] (110) at (1.5, -0.75) {$#4$};
            \node [style=none] (111) at (2.5, -0.75) {};
        \end{pgfonlayer}
        \begin{pgfonlayer}{edgelayer}
            \draw [fill=white] (51.center)
                 to (46.center)
                 to (47.center)
                 to (50.center)
                 to cycle;
            \draw (53.center) to (48.center);
            \draw (57.center) to (49.center);
            \draw [fill=white] (86.center) to (81.center);
            \draw [fill=white] (81.center) to (82.center);
            \draw [fill=white] (82.center) to (85.center);
            \draw [fill=white] (85.center) to (86.center);
            \draw (88.center) to (83.center);
            \draw [fill=white] (94.center)
                 to (89.center)
                 to (90.center)
                 to (93.center)
                 to cycle;
            \draw (91.center) to (84.center);
            \draw [fill=white] (109.center)
                 to (104.center)
                 to (105.center)
                 to (108.center)
                 to cycle;
            \draw (107.center) to (111.center);
            \draw (92.center) to (103.center);
        \end{pgfonlayer}
    \end{tikzpicture}     
}

\newcommand{\EntryDxU}[3]{
    \begin{tikzpicture}
        \begin{pgfonlayer}{nodelayer}
            \node [style=none] (37) at (-1.25, 1.25) {};
            \node [style=none] (38) at (-1.25, 0.25) {};
            \node [style=none] (39) at (-1.25, 0.75) {};
            \node [style=none] (40) at (-0.25, 0.75) {};
            \node [style=none] (41) at (-0.25, 0.25) {};
            \node [style=none] (42) at (-0.25, 1.25) {};
            \node [style=none, scale=0.9] (43) at (-0.75, 0.75) {$#1$};
            \node [style=none] (44) at (-2.5, 0.75) {};
            \node [style=none] (46) at (-0.5, -0.25) {};
            \node [style=none] (47) at (-0.5, -1.25) {};
            \node [style=none] (48) at (-0.5, -0.75) {};
            \node [style=none] (49) at (0.5, -0.75) {};
            \node [style=none] (50) at (0.5, -1.25) {};
            \node [style=none] (51) at (0.5, -0.25) {};
            \node [style=none, scale=0.9] (52) at (0, -0.75) {$#3$};
            \node [style=none] (53) at (-2.5, -0.75) {};
            \node [style=none] (57) at (2.5, -0.75) {};
            \node [style=none] (64) at (0.25, 1.25) {};
            \node [style=none] (65) at (0.25, 0.25) {};
            \node [style=none] (66) at (0.25, 0.75) {};
            \node [style=none] (67) at (1.25, 0.75) {};
            \node [style=none] (68) at (1.25, 0.25) {};
            \node [style=none] (69) at (1.25, 1.25) {};
            \node [style=none, scale=0.9] (70) at (0.75, 0.75) {$#2$};
            \node [style=none] (72) at (2.5, 0.75) {};
        \end{pgfonlayer}
        \begin{pgfonlayer}{edgelayer}
            \draw [fill=white] (42.center)
                 to (37.center)
                 to (38.center)
                 to (41.center)
                 to cycle;
            \draw (44.center) to (39.center);
            \draw [fill=white] (51.center)
                 to (46.center)
                 to (47.center)
                 to (50.center)
                 to cycle;
            \draw (53.center) to (48.center);
            \draw [fill=white] (69.center)
                 to (64.center)
                 to (65.center)
                 to (68.center)
                 to cycle;
            \draw (72.center) to (67.center);
            \draw (40.center) to (66.center);
            \draw (57.center) to (49.center);
        \end{pgfonlayer}
    \end{tikzpicture}      
}

\newcommand{\EntryDxD}[4]{
    \begin{tikzpicture}
        \begin{pgfonlayer}{nodelayer}
            \node [style=none] (37) at (-1.25, 1.25) {};
            \node [style=none] (38) at (-1.25, 0.25) {};
            \node [style=none] (39) at (-1.25, 0.75) {};
            \node [style=none] (40) at (-0.25, 0.75) {};
            \node [style=none] (41) at (-0.25, 0.25) {};
            \node [style=none] (42) at (-0.25, 1.25) {};
            \node [style=none, scale=0.9] (43) at (-0.75, 0.75) {$#1$};
            \node [style=none] (44) at (-2.5, 0.75) {};
            \node [style=none] (46) at (-1.25, -0.25) {};
            \node [style=none] (47) at (-1.25, -1.25) {};
            \node [style=none] (48) at (-1.25, -0.75) {};
            \node [style=none] (49) at (-0.25, -0.75) {};
            \node [style=none] (50) at (-0.25, -1.25) {};
            \node [style=none] (51) at (-0.25, -0.25) {};
            \node [style=none, scale=0.9] (52) at (-0.75, -0.75) {$#3$};
            \node [style=none] (53) at (-2.5, -0.75) {};
            \node [style=none] (55) at (0.25, -0.25) {};
            \node [style=none] (56) at (0.25, -1.25) {};
            \node [style=none] (57) at (0.25, -0.75) {};
            \node [style=none] (58) at (1.25, -0.75) {};
            \node [style=none] (59) at (1.25, -1.25) {};
            \node [style=none] (60) at (1.25, -0.25) {};
            \node [style=none, scale=0.9] (61) at (0.75, -0.75) {$#4$};
            \node [style=none] (63) at (2.5, -0.75) {};
            \node [style=none] (64) at (0.25, 1.25) {};
            \node [style=none] (65) at (0.25, 0.25) {};
            \node [style=none] (66) at (0.25, 0.75) {};
            \node [style=none] (67) at (1.25, 0.75) {};
            \node [style=none] (68) at (1.25, 0.25) {};
            \node [style=none] (69) at (1.25, 1.25) {};
            \node [style=none, scale=0.9] (70) at (0.75, 0.75) {$#2$};
            \node [style=none] (72) at (2.5, 0.75) {};
        \end{pgfonlayer}
        \begin{pgfonlayer}{edgelayer}
            \draw [fill=white] (42.center)
                 to (37.center)
                 to (38.center)
                 to (41.center)
                 to cycle;
            \draw (44.center) to (39.center);
            \draw [fill=white] (51.center)
                 to (46.center)
                 to (47.center)
                 to (50.center)
                 to cycle;
            \draw (53.center) to (48.center);
            \draw [fill=white] (60.center)
                 to (55.center)
                 to (56.center)
                 to (59.center)
                 to cycle;
            \draw (63.center) to (58.center);
            \draw [fill=white] (69.center)
                 to (64.center)
                 to (65.center)
                 to (68.center)
                 to cycle;
            \draw (72.center) to (67.center);
            \draw (40.center) to (66.center);
            \draw (57.center) to (49.center);
        \end{pgfonlayer}
    \end{tikzpicture}    
}

\newcommand{\EntryDxT}[5]{
    \begin{tikzpicture}
        \begin{pgfonlayer}{nodelayer}
            \node [style=none] (37) at (-1.25, 1.25) {};
            \node [style=none] (38) at (-1.25, 0.25) {};
            \node [style=none] (39) at (-1.25, 0.75) {};
            \node [style=none] (40) at (-0.25, 0.75) {};
            \node [style=none] (41) at (-0.25, 0.25) {};
            \node [style=none] (42) at (-0.25, 1.25) {};
            \node [style=none, scale=0.9] (43) at (-0.75, 0.75) {$#1$};
            \node [style=none] (44) at (-2.5, 0.75) {};
            \node [style=none] (46) at (-2, -0.25) {};
            \node [style=none] (47) at (-2, -1.25) {};
            \node [style=none] (48) at (-2, -0.75) {};
            \node [style=none] (49) at (-1, -0.75) {};
            \node [style=none] (50) at (-1, -1.25) {};
            \node [style=none] (51) at (-1, -0.25) {};
            \node [style=none, scale=0.9] (52) at (-1.5, -0.75) {$#3$};
            \node [style=none] (53) at (-2.5, -0.75) {};
            \node [style=none] (55) at (-0.5, -0.25) {};
            \node [style=none] (56) at (-0.5, -1.25) {};
            \node [style=none] (57) at (-0.5, -0.75) {};
            \node [style=none] (58) at (0.5, -0.75) {};
            \node [style=none] (59) at (0.5, -1.25) {};
            \node [style=none] (60) at (0.5, -0.25) {};
            \node [style=none, scale=0.9] (61) at (0, -0.75) {$#4$};
            \node [style=none] (64) at (0.25, 1.25) {};
            \node [style=none] (65) at (0.25, 0.25) {};
            \node [style=none] (66) at (0.25, 0.75) {};
            \node [style=none] (67) at (1.25, 0.75) {};
            \node [style=none] (68) at (1.25, 0.25) {};
            \node [style=none] (69) at (1.25, 1.25) {};
            \node [style=none, scale=0.9] (70) at (0.75, 0.75) {$#2$};
            \node [style=none] (72) at (2.5, 0.75) {};
            \node [style=none] (73) at (1, -0.25) {};
            \node [style=none] (74) at (1, -1.25) {};
            \node [style=none] (75) at (2, -0.75) {};
            \node [style=none] (76) at (2, -1.25) {};
            \node [style=none] (77) at (2, -0.25) {};
            \node [style=none, scale=0.9] (78) at (1.5, -0.75) {$#5$};
            \node [style=none] (79) at (2.5, -0.75) {};
            \node [style=none] (80) at (1, -0.75) {};
        \end{pgfonlayer}
        \begin{pgfonlayer}{edgelayer}
            \draw [fill=white] (42.center)
                 to (37.center)
                 to (38.center)
                 to (41.center)
                 to cycle;
            \draw (44.center) to (39.center);
            \draw [fill=white] (51.center)
                 to (46.center)
                 to (47.center)
                 to (50.center)
                 to cycle;
            \draw (53.center) to (48.center);
            \draw [fill=white] (60.center)
                 to (55.center)
                 to (56.center)
                 to (59.center)
                 to cycle;
            \draw [fill=white] (69.center)
                 to (64.center)
                 to (65.center)
                 to (68.center)
                 to cycle;
            \draw (72.center) to (67.center);
            \draw (40.center) to (66.center);
            \draw (57.center) to (49.center);
            \draw [fill=white] (77.center)
                 to (73.center)
                 to (74.center)
                 to (76.center)
                 to cycle;
            \draw (79.center) to (75.center);
            \draw (58.center) to (80.center);
        \end{pgfonlayer}
    \end{tikzpicture}    
}

\newcommand{\EntryTxU}[4]{
    \begin{tikzpicture}
        \begin{pgfonlayer}{nodelayer}
            \node [style=none] (44) at (-2.5, -0.75) {};
            \node [style=none] (46) at (-2, 0.25) {};
            \node [style=none] (47) at (-2, 1.25) {};
            \node [style=none] (48) at (-2, 0.75) {};
            \node [style=none] (49) at (-1, 0.75) {};
            \node [style=none] (50) at (-1, 1.25) {};
            \node [style=none] (51) at (-1, 0.25) {};
            \node [style=none, scale=0.9] (52) at (-1.5, 0.75) {$#1$};
            \node [style=none] (53) at (-2.5, 0.75) {};
            \node [style=none] (55) at (-0.5, 0.25) {};
            \node [style=none] (56) at (-0.5, 1.25) {};
            \node [style=none] (57) at (-0.5, 0.75) {};
            \node [style=none] (58) at (0.5, 0.75) {};
            \node [style=none] (59) at (0.5, 1.25) {};
            \node [style=none] (60) at (0.5, 0.25) {};
            \node [style=none, scale=0.9] (61) at (0, 0.75) {$#2$};
            \node [style=none] (64) at (-0.5, -1.25) {};
            \node [style=none] (65) at (-0.5, -0.25) {};
            \node [style=none] (66) at (-0.5, -0.75) {};
            \node [style=none] (67) at (0.5, -0.75) {};
            \node [style=none] (68) at (0.5, -0.25) {};
            \node [style=none] (69) at (0.5, -1.25) {};
            \node [style=none, scale=0.9] (70) at (0, -0.75) {$#4$};
            \node [style=none] (72) at (2.5, -0.75) {};
            \node [style=none] (73) at (1, 0.25) {};
            \node [style=none] (74) at (1, 1.25) {};
            \node [style=none] (75) at (2, 0.75) {};
            \node [style=none] (76) at (2, 1.25) {};
            \node [style=none] (77) at (2, 0.25) {};
            \node [style=none, scale=0.9] (78) at (1.5, 0.75) {$#3$};
            \node [style=none] (79) at (2.5, 0.75) {};
            \node [style=none] (80) at (1, 0.75) {};
        \end{pgfonlayer}
        \begin{pgfonlayer}{edgelayer}
            \draw [fill=white] (51.center)
                 to (46.center)
                 to (47.center)
                 to (50.center)
                 to cycle;
            \draw (53.center) to (48.center);
            \draw [fill=white] (60.center)
                 to (55.center)
                 to (56.center)
                 to (59.center)
                 to cycle;
            \draw [fill=white] (69.center)
                 to (64.center)
                 to (65.center)
                 to (68.center)
                 to cycle;
            \draw (72.center) to (67.center);
            \draw (57.center) to (49.center);
            \draw [fill=white] (77.center)
                 to (73.center)
                 to (74.center)
                 to (76.center)
                 to cycle;
            \draw (79.center) to (75.center);
            \draw (58.center) to (80.center);
            \draw (44.center) to (66.center);
        \end{pgfonlayer}
    \end{tikzpicture}    
}

\newcommand{\EntryTxD}[5]{
    \begin{tikzpicture}
        \begin{pgfonlayer}{nodelayer}
            \node [style=none] (37) at (-1.25, -1.25) {};
            \node [style=none] (38) at (-1.25, -0.25) {};
            \node [style=none] (39) at (-1.25, -0.75) {};
            \node [style=none] (40) at (-0.25, -0.75) {};
            \node [style=none] (41) at (-0.25, -0.25) {};
            \node [style=none] (42) at (-0.25, -1.25) {};
            \node [style=none, scale=0.9] (43) at (-0.75, -0.75) {$#4$};
            \node [style=none] (44) at (-2.5, -0.75) {};
            \node [style=none] (46) at (-2, 0.25) {};
            \node [style=none] (47) at (-2, 1.25) {};
            \node [style=none] (48) at (-2, 0.75) {};
            \node [style=none] (49) at (-1, 0.75) {};
            \node [style=none] (50) at (-1, 1.25) {};
            \node [style=none] (51) at (-1, 0.25) {};
            \node [style=none, scale=0.9] (52) at (-1.5, 0.75) {$#1$};
            \node [style=none] (53) at (-2.5, 0.75) {};
            \node [style=none] (55) at (-0.5, 0.25) {};
            \node [style=none] (56) at (-0.5, 1.25) {};
            \node [style=none] (57) at (-0.5, 0.75) {};
            \node [style=none] (58) at (0.5, 0.75) {};
            \node [style=none] (59) at (0.5, 1.25) {};
            \node [style=none] (60) at (0.5, 0.25) {};
            \node [style=none, scale=0.9] (61) at (0, 0.75) {$#2$};
            \node [style=none] (64) at (0.25, -1.25) {};
            \node [style=none] (65) at (0.25, -0.25) {};
            \node [style=none] (66) at (0.25, -0.75) {};
            \node [style=none] (67) at (1.25, -0.75) {};
            \node [style=none] (68) at (1.25, -0.25) {};
            \node [style=none] (69) at (1.25, -1.25) {};
            \node [style=none, scale=0.9] (70) at (0.75, -0.75) {$#5$};
            \node [style=none] (72) at (2.5, -0.75) {};
            \node [style=none] (73) at (1, 0.25) {};
            \node [style=none] (74) at (1, 1.25) {};
            \node [style=none] (75) at (2, 0.75) {};
            \node [style=none] (76) at (2, 1.25) {};
            \node [style=none] (77) at (2, 0.25) {};
            \node [style=none, scale=0.9] (78) at (1.5, 0.75) {$#3$};
            \node [style=none] (79) at (2.5, 0.75) {};
            \node [style=none] (80) at (1, 0.75) {};
        \end{pgfonlayer}
        \begin{pgfonlayer}{edgelayer}
            \draw [fill=white] (42.center)
                 to (37.center)
                 to (38.center)
                 to (41.center)
                 to cycle;
            \draw (44.center) to (39.center);
            \draw [fill=white] (51.center)
                 to (46.center)
                 to (47.center)
                 to (50.center)
                 to cycle;
            \draw (53.center) to (48.center);
            \draw [fill=white] (60.center)
                 to (55.center)
                 to (56.center)
                 to (59.center)
                 to cycle;
            \draw [fill=white] (69.center)
                 to (64.center)
                 to (65.center)
                 to (68.center)
                 to cycle;
            \draw (72.center) to (67.center);
            \draw (40.center) to (66.center);
            \draw (57.center) to (49.center);
            \draw [fill=white] (77.center)
                 to (73.center)
                 to (74.center)
                 to (76.center)
                 to cycle;
            \draw (79.center) to (75.center);
            \draw (58.center) to (80.center);
        \end{pgfonlayer}
    \end{tikzpicture}    
}

\newcommand{\EntryTxT}[6]{
    \begin{tikzpicture}
        \begin{pgfonlayer}{nodelayer}
            \node [style=none] (46) at (-2, 0.25) {};
            \node [style=none] (47) at (-2, 1.25) {};
            \node [style=none] (48) at (-2, 0.75) {};
            \node [style=none] (49) at (-1, 0.75) {};
            \node [style=none] (50) at (-1, 1.25) {};
            \node [style=none] (51) at (-1, 0.25) {};
            \node [style=none, scale=0.9] (52) at (-1.5, 0.75) {$#1$};
            \node [style=none] (53) at (-2.5, 0.75) {};
            \node [style=none] (55) at (-0.5, 0.25) {};
            \node [style=none] (56) at (-0.5, 1.25) {};
            \node [style=none] (57) at (-0.5, 0.75) {};
            \node [style=none] (58) at (0.5, 0.75) {};
            \node [style=none] (59) at (0.5, 1.25) {};
            \node [style=none] (60) at (0.5, 0.25) {};
            \node [style=none, scale=0.9] (61) at (0, 0.75) {$#2$};
            \node [style=none] (73) at (1, 0.25) {};
            \node [style=none] (74) at (1, 1.25) {};
            \node [style=none] (75) at (2, 0.75) {};
            \node [style=none] (76) at (2, 1.25) {};
            \node [style=none] (77) at (2, 0.25) {};
            \node [style=none, scale=0.9] (78) at (1.5, 0.75) {$#3$};
            \node [style=none] (79) at (2.5, 0.75) {};
            \node [style=none] (80) at (1, 0.75) {};
            \node [style=none] (81) at (-2, -1.25) {};
            \node [style=none] (82) at (-2, -0.25) {};
            \node [style=none] (83) at (-2, -0.75) {};
            \node [style=none] (84) at (-1, -0.75) {};
            \node [style=none] (85) at (-1, -0.25) {};
            \node [style=none] (86) at (-1, -1.25) {};
            \node [style=none, scale=0.9] (87) at (-1.5, -0.75) {$#4$};
            \node [style=none] (88) at (-2.5, -0.75) {};
            \node [style=none] (89) at (-0.5, -1.25) {};
            \node [style=none] (90) at (-0.5, -0.25) {};
            \node [style=none] (91) at (-0.5, -0.75) {};
            \node [style=none] (92) at (0.5, -0.75) {};
            \node [style=none] (93) at (0.5, -0.25) {};
            \node [style=none] (94) at (0.5, -1.25) {};
            \node [style=none, scale=0.9] (95) at (0, -0.75) {$#5$};
            \node [style=none] (96) at (1, -1.25) {};
            \node [style=none] (97) at (1, -0.25) {};
            \node [style=none] (98) at (2, -0.75) {};
            \node [style=none] (99) at (2, -0.25) {};
            \node [style=none] (100) at (2, -1.25) {};
            \node [style=none, scale=0.9] (101) at (1.5, -0.75) {$#6$};
            \node [style=none] (102) at (2.5, -0.75) {};
            \node [style=none] (103) at (1, -0.75) {};
        \end{pgfonlayer}
        \begin{pgfonlayer}{edgelayer}
            \draw [fill=white] (51.center)
                 to (46.center)
                 to (47.center)
                 to (50.center)
                 to cycle;
            \draw (53.center) to (48.center);
            \draw [fill=white] (60.center)
                 to (55.center)
                 to (56.center)
                 to (59.center)
                 to cycle;
            \draw (57.center) to (49.center);
            \draw [fill=white] (77.center)
                 to (73.center)
                 to (74.center)
                 to (76.center)
                 to cycle;
            \draw (79.center) to (75.center);
            \draw (58.center) to (80.center);
            \draw [fill=white] (86.center) to (81.center);
            \draw [fill=white] (81.center) to (82.center);
            \draw [fill=white] (82.center) to (85.center);
            \draw [fill=white] (85.center) to (86.center);
            \draw (88.center) to (83.center);
            \draw [fill=white] (94.center)
                 to (89.center)
                 to (90.center)
                 to (93.center)
                 to cycle;
            \draw (91.center) to (84.center);
            \draw [fill=white] (100.center) to (96.center);
            \draw [fill=white] (96.center) to (97.center);
            \draw [fill=white] (97.center) to (99.center);
            \draw [fill=white] (99.center) to (100.center);
            \draw (102.center) to (98.center);
            \draw (92.center) to (103.center);
        \end{pgfonlayer}
    \end{tikzpicture}    
}

\newcommand{\EntryU}[1]{
    \begin{tikzpicture}
        \begin{pgfonlayer}{nodelayer}
            \node [style=none] (46) at (-0.5, -0.5) {};
            \node [style=none] (47) at (-0.5, 0.5) {};
            \node [style=none] (48) at (-0.5, 0) {};
            \node [style=none] (49) at (0.5, 0) {};
            \node [style=none] (50) at (0.5, 0.5) {};
            \node [style=none] (51) at (0.5, -0.5) {};
            \node [style=none, scale=0.9] (52) at (0, 0) {$#1$};
            \node [style=none] (53) at (-2.5, 0) {};
            \node [style=none] (57) at (2.5, 0) {};
        \end{pgfonlayer}
        \begin{pgfonlayer}{edgelayer}
            \draw [fill=white] (51.center)
                 to (46.center)
                 to (47.center)
                 to (50.center)
                 to cycle;
            \draw (53.center) to (48.center);
            \draw (57.center) to (49.center);
        \end{pgfonlayer}
    \end{tikzpicture}    
}

\newcommand{\EntryD}[2]{
    \begin{tikzpicture}
        \begin{pgfonlayer}{nodelayer}
            \node [style=none] (46) at (-1.25, -0.5) {};
            \node [style=none] (47) at (-1.25, 0.5) {};
            \node [style=none] (48) at (-1.25, 0) {};
            \node [style=none] (49) at (-0.25, 0) {};
            \node [style=none] (50) at (-0.25, 0.5) {};
            \node [style=none] (51) at (-0.25, -0.5) {};
            \node [style=none, scale=0.9] (52) at (-0.75, 0) {$#1$};
            \node [style=none] (53) at (-2.5, 0) {};
            \node [style=none] (55) at (0.25, -0.5) {};
            \node [style=none] (56) at (0.25, 0.5) {};
            \node [style=none] (57) at (0.25, 0) {};
            \node [style=none] (58) at (1.25, 0) {};
            \node [style=none] (59) at (1.25, 0.5) {};
            \node [style=none] (60) at (1.25, -0.5) {};
            \node [style=none, scale=0.9] (61) at (0.75, 0) {$#2$};
            \node [style=none] (80) at (2.5, 0) {};
        \end{pgfonlayer}
        \begin{pgfonlayer}{edgelayer}
            \draw [fill=white] (51.center)
                 to (46.center)
                 to (47.center)
                 to (50.center)
                 to cycle;
            \draw (53.center) to (48.center);
            \draw [fill=white] (60.center)
                 to (55.center)
                 to (56.center)
                 to (59.center)
                 to cycle;
            \draw (57.center) to (49.center);
            \draw (58.center) to (80.center);
        \end{pgfonlayer}
    \end{tikzpicture}    
}

\newcommand{\EntryT}[3]{
    \begin{tikzpicture}
        \begin{pgfonlayer}{nodelayer}
            \node [style=none] (46) at (-2, -0.5) {};
            \node [style=none] (47) at (-2, 0.5) {};
            \node [style=none] (48) at (-2, 0) {};
            \node [style=none] (49) at (-1, 0) {};
            \node [style=none] (50) at (-1, 0.5) {};
            \node [style=none] (51) at (-1, -0.5) {};
            \node [style=none, scale=0.9] (52) at (-1.5, 0) {$#1$};
            \node [style=none] (53) at (-2.5, 0) {};
            \node [style=none] (55) at (-0.5, -0.5) {};
            \node [style=none] (56) at (-0.5, 0.5) {};
            \node [style=none] (57) at (-0.5, 0) {};
            \node [style=none] (58) at (0.5, 0) {};
            \node [style=none] (59) at (0.5, 0.5) {};
            \node [style=none] (60) at (0.5, -0.5) {};
            \node [style=none, scale=0.9] (61) at (0, 0) {$#2$};
            \node [style=none] (73) at (1, -0.5) {};
            \node [style=none] (74) at (1, 0.5) {};
            \node [style=none] (75) at (2, 0) {};
            \node [style=none] (76) at (2, 0.5) {};
            \node [style=none] (77) at (2, -0.5) {};
            \node [style=none, scale=0.9] (78) at (1.5, 0) {$#3$};
            \node [style=none] (79) at (2.5, 0) {};
            \node [style=none] (80) at (1, 0) {};
        \end{pgfonlayer}
        \begin{pgfonlayer}{edgelayer}
            \draw [fill=white] (51.center)
                 to (46.center)
                 to (47.center)
                 to (50.center)
                 to cycle;
            \draw (53.center) to (48.center);
            \draw [fill=white] (60.center)
                 to (55.center)
                 to (56.center)
                 to (59.center)
                 to cycle;
            \draw (57.center) to (49.center);
            \draw [fill=white] (77.center)
                 to (73.center)
                 to (74.center)
                 to (76.center)
                 to cycle;
            \draw (79.center) to (75.center);
            \draw (58.center) to (80.center);
        \end{pgfonlayer}
    \end{tikzpicture}    
}

\newcommand{\Tcopier}[1]{\copier{#1}
}

\newcommand{\Tcocopier}[1]{
    \cocopier{#1}
}

\newcommand{\Tdischarger}[1]{\discharger{#1}
}

\newcommand{\Tcodischarger}[1]{\codischarger{#1}
}

\newcommand{\qEffect}[1]{
    \begin{tikzpicture}
        \begin{pgfonlayer}{nodelayer}
            \node [style=none] (81) at (-1, 0) {};
            \node [style=none] (112) at (0, 0) {};
            \node [style=none] (113) at (0, 0.5) {};
            \node [style=none] (114) at (0, -0.5) {};
            \node [style=none] (115) at (1, 0) {};
            \node [scale=0.6] (116) at (0.35, 0) {$#1$};
        \end{pgfonlayer}
        \begin{pgfonlayer}{edgelayer}
            \draw (81.center) to (112.center);
            \draw (114.center) to (115.center);
            \draw (113.center) to (112.center);
            \draw (112.center) to (114.center);
            \draw (113.center) to (115.center);
        \end{pgfonlayer}
    \end{tikzpicture}    
}

\newcommand{\qState}[1]{
    \begin{tikzpicture}
        \begin{pgfonlayer}{nodelayer}
            \node [style=none] (81) at (1, 0) {};
            \node [style=none] (112) at (0, 0) {};
            \node [style=none] (113) at (0, 0.5) {};
            \node [style=none] (114) at (0, -0.5) {};
            \node [style=none] (115) at (-1, 0) {};
            \node [scale=0.6] (116) at (-0.35, 0) {$#1$};
        \end{pgfonlayer}
        \begin{pgfonlayer}{edgelayer}
            \draw (81.center) to (112.center);
            \draw (114.center) to (115.center);
            \draw (113.center) to (112.center);
            \draw (112.center) to (114.center);
            \draw (113.center) to (115.center);
        \end{pgfonlayer}
    \end{tikzpicture}       
}

\newcommand{\Trel}[1]{
    \begin{tikzpicture}
        \begin{pgfonlayer}{nodelayer}
            \node [style=none] (55) at (1.5, -0.625) {};
            \node [style=none] (56) at (1.5, 0.625) {};
            \node [style=none] (57) at (-1.5, 0.625) {};
            \node [style=none] (58) at (-1.5, -0.625) {};
            \node [style=none] (60) at (-1.5, 0) {};
            \node [style=none] (61) at (1.5, 0) {};
            \node [style=box] (62) at (0, 0) {$#1$};
        \end{pgfonlayer}
        \begin{pgfonlayer}{edgelayer}
            \draw[style=tape] (56.center)
                 to (57.center)
                 to (58.center)
                 to (55.center)
                 to cycle;
            \draw (60.center) to (62);
            \draw (62) to (61.center);
        \end{pgfonlayer}
    \end{tikzpicture}    
}

\newcommand{\CBcopier}[1]{
    \begin{tikzpicture}
        \begin{pgfonlayer}{nodelayer}
            \node [style=none] (6) at (-0.75, 0) {};
            \node [style=black] (7) at (0, 0) {};
            \node [style=label] (8) at (-1.25, 0) {$#1$};
            \node [style=label] (9) at (1.25, 0.325) {$#1$};
            \node [style=none] (10) at (0.5, 0.325) {};
            \node [style=none] (11) at (0.5, -0.325) {};
            \node [style=label] (12) at (1.25, -0.325) {$#1$};
            \node [style=none] (13) at (0.75, 0.325) {};
            \node [style=none] (14) at (0.75, -0.325) {};
        \end{pgfonlayer}
        \begin{pgfonlayer}{edgelayer}
            \draw (6.center) to (7);
            \draw [bend left] (7) to (10.center);
            \draw [bend right] (7) to (11.center);
            \draw (10.center) to (13.center);
            \draw (14.center) to (11.center);
        \end{pgfonlayer}
    \end{tikzpicture}    
}

\newcommand{\CBcocopier}[1]{
    \begin{tikzpicture}
        \begin{pgfonlayer}{nodelayer}
            \node [style=none] (6) at (0.75, 0) {};
            \node [style=black] (7) at (0, 0) {};
            \node [style=label] (8) at (1.25, 0) {$#1$};
            \node [style=label] (9) at (-1.25, 0.325) {$#1$};
            \node [style=none] (10) at (-0.5, 0.325) {};
            \node [style=none] (11) at (-0.5, -0.325) {};
            \node [style=label] (12) at (-1.25, -0.325) {$#1$};
            \node [style=none] (13) at (-0.75, 0.325) {};
            \node [style=none] (14) at (-0.75, -0.325) {};
        \end{pgfonlayer}
        \begin{pgfonlayer}{edgelayer}
            \draw (6.center) to (7);
            \draw [bend right] (7) to (10.center);
            \draw [bend left] (7) to (11.center);
            \draw (10.center) to (13.center);
            \draw (14.center) to (11.center);
        \end{pgfonlayer}
    \end{tikzpicture}          
}

\newcommand{\CBdischarger}[1]{
    \begin{tikzpicture}
        \begin{pgfonlayer}{nodelayer}
            \node [style=none] (6) at (-0.25, 0) {};
            \node [style=black] (7) at (0.5, 0) {};
            \node [style=label] (8) at (-0.75, 0) {$#1$};
        \end{pgfonlayer}
        \begin{pgfonlayer}{edgelayer}
            \draw (6.center) to (7);
        \end{pgfonlayer}
    \end{tikzpicture}    
}

\newcommand{\CBcodischarger}[1]{
    \begin{tikzpicture}
        \begin{pgfonlayer}{nodelayer}
            \node [style=none] (6) at (0.75, 0) {};
            \node [style=black] (7) at (0, 0) {};
            \node [style=label] (8) at (1.25, 0) {$#1$};
        \end{pgfonlayer}
        \begin{pgfonlayer}{edgelayer}
            \draw (6.center) to (7);
        \end{pgfonlayer}
    \end{tikzpicture}    
}

\newcommand{\CBRel}[3]{
    \begin{tikzpicture}
        \begin{pgfonlayer}{nodelayer}
            \node [style=none] (60) at (-1, 0) {};
            \node [style=none] (61) at (1, 0) {};
            \node [style=box] (62) at (0, 0) {$#1$};
            \node [style=label] (63) at (-1.5, 0) {$#2$};
            \node [style=label] (64) at (1.5, 0) {$#3$};
        \end{pgfonlayer}
        \begin{pgfonlayer}{edgelayer}
            \draw (60.center) to (62);
            \draw (62) to (61.center);
        \end{pgfonlayer}
    \end{tikzpicture}      
}

\newcommand{\TRelCodiagDiag}[1]{\begin{tikzpicture}
	\begin{pgfonlayer}{nodelayer}
		\node [style=none] (0) at (-1.5, 0) {};
		\node [style=none] (1) at (-0.25, 1.25) {};
		\node [style=none] (2) at (-0.25, -1.25) {};
		\node [style=none] (8) at (-0.25, -1.75) {};
		\node [style=none] (9) at (-0.25, 1.75) {};
		\node [style=none] (10) at (-0.25, 0.75) {};
		\node [style=none] (11) at (-1, 0) {};
		\node [style=none] (12) at (-0.25, -0.75) {};
		\node [style=none] (13) at (-2, -0.5) {};
		\node [style=none] (14) at (-2, 0.5) {};
		\node [style=none] (24) at (-3, 0) {};
		\node [style=none] (35) at (-3, -0.5) {};
		\node [style=none] (36) at (-3, 0.5) {};
		\node [style=none] (50) at (-0.25, 1.25) {};
		\node [style=none] (80) at (1.5, 0) {};
		\node [style=none] (81) at (0.25, 1.25) {};
		\node [style=none] (82) at (0.25, -1.25) {};
		\node [style=none] (83) at (0.25, -1.75) {};
		\node [style=none] (84) at (0.25, 1.75) {};
		\node [style=none] (85) at (0.25, 0.75) {};
		\node [style=none] (86) at (1, 0) {};
		\node [style=none] (87) at (0.25, -0.75) {};
		\node [style=none] (88) at (2, -0.5) {};
		\node [style=none] (89) at (2, 0.5) {};
		\node [style=none] (90) at (3, 0) {};
		\node [style=none] (91) at (3, -0.5) {};
		\node [style=none] (92) at (3, 0.5) {};
		\node [style=none] (93) at (0.25, 1.25) {};
		\node [style=label] (94) at (-3.5, 0) {$#1$};
		\node [style=label] (95) at (3.5, 0) {$#1$};
	\end{pgfonlayer}
	\begin{pgfonlayer}{edgelayer}
		\draw [style=tape] (83.center)
			 to (8.center)
			 to [bend left] (13.center)
			 to (35.center)
			 to (36.center)
			 to (14.center)
			 to [bend left] (9.center)
			 to (84.center)
			 to [bend left] (89.center)
			 to (92.center)
			 to (91.center)
			 to (88.center)
			 to [bend left] cycle;
		\draw [style=tapeNoFill, fill=white] (87.center)
			 to (12.center)
			 to [bend left=45] (11.center)
			 to [bend left=45] (10.center)
			 to (85.center)
			 to [bend left=45] (86.center)
			 to [bend left=45] cycle;
		\draw [bend right=45] (0.center) to (2.center);
		\draw [bend left=45] (0.center) to (1.center);
		\draw (24.center) to (0.center);
		\draw [bend left=45] (80.center) to (82.center);
		\draw [bend right=45] (80.center) to (81.center);
		\draw (90.center) to (80.center);
		\draw (50.center) to (93.center);
		\draw (82.center) to (2.center);
	\end{pgfonlayer}
\end{tikzpicture}
}

\newcommand{\TapedMonoid}{
    \begin{tikzpicture}
        \begin{pgfonlayer}{nodelayer}
            \node [style=none] (81) at (1.25, 0) {};
            \node [style=black] (107) at (-0.25, 0) {};
            \node [style=none] (108) at (-0.75, -0.375) {};
            \node [style=none] (109) at (-0.75, 0.375) {};
            \node [style=none] (110) at (-1.25, -0.375) {};
            \node [style=none] (111) at (-1.25, 0.375) {};
            \node [style=none] (115) at (-1.25, 0.625) {};
            \node [style=none] (116) at (-1.25, -0.6) {};
            \node [style=none] (117) at (1.25, -0.6) {};
            \node [style=none] (118) at (1.25, 0.625) {};
        \end{pgfonlayer}
        \begin{pgfonlayer}{edgelayer}
            \draw [style=tape] (115.center)
                 to (118.center)
                 to (117.center)
                 to (116.center)
                 to cycle;
            \draw [bend left] (109.center) to (107);
            \draw [bend left] (107) to (108.center);
            \draw (81.center) to (107);
            \draw (111.center) to (109.center);
            \draw (108.center) to (110.center);
        \end{pgfonlayer}
    \end{tikzpicture}        
}

\newcommand{\TapedComonoid}{
    \begin{tikzpicture}
        \begin{pgfonlayer}{nodelayer}
            \node [style=none] (81) at (-1.25, 0) {};
            \node [style=black] (107) at (0.25, 0) {};
            \node [style=none] (108) at (0.75, -0.375) {};
            \node [style=none] (109) at (0.75, 0.375) {};
            \node [style=none] (110) at (1.25, -0.375) {};
            \node [style=none] (111) at (1.25, 0.375) {};
            \node [style=none] (115) at (1.25, 0.625) {};
            \node [style=none] (116) at (1.25, -0.6) {};
            \node [style=none] (117) at (-1.25, -0.6) {};
            \node [style=none] (118) at (-1.25, 0.625) {};
        \end{pgfonlayer}
        \begin{pgfonlayer}{edgelayer}
            \draw [style=tape] (115.center)
                 to (118.center)
                 to (117.center)
                 to (116.center)
                 to cycle;
            \draw [bend right] (109.center) to (107);
            \draw [bend right] (107) to (108.center);
            \draw (81.center) to (107);
            \draw (111.center) to (109.center);
            \draw (108.center) to (110.center);
        \end{pgfonlayer}
    \end{tikzpicture}             
} 
\AtBeginDocument{}

\setcopyright{acmcopyright}
\copyrightyear{2018}
\acmYear{2018}
\acmDOI{XXXXXXX.XXXXXXX}

\acmConference[Conference acronym 'XX]{Make sure to enter the correct
	conference title from your rights confirmation emai}{June 03--05,
	2018}{Woodstock, NY}
\acmPrice{15.00}
\acmISBN{978-1-4503-XXXX-X/18/06}

\begin{document}

\title{Deconstructing the Calculus of Relations with Tape~Diagrams}

\thanks{\textcopyright 2022 Filippo Bonchi, Alessandro Di Giorgio, Alessio Santamaria. This preprint article is available under licence CC-BY 4.0 \url{https://creativecommons.org/licenses/by/4.0/}}

\author{Filippo Bonchi}
\email{filippo.bonchi@unipi.it}
\orcid{0000-0002-3433-723X}             \affiliation{
	\department{Dipartimento di Informatica}              \institution{University of Pisa}            \city{Pisa}
\country{Italy}                    }
\author{Alessandro Di Giorgio}
\email{alessandro.digiorgio@phd.unipi.it}
\orcid{0000-0002-6428-6461}
\affiliation{
	\department{Dipartimento di Informatica}              \institution{University of Pisa}            \city{Pisa}
\country{Italy}                    }
\author{Alessio Santamaria}
\email{alessio.santamaria@di.unipi.it}
\orcid{0000-0001-7683-5221}   
\affiliation{
	\department{Dipartimento di Informatica}              \institution{University of Pisa}            \city{Pisa}
\country{Italy}                    }
\affiliation{
	\department{Department of Informatics}
	\institution{University of Sussex}
	\city{Brighton}
	\country{United Kingdom}
}

\begin{abstract}
  Rig categories with finite biproducts are categories with two monoidal products, where one is a biproduct and the other distributes over it. In this work we present tape diagrams, a sound and complete diagrammatic language for these categories, that can be intuitively thought as string diagrams of string diagrams.
  We test the effectiveness of our approach against the positive fragment of Tarski's calculus of relations.
\end{abstract}



\maketitle

\section{Introduction}

Diagrammatic notations have been used in computer science since its early stages. A famous example is the proof of the structured program theorem~\cite{bohm1966flow} by B\"ohm and Jacopini: they rely on a syntax of flow diagrams and, by means of several transformations preserving the semantics, prove the existence of a normal form.
In physics, diagrams by Feynman~\cite{kaiser2009drawing} and Penrose~\cite{penrose1971applications} became essential linguistic tools: on the one hand they provide intuitive visualisations for otherwise arcane formulas; on the other, they allow a critical simplification of calculations, pretty much like adding two Hindu-Arabic numerals is far easier than two Roman ones~\cite{fibonacci_liber_2020}.
 
In general, well-behaved diagrammatic languages share some desirable features: (a) diagrams can be composed, like one composes formulas in mathematics, programs in a programming language or sentences in English; (b) they are equipped with a compositional semantics: the meaning of a compound diagram is given by the meaning of its components; (c) some basic laws allow us to transform diagrams into semantically equivalent ones, like the laws of arithmetic allow safe manipulation of expressions.
 
Motivated by the interest in dealing with diagrammatic languages that enjoy such features, a growing number of works~\cite{BaezErbele-CategoriesInControl,DBLP:journals/pacmpl/BonchiHPSZ19,Bonchi2015,coecke2011interacting, Fong2015,DBLP:journals/corr/abs-2009-06836,Ghica2016,DBLP:conf/lics/MuroyaCG18,Piedeleu2021,DBLP:journals/jacm/BonchiGKSZ22}  exploits \emph{string diagrams}~\cite{joyal1991geometry,selinger2010survey}, a graphical notation that emerged in the field of category theory. Formally, string diagrams are arrows of a strict symmetric monoidal category freely generated by a \emph{monoidal signature}. A symbol $s$ in the signature is represented as a box
$\Cgen{\gen}{}{}$, and arbitrary string diagrams are depicted by composing horizontally ($;$) and vertically $(\per)$ such symbols (plus some wiring technology that we are going to ignore in this introduction). The following is our first example of a string diagram:
\[ 
    \InputIfFileExists{functoriality.tikz}{}{\input{./tikz/functoriality.tikz}}
 \]
Observe that it can be regarded as both $(f_1;g_1) \per (f_2;g_2)$ and $(f_1 \per f_2); (g_1 \per g_2)$. This ambiguity is not an issue, since in any monoidal category the law $(f_1;g_1) \per (f_2;g_2) = (f_1 \per f_2); (g_1 \per g_2)$ holds for all arrows $f_1,f_2,g_1,g_2$.
More generally, the main result in~\cite{joyal1991geometry} states that the diagrammatic representation identifies exactly \emph{all and only} the laws of strict monoidal categories. This is  the key feature of string diagrams.
Indeed, by virtue of this fact, one can safely exploit diagrams to make proofs, which in this way often amount to suggestive manipulations of diagrams.

By carefully crafting the monoidal signature, hereafter denoted as $\Sigma$, one obtains the syntaxes of several languages specifying a large variety of systems: quantum processes~\cite{Coecke2017}, linear dynamical systems~\cite{Bonchi2015}, Petri nets~\cite{DBLP:journals/pacmpl/BonchiHPSZ19}, concurrent connectors~\cite{Bruni2006}, digital circuits~\cite{Ghica2016}, automata~\cite{Piedeleu2021}, or conjunctive queries~\cite{GCQ}. In these approaches, the semantics are defined by monoidal functors
\begin{equation}\label{eq:functorSemantics}
\begin{tikzpicture}
	\begin{pgfonlayer}{nodelayer}
		\node [style=none] (0) at (-2.5, 0) {$\CatString\;\;$};
		\node [style=none] (1) at (2.5, 0) {$\;\;\Cat{D}$};
	\end{pgfonlayer}
	\begin{pgfonlayer}{edgelayer}
		\draw [->] (0) -- node[above] {\scriptsize $\dsem{\cdot}$} ++ (1);
	\end{pgfonlayer}
\end{tikzpicture}    
\end{equation}
going from the category of string diagrams $\CatString$ to some monoidal category $\Cat{D}$ representing the semantic domain. Since $\dsem{\cdot}$ is a monoidal functor, it preserves $;$ and $\per$, and thus the semantics is guaranteed to be \emph{compositional}. Typically, the languages come with a set of axioms, namely equalities or inequalities between string diagrams, that are sound with respect to the semantics interpretation. Interestingly, the \emph{same} algebraic structures seem to appear in many different contexts, e.g.\ commutative monoids and comonoids, Frobenius algebras, bialgebras, etc.

However, in several occasions the string diagrammatic syntax seems to be too restrictive. For instance, in the context of the ZX-calculus~\cite{Coecke2008,Coecke2017}, a well known quantum diagrammatic language,
several works~\cite{DBLP:journals/corr/abs-2103-07960,zhao2021analyzing,stollenwerk2022diagrammatic} make use of a mixture of diagrammatic and algebraic syntax to represent, for example,  addition of diagrams. Similarly, in~\cite{DBLP:conf/fossacs/BoisseauP22} $\sqcup$-props have been introduced in order to enrich string diagrams with a join operation. Sometimes, the structure of monoidal category is not enough and one needs to depict arrows  of \emph{rig categories}~\cite{laplaza_coherence_1972, johnson2021bimonoidal}, roughly categories equipped with two monoidal products: $\per$ and $\piu$. In these cases, the authors often introduce novel kinds of diagrams \cite{duncan2009generalised, james2012information, staton2015algebraic} to convey the intuition to the reader, but without a soundness and completeness theorem like in the aforementioned~\cite{joyal1991geometry} for string diagrams.  
The main challenge in depicting arrows of a rig category is given by the possibility of composing them not only with $;$ (horizontally) and  $\per$ (vertically), but also with the novel monoidal product $\piu$. The natural solution consists in exploiting 3 dimensions. This is the approach taken by \emph{sheet diagrams}~\cite{comfort2020sheet}, certain topological manifolds that, modulo a notion of isotopy, capture exactly the laws of rig categories.

In this paper, we  introduce \emph{tape diagrams} as a way to depict arrows not of  arbitrary rig categories but only of those where $\piu$ is a \emph{biproduct}~\cite{mac_lane_categories_1978, coecke2017two}. The payoff of this restriction in expressiveness is a better usability: tape diagrams are two dimensional pictures and for this reason they are, in our opinion, more intuitive and more easily drawable than three dimensional diagrams. A second important novelty is that we do not need to define ad-hoc topological structures and transformations, since tape diagrams are, literally, \emph{string diagrams of string diagrams}.

Our main result, Theorem~\ref{thm:Tapes is free sesquistrict generated by sigma}, is analogous to the one of~\cite{joyal1991geometry}: it states  that  the category of tape diagrams, hereafter referred to as $\CatTape$, is the rig category with finite biproducts freely generated by a \emph{monoidal} signature. 
Another result, Theorem~\ref{thm:equivalentsignature}, states that  for finite biproduct rig categories considering only monoidal signatures, rather than the more general \emph{rig} signatures, does not affect the expressivity, in the sense that for every rig signature $\Sigma$ one can find a monoidal signature $\Sigma_M$ such that the finite biproduct rig category generated by $\Sigma$ is isomorphic to the one generated by $\Sigma_M$. So, Theorems~\ref{thm:Tapes is free sesquistrict generated by sigma} and \ref{thm:equivalentsignature} together allow us to state that tape diagrams form a universal diagrammatic language for rig categories with finite biproducts (see Remark~\ref{rem:tapes universal language}).

A useful  consequence of Theorem~\ref{thm:Tapes is free sesquistrict generated by sigma} is Corollary~\ref{co:rig semantics}, which states that whenever the semantic domain $\Cat{D}$ of a string diagrammatic language as in~\eqref{eq:functorSemantics} carries the structure of a finite biproduct rig category, the semantics map $\dsem{\cdot}$ can be extended to tape diagrams as follows.
\[
\begin{tikzcd}
            \CatString \ar[r,"\dsem{\cdot}"] \ar[d,hook] & \Cat D \\
            \CatTape \ar[ur,"\dsem{\cdot}^\sharp"']
\end{tikzcd}
\]
In Example \ref{ex:zxcalculus}, we quickly show that applying the above construction to the ZX-calculus \cite{Coecke2008}, one can easily express through a tape diagram a quantum Controlled Unitary gate. In Example~\ref{ex:mu-props}, we show that, with the help of four ``adjointness'' axioms (in Figure~\ref{fig:rel axioms}), $\sqcup$-props from~\cite{DBLP:conf/fossacs/BoisseauP22} can be comfortably translated into tape diagrams so to obtain a purely graphical calculus that avoids the use of algebraic operators. 
Finally, by taping the calculus of graphical conjunctive queries from~\cite{GCQ}, one obtains a complete axiomatisation for the calculus of relations by Tarski~\cite{tarski1941calculus}. This is our main application, which we will illustrate in the next section.

\paragraph{Structure of the paper.} 
In Section \ref{sc:monoidal} we recall string diagrams and monoidal categories. In particular, we show that $\Rel$, the category of sets and relations, carries two  monoidal structures satisfying distinct algebraic properties: $(\Rel, \piu, \zero)$ is a finite biproduct (fb) category, while $(\Rel, \per, \uno)$ is a cartesian bicategory (cb). In Section~\ref{sc:rig} we recall rig categories and we introduce a novel (to the best of our knowledge) notion of strictness that is useful to simplify the presentation. We also illustrate rig signatures, finite biproduct rig categories and Theorem~\ref{thm:equivalentsignature}.

In Section~\ref{sc:tape} we introduce tape diagrams and we prove Theorem~\ref{thm:Tapes is free sesquistrict generated by sigma}. The key step of its proof consists in showing that tapes form a rig category (Theorem~\ref{thm:taperig}): this is carefully done by inductively defining left and right whiskerings (Definition~\ref{def:tape:whiskG}) that enjoy beautiful algebraic properties (Table~\ref{table:whisk}). Differently from $\piu$, the representation of the product $\otimes$ of two tapes involves some computations. However, a further result, Theorem~\ref{thm:contextual}, allows us to avoid this issue: in diagrammatic proofs one can safely forget about $\per$ of tapes.

In Section~\ref{sc:matrix} we investigate the matrix calculus for $\CatTape$ that is provided by its finite biproduct structure.
Corollary~\ref{thm:tapes as matrices} characterises tape diagrams as matrices of multisets of string diagrams. Interestingly enough $\piu$ of tapes corresponds to direct sum of matrices, while $\per$ to  their Kronecker product. Such correspondence is then extended to a poset enriched setting: the category of tapes resulting from the four aforementioned adjointness axioms is isomorphic to matrices of downsets of string diagrams (Theorem~\ref{thm:tapes as matrices poset version}). From this follows a characterisation of the induced poset (Corollary~\ref{cor:poset}) that is fundamental for the completeness result in Section~\ref{sec:CBPOPL}: Theorem~\ref{thm:completeness}. 

This result is analogous to Theorem 2.2 in~\cite{selinger2012finite} stating that finite dimensional Hilbert spaces are complete for dagger compact closed categories. Theorem~\ref{thm:completeness} shows that $\Rel$ is complete for 
\emph{fb-cb rig categories} (Definition~\ref{def:fbcbrig}), namely rig categories where $\per$ forms a cartesian bicategory and $\piu$ a finite biproduct category enjoying the aforementioned adjointness axioms.
From the completeness theorem, one can easily show (Corollary~\ref{crlFinal}) that the laws of fb-cb categories provide a sound and complete axiomatisation for the positive fragment of Tarski's calculus of relations.

This article comes with a series of appendixes. All the coherence conditions are in Appendix~\ref{app:coherence axioms}. Appendix~\ref{app:morefigures} contains some optional figures which may interest the curious reader. Indeed, apart from Figures~\ref{fig:tapesax},~\ref{fig:rel axioms} and~\ref{fig:cb axioms} which display all the tape axioms relevant to this paper, all the remaining figures are in Appendixes~\ref{app:coherence axioms} or \ref{app:morefigures}. Finally, every technical section comes with its own appendix providing additional results and detailed proofs (those in the main text are just sketches).

 \section{Leading example: the calculus of relations}\label{sc:leading example}

In order to provide a preliminary intuition about tape diagrams and, meanwhile, explain their main application investigated in this paper, we recall now
the positive fragment of the calculus of binary relations by Tarski~\cite{tarski1941calculus}.
Its syntax is specified by the following context free grammar
\begin{equation}\tag{$\CRS$} \label{eq:calculusofrelation}
\begin{array}{rcl}
E & ::=&\; R  \; \mid \; 1 \; \mid  \;  E ; E \; \mid \; \bot \; \mid \;  E \cup E  \; \mid \;  \top \; \mid \; E \cap E \; \mid \;  \op{E}  
\end{array}
\end{equation}    
where $R$ is taken from a given set $\Sigma$ of generating symbols. The expression $1$ denotes the identity relation, $;$  relational composition, $\op{\cdot}$ the opposite relation and the remaining expressions are the usual set-theoretic union $\cup$ and intersection $\cap$, together with their units $\bot$ and $\top$ being, respectively, the empty relation and the total relation. Formally, the semantics of $\CRS$ is defined w.r.t. a  \emph{relational interpretation} $\interpretation$, that is, a set $X$ together with a binary relation $\rho(R)\subseteq X \times X$ for each $R\in \Sigma$.
\begin{equation}\label{eq:sematicsExpr}
\setlength{\tabcolsep}{3pt}
        \begin{tabular}{rclrcl}
            $\dsemRel{R}$& \!\!=\!\! &$\rho(R)$ & $\dsemRel{ \op{E}} $&\!\!=\!\!&$ \{(y,x) \mid (x,y) \!\in\! \dsemRel{E}\}$ \\
            $\dsemRel{\bot} $&\!\!=\!\!&$ \{ \; \}$ &$\dsemRel{ E_1 \cup E_2} $&\!\!=\!\!&$ \dsemRel{ E_1} \cup \dsemRel{ E_2}$ \\
            $\dsemRel{\top} $&\!\!=\!\!&$ \{ (x,y) \mid x,y \!\in\! X \}$ &$\dsemRel{ E_1 \cap E_2} $&\!\!=\!\!&$ \dsemRel{ E_1} \cap \dsemRel{ E_2}$ \\
            $\dsemRel{1} $&\!\!=\!\!&$ \id{X} \!=\!\{(x,x) \mid x \!\in\! X\}$ & $\dsemRel{ E_1 ; E_2} $&\!\!=\!\!&$ \{ (x,z) \mid \exists y. \; (x,y) \!\in\! \dsemRel{ E_1} \wedge (y,z) \!\in\! \dsemRel{ E_2} \}$ \\
        \end{tabular}
\end{equation}
Two expressions $E_1$, $E_2$ are said to be \emph{equivalent}, written $E_1 \equiv_{\CR}E_2$, if and only if $\dsemRel{E_1} = \dsemRel{E_2}$, for all interpretations $\interpretation$. Inclusion, denoted by $\minorExpression$, is defined analogously by replacing $=$ with $\subseteq$. For instance, the following two laws assert that $;$ distributes over $\cup$ but only laxly over $\cap$.
\begin{equation}\label{eq:distributivityExpres}
R;(S\cup T) \equiv_{\CR}  (R;S)\cup(R;T) \qquad R;(S\cap T) \minorExpression  (R;S)\cap(R;T) 
\end{equation}
The question left open by Tarski is whether or not $\equiv_{\CR}$ can be axiomatised: is there a basic set of laws from which one can prove all the valid equivalences? Unfortunately, many negative results show that there are no finite complete axiomatisations for the whole calculus~\cite{monk}, for the positive fragment~\cite{hodkinson2000axiomatizability} and several other fragments, e.g.\ \cite{redko1964defining,freyd1990categories,doumane2020non}. See \cite{DBLP:conf/stacs/Pous18} for a recent overview.

In this paper we propose a solution to the same problem, but from a radically different perspective: we encode the calculus of relations into a novel calculus that is based on, in our opinion, more primitive linguistic constructions. Our language, named $\CatTapeCB$, is strictly more expressive than $\CRS$ but allows to obtain a complete axiomatisation of equivalence and inclusion.

The syntax of $\CatTapeCB$ is based on \emph{circuits} and \emph{tapes}. Circuits are obtained by composing horizontally and vertically the following set of basic gates, where $R$ is a symbol in $\sign$.
\[  \comonoid{} \qquad \counit{} \qquad \boxCirc{R} \qquad \unit{}  \qquad \monoid{} \]
To abbreviate, we will denote $\comonoid{}, \counit{}, \unit{}$ and $\monoid{}$ respectively as $\copier{}, \discharger{}, \codischarger{}$ and $\cocopier{}$.
Intuitively, $\copier{}$ acts as a \emph{copier}: it receives a signal (i.e.\ a value from some set $X$) on its left wire and sends it to \emph{both} its wires on the right. Instead, $\discharger{}$ is a \emph{discharger}: it throws away the signal coming from its only wire on the left. Formally, $\copier{}$ is the pairing function $\langle \id{X}, \id{X}\rangle$ going from $X$ to $X \times X$, namely the cartesian product of $X$ with itself, while $\discharger{}$ is the only function going from a set $X$ to a singleton set $\uno$. The gates $\cocopier{}$ and $\codischarger{}$ are interpreted as the opposite relations of $\copier{}$ and $\discharger{}$, respectively. Finally $\boxCirc{R}$ simply denotes an arbitrary binary relation $R$ on $X$.

It is worth emphasising here that  signal flows through circuits as a \emph{wave}, i.e.\ it passes through \emph{all} the vertical components at the same time. 
For instance, the circuit 
\begin{tikzpicture}
	\begin{pgfonlayer}{nodelayer}
		\node [style=none] (81) at (1.75, 0) {};
		\node [style=black] (107) at (1, 0) {};
		\node [style=none] (112) at (-1.75, 0) {};
		\node [style=black] (113) at (-1, 0) {};
		\node [style=box] (114) at (0, -0.625) {$S$};
		\node [style=box] (115) at (0, 0.625) {$R$};
	\end{pgfonlayer}
	\begin{pgfonlayer}{edgelayer}
		\draw (81.center) to (107);
		\draw [bend right] (115) to (113);
		\draw [bend right] (113) to (114);
		\draw (112.center) to (113);
		\draw [bend left] (115) to (107);
		\draw [bend left] (107) to (114);
	\end{pgfonlayer}
\end{tikzpicture}
denotes the relation $R \cap S$, i.e.\ the set of all pairs of signals $(x,y)$ such that, at the same time, $x\,R\,y$ and $x\,S\,y$.

Tapes are obtained by composing horizontally and vertically the following generators
\[ 
\Tcomonoid{} \qquad \Tcounit{} \qquad \Tcirc{c}{}{}  \qquad \Tunit{} \qquad \Tmonoid{}
\]
where the middle one represents a circuit $c$ wrapped inside a tape. To abbreviate, we will denote the first two generators on the left as $\diag{}$ and $\bang{}$ and the last two on the right as $\cobang{}$ and $\codiag{}$.
Intuitively, $\codiag{}$ lets pass to the right the signal coming from either the top \emph{or} the bottom branch on the left, while $\cobang{}$ simply closes a branch. Formally, $\codiag{}$ is the coparing function $\left[ \id X, \id X \right]$ going from $X + X$, i.e.\ the disjoint union of $X$ with itself, to $X$, while $\cobang{}$ is the only function going from the the empty set $\zero$ to $X$. The generators $\diag{}$ and $\bang{}$ are interpreted as the opposite relations of $\codiag{}$ and $\cobang{}$, respectively.

Differently from circuits, a signal flows through a tape as a \emph{particle}, i.e.\ it passes through \emph{only one} of the vertical components at a time. For instance, the tape in~\eqref{eq:union bottom tapes} denotes the relation 
$R \cup S$, i.e.\ the set of all pairs $(x,y)$ such that either $x\,R\,y$ or $x\,S\,y$.

\begin{minipage}{0.5\textwidth}
    \begin{equation}\label{eq:union bottom tapes}
        \begin{tikzpicture}
            \begin{pgfonlayer}{nodelayer}
                \node [style=none] (0) at (-1.5, 0) {};
                \node [style=none] (1) at (-0.25, 1.25) {};
                \node [style=none] (2) at (-0.25, -1.25) {};
                \node [style=none] (8) at (-0.5, -1.9) {};
                \node [style=none] (9) at (-0.5, 1.9) {};
                \node [style=none] (10) at (-0.25, 0.6) {};
                \node [style=none] (11) at (-1, 0) {};
                \node [style=none] (12) at (-0.25, -0.6) {};
                \node [style=none] (13) at (-2, -0.65) {};
                \node [style=none] (14) at (-2, 0.65) {};
                \node [style=none] (24) at (-3, 0) {};
                \node [style=none] (35) at (-3, -0.65) {};
                \node [style=none] (36) at (-3, 0.65) {};
                \node [style=none] (50) at (-0.25, 1.25) {};
                \node [style=none] (80) at (1.5, 0) {};
                \node [style=none] (81) at (0.25, 1.25) {};
                \node [style=none] (82) at (0.25, -1.25) {};
                \node [style=none] (83) at (0.5, -1.9) {};
                \node [style=none] (84) at (0.5, 1.9) {};
                \node [style=none] (85) at (0.25, 0.6) {};
                \node [style=none] (86) at (1, 0) {};
                \node [style=none] (87) at (0.25, -0.6) {};
                \node [style=none] (88) at (2, -0.65) {};
                \node [style=none] (89) at (2, 0.575) {};
                \node [style=none] (90) at (3, 0) {};
                \node [style=none] (91) at (3, -0.65) {};
                \node [style=none] (92) at (3, 0.575) {};
                \node [style=none] (93) at (0.25, 1.25) {};
                \node [style=box, scale=0.9] (96) at (0, 1.25) {$R$};
                \node [style=box, scale=0.9] (97) at (0, -1.25) {$S$};
            \end{pgfonlayer}
            \begin{pgfonlayer}{edgelayer}
                \draw [style=tape] (83.center)
                     to (8.center)
                     to [bend left] (13.center)
                     to (35.center)
                     to (36.center)
                     to (14.center)
                     to [bend left] (9.center)
                     to (84.center)
                     to [bend left] (89.center)
                     to (92.center)
                     to (91.center)
                     to (88.center)
                     to [bend left] cycle;
                \draw [style=tapeNoFill, fill=white] (87.center)
                     to (12.center)
                     to [bend left=45] (11.center)
                     to [bend left=45] (10.center)
                     to (85.center)
                     to [bend left=45] (86.center)
                     to [bend left=45] cycle;
                \draw [bend right=45] (0.center) to (2.center);
                \draw [bend left=45] (0.center) to (1.center);
                \draw (24.center) to (0.center);
                \draw [bend left=45] (80.center) to (82.center);
                \draw [bend right=45] (80.center) to (81.center);
                \draw (90.center) to (80.center);
                \draw (50.center) to (96);
                \draw (93.center) to (96);
                \draw (2.center) to (97);
                \draw (82.center) to (97);
            \end{pgfonlayer}
        \end{tikzpicture}        
    \end{equation}
\end{minipage}
\hfill
\begin{minipage}{0.4\textwidth}
    \begin{equation}\label{eq:R or (S and T)}
        
    \InputIfFileExists{relations/RoSaT.tikz}{}{\input{./tikz/relations/RoSaT.tikz}}

    \end{equation}
\end{minipage}

\begin{figure}[p]
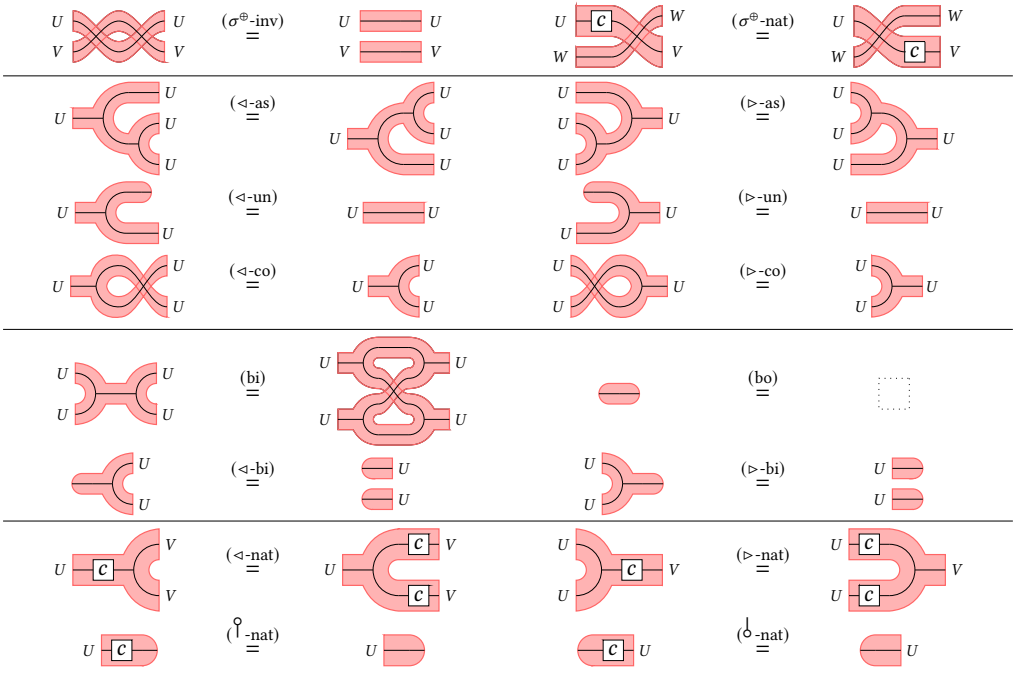

    \mylabel{ax:symmpinv}{$\symmp$-inv}
    \mylabel{ax:symmpnat}{$\symmp$-nat}
    \mylabel{ax:diagas}{$\diag{}$-as}
    \mylabel{ax:diagun}{$\diag{}$-un}
    \mylabel{ax:diagco}{$\diag{}$-co}
    \mylabel{ax:codiagas}{$\codiag{}$-as}
    \mylabel{ax:codiagun}{$\codiag{}$-un}
    \mylabel{ax:codiagco}{$\codiag{}$-co}
    \mylabel{ax:bi}{bi}
    \mylabel{ax:bo}{bo}
    \mylabel{ax:diagbi}{$\diag{}$-bi}
    \mylabel{ax:codiagbi}{$\codiag{}$-bi}
    \mylabel{ax:diagnat}{$\diag{}$-nat}
    \mylabel{ax:bangnat}{$\bang{}$-nat}
    \mylabel{ax:codiagnat}{$\codiag{}$-nat}
    \mylabel{ax:cobangnat}{$\cobang{}$-nat}
    \scalebox{0.9}{
    \begin{tabular}{C{1px} C{2.3cm} c C{2.5cm} C{5px} C{2.5cm} c C{2.3cm} C{1px}}
        &
    \InputIfFileExists{tapes/ax/symminv_left.tikz}{}{\input{./tikz/tapes/ax/symminv_left.tikz}}
 &$\stackrel{(\symmp\text{-inv})}{=}$& 
    \InputIfFileExists{tapes/ax/symminv_right.tikz}{}{\input{./tikz/tapes/ax/symminv_right.tikz}}
  && 
    \InputIfFileExists{tapes/ax/symmnat_left.tikz}{}{\input{./tikz/tapes/ax/symmnat_left.tikz}}
  &$\stackrel{(\symmp\text{-nat})}{=}$&  
    \InputIfFileExists{tapes/ax/symmnat_right.tikz}{}{\input{./tikz/tapes/ax/symmnat_right.tikz}}
 & \\[-1ex]
\hline
&
    \InputIfFileExists{tapes/whiskered_ax/comonoid_assoc_left.tikz}{}{\input{./tikz/tapes/whiskered_ax/comonoid_assoc_left.tikz}}
  &$\stackrel{(\diag{}\text{-as})}{=}$&  
    \InputIfFileExists{tapes/whiskered_ax/comonoid_assoc_right.tikz}{}{\input{./tikz/tapes/whiskered_ax/comonoid_assoc_right.tikz}}
 && 
    \InputIfFileExists{tapes/whiskered_ax/monoid_assoc_left.tikz}{}{\input{./tikz/tapes/whiskered_ax/monoid_assoc_left.tikz}}
 &$\stackrel{(\codiag{}\text{-as})}{=}$& 
    \InputIfFileExists{tapes/whiskered_ax/monoid_assoc_right.tikz}{}{\input{./tikz/tapes/whiskered_ax/monoid_assoc_right.tikz}}
 & \\
        &
    \InputIfFileExists{tapes/whiskered_ax/comonoid_unit_left.tikz}{}{\input{./tikz/tapes/whiskered_ax/comonoid_unit_left.tikz}}
 &$\stackrel{(\diag{}\text{-un})}{=}$& \Twire{U} && 
    \InputIfFileExists{tapes/whiskered_ax/monoid_unit_left.tikz}{}{\input{./tikz/tapes/whiskered_ax/monoid_unit_left.tikz}}
 &$\stackrel{(\codiag{}\text{-un})}{=}$& \Twire{U} & \\
        &
    \InputIfFileExists{tapes/whiskered_ax/comonoid_comm_left.tikz}{}{\input{./tikz/tapes/whiskered_ax/comonoid_comm_left.tikz}}
 &$\stackrel{(\diag{}\text{-co})}{=}$& \Tcomonoid{U} && 
    \InputIfFileExists{tapes/whiskered_ax/monoid_comm_left.tikz}{}{\input{./tikz/tapes/whiskered_ax/monoid_comm_left.tikz}}
 &$\stackrel{(\codiag{}\text{-co})}{=}$& \Tmonoid{U} &\\[-0.5ex]
\hline
        \\[-2.2ex]
        &
    \InputIfFileExists{tapes/whiskered_ax/bialg1_left.tikz}{}{\input{./tikz/tapes/whiskered_ax/bialg1_left.tikz}}
 &$\stackrel{(\text{bi})}{=}$& 
    \InputIfFileExists{tapes/whiskered_ax/bialg1_right.tikz}{}{\input{./tikz/tapes/whiskered_ax/bialg1_right.tikz}}
 && 
    \InputIfFileExists{tapes/whiskered_ax/bialg2_left.tikz}{}{\input{./tikz/tapes/whiskered_ax/bialg2_left.tikz}}
  &$\stackrel{(\text{bo})}{=}$&  
    \InputIfFileExists{empty.tikz}{}{\input{./tikz/empty.tikz}}
 & \\[-1ex]
        &
    \InputIfFileExists{tapes/whiskered_ax/bialg3_left.tikz}{}{\input{./tikz/tapes/whiskered_ax/bialg3_left.tikz}}
  &$\stackrel{(\diag{}\text{-bi})}{=}$&  
    \InputIfFileExists{tapes/whiskered_ax/bialg3_right.tikz}{}{\input{./tikz/tapes/whiskered_ax/bialg3_right.tikz}}
 && 
    \InputIfFileExists{tapes/whiskered_ax/bialg4_left.tikz}{}{\input{./tikz/tapes/whiskered_ax/bialg4_left.tikz}}
  &$\stackrel{(\codiag{}\text{-bi})}{=}$&  
    \InputIfFileExists{tapes/whiskered_ax/bialg4_right.tikz}{}{\input{./tikz/tapes/whiskered_ax/bialg4_right.tikz}}
 & \\
\hline
        \\[-2.2ex]
        &
    \InputIfFileExists{tapes/ax/comonoidnat_left.tikz}{}{\input{./tikz/tapes/ax/comonoidnat_left.tikz}}
 &$\stackrel{(\diag{}\text{-nat})}{=}$& 
    \InputIfFileExists{tapes/ax/comonoidnat_right.tikz}{}{\input{./tikz/tapes/ax/comonoidnat_right.tikz}}
 && 
    \InputIfFileExists{tapes/ax/monoidnat_left.tikz}{}{\input{./tikz/tapes/ax/monoidnat_left.tikz}}
 &$\stackrel{(\codiag{}\text{-nat})}{=}$& 
    \InputIfFileExists{tapes/ax/monoidnat_right.tikz}{}{\input{./tikz/tapes/ax/monoidnat_right.tikz}}
 & \\[-1ex]
        &
    \InputIfFileExists{tapes/ax/counitnat_left.tikz}{}{\input{./tikz/tapes/ax/counitnat_left.tikz}}
  &$\stackrel{(\bang{}\text{-nat})}{=}$& 
    \InputIfFileExists{tapes/ax/counitnat_right.tikz}{}{\input{./tikz/tapes/ax/counitnat_right.tikz}}
 && 
    \InputIfFileExists{tapes/ax/unitnat_left.tikz}{}{\input{./tikz/tapes/ax/unitnat_left.tikz}}
  &$\stackrel{(\cobang{}\text{-nat})}{=}$& 
    \InputIfFileExists{tapes/ax/unitnat_right.tikz}{}{\input{./tikz/tapes/ax/unitnat_right.tikz}}
 &
    \end{tabular}}
    \caption{Axioms for tape diagrams}
    \label{fig:tapesax}
\end{figure}

Since circuits can be nested inside tapes, an expression such as $R \cup (S \cap T)$ can be represented as the tape in~\eqref{eq:R or (S and T)}. A formal semantics of $\CatTapeCB$, as well as an encoding of $\CRS$ into   $\CatTapeCB$, will be given in Section~\ref{sec:CBPOPL}. Moreover,  Theorem~\ref{thm:completeness} will state that  the axioms in Figures~\ref{fig:tapesax},~\ref{fig:rel axioms} and~\ref{fig:cb axioms} are complete. Figure~\ref{fig:tapesax} features the axioms for ``plain'' tape diagrams and Figure~\ref{fig:rel axioms} for their partial order enrichment (discussed in Sections~\ref{sec:preorderenrichment} and~\ref{sec:CBPOPL}); Figure~\ref{fig:cb axioms} lists axioms to be used specifically for $\CatTapeCB$.

Our axiomatisation is far from those proposed in more traditional approaches to $\CR$, but it elegantly features some well-known algebraic structures that occur frequently in various fields \cite{lafont2003towards,Bruni2006,coecke2011interacting,BaezErbele-CategoriesInControl,Bonchi2015,DBLP:journals/pacmpl/BonchiHPSZ19,BonchiPSZ19,Fritz_stochasticmatrices,DBLP:journals/jacm/BonchiGKSZ22}. The second group of axioms in Figures~\ref{fig:tapesax} and~\ref{fig:cb axioms} state that $(\diag{}, \bang{})$ and $(\copier{}, \discharger{})$ are cocommutative comonoids while $(\codiag{}, \cobang{})$ and $(\cocopier{}, \codischarger{})$ are commutative monoids. The third group expresses the facts that $(\diag{}, \bang{}, \codiag{}, \cobang{})$ form a bialgebra and $(\copier{}, \discharger{}, \cocopier{}, \codischarger{})$ a special Frobenius bimonoid.
The axioms in Figure~\ref{fig:rel axioms} assert that the monoid $(\codiag{}, \cobang{})$ is \emph{left} adjoint to the comonoid $(\diag{}, \bang{})$, while  those in the last group in Figure~\ref{fig:cb axioms} state the the monoid $(\cocopier{}, \codischarger{})$ is \emph{right} adjoint to the monoid $(\copier{}, \discharger{})$. 

The formal justification to these axioms will be clarified through the paper but it is worth establishing now a preliminary intuition: consider the axioms \eqref{ax:relCodiagDiag} and \eqref{ax:copieradj1}. In the leftmost tape of \eqref{ax:relCodiagDiag}, the signal flows either in the top branch or in the bottom one, while in the rightmost tape, the signal coming from any of the two branches on the left may go through any of the branches on the right. In the two tapes of \eqref{ax:copieradj1} the signal flows, at the same time, through the top and bottom wires: in the leftmost such signals must be equal, in the rightmost they may be different.

The remaining axioms are those in the first and fourth groups in Figures~\ref{fig:tapesax} and~\ref{fig:cb axioms}. The laws in the first group assert that crossings of tapes and wires behave like symmetries. The axioms in 
the fourth group force naturality for $\diag{}, \bang{}, \codiag{}$ and $\cobang{}$, while for $\copier{}$ and $\discharger{}$ naturality only holds laxly (and from these one can easily derive that $\cocopier{}$ and $\codischarger{}$ are colax natural). With these naturality axioms one can immediately derive the laws in \eqref{eq:distributivityExpres} as follows:
\[

    \InputIfFileExists{relations/distrU1.tikz}{}{\input{./tikz/relations/distrU1.tikz}}
 \axeq{\refeq{ax:diagnat}} 
    \InputIfFileExists{relations/distrU2.tikz}{}{\input{./tikz/relations/distrU2.tikz}}

\qquad  \qquad 
 
    \InputIfFileExists{relations/distrA1.tikz}{}{\input{./tikz/relations/distrA1.tikz}}
 \axsubeq{\refeq{ax:copiernat}} 
    \InputIfFileExists{relations/distrA2.tikz}{}{\input{./tikz/relations/distrA2.tikz}}
 \]
By means of simple graphical manipulations prescribed by the laws in Figures~\ref{fig:tapesax},~\ref{fig:rel axioms} and~\ref{fig:cb axioms}, one can prove all the valid equivalences in $\CR$. The curious reader may have a look to Figure~\ref{fig:R or R and S} illustrating a graphical derivation for $R\cup (R\cap S) \equiv_{\CR} R$. We conclude this preliminary section by remarking that the discussed axiomatisation has some redundancies, e.g.\  \eqref{ax:relCobangBang} evidently follows from \eqref{ax:bo}. We have however chosen this presentation to emphasise the various elegant dualities.

\begin{figure}[p]
    \mylabel{ax:relCobangBang}{$\cobang{}\bang{}$}
    \mylabel{ax:relBangCobang}{$\bang{}\cobang{}$}
    \mylabel{ax:relCodiagDiag}{$\codiag{}\diag{}$}
    \mylabel{ax:relDiagCodiag}{$\diag{}\codiag{}$}
    \scalebox{0.9}{
    \begin{tabular}{C{1px} C{2.5cm} c C{2.5cm} C{5px} C{3cm} c C{2.5cm} C{1px}}
        &
    \InputIfFileExists{empty.tikz}{}{\input{./tikz/empty.tikz}}
 & $\axsubeq{\cobang{}\bang{}}$ & 
    \InputIfFileExists{tapes/whiskered_ax/bialg2_left.tikz}{}{\input{./tikz/tapes/whiskered_ax/bialg2_left.tikz}}
  && \Tcounit{U}\Tunit{U} & $\axsubeq{\bang{}\cobang{}}$ & \Twire{U} & \\[2ex]
        &$\begin{aligned}\begin{gathered}\Twire{U} \\ \Twire{U} \end{gathered}\end{aligned}$ & $\axsubeq{\codiag{}\diag{}}$ & 
    \InputIfFileExists{tapes/whiskered_ax/bialg1_left.tikz}{}{\input{./tikz/tapes/whiskered_ax/bialg1_left.tikz}}
 && \TRelCodiagDiag{U} & $\axsubeq{\diag{}\codiag{}}$ & \Twire{U} &
    \end{tabular}
    }
    \caption{Axioms for  $\cobang{}  \dashv \bang{}$ and $\codiag{} \dashv \diag{}$}
    \label{fig:rel axioms}
\end{figure}
\begin{figure}[p]
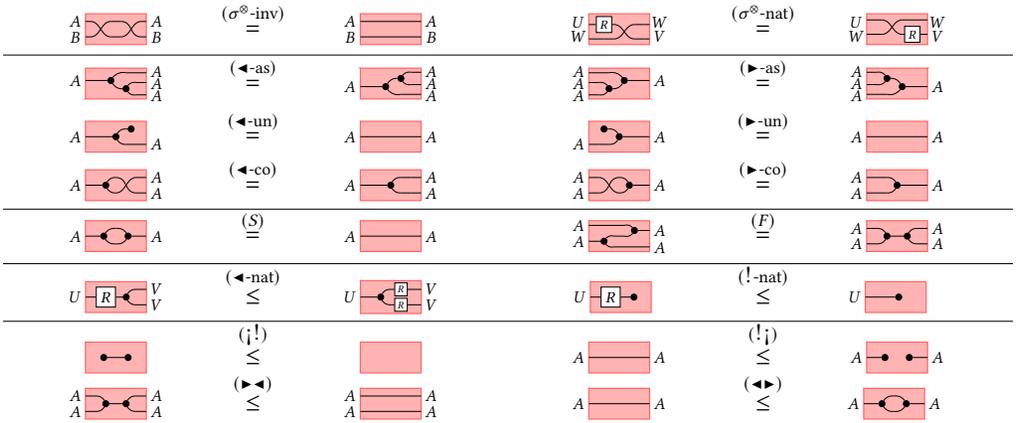

    \mylabel{ax:symmtinv}{$\sigma^{\per}$-inv}
    \mylabel{ax:symmtnat}{$\sigma^{\per}$-nat}
    \mylabel{ax:copieras}{$\copier{}$-as}
    \mylabel{ax:copierun}{$\copier{}$-un}
    \mylabel{ax:copierco}{$\copier{}$-co}
    \mylabel{ax:cocopieras}{$\cocopier{}$-as}
    \mylabel{ax:cocopierun}{$\cocopier{}$-un}
    \mylabel{ax:cocopierco}{$\cocopier{}$-co}
    \mylabel{ax:specfrob}{S}
    \mylabel{ax:frob}{F}
    \mylabel{ax:copiernat}{$\copier{}$-nat}
    \mylabel{ax:dischargernat}{$\discharger{}$-nat}
    \mylabel{ax:dischargeradj1}{$\codischarger{}\discharger{}$}
    \mylabel{ax:dischargeradj2}{$\discharger{}\codischarger{}$}
    \mylabel{ax:copieradj1}{$\cocopier{}\copier{}$}
    \mylabel{ax:copieradj2}{$\copier{}\cocopier{}$}
    \scalebox{0.9}{
    \begin{tabular}{C{1px} C{2.3cm} c C{2.5cm} C{5px} C{2.5cm} c C{2.3cm} C{1px}}
        &
    \InputIfFileExists{cb/symm_inv_left.tikz}{}{\input{./tikz/cb/symm_inv_left.tikz}}
 & $\axeq{\sigma^{\per}\text{-inv}}$ & 
    \InputIfFileExists{cb/symm_inv_right.tikz}{}{\input{./tikz/cb/symm_inv_right.tikz}}
 && 
    \InputIfFileExists{cb/symm_nat_left.tikz}{}{\input{./tikz/cb/symm_nat_left.tikz}}
 & $\axeq{\sigma^{\per}\text{-nat}}$ & 
    \InputIfFileExists{cb/symm_nat_right.tikz}{}{\input{./tikz/cb/symm_nat_right.tikz}}
& \\
        \hline
        &
    \InputIfFileExists{cb/monoid_assoc_left.tikz}{}{\input{./tikz/cb/monoid_assoc_left.tikz}}
 & $\axeq{\copier{}\text{-as}}$ & 
    \InputIfFileExists{cb/monoid_assoc_right.tikz}{}{\input{./tikz/cb/monoid_assoc_right.tikz}}
 && 
    \InputIfFileExists{cb/comonoid_assoc_left.tikz}{}{\input{./tikz/cb/comonoid_assoc_left.tikz}}
 & $\axeq{\cocopier{}\text{-as}}$ & 
    \InputIfFileExists{cb/comonoid_assoc_right.tikz}{}{\input{./tikz/cb/comonoid_assoc_right.tikz}}
& \\
        &
    \InputIfFileExists{cb/monoid_unit_left.tikz}{}{\input{./tikz/cb/monoid_unit_left.tikz}}
 & $\axeq{\copier{}\text{-un}}$ & 
    \InputIfFileExists{cb/monoid_unit_right.tikz}{}{\input{./tikz/cb/monoid_unit_right.tikz}}
 && 
    \InputIfFileExists{cb/comonoid_unit_left.tikz}{}{\input{./tikz/cb/comonoid_unit_left.tikz}}
 & $\axeq{\cocopier{}\text{-un}}$ & 
    \InputIfFileExists{cb/comonoid_unit_right.tikz}{}{\input{./tikz/cb/comonoid_unit_right.tikz}}
& \\
        &
    \InputIfFileExists{cb/monoid_comm_left.tikz}{}{\input{./tikz/cb/monoid_comm_left.tikz}}
 & $\axeq{\copier{}\text{-co}}$ & 
    \InputIfFileExists{cb/monoid_comm_right.tikz}{}{\input{./tikz/cb/monoid_comm_right.tikz}}
 && 
    \InputIfFileExists{cb/comonoid_comm_left.tikz}{}{\input{./tikz/cb/comonoid_comm_left.tikz}}
 & $\axeq{\cocopier{}\text{-co}}$ & 
    \InputIfFileExists{cb/comonoid_comm_right.tikz}{}{\input{./tikz/cb/comonoid_comm_right.tikz}}
& \\
        \hline
        &
    \InputIfFileExists{cb/spec_frob_left.tikz}{}{\input{./tikz/cb/spec_frob_left.tikz}}
 & $\axeq{S}$ & 
    \InputIfFileExists{cb/monoid_unit_right.tikz}{}{\input{./tikz/cb/monoid_unit_right.tikz}}
 && 
    \InputIfFileExists{cb/frob_left.tikz}{}{\input{./tikz/cb/frob_left.tikz}}
 & $\axeq{F}$ & 
    \InputIfFileExists{cb/frob_center.tikz}{}{\input{./tikz/cb/frob_center.tikz}}
& \\
        \hline
        &
    \InputIfFileExists{cb/copier_nat_left.tikz}{}{\input{./tikz/cb/copier_nat_left.tikz}}
 & $\axsubeq{\copier{}\text{-nat}}$ & 
    \InputIfFileExists{cb/copier_nat_right.tikz}{}{\input{./tikz/cb/copier_nat_right.tikz}}
 && 
    \InputIfFileExists{cb/discharger_nat_left.tikz}{}{\input{./tikz/cb/discharger_nat_left.tikz}}
 & $\axsubeq{\discharger{}\text{-nat}}$ & 
    \InputIfFileExists{cb/discharger_nat_right.tikz}{}{\input{./tikz/cb/discharger_nat_right.tikz}}
 &\\
        \hline
        &
    \InputIfFileExists{cb/adjoint1_bangs_left.tikz}{}{\input{./tikz/cb/adjoint1_bangs_left.tikz}}
 & $\axsubeq{\codischarger{}\discharger{}}$ & 
    \InputIfFileExists{cb/empty.tikz}{}{\input{./tikz/cb/empty.tikz}}
 && 
    \InputIfFileExists{cb/monoid_unit_right.tikz}{}{\input{./tikz/cb/monoid_unit_right.tikz}}
 & $\axsubeq{\discharger{}\codischarger{}}$ & 
    \InputIfFileExists{cb/adjoint2_bangs_right.tikz}{}{\input{./tikz/cb/adjoint2_bangs_right.tikz}}
& \\
        &
    \InputIfFileExists{cb/frob_center.tikz}{}{\input{./tikz/cb/frob_center.tikz}}
 & $\axsubeq{\cocopier{}\copier{}}$ & 
    \InputIfFileExists{cb/adjoint1_diags_right.tikz}{}{\input{./tikz/cb/adjoint1_diags_right.tikz}}
 && 
    \InputIfFileExists{cb/monoid_unit_right.tikz}{}{\input{./tikz/cb/monoid_unit_right.tikz}}
 & $\axsubeq{\copier{}\cocopier{}}$ & 
    \InputIfFileExists{cb/spec_frob_left.tikz}{}{\input{./tikz/cb/spec_frob_left.tikz}}
 &
    \end{tabular}
    }
    \caption{Axioms of cartesian bicategories}
    \label{fig:cb axioms}
\end{figure}

\section{String diagrams and monoidal categories}\label{sc:monoidal}

We begin our exposition by regarding string diagrams \cite{joyal1991geometry,selinger2010survey} as terms of a typed language. Given a set $\sort$ of basic \emph{sorts}, hereafter denoted by $A,B\dots$, types are elements of $\sort^\star$, i.e.\ words over $\sort$. Terms are defined by the following context free grammar
\begin{equation}\label{eq:syntaxsymmetricstrict}
\begin{array}{rcl}
f & ::=& \; \id{A} \; \mid \; \id{I} \; \mid \; \gen \; \mid \; \sigma_{A,B}^{\perG} \; \mid \;   f ; f   \; \mid \;  f \perG f \\
\end{array}
\end{equation}  
where $s$ belongs to a fixed set $\sign$ of \emph{generators} and $I$ is the empty word. Each $s\in \sign$ comes with two types: arity $\ar(s)$ and coarity $\coar(s)$. The tuple $(\sort, \sign, \ar, \coar)$, $\sign$ for short, forms a \emph{monoidal signature}. Amongst the terms generated by  \eqref{eq:syntaxsymmetricstrict} we consider only those that can be typed according to the inference rules in Table \ref{fig:freestricmmoncatax}. String diagrams are such terms modulo the axioms in Table \ref{fig:freestricmmoncatax} where,  for any $X,Y\in \sort^\star$,  $\id{X}$ and $\sigma_{X,Y}^{\perG}$ can be easily built using $\id I$, $\id{A}$, $\sigma_{A,B}^{\perG}$, $\perG$ and $;$ (see e.g.~\cite{ZanasiThesis}).
\begin{table}[t]
\tiny{
\begin{center}
\begin{tabular}{c  c}
\begin{tabular}{c}
    \toprule
Objects ($A\in \sort$)\\
\midrule
$X \; ::=\; \; A \; \mid \; \unoG \; \mid \;  X \perG X \vphantom{\sigma_{A,B}^{\perG}}$\\
\midrule
\makecell{
    \\[-2pt] $(X\perG Y)\perG Z=X \perG (Y \perG Z)$ \\ $X \perG \unoG = X $ \\ $\unoG \perG X = X$ \\[2pt]
} \\[13pt]
\bottomrule
\end{tabular}
&
\begin{tabular}{cc}
\toprule
\multicolumn{2}{c}{Arrows ($A\in \sort$, $s\in \sign$)} \\
\midrule
\multicolumn{2}{c}{$f \; ::=\; \; \id{A} \; \mid \; \id{\unoG} \; \mid \; \gen  \; \mid \;   f ; f   \; \mid \;  f \perG f \; \mid \; \sigma_{A,B}^{\perG}$} \\
\midrule
$(f;g);h=f;(g;h)$ & $id_X;f=f=f;id_Y$\\

\multicolumn{2}{c}{$(f_1\perG f_2) ; (g_1 \perG g_2) = (f_1;g_1) \perG (f_2;g_2)$} \\

$id_{\unoG}\perG f = f = f \perG id_{\unoG}$ & $(f \perG g)\, \perG h = f \perG \,(g \perG h)$ \\

$\sigma_{A, B}^{\perG}; \sigma_{B, A}^{\perG}= id_{A \perG B}$ & $(\gen \perG id_Z) ; \sigma_{Y, Z}^{\perG} = \sigma_{X,Z}^{\perG} ; (id_Z \perG \gen)$ \\
\bottomrule
\end{tabular}
\end{tabular}

\vspace{1em}

\begin{tabular}{c}
    \toprule
    Typing rules \\
    \midrule
    $
    {id_A \colon A \!\to\! A} \qquad  {id_\unoG \colon \unoG \!\to\! \unoG} \qquad {\sigma_{A, B}^{\perG} \colon A \perG B \!\to\! B \perG A} \qquad 
        \inferrule{\gen \colon \ar(s) \!\to\! \coar(s) \in \sign}{\gen \colon \ar(s) \!\to\! \coar(s)} \qquad
        \inferrule{f \colon X_1 \!\to\! Y_1 \and g \colon X_2 \!\to\! Y_2}{f \perG g \colon X_1 \perG X_2 \!\to\! Y_1 \perG Y_2}  \qquad
        \inferrule{f \colon X \!\to\! Y \and g \colon Y \!\to\! Z}{f ; g \colon X \!\to\! Z}
    $\\    
\bottomrule  
\end{tabular}
\end{center}
 \caption{Axioms for $\CatString$}
    \label{fig:freestricmmoncatax}
    }
\end{table}

String diagrams enjoy an elegant graphical representation: a generator $\gen$ in $\sign$ with arity $X$ and coarity $Y$ is depicted as  a \emph{box} having \emph{labelled wires} on the left and on the right representing, respectively, the words $X$ and $Y$. For instance $\gen \colon AB \to C$ in $\sign$ is depicted as the leftmost diagram below. Moreover, $\id{A}$ is displayed as one wire,  $id_{\unoG} $ as the empty diagram and $\sigma_{A,B}^{\perG}$ as a crossing:
\[
    \InputIfFileExists{generator.tikz}{}{\input{./tikz/generator.tikz}}
 \qquad \qquad 
    \InputIfFileExists{id.tikz}{}{\input{./tikz/id.tikz}}
 \qquad  \qquad     
    \InputIfFileExists{empty.tikz}{}{\input{./tikz/empty.tikz}}
 \qquad  \qquad   
    \InputIfFileExists{symm.tikz}{}{\input{./tikz/symm.tikz}}
\]
Finally, composition $f;g$ is represented by connecting the right wires 
of $f$ with the left wires of $g$ when their labels match, 
while the monoidal product $f \perG g$ is depicted by stacking the corresponding 
diagrams on top of each other: \[
    \InputIfFileExists{seq_comp.tikz}{}{\input{./tikz/seq_comp.tikz}}
 \qquad \qquad \qquad  
    \InputIfFileExists{par_comp.tikz}{}{\input{./tikz/par_comp.tikz}}
 \]
The first three rows of axioms for arrows in Table~\ref{fig:freestricmmoncatax}
are implicit in the 
graphical representation while the axioms in the last row  are displayed as 
\[ 
    \InputIfFileExists{stringdiag_ax1_left.tikz}{}{\input{./tikz/stringdiag_ax1_left.tikz}}
 = 
    \InputIfFileExists{stringdiag_ax1_right.tikz}{}{\input{./tikz/stringdiag_ax1_right.tikz}}
 \quad\qquad 
    \InputIfFileExists{stringdiag_ax2_left.tikz}{}{\input{./tikz/stringdiag_ax2_left.tikz}}
 = 
    \InputIfFileExists{stringdiag_ax2_right.tikz}{}{\input{./tikz/stringdiag_ax2_right.tikz}}
 \]

Hereafter, we call  $\CatString$ the category having as objects words in $\sort^\star$ and as arrows string diagrams. Theorem 2.3 in~\cite{joyal1991geometry} states that  $\Cat{C}_\sign$ is a \emph{symmetric strict monoidal category freely generated} by $\sign$.
\begin{definition} A \emph{symmetric monoidal category}  consists 
of a category $\Cat{C}$, a bifunctor $\perG \colon \Cat{C} \times \Cat{C} \to \Cat{C}$,
an object $\unoG$  and natural isomorphisms \[ \alpha_{X, Y, Z} \colon (X \perG Y) \perG Z \to X \perG (Y \perG Z) \qquad \lambda_X \colon \unoG \perG X \to X \qquad \rho_X \colon X \perG \unoG \to X \qquad \sigma_{X, Y}^{\perG} \colon X \perG Y \to Y \perG X \]
satisfying some coherence axioms (in Figures~\ref{fig:moncatax} and~\ref{fig:symmmoncatax}).
A monoidal category is said to be \emph{strict} when $\alpha$, $\lambda$ and $\rho$ are all identity natural isomorphisms. A \emph{strict symmetric monoidal functor} is a functor $F \colon \Cat{C} \to \Cat{D}$  preserving $\perG$, $\unoG$ and $\sigma^{\perG}$.
\end{definition}
\begin{remark}\label{rmk:symstrict}
In \emph{strict}  symmetric monoidal (ssm) categories  the symmetry $\sigma$ is not forced to be the identity, since this would raise some problems: for instance, $(f_1;g_1) \perG (f_2;g_2) = (f_1;g_2) \perG (f_2;g_1)$ for all $f_1,f_2\colon A \to B$ and $g_1,g_2\colon B \to C$. As we will see in  Section~\ref{sc:rig}, this fact will make the issue of strictness for rig categories rather subtle.
\end{remark}

To illustrate in which sense $\Cat{C}_\sign$ is freely generated, it is convenient to introduce \emph{interpretations} in a fashion similar to~\cite{selinger2010survey}: an interpretation $\interpretation$ of $\sign$ into an ssm category $\Cat{D}$ consists of two functions $\alpha_{\sort} \colon \sort \to Ob(\Cat{D})$ and $\alpha_{\sign}\colon \sign \to Ar(\Cat{D})$ such that, for all $s\in \sign$, $\alpha_{\sign}(s)$ is an arrow having as domain $\alpha_{\sort}^\sharp(\ar(s))$ and codomain $\alpha_{\sort}^\sharp(\coar(s))$, for $\alpha_{\sort}^\sharp\colon \sort^\star \to Ob(\Cat{D})$ the inductive extension of  $\alpha_{\sort}$.  $\CatString$ is freely generated by $\sign$ in the sense that, for all symmetric strict monoidal categories $\Cat{D}$ and  all interpretations $\interpretation$ of $\sign$ in $\Cat{D}$, there exists a unique ssm-functor $\dsem{-}_{\interpretation}\colon \Cat{C}_\sign \to \Cat{D}$ extending $\interpretation$ (i.e. $\dsem{s}_\interpretation=\alpha_{\sign}(s)$ for all $s\in \sign$).

One can easily extend the notion of interpretation of $\sign$ into a symmetric monoidal category $\Cat D$ that is not necessarily strict. In this case we set $\alpha^\sharp_\sort \colon \sort^\star \to Ob(\Cat D)$ to be the \emph{right bracketing} of the inductive extension of $\alpha_\sort$. For instance, $\alpha^\sharp_\sort(ABC) = \alpha_\sort(A) \perG (\alpha_\sort(B) \perG \alpha_\sort(C))$.

\begin{example}\label{ex:signature}The set $\Sigma$ in $\CRS$ can be regarded as a monoidal signature. The set of sorts $\sort$ is the singleton set $\{A\}$, while the set of generators is exactly $\sign$. Each generator $R \in \Sigma$ has both arity and coarity $A$. Now, interpretations $\interpretation=(\alpha_{\sort},\alpha_{\sign})$ of this monoidal signature into $\Rel$, the category of sets and relations, are exactly relational interpretations as intended for $\CR$: a set $\alpha_{\sort}(A)$ (named $X$ in $\CR$) together with, for all $R\in \sign$, a relation $\alpha_{\sign}(R)\colon \alpha_{\sort}(A) \to \alpha_{\sort}(A)$ (named $\rho(R)$ in $\CR$). 
\end{example}

\subsection{The two monoidal structures of $\Rel$}\label{sec:2monREL} It is often the case that the same category carries more than one monoidal product. An example relevant to this work is $\Rel$ which exhibits two monoidal structures:  $(\Rel, \per, \uno)$ and $(\Rel, \piu, \zero)$.
In the former, $\per$ is given by the cartesian product, i.e.  $R \per S \defeq \{(\,(x_1,x_2),\, (y_1,y_2)\,) \mid (x_1,y_1)\in R \text{ and } (x_2,y_2)\in S  \}$ for all relations $R,S$, and the monoidal unit is the singleton set $1=\{\bullet\}$.
In the latter, $\piu$ is given by disjoint union, i.e.\ $R \piu S \defeq \{(\,(x,0),\,(y,0)\,) \mid (x,y)\in R \} \cup \{(\,(x,1),\,(y,1)\,) \mid (x,y)\in S \}$, and the monoidal unit $0$ is the empty set. It is worth recalling that in
$\Rel$ the empty set is both an initial and final object, i.e.\ a zero object, and that the disjoint union is both a coproduct and product, in fact a biproduct. Indeed  $(\Rel, \piu, \zero)$ is our first example of a finite biproduct category.

\begin{definition}\label{def:biproduct category}
    A \emph{finite biproduct category} is a symmetric monoidal category $(\Cat{C}, \perG, \unoG)$ where for every object $X$ there are morphisms $\codiag{X} \colon X \perG X \!\to\! X, \;\; \cobang{X} \colon \unoG \!\to\! X, \;\; \diag{X} \colon X \!\to\!  X \perG X, \;\; \bang{X} \colon X \!\to\! \unoG$ s.t.\
    \begin{enumerate}
\item $(\codiag{X}, \cobang{X})$ is a commutative monoid and $(\diag{X}, \bang{X})$ is a cocommutative comonoid, satisfying the coherence axioms in Figure~\ref{fig:fbcoherence},
        \item every arrow $f \colon X \to Y$ is both a monoid and a comonoid homomorphism, i.e.\
        \[
        (f\perG f) ; \codiag{Y} = \codiag{X}; f, \quad \cobang{X}; f = \cobang{Y}, \quad f; \diag{Y}= \diag{X}; (f\perG f) \quad \text{and} \quad f;\bang{Y}= \bang{X}.
        \] 
\end{enumerate}
    A \emph{morphism of finite biproduct categories} is a symmetric monoidal functor preserving  $\codiag{X},\cobang{X},\diag{X},\cobang{X}$. \end{definition}
Observe that the second condition simply amounts to naturality of monoids and comonoids. More generally, the reader who does not recognise the familiar definition of finite biproduct (fb) category may have a look at Appendix~\ref{appendix:rig}.  Monoids and comonoids in the monoidal category $(\Rel, \piu, \zero)$ are illustrated in the first column below: \begin{equation}\label{eq:comonoidsREL}
		\begin{tabular}{rcl c rcl}
			$\codiag{X}$ & $\!\!\defeq\!\!$ & $\{((x, 0), \; x) \mid x\in X\} \cup \{((x, 1), \; x) \mid x\in X\}$ && $\cocopier{X}$ & $\!\!\defeq\!\!$ & $\op{\copier{X}}$ \\[0.5em]
			$\cobang{X}$ & $\!\!\defeq\!\!$ & $\{\}$ && $\codischarger{X}$ & $\!\!\defeq\!\!$ &  $\op{\discharger{X}} $ \\[0.5em]
			$\diag{X}$ & $\!\!\defeq\!\!$ & $\op{\codiag{X}}$ && $\copier{X}$ & $\!\!\defeq\!\!$ & $\{(x, \; (x,x)) \mid x\in X\} $ \\[0.5em]
	     	$\bang{X}$ & $\!\!\defeq\!\!$ & $\op{\cobang{X}}$ && $\discharger{X}$ & $\!\!\defeq\!\!$ & $\{(x, \bullet) \mid x\in X\}\subseteq X \times 1$
		\end{tabular}
\end{equation}
Also $(\Rel, \per, \uno)$ has monoids and comonoids, illustrated in the second column above. However, they fail to be natural and, for this reason, $(\Rel, \per, \uno)$ is not an fb category. It is instead the archetypical example of a cartesian bicategory.

\begin{definition}\label{def:cartesian bicategory}
    A \emph{cartesian bicategory}, in the sense of~\cite{Carboni1987}, is a symmetric monoidal category $(\Cat{C}, \perG, \unoG)$ enriched over the category of posets where for every object $X$ there are morphisms $\cocopier{X} \colon X \perG X \to X, \;\; \codischarger{X} \colon \unoG \to X, \;\; \copier{X} \colon X \to X \perG X, \;\; \discharger{X} \colon X \to \unoG$ such that
    \begin{enumerate}
        \item $(\cocopier{X}, \codischarger{X})$ is a commutative monoid and $(\copier{X}, \discharger{X})$ is a cocommutative comonoid, satisfying the coherence axioms in Figure~\ref{fig:fbcoherence},
        \item every arrow $f \colon X \to Y$ is a lax comonoid homomorphism, i.e.\
        \[
        f;\copier{Y}\leq \copier{X}; (f \perG f) \quad \text{and} \quad f;\discharger{Y} \leq \discharger{X},
        \] 
\item monoids and comonoids form special Frobenius bimonoids (see e.g.\ \cite{Lack2004a}), 
        \item the comonoid $(\copier{X}, \discharger{X})$ is left adjoint to the monoid $(\cocopier{X}, \codischarger{X})$, i.e.\ :
        \[ \codischarger{X} ; \discharger{X} \leq \id{\unoG} \qquad \cocopier{X};\copier{X} \leq \id X \perG \id X \qquad \id X \leq \discharger{X} ; \codischarger{X} \qquad \id{X} \leq \copier{X};\cocopier{X} \]
    \end{enumerate}
A \emph{morphism of cartesian bicategories} is a poset enriched symmetric monoidal functor preserving monoids and comonoids.
\end{definition}

In~\cite{GCQ}, the authors introduced a string diagrammatic language, named $\CB$, expressing the cartesian bicategory structure of  $(\Rel, \per, \uno)$. One can similarly define a language for $(\Rel, \piu, \zero)$, however combining them would require a diagrammatic language that is able to express two different monoidal products at once. The appropriate categorical structures for this are rig categories, discussed in the next section.

 \section{Rig categories}\label{sc:rig}

Rig categories~\cite{laplaza_coherence_1972,johnson2021bimonoidal}, also known as \emph{bimonoidal categories}, involve two monoidal structures where one distributes over the other. 
The structures introduced below are sometimes referred to as \emph{symmetric} rig categories, but for the sake of brevity we will just call them rig categories.

\begin{definition}\label{def:rig}
    A \emph{rig category} is a category $\Cat{C}$ with 
    two symmetric monoidal structures $(\Cat{C}, \per, \uno, \symmt)$ and 
    $(\Cat{C}, \piu, \zero, \symmp)$ and natural isomorphisms 
    \[ \dl{X}{Y}{Z} \colon X \per (Y \piu Z) \to (X \per Y) \piu (X \per Z) \qquad  \annl{X} \colon \zero \per X \to \zero \]
    \[ \dr{X}{Y}{Z} \colon (X \piu Y) \per Z \to (X \per Z) \piu (Y \per Z) \qquad \annr{X} \colon X \per \zero \to \zero \]
satisfying the coherence axioms in Figure~\ref{fig:rigax}.    A rig category is said to be \emph{right} (respectively \emph{left}) \emph{strict} when both its monoidal structures are 
    strict and $\lambda^\bullet, \rho^\bullet$ and $\delta^r$ (respectively $\delta^l$) are all identity 
    natural isomorphisms.
\end{definition}
The natural isomorphism $\delta^l$ ($\delta^r$) is called \emph{left} (\emph{right}) \emph{distributor}.
The reader may wonder why only one of the two distributors is forced to be the identity within a strict rig category.
This can intuitively be explained as follows: imagine requiring that both distributors are identities. This would imply that both equations below should hold for all objects $X,Y,Z$ of any such strict category:
\[ (X \piu Y) \per Z = (X \per Z) \piu (Y \per Z)  \quad \text{and} \quad X \per (Y \piu Z) = (X \per Y) \piu (X \per Z). \]
The coexistence of the above laws would raise the same problems of strictification of symmetries (see Remark \ref{rmk:symstrict}). Indeed it would hold at once that \begin{align*}
(\!A\piu B\!) \per (\!C \piu D\!)  = (\!(A\piu B) \per C) \piu (\!(A\piu B) \per D) 
 = (\!(A \per C)\piu (B\per C)\!) \piu (\!(A \per D)\piu (B \per D)\!) 
\end{align*}
and 
\begin{align*}
(\!A\piu B\!) \per (\!C \piu D\!)  = (A \per (C \piu D)\!) \piu  (B \per (C \piu D)\!) 
 = (\!(A \per C)\piu (A\per D)\!) \piu (\!(B \per C)\piu (B \per D)\!)
\end{align*}
Note that $(B\per C)$ and $(A \per D)$ are in the opposite order in the two terms.

The traditional approach to strictness is however unsatisfactory when studying freely generated categories. To illustrate our concerns, consider a right strict rig category freely generated by a signature $\sign$ with sorts $\sort$. The objects of this category are terms generated by the grammar in Table~\ref{table:eq objects fsr} modulo the equations in the first three rows of the same table. These equivalence classes of terms do not come with a very handy form, unlike, for instance, the objects of a strict monoidal category, which are words. For this reason several authors, like~\cite{comfort2020sheet,johnson2021bimonoidal}, prefer to take as objects polynomials in $\sort$ at the price of working with a category that is not freely generated but only equivalent to a freely generated one. This fact forces one to consider functors that are not necessarily strict,  thus most of the constructions need to properly deal with the tedious natural isomorphisms.

\begin{table}
	\begin{center}
	\tiny{	
		\begin{subtable}{0.52\textwidth}
            \begin{tabular}{c}
				\toprule
				$X \; ::=\; \; A \; \mid \; \uno \; \mid \; \zero \; \mid \;  X \per X \; \mid \;  X \piu X$\\
				\midrule
				\begin{tabular}{ccc}
					$ (X \per Y) \per Z = X \per (Y \per Z)$ & $\uno \per X = X$ &  $X \per \uno = X $\\
					$(X \piu Y) \piu Z = X \piu (Y \piu Z)$ & $\zero \piu X = X$  & $X \piu \zero =X  $\\
					$(X \piu Y) \per Z = (X \per Z) \piu (Y \per Z)$ &  $\!\zero \per X = \zero$ & $X \per \zero=X $\\
				\end{tabular}\\
				$A \per (Y \piu Z) = (A \per Y) \piu (A \per Z)$\\
				\bottomrule
			\end{tabular}
            \caption{}
            \label{table:eq objects fsr}
        \end{subtable}
		\hfill
		\begin{subtable}{0.45\textwidth}
            \begin{tabular}{c}
				\toprule
				$n$-ary sums and products $\vphantom{\mid}$\\
				\midrule
				\makecell{
					\\[-5pt]
					$\Piu[i=1][0]{X_i} = \zero \; \Piu[i=1][1]{X_i}= X_1 \; \Piu[i=1][n+1]{X_i} = X_1 \piu (\Piu[i=1][n]{X_{i+1}} )$ \\[1em]
					$\Per[i=1][0]{X_i} = \uno \; \Per[i=1][1]{X_i}= X_1 \; \Per[i=1][n+1]{X_i} = X_1 \per (\Per[i=1][n]{X_{i+1}} )$ \\[3pt]
				} \\
				\bottomrule
			\end{tabular}
            \caption{}
            \label{table:n-ary sums and prod}
        \end{subtable}
	}
	\end{center}
	\caption{Equations for the objects of a free sesquistrict rig category}\label{tab:equationsonobject}
\end{table}

Here we propose a solution that might look a bit technical but, in our opinion, is rewarding. We focus on freely generated rig categories that we call \emph{sesquistrict}, i.e.\ right strict but only partially left strict: namely the left distributor  $\dl{X}{Y}{Z} \colon X \per (Y \piu Z) \to (X \per Y) \piu (X \per Z)$ is the identity only when $X$ is a basic sort $A\in \sort$. In terms of the equations to impose on objects, this amounts to the one in the fourth row in Table~\ref{table:eq objects fsr} for each $A\in \sort$. It is useful to observe that the addition of these equations avoids the problem of using left and right strictness at the same time. Indeed $(A\piu B) \per (C \piu D)$ turns out to be equal to $(A \per C) \piu (A \per D) \piu (B \per C) \piu (B \per D)$ but not to $(A \per C) \piu (B \per C) \piu (A \per D)  \piu (B \per D)$. Moreover, by orienting all the equations in Table~\ref{table:eq objects fsr} from left to right, one obtains a rewriting system that is confluent and terminating and, most importantly, the unique normal forms are exactly polynomials.

\begin{definition}
A term $X$ of the grammar in Table~\ref{table:eq objects fsr} is said to be in \emph{polynomial} form if there exist $n$, $m_i$ and $A_{i,j}\in \sort$ for $i=1 \dots n$ and $j=1 \dots m_i$ such that $X=\Piu[i=1][n]{\Per[j=1][m_i]{A_{i,j}}}$ (for $n$-ary sums and products as in Table~\ref{table:n-ary sums and prod}).\end{definition}
We will always refer to terms in polynomial form as \emph{polynomials} and, for a polynomial like $X$ above, we will call \emph{monomials} of $X$ the $n$ terms $\Per[j=1][m_i]{A_{i,j}}$. For instance the monomials of $(A \per B) \piu \uno$ are $A \per B$ and $1$. Note that, differently from the polynomials we are used to dealing with, here neither $\piu$ nor $\per$ is commutative so, for instance, $(A \per B) \piu \uno$ is different from both $\uno \piu (A \per B)$ and $(B \per A) \piu \uno$. Note that non-commutative polynomials are in one to one correspondence with \emph{words of words} over $\sort$, while monomials are words over $\sort$. 
\begin{notation}
Through the whole paper, we will denote by $A,B,C\dots$ the sorts in $\sort$, by $U,V,W \dots$ the words in $\sort^\star$ and by $P,Q,R,S \dots$ the words of words in $(\sort^\star)^\star$. Given two words $U,V\in \sort^\star$, we will write $UV$ for their concatenation and $1$ for the empty word. Given two words of words $P,Q\in (\sort^\star)^\star$, we will write $P\piu Q$ for their concatenation and $\zero$ for the empty word of words. Given a word of words $P$, we will write $\pi P$ for the corresponding term in polynomial form, for instance $\pi(A \piu BCD\piu 1 )$ is the term $A \piu ((B \per (C \per D)) \piu \uno)$. Throughout this paper  we  often omit $\pi$, thus we implicitly identify words of words with polynomials.
\end{notation}

Beyond concatenation ($\piu$), one can define on $(\sort^\star)^\star$ a product operation $\per$ by taking the unique normal form of $\pi(P) \per \pi(Q)$ for any $P,Q\in (\sort^\star)^\star$. More explicitly for 
$P = \Piu[i]{U_i}$ and $Q = \Piu[j]{V_j}$, 
\begin{equation}\label{def:productPolynomials} P \per Q \defeq \Piu[i]{\Piu[j]{U_iV_j}}.
\end{equation}
Observe that, if both $P$ and $Q$ are monomials, namely, $P=U$ and $Q=V$ for some $U,V\in \sort^\star$, then $P\per Q = UV$. We can thus safely write $PQ$ in place of $P\per Q$ without the risk of any confusion.

Words of words are also exploited in the traditional \emph{strictification} theorem for rig categories (see e.g.\ Theorem 5.4.6 in~\cite{johnson2021bimonoidal}):
given a rig category $\Cat C$, one can build a right strict rig category $\sCatT C$ that is equivalent to it. The objects of the strictification $\sCatT C$ are words of words of objects of $\Cat C$, while $\sCatT C[P,Q] = \Cat C[\pi P, \pi Q]$. It is easy to see that $\Cat C$ embeds into $\sCatT C$: a morphism $f \!\colon\! X \!\to\! Y$ in $\Cat C$ can be seen as a morphism in $\sCatT C[X,Y]$, where $X$ and $Y$ are considered as unary words made of a single unary word. This embedding forms an equivalence of rig categories $\Cat C \simeq \sCatT C$. See~\cite{johnson2021bimonoidal} for further details.

It turns out that $\sCatT C$ does actually satisfy a partial left distribution, with $Ob(\Cat C)$ playing the role of $\sort$, as it were. To make this statement precise, we introduce the following notion.

\begin{definition}\label{def:sesquistrict rig category}
	A \emph{sesquistrict rig category} is a functor $H \colon \Cat S \to \Cat C$, where $\Cat S$ is a discrete category and $\Cat C$ is a strict rig category, such that for all $A \in \Cat S$
	\[
	\dl{H(A)}{X}{Y} \colon H(A) \per (X \piu Y) \to (H(A) \per X) \piu (H(A) \per Y)
	\]
	is an identity morphism. We will also say, in this case, that $\Cat C$ is a $\Cat S$-sesquistrict rig category.
	
	Given $H \!\colon\! \Cat S \!\to\! \Cat C$ and $H' \!\colon\! \Cat S' \!\to\! \Cat C'$ two sesquistrict rig categories, a \emph{sesquistrict rig functor} from $H$ to $H'$ is a pair $(\alpha \!\colon\! \Cat S \!\to\! \Cat S', \beta \!\colon\! \Cat C \!\to\! \Cat C')$, with $\alpha$ a functor and $\beta$ a strict rig functor, such that $\alpha; H' = H; \beta$.
\end{definition}

\begin{theorem}\label{thm:strict(C) is sesquistrict}
Let $\Cat C$ be a rig category. Then its strictification $\sCatT C$ is a $Ob(\Cat C)$-sesquistrict rig category.\end{theorem}
\begin{proof}
	We recall from~\cite{johnson2021bimonoidal} some facts regarding the strictification $\sCatT C$ of $\Cat C$.
	
	Let $P$ and $Q$ be objects of $\sCatT C$. Then $P \per Q$ is given as in~\eqref{def:productPolynomials} and it is immediate to see that $A \per (P \piu Q) = (A \per P) \piu (A \per Q)$ when $A \in Ob(\Cat C)$. In order to prove that $\id{A \per (P \piu Q)} = \dl{A}{P}{Q}$ in $\sCatT C$, one needs to expand the definition of the left distributor $\delta^l$ and the symmetries $\sigma^\otimes$ in $\sCatT C$, which involve structural isomorphisms of $\Cat C$, and use the coherence theorem for rig categories, as stated in Theorem~3.9.1 of~\cite{johnson2021bimonoidal}. For this it is enough to observe that $\pi(A \per (P \piu Q))$ is \emph{regular} in the sense of~\cite[Definition 3.1.25]{johnson2021bimonoidal}. 
\end{proof}
\begin{corollary}[Sesquistrictification]
Any rig category is rig equivalent to a sesquistrict rig category.
\end{corollary}

In light of the corollary above, from now on we will only consider strictified rig categories. Moreover, in the rest of the article, whenever we will be considering strict symmetric monoidal functors $F \colon \Cat C \to \Cat D$ where $\Cat D$ is a rig category, we will actually mean that $F \colon \Cat C \to \sCatT D$ and we will assume that $F$ sends objects of $\Cat C$ to objects of $\Cat D$ (embedded in $\sCatT D$). This assumption is not restrictive, as witnessed by the following lemma.

\begin{lemma}\label{lemma:every functor in strict(D) is well behaved}
	Let $\Cat D$ be a rig category, $(\Cat C,\per,\uno)$ a strict symmetric monoidal category and $F \colon \Cat C \to (\sCatT D,\per,\uno)$ a strict symmetric monoidal functor. Then there exists a ssm functor $F' \colon \Cat C \to \sCatT D$, monoidally-naturally isomorphic to $F$, such that $F(Ob(\Cat C)) \subseteq Ob(\Cat D)$.
\end{lemma}

\subsection{Finite biproduct rig categories}

On many occasions, one is interested in rig categories where $\piu$ has some additional structure. For instance, distributive monoidal categories are rig categories where  $\piu$ is a coproduct. In this paper we will focus on rig categories where $\piu$ is a biproduct.
\begin{definition}
	A \emph{finite biproduct (fb) rig category} is a rig category $(\Cat C, \piu, \zero, \per, \uno)$ such that  $(\Cat C, \piu, \zero)$ is a finite biproduct category.
\end{definition}
It is worth remarking that, without the necessity of any extra coherence axioms, monoids and comonoids given by the fb structure of $(\Cat C, \piu, \zero)$ interact well with $\per$. See Appendix~\ref{appendix:rig} for all details.

$\Rel$, with the two monoidal structures defined as in Section~\ref{sec:2monREL}, is a fb rig category. Another example relevant to this paper is $\Mat{\CNum}$, the category of complex matrices, with $\per$ and $\piu$ being, respectively, the Kronecker product and the direct sum of matrices. Beyond these examples, our interest in fb rig categories is motivated by Theorem~\ref{thm:equivalentsignature}, which we are going to illustrate now.

Given $\sort$ a set of sorts, a \emph{rig signature} is a tuple $(\sort,\sign,\ar,\coar)$ where $\ar$ and $\coar$ assign to each $\gen \in \sign$ an arity and a coarity respectively, which are terms in the grammar specified in Table~\ref{table:eq objects fsr} modulo the equations underneath it. (Notice that any monoidal signature is in particular a rig signature.) To define the notion of free sesquistrict fb rig category, we need to extend interpretations of monoidal signatures to the fb rig case. An \emph{interpretation} of a rig signature $(\sort,\sign,\ar,\coar)$ in a sesquistrict fb rig category $H \colon \Cat M \to \Cat D$ is a pair of functions $(\alpha_{\sort} \colon \sort \to Ob(\Cat M), \alpha_\sign \colon \sign \to Ar(\Cat D))$ such that, for all $\gen \in \sign$, $\alpha_{\sign}(s)$ is an arrow having as domain and codomain $(\alpha_{\sort};H)^\sharp(\ar(s))$ and  $(\alpha_{\sort};H)^\sharp(\coar(s))$.

\begin{definition}
	Let $(\sort,\sign,\ar,\coar)$ (simply $\sign$ for short) be a rig signature. A sesquistrict fb rig category $H \colon \Cat M \to \Cat D$ is said to be \emph{freely generated} by $\sign$ if there is an interpretation $(\alpha_S,\alpha_\sign)$ of $\sign$ in $H$ such that for every sesquistrict rig category $H' \colon \Cat M' \to \Cat D'$ and every interpretation $(\alpha_\sort' \colon \sort \to Ob(\Cat M'), \alpha_\sign' \colon \sign \to Ar(\Cat D'))$ there exists a unique sesquistrict rig functor $(\alpha \colon \Cat M \to \Cat M', \beta \colon \Cat D \to \Cat D')$ such that $\alpha_\sort ; \alpha = \alpha_\sort'$ and $\alpha_\sign ; \beta = \alpha_{\sign}'$. 
\end{definition}

Sesquistrict fb rig categories generated by a given signature are isomorphic to each other, hence we will refer to ``the'' free sesquistrict rig category generated by a signature.

\begin{theorem}\label{thm:equivalentsignature}
	For every rig signature $(\sort,\sign)$ there exists a monoidal signature $(\sort, \sign_M)$ such that the free sesquistrict fb rig categories generated by $(\sort,\sign)$ and by $(\sort, \sign_M)$ are isomorphic.
\end{theorem}
The proof of the above theorem follows the same pattern as the proof of the well known fact that one can reduce monoidal signatures to standard cartesian signatures when generating the free finite product category. In particular, $\sign_M$ is obtained from $\sign$ by replacing each generator $\gen \colon \Piu[i=1][n]{U_i} \to \Piu[j=1][m]{V_j}$ in $\sign$ with new, formal symbols $\gen_{j,i}$ of type $U_i \to V_j$.

 \section{Tape Diagrams}\label{sc:tape}

\begin{table}[t]
\tiny{
\begin{center}

\begin{tabular}{cccc}
    \toprule
    \multicolumn{2}{c}{$\diag{ A}; (\id{ A}\perG \diag{ A}) = \diag{ A};(\diag{ A}\perG \id{ A})$} & $\diag{ A} ; (\bang{ A}\perG \id{ A}) = \id{ A} $ & $ \diag{ A};\sigma_{ A, A}=\diag{ A}$ \\
    \multicolumn{2}{c}{$(\id{ A}\perG \codiag{ A}) ; \codiag{ A} = (\codiag{ A}\perG \id{ A}) ; \codiag{ A}$} & $(\cobang{ A}\perG \id{ A}) ; \codiag{ A}  = \id{ A} $ & $ \sigma_{ A, A};\codiag{ A}=\codiag{ A}$ \\
    $\codiag{ A};\diag{ A} = \diag{ A\perG  A} ; (\codiag{ A} \perG \codiag{ A}) $ & $ \cobang{ A}; \bang{ A}= \id{\unoG} $ & $ \cobang{ A}; \diag{ A}= \cobang{ A\perG  A} $ & $ \diag{ A}; \bang{ A}= \bang{ A\perG  A}$\\
    $\tapeFunct{c}; \bang{B}=\bang{A} $ & $ \tapeFunct{c}; \diag{B}=\diag{A}; (\tapeFunct{c} \perG \tapeFunct{c}) $ & $ \cobang{A};\tapeFunct{c} =\cobang{B} $ & $ \codiag{A};\tapeFunct{c} =(\tapeFunct{c} \perG \tapeFunct{c}); \codiag{B}$ \\
    \multicolumn{2}{c}{$\tape{\id{A}} = \id{A}$} & \multicolumn{2}{c}{$\tape{c ; d} = \tape{c} ; \tape{d}$} \\
    \bottomrule
    \end{tabular}
\end{center}
}
\caption{Additional axioms for $F_2(\Cat{C})$. Above, $c\colon A \to B$ is an arbitrary arrow of $\Cat{C}$}\label{fig:freestrictfbcat}
\end{table}

We have seen in Section~\ref{sc:monoidal} that string diagrams provide a convenient graphical language for strict monoidal categories.
In this section, we introduce tape diagrams to graphically represent arrows of sesquistrict rig categories.

In order to illustrate our main idea, it is convenient to recall the following functors amongst $\CAT$, the category  of categories and functors, $\SMC$, the category 
 of ssm categories and functors, and $\FBC$, the category of strict finite biproduct categories and functors. 
\begin{equation}\label{eq:adjunction} \begin{tikzcd}
		\SMC \ar[r,"U_1"] & \CAT \ar[r,bend left,"F_2"] \ar[r,draw=none,"\bot"description] & \FBC \ar[l,bend left,"U_2"]
\end{tikzcd}
\end{equation}
The functors $U_1$ and $U_2$ are the obvious forgetful functors. The functor $F_2$ is the left adjoint to $U_2$, and can be described as follows.

\begin{definition}\label{def:strict fb freely generated by C}
Let $\Cat{C}$ be a category. The  strict fb category freely generated by $\Cat{C}$, hereafter denoted by $F_2(\Cat{C})$, has as objects words of objects of $\Cat{C}$. Arrows are terms inductively generated by the following grammar, where $A,B$ and $c$ range over arbitrary objects and arrows of $\Cat{C}$,\begin{equation}
    \begin{array}{rcl}
        f & ::=& \; \id{A} \; \mid \; \id{I} \; \mid \; \tapeFunct{c} \; \mid \; \sigma_{A,B}^{\perG} \; \mid \;   f ; f   \; \mid \;  f \perG f  \; \mid \; \bang{A} \; \mid \; \diag{A} \;  \mid \; \cobang{A}\; \mid \; \codiag{A}\\
        \end{array}
\end{equation}  
modulo the axioms in  Tables~\ref{fig:freestricmmoncatax} and~\ref{fig:freestrictfbcat}. Notice in particular the last two from Table~\ref{fig:freestrictfbcat}:
\begin{equation}\label{ax:tape}\tag{Tape}
\tape{\id{A}} = \id{A} \qquad \tape{c;d}= \tape{c} ; \tape{d}
\end{equation}
\end{definition}
The assignment $\Cat{C} \mapsto F_2({\Cat{C}})$ easily extends to functors $H\colon \Cat{C} \to \Cat{D}$.
The unit of the adjunction $\eta \colon Id_{\CAT} \Rightarrow F_2U_2$ is defined for each category $\Cat{C}$ as the functor ${\tapeFunct{\cdot} \; \colon \Cat{C} \to U_2F_2(\Cat{C})}$ which is the identity on objects and maps every arrow $c$ in $\Cat{C}$ into the arrow $\tapeFunct{c}$ of $U_2F_2(\Cat{C})$. Observe that $\tapeFunct{\cdot}$ is indeed a functor, namely an arrow in $\CAT$, thanks to the axioms \eqref{ax:tape}. We will refer hereafter to this functor as the \emph{taping functor}.

\begin{lemma}\label{lemma:F2 left adjoint to U2}
$F_2 \colon \CAT \to \FBC$ is left adjoint to $U_2 \colon \FBC \to \CAT$.
\end{lemma}

The main result of this section (Theorem~\ref{thm:Tapes is free sesquistrict generated by sigma}) states that the  sesquistrict fb rig category freely generated by a monoidal signature $\sign$ is $F_2U_1(\CatString)$, hereafter referred to as $\CatTape$. 
This is somehow the fb rig category analogue to Theorem 2.3 in \cite{joyal1991geometry} mentioned in Section~\ref{sc:monoidal}. Let us see why.

Recall that the set of objects of 
$\CatString$ is $\sort^\star$, i.e.\ words of sorts in $\sort$. The set of objects of $\CatTape$ is thus $(\sort^\star)^\star$, namely words of words of sorts in $\sort$. For arrows, consider the following two-layer grammar where $s \in \sign$, $A,B \in \sort$ and $U,V \in \sort^\star$.
\begin{equation}\label{tapesGrammar}
    \begin{tabular}{rc ccccccccccccccccccc}\setlength{\tabcolsep}{0.0pt}
        $c$  & ::= & $\id{A}$ & $\!\!\! \mid \!\!\!$ & $ \id{\uno} $ & $\!\!\! \mid \!\!\!$ & $ \gen $ & $\!\!\! \mid \!\!\!$ & $ \sigma_{A,B} $ & $\!\!\! \mid \!\!\!$ & $   c ; c   $ & $\!\!\! \mid \!\!\!$ & $  c \per c$ & \multicolumn{8}{c}{\;} \\
        $\t$ & ::= & $\id{U}$ & $\!\!\! \mid \!\!\!$ & $ \id{\zero} $ & $\!\!\! \mid \!\!\!$ & $ \tapeFunct{c} $ & $\!\!\! \mid \!\!\!$ & $ \sigma_{U,V}^{\piu} $ & $\!\!\! \mid \!\!\!$ & $   \t ; \t   $ & $\!\!\! \mid \!\!\!$ & $  \t \piu \t  $ & $\!\!\! \mid \!\!\!$ & $ \bang{U} $ & $\!\!\! \mid \!\!\!$ & $\diag{U}$ & $\!\!\! \mid \!\!\!$ & $\cobang{U}$ & $\!\!\! \mid \!\!\!$ & $\codiag{U}$    
    \end{tabular}
\end{equation}  
The terms of the first row, denoted by $c$, are called \emph{circuits}. Modulo the axioms in Table~\ref{fig:freestricmmoncatax} (after replacing $\perG$ with $\per$), these are exactly the arrows of $\CatString$.
The terms of the second row, denoted by $\t$, are called \emph{tapes}. Modulo the axioms in Tables \ref{fig:freestricmmoncatax} and \ref{fig:freestrictfbcat} (after replacing $\perG$ with $\piu$ and $A,B$ with $U,V$), these are exactly the arrows of $F_2U_1(\CatString)$, i.e.\  $\CatTape$. 

Since circuits are arrows of $\CatString$, these can be graphically represented as string diagrams. Also tapes can be represented as string diagrams, since they satisfy all of the axioms of ssmc. Note however that \emph{inside} tapes, there are string diagrams: this justifies the motto \emph{tape diagrams are string diagrams of string diagrams}. We can thus render graphically the grammar in~\eqref{tapesGrammar}: 
\begin{equation*}\label{tapesDiagGrammar}
    \setlength{\tabcolsep}{2pt}
    \begin{tabular}{rc cccccccccccc}
        $c$  & ::= &  $\wire{A}$ & $\mid$ & $ 
    \InputIfFileExists{empty.tikz}{}{\input{./tikz/empty.tikz}}
 $ & $\mid$ & $ \Cgen{\gen}{A}{B}  $ & $\mid$ & $ \Csymm{A}{B} $ & $\mid$ & $ 
    \InputIfFileExists{seq_compC.tikz}{}{\input{./tikz/seq_compC.tikz}}
   $ & $\mid$ & $  
    \InputIfFileExists{par_compC.tikz}{}{\input{./tikz/par_compC.tikz}}
$ & \\
        $\t$ & ::= & $\Twire{U}$ & $\mid$ & $ 
    \InputIfFileExists{empty.tikz}{}{\input{./tikz/empty.tikz}}
 $ & $\mid$ & $ \Tcirc{c}{U}{V}  $ & $\mid$ & $ \Tsymmp{U}{V} $ & $\mid$ & $ 
    \InputIfFileExists{tapes/seq_comp.tikz}{}{\input{./tikz/tapes/seq_comp.tikz}}
  $ & $\mid$ & $  
    \InputIfFileExists{tapes/par_comp.tikz}{}{\input{./tikz/tapes/par_comp.tikz}}
$ & $\mid$ \\
             &     & \multicolumn{12}{l}{$\Tcounit{U} \; \mid \; \Tcomonoid{U} \; \mid \; \Tunit{U} \; \mid \; \Tmonoid{U}$}
    \end{tabular}
\end{equation*}  
The identity $\id\zero$ is rendered as the empty tape $
    \InputIfFileExists{empty.tikz}{}{\input{./tikz/empty.tikz}}
$, while $\id\uno$ is $
    \InputIfFileExists{tapes/empty.tikz}{}{\input{./tikz/tapes/empty.tikz}}
$: a tape filled with the empty circuit. 
For a monomial $U \!=\! A_1\dots A_n$, $\id U$ is depicted as a tape containing  $n$ wires labelled by $A_i$. For instance, $\id{AB}$ is rendered as $\TRwire{A}{B}$. When clear from the context, we will simply represent it as a single wire  $\Twire{U}$ with the appropriate label.
Similarly, for a polynomial $P = \Piu[i=1][n]{U_i}$, $\id{P}$ is obtained as a vertical composition of tapes, as illustrated below on the left. \[ \id{AB \piu \uno \piu C} = \begin{aligned}\begin{gathered} \TRwire{A}{B} \\[-1.5mm] \Twire{\uno} \\[-2mm] \Twire{C} \end{gathered}\end{aligned} \qquad \qquad \tapesymm{AB}{C} = 
    \InputIfFileExists{tapes/examples/tapesymmABxC.tikz}{}{\input{./tikz/tapes/examples/tapesymmABxC.tikz}}
 \qquad\qquad \symmp{AB \piu \uno}{C} = 
    \InputIfFileExists{tapes/examples/symmpABp1pC.tikz}{}{\input{./tikz/tapes/examples/symmpABp1pC.tikz}}
 \]
We can render graphically the symmetries $\tapesymm{U}{V} \colon UV \!\to\! VU$ and $\symmp{P}{Q} \colon P \piu Q \!\to\! Q \piu P$ as crossings of wires and crossings of tapes,  see the two rightmost diagrams above.
The diagonal $\diag{U} \colon U \!\to\! U \piu U$ is represented as a splitting of tapes, while the bang $\bang{U} \colon U \!\to\! \zero$ is a tape closed on its right boundary. 
Codiagonals and cobangs are represented in the same way but mirrored along the y-axis. Exploiting the coherence axioms in Figure~\ref{fig:fbcoherence}, we can construct (co)diagonals and (co)bangs for arbitrary polynomials $P$.
For example, $\diag{AB}$, $\bang{CD}$, $\codiag{A\piu B \piu C}$ and $\cobang{AB \piu B \piu C}$ are depicted as:
\[ 
    \diag{AB} = \TLcomonoid{B}{A} \qquad \bang{CD} = \TLcounit{D}{C} \qquad  \codiag{A\piu B \piu C} = 
    \InputIfFileExists{tapes/examples/codiagApBpC.tikz}{}{\input{./tikz/tapes/examples/codiagApBpC.tikz}}
 \qquad   \cobang{AB \piu B \piu C} = 
    \InputIfFileExists{tapes/examples/cobangABpBpC.tikz}{}{\input{./tikz/tapes/examples/cobangABpBpC.tikz}}
 
\]

When the structure inside a tape is not relevant the graphical language can be ``compressed'' in order to simplify the diagrammatic reasoning. For example, for arbitrary polynomials $P, Q$ we represent $\id{P}, \symmp{P}{Q}, \diag{P}, \bang{P}, \codiag{P}, \cobang{P}$ as follows:
\[ \TPolyWire{P} \qquad \TPolySymmp{P}{Q} \qquad \TPolyDiag{P} \qquad \TPolyCounit{P} \qquad \TPolyCodiag{P} \qquad \TPolyUnit{P} \]
Moreover, for an arbitrary tape diagram $\t \colon P \to Q$ we write $\Tbox{\t}{P}{Q}$.

It is important to observe that the graphical representation takes care of the two axioms in \eqref{ax:tape}: both sides of the leftmost axiom are depicted as $\Twire{A}$ while both sides of the rightmost axiom as
$
    \InputIfFileExists{tapes/ax/c_poi_d.tikz}{}{\input{./tikz/tapes/ax/c_poi_d.tikz}}
$. The axioms of monoidal categories are also implicit in the graphical representation, while those for symmetries and the fb-structure (in Table~\ref{fig:freestrictfbcat}) have to be depicted explicitly as in Figure~\ref{fig:tapesax}. In particular, the diagrams in the first row express the inverse law and naturality of $\symmp$. In the second group there are the (co)monoid axioms and in the third group the bialgebra ones. Finally, the last group depicts naturality of the (co)diagonals and (co)bangs.

\subsection{The finite biproduct rig structure of tapes} By definition $\CatTape$ is equipped with a monoidal product $\piu$ that is a biproduct. We now show that
$\CatTape$ carries a second monoidal product $\per$ which makes it a sesquistrict rig category.

On objects, $\per$ is defined as in \eqref{def:productPolynomials}.
For instance, $(A \piu B) \per (C \piu D)$ is $AC \piu AD \piu BC \piu BD$ and not $AC \piu BC \piu AD \piu BD$. This is justified by the fact that we want $\CatTape$ to be a sesquistrict rig category and thus $\per$ distributes over $\piu$ on the right (and only partially on the left). 
\begin{remark}\label{rem:per tape is right distributive}
Observe that left distributivity holds not only for $A\in \sort$, but for all monomials $U$: $U \per (P \piu Q) = (U \per P) \piu (U \per Q)$ for all polynomials $P,Q$.
\end{remark}

In general, distributivity on the left is possible, but has to be made explicit through left distributors built just from identities and $\piu$-symmetries, as shown in Table~\ref{table:def dl}. Similarly, arbitrary $\per$-symmetries can be built, as shown in Table~\ref{table:def symmt}, from left distributors and symmetries within tapes $\tapesymm{U}{V}$.

\begin{table}
    \begin{center}
    \tiny{
        \hfill
        \begin{subtable}{0.60\textwidth}
            \begin{tabular}{l}
                \toprule
                $\dl{P}{Q}{R} \colon P \per (Q\piu R)  \to (P \per Q) \piu (P\per R) \vphantom{\symmt{P}{Q}}$ \\
                \midrule
                $\dl{\zero}{Q}{R} \defeq \id{\zero} \vphantom{\symmt{P}{\zero} \defeq \id{\zero}}$ \\
                $\dl{U \piu P'}{Q}{R} \defeq (\id{U\per (Q \piu R)} \piu \dl{P'}{Q}{R});(\id{U\per Q} \piu \symmp{U\per R}{P'\per Q} \piu \id{P'\per R}) \vphantom{\symmt{P}{V \piu Q'} \defeq \dl{P}{V}{Q'} ; (\Piu[i]{\tapesymm{U_i}{V}} \piu \symmt{P}{Q'})}$ \\
                \bottomrule
            \end{tabular}
            \caption{}
            \label{table:def dl}
        \end{subtable}
        \hfill
        \begin{subtable}{0.35\textwidth}
            \begin{tabular}{l}
                \toprule
                $\symmt{P}{Q} \colon P\per Q \to Q \per P$, with $P = \Piu[i]{U_i}$ \\
                \midrule
                $\symmt{P}{\zero} \defeq \id{\zero}$ \\
                $\symmt{P}{V \piu Q'} \defeq \dl{P}{V}{Q'} ; (\Piu[i]{\tapesymm{U_i}{V}} \piu \symmt{P}{Q'})$ \\
                \bottomrule
            \end{tabular}
            \caption{}
            \label{table:def symmt}
        \end{subtable}
        \hfill
        \caption{Inductive definition of $\delta^l$ and $\symmt$}
        \label{table:def dl symmt}
    }
    \end{center}
\end{table}

\begin{example}\label{ex:diag:dl symmt}
By means of the inductive definition in Table~\ref{table:def dl symmt},  $\dl{A\piu B}{C}{D} \colon (A \piu B)(C \piu D) \to (A \piu B)C \piu (A \piu B)D$ and $\symmt{A\piu B}{C \piu D} \colon (A \piu B)(C \piu D) \to (C \piu D)(A \piu B)$ are depicted as follows:
{\setlength{\abovedisplayskip}{4pt}
    \setlength{\belowdisplayskip}{4pt}
    \begin{equation*}
        \dl{A \piu B}{C}{D} = \Tdl{A}{B}{C}{D} \qquad \qquad \symmt{A\piu B}{C \piu D} = \Tsymmt{A}{B}{C}{D}
    \end{equation*}}
\end{example}
We can now introduce left and right whiskering for those objects $U$ that are monomials. \begin{definition}\label{def:tape:whisk} Let $U$ be a monomial. The \emph{left} and \emph{right whiskering} (with respect to $U$) are two functors $L_U, R_U \colon \CatTape \to \CatTape$ 
    which are defined on objects as $\LW{U}{P} \defeq U \per P$ and $\RW{U}{P} \defeq P\per U$
    and on arrows as:
\begin{align*}
        &\LW U {\id\zero} \defeq \id\zero &&\RW U {\id\zero} \defeq \id\zero \\
        &\LW U {\tapeFunct{c}} \defeq \tape{\id U \per c} \,\;\quad \LW U {\symmp{V}{W}} \defeq \symmp{UV}{UW} &&\RW U {\tapeFunct{c}} \defeq \tape{c \per \id U} \,\;\quad \RW U {\symmp{V}{W}} \defeq \symmp{VU}{WU} \\
        &\LW U {\diag V} \defeq \diag{UV} \quad\qquad \LW U {\bang V} \defeq \bang{UV} &&\RW U {\diag V} \defeq \diag{VU} \quad\qquad \RW U {\bang V} \defeq \bang{VU} \\
        &\LW U {\codiag V} \defeq \codiag{UV} \quad\qquad \LW U {\cobang V} \defeq \cobang{UV} &&\RW U {\codiag V} \defeq \codiag{VU} \quad\qquad \RW U {\cobang V} \defeq \cobang{VU} \\
        &\LW U {\t_1 ; \t_2} \defeq \LW U {\t_1} ; \LW U {\t_2} &&\RW U {\t_1 ; \t_2} \defeq \RW U {\t_1} ; \RW U {\t_2} \\
        &\LW U {\t_1 \piu \t_2} \defeq \LW U {\t_1} \piu \LW U {\t_2} &&\RW U {\t_1 \piu \t_2} \defeq \RW U {\t_1} \piu \RW U {\t_2}
    \end{align*}
\end{definition}

\begin{example}\label{ex:diagWhisk}The meaning of the monomial whiskering is quite immediate in graphical terms: below we draw the left whiskering of a tape $\s$ and the right whiskering of a tape $\t$. $\LW{U}{\s}$ \emph{thickens} the tapes which $\s$ is made of, by stacking the wires of $\id U$ inside them; the right whiskering works analogously except that the additional wires are stacked at the bottom of the single tapes:
\[\LW{U}{
    \InputIfFileExists{tapes/examples/t2.tikz}{}{\input{./tikz/tapes/examples/t2.tikz}}
} = 
    \InputIfFileExists{tapes/examples/L_Ut2.tikz}{}{\input{./tikz/tapes/examples/L_Ut2.tikz}}
 \quad \RW{W'}{
    \InputIfFileExists{tapes/examples/t1.tikz}{}{\input{./tikz/tapes/examples/t1.tikz}}
} = 
    \InputIfFileExists{tapes/examples/R_Wpt1.tikz}{}{\input{./tikz/tapes/examples/R_Wpt1.tikz}}
\]
\end{example}

We can extend Definition~\ref{def:tape:whisk} to arbitrary polynomials $S$ as follows.

\begin{definition}\label{def:tape:whiskG} 
Let $S$ be a polynomial. $L_{S}, R_{S} \colon \CatTape \to \CatTape$ are defined on objects as $\LW{S}{P} \defeq S \per P$ and $\RW{S}{P} \defeq P\per S$ and on arrows $\t \colon P \to Q$ by induction on $S$:
\begin{align*}
    &\LW{\zero}{\t} \defeq \id{\zero} && \RW{\zero}{\t} \defeq \id{\zero} \\
    & \LW{W\piu S'}{\t} \defeq \LW{W}{\t} \piu \LW{S'}{\t} && \RW{W \piu S'}{\t} \defeq \dl{P}{W}{S'} ; (\RW{W}{\t} \piu \RW{S'}{\t}) ; \Idl{Q}{W}{S'}
\end{align*}
\end{definition}
Observe that there is an asymmetry in the definition of left and right whiskerings for polynomials: again, this is justified by the fact that $\per$ distributes over $\piu$ on the right. The whiskering of  a tape $\t \colon A\piu B \to A' \piu B'$ on the right with the polynomial $C \piu D$, that is $\RW{C \piu D}{\t}$, should go from $(A\piu B) \per (C \piu D) = AC \piu AD \piu BC \piu BD$ to $(A'\piu B') \per (C \piu D) = A'C \piu A'D \piu B'C \piu B'D$, whereas $\RW{C}{\t} \piu \RW{D}{\t}$ goes from  $AC \piu BC \piu AD \piu BD$ to $A'C \piu B'C \piu A'D \piu B'D$.

With Definition \ref{def:tape:whiskG} we can finally introduce $\per$ on arrows  $\t_1 \colon P_1 \to Q_1, \t_2 \colon P_2 \to Q_2$ as
\begin{equation}\label{eq:def tensore tape}
\t_1 \per \t_2 \defeq \LW{P_1}{\t_2} ; \RW{Q_2}{\t_1}\text{.}
\end{equation}

\begin{example}\label{ex:diagPer}
Consider $\t\colon U\piu V \to W\piu Z$ and $\s\colon U'\piu V' \to W'\piu Z'$ from Example~\ref{ex:diagWhisk}, then $\t \per \s$ is simply the sequential composition of $\LW{U\piu V}{\s}$ and $\RW{W' \piu Z'}{\t}$:
    \[  
    \InputIfFileExists{tapes/examples/t1.tikz}{}{\input{./tikz/tapes/examples/t1.tikz}}
 \per 
    \InputIfFileExists{tapes/examples/t2.tikz}{}{\input{./tikz/tapes/examples/t2.tikz}}
 = 
    \InputIfFileExists{tapes/examples/t1xt2_line.tikz}{}{\input{./tikz/tapes/examples/t1xt2_line.tikz}}
 \]
    The dashed line highlights the boundary between left and right polynomial whiskerings: $\LW{U \piu V}{\s}$, on the left, is simply the vertical composition of the monomial whiskerings $\LW{U}{\s}$ and $\LW{V}{\s}$ while, on the right, $\RW{W' \piu Z'}{\t}$ is rendered as the vertical composition of $\RW{W'}{\t}$ and $\RW{Z'}{\t}$, precomposed and postcomposed with left distributors.
\end{example}

\begin{table}[t]
    \tiny{\begin{tabular}{cc cc}
        \toprule
        \multicolumn{2}{c}{$\LW S {\id{P}} = \id{SP}$} & $\RW S {\id{P}} = \id{PS}$ & (\newtag{W1}{eq:whisk:id})\\[0.3em]
        \multicolumn{2}{c}{$\LW{S}{\t ; \s} = \LW{S}{\t} ; \LW{S}{\s}$} & $\RW{S}{\t ; \s} = \RW{S}{\t} ; \RW{S}{\s}$ & (\newtag{W2}{eq:whisk:funct})\\[0.3em]
        \multicolumn{2}{c}{$\LW{\uno}{\t} = \t$} & $\RW{\uno}{\t} = \t$ & (\newtag{W3}{eq:whisk:uno}) \\[0.3em]
        \multicolumn{2}{c}{$\LW{\zero}{\t} = \id{\zero}$} & $\RW{\zero}{\t} = \id{\zero}$ & (\newtag{W4}{eq:whisk:zero}) \\[0.3em]
\multicolumn{2}{c}{$\LW{S}{\t_1 \piu \t_2} = \dl{S}{P_1}{P_2} ; (\LW{S}{\t_1} \piu \LW{S}{\t_2}) ; \Idl{S}{Q_1}{Q_2}$}  & $\RW{S}{\t_1 \piu \t_2} = \RW{S}{\t_1} \piu \RW{S}{\t_2}$ & (\newtag{W5}{eq:whisk:funct piu}) \\[0.3em]
        \multicolumn{2}{c}{$\LW{S \piu T}{\t} = \LW{S}{\t} \piu \LW{T}{\t}$}  & $\RW{S \piu T}{\t} = \dl{P}{S}{T} ; ( \RW{S}{\t} \piu \RW{T}{\t} ) ; \Idl{Q}{S}{T}$ & (\newtag{W6}{eq:whisk:sum}) \\[0.3em]
        \multicolumn{3}{c}{$\LW{P_1}{\t_2} ; \RW{Q_2}{\t_1} = \RW{P_2}{\t_1} ; \LW{Q_1}{\t_2}$} & (\newtag{W7}{eq:tape:LexchangeR}) \\[0.3em]
        $\RW S {\diag{U}} = \diag{US} \qquad \RW S {\codiag{U}} = \codiag{US}$ & (\newtag{W8}{eq:whisk:diag}) &  $\RW S {\bang{U}} = \bang{US} \qquad \RW S {\cobang{U}} = \cobang{US}$ & (\newtag{W9}{eq:whisk:bang}) \\[0.3em]
        $\RW S {\symmp{P}{Q}} = \symmp{PS}{QS}$ & (\newtag{W10}{eq:whisk:symmp}) & $\symmt{PQ}{S} = \LW{P}{ \symmt{Q}{S}} ; \RW{Q}{\symmt{P}{S}}$ & (\newtag{W11}{eq:symmper}) \\[0.3em]
        $\RW{S}{\t} ; \symmt{Q}{S} = \symmt{P}{S} ; \LW{S}{\t}$ & (\newtag{W12}{eq:LRnatsym}) & $\LW{S}{\RW{T}{\t}} = \RW{T}{\LW{S}{\t}}$ & (\newtag{W13}{eq:tape:LR}) \\[0.3em]
        $\LW{ST}{\t} = \LW{S}{\LW{T}{\t}}$ & (\newtag{W14}{eq:tape:LL}) & $\RW{TS}{\t} = \RW{S}{\RW{T}{\t}}$ & (\newtag{W15}{eq:tape:RR}) \\[0.3em]
        $\RW S {\dl{P}{Q}{R}} = \dl{P}{QS}{RS}$ & (\newtag{W16}{eq:whisk:dl}) & $\LW S {\dl{P}{Q}{R}} = \dl{SP}{Q}{R} ; \Idl{S}{PQ}{PR}$ & (\newtag{W17}{eq:whisk:Ldl}) \\
        \bottomrule
    \end{tabular}}
    \caption{The algebra of whiskerings}  
    \label{table:whisk}
\end{table}

\begin{lemma}\label{lm:whisk properties}
    The laws in Table~\ref{table:whisk} hold for any $\t \colon P \to Q, \s \colon Q \to R, \t_1 \colon P_1 \to Q_1, \t_2 \colon P_2 \to Q_2$.
\end{lemma}
The laws in Table~\ref{table:whisk} are useful in several occasions. In particular, they make it possible to easily prove that $\CatTape$ is a fb rig category. For instance, functoriality of $\per$ immediately follows from \eqref{eq:whisk:funct} and 
\eqref{eq:tape:LexchangeR}:  for all  $\t_1 \colon P \to Q, \t_2 \colon Q \to S, \t_3 \colon P' \to Q', \t_4 \colon Q' \to S'$, 
\begin{align*}
    (\t_1 ; \t_2) \per (\t_3 ; \t_4) &=\tag{Def. $\per$}   \LW {P} {\t_3 ; \t_4} ; \RW {S'} {\t_1 ; \t_2} \\
    &=\tag{\ref{eq:whisk:funct}} \LW {P} {\t_3} ; \LW {P} {\t_4} ; \RW {S'} {\t_1} ; \RW {S'} {\t_2} \\
    &=\tag{\ref{eq:tape:LexchangeR}} \LW {P} {\t_3} ; \RW {Q'} {\t_1} ; \LW {Q} {\t_4}   ; \RW {S'} {\t_2} \\
    &=\tag{Def. $\per$} (\t_1 \per \t_3) ; (\t_2 \per \t_4)
\end{align*}

\begin{theorem}\label{thm:taperig}
$\CatTape$ is a $\sort$-sesquistrict finite biproduct rig category.
\end{theorem}

In the future we will refer to the sesquistrict fb rig category $\sort \to \CatTape$ simply as $\CatTape$. We can now state the main result of this section.

\begin{theorem}\label{thm:Tapes is free sesquistrict generated by sigma}
	$\CatTape$ is the free sesquistrict fb rig category generated by the monoidal signature $(\sort, \sign)$.
\end{theorem}
\begin{proof}
	We have a trivial interpretation of $(\sort,\sign)$ in $\sort \to \CatTape$, given by $(\id\sort,\tapeFunct{\cdot})$. Suppose now that $H \colon \Cat M \to \Cat D$ is a sesquistrict fb rig category with an interpretation $\interpretation=(\alpha_\sort \colon \sort \to {Ob}(\Cat M), \alpha_\sign \colon \sign \to {Ar}(\Cat D))$. We aim to find a sesquistrict fb rig functor $(\alpha \colon \sort \to \Cat M, \beta \colon \CatTape \to \Cat D)$ such that $\id\sort ; \alpha = \alpha_{\sort}$ and $\tapeFunct{\cdot} ; \beta = \alpha_{\sign}$.
	Since $(\sort,\sign)$ is a monoidal signature, $\interpretation$ is in fact a monoidal interpretation of $(\sort,\sign)$ into the ssm category $(\Cat D,\per,\uno)$, hence by freeness of $\CatString$ there exists a unique ssm functor $\dsem{-}_{\interpretation} \colon \CatString \to \Cat D$ that extends $\interpretation$. Now, because $\CatTape = F_2(\CatString)$ is by definition the free fb category generated by $\CatString$ and $\Cat D$ is also a fb category, we have that there is a unique fb functor $\beta \colon \CatTape \to \Cat D$ that extends $\dsem{-}_{\interpretation}$. The fact that $\beta$ preserves the rest of the rig structure can be proved using the inductive definitions of the symmetries, distributors and the tensor of $\CatTape$.
\end{proof}

\begin{corollary}\label{co:rig semantics}
	Let $\Cat D$ be a fb rig category and $F \colon \CatString \to (\sCat D,\otimes,1)$ be a strict symmetric monoidal functor. Then there exists a unique sesquistrict rig functor $F^\sharp \colon \CatTape \to \sCat D$ such that $F = F^\sharp \circ \tapeFunct{\cdot}$.
\end{corollary}

The above corollary is rather effective: take $\CatString$ to be the syntax of a string diagrammatic language and the functor $F \colon \CatString \to \sCat D$ to be its semantics. Whenever $\sCat D$ carries the structure of an fb rig category, then one can extend the semantics $F$ to the language of tape diagrams $\CatTape$.

\begin{example}[ZX-calculus]\label{ex:zxcalculus}
    Consider the ZX-calculus with scalars and let $\ZX$ be the smc of ZX diagrams and $\dsem{-} \colon \ZX \to \Mat{\CNum}$ the semantic functor given in~\cite[Definition 3.1.1]{backens2016completeness}.    
    Since $\Mat{\CNum}$ is a fb rig category, one can easily derive the semantics $\dsem{-}^{\sharp} \colon \TZX \to \Mat{\CNum}$ of ZX \emph{tapes} (i.e.\ tape diagrams of ZX diagrams) as suggested in Corollary~\ref{co:rig semantics} above.
    
    We show that this construction allows us to obtain an immediate graphical representation of control. Let $
    \InputIfFileExists{zx/unitary.tikz}{}{\input{./tikz/zx/unitary.tikz}}
$ be a ZX diagram for a unitary $U$ and $\qState{0}, \qState{1}, \qEffect{0}, \qEffect{1}$ be shorthands for the computational basis states and effects given in~\cite[Equation 3.72]{backens2016completeness}. For instance, $\dsem{\qState{0}} = \ket{0} = \left(\begin{smallmatrix}1 \\ 0\end{smallmatrix}\right)$ and $\dsem{\qEffect{0} \qState{0}} = \ket{0} \bra{0} = \left(\begin{smallmatrix}1 \\ 0\end{smallmatrix}\right)  \left(\begin{smallmatrix}1 & 0\end{smallmatrix}\right) = \left(\begin{smallmatrix}1 & 0 \\ 0 & 0\end{smallmatrix}\right)$.
    Consider the following diagram:
    \[\dsem{
    \InputIfFileExists{zx/control.tikz}{}{\input{./tikz/zx/control.tikz}}
}^\sharp = \ket{00}\bra{00} + \ket{01}\bra{01} + (\ket{1} \per U\ket{0})\bra{10} + (\ket{1} \per U\ket{1})\bra{11} \]
As witnessed by the semantics the first wire acts as a control qubit on the unitary $U$ and thus the whole diagram can be regarded as a rendering of the controlled unitary $CU$.
    The correspondence between the diagrammatic notation and the sum of matrices ($+$) will be clear in Section~\ref{sc:matrix}.
\end{example}

\subsection{Diagrammatic reasoning with tapes}\label{sc:contextual}

String diagrams allow graphical reasoning on arrows of (symmetric) monoidal categories. In particular,  graphical proofs are significantly simpler, since the diagrammatic representation implicitly embodies the laws of strict monoidal categories (Table~\ref{fig:freestricmmoncatax}). In this section, we show that tape diagrams allow  the same kind of graphical reasoning as string diagrams. However, this fact is not completely obvious because of the peculiar role played by $\per$ in tape diagrams.

The usual way of reasoning through string diagrams is based on monoidal theories, namely a signature plus a set of axioms: either equations or inequations. Similarly a \emph{tape theory} is a pair $(\sign, \basicR)$ where $\sign$ is a monoidal signature (or by Theorem~\ref{thm:equivalentsignature} even a rig signature) and $\basicR$ is a set of axioms, namely a set of pairs of tapes with same domain and codomain. Hereafter, we think of each pair $(\t_1,\t_2)$ as an inequation $\t_1 \leq \t_2$, but the results that we develop in this section trivially hold  also for equations: it is enough to add in $\basicR$ a pair $(\t_2,\t_1)$ for each $(\t_1, \t_2) \in \basicR$.

In the following, we write $\t_1 \basicR \t_2$ for $(\t_1,\t_2)\in \basicR$ and $\precongB$ for the smallest precongruence (w.r.t. $\piu$, $\per$ and $;$) generated by $\basicR$, i.e.\ the relation inductively generated as 
\[
\begin{array}{ccc}
\inferrule*[right=($\basicR$)]{\t_1 \basicR \t_2}{\t_1 \precongB \t_2}
&
\inferrule*[right=($r$)]{-}{\t \precongB \t}
&
\inferrule*[right=($t$)]{\t_1 \precongB \t_2 \quad \t_2 \precongB \t_3}{\t_1 \precongB \t_3} \\
\inferrule*[right=($\piu$)]{\t_1 \precongB \t_2 \quad \s_1 \precongB \s_2}{\t_1\piu\s_1 \precongB \t_2 \piu \s_2} &
\inferrule*[right=($\per $)]{\t_1 \precongB \t_2 \quad \s_1 \precongB \s_2}{\t_1\per \s_1 \precongB \t_2 \per \s_2}
&
\inferrule*[right=($;$)]{\t_1 \precongB \t_2 \quad \s_1 \precongB \s_2}{\t_1;\s_1 \precongB \t_2;\s_2}
\end{array}\]

By enriching $\CatTape$ with $\precongB$, we obtain a \emph{preorder enriched rig category} (namely, $\piu$, $\per$ and $;$ are monotone) that we denote hereafter by 
$\PreCatTapeI{\basicR}$.
For an arbitrary category $\Cat{C}$ enriched over a preorder $\leq$, one can define a corresponding poset enriched category $\posetification{\Cat{C}}$ by quotienting the homsets of $\Cat{C}$ by the equivalence relation $\sim$ defined as $\sim  \defeq  \leq \cap \geq$. Moreover, if $\Cat{C}$ is a preorder enriched rig category, then $\posetification{\Cat{C}}$ is a \emph{poset enriched rig category}. Particularly relevant for this paper will be $\CatTapeI{\basicR}$. Similarly, for a monoidal theory $(\sign, \basicR)$, one can construct preorder and poset enriched monoidal categories, hereafter denoted by $\Cat{C}_{\sign,\basicR}$ and $\posetification{\Cat{C}_{\sign,\basicR}}$, respectively.

Now, given two tape diagrams $\s$ and $\t$, one would like to prove that $\s \precongB \t$ through some graphical manipulation involving the axioms in $\basicR$ and the one in Figure~\ref{fig:tapesax}.  Unfortunately, this is not completely obvious with tapes, as illustrated by the following example.

\begin{example}\label{ex:zxcontext} Consider the ZX-calculus mentioned in Example~\ref{ex:zxcalculus} and let $\basicR$ be the set consisting of the following axioms
    (which are part of a larger set of axioms stating that $\qState{0}$ and $\qState{1}$ form an orthonormal basis, see Figure \ref{fig:orthonormal}).

\begin{center}
        \mylabel{ax:onb1}{id}
        \mylabel{ax:onb2}{$\emptyset$}
        \begin{tabular}{ccc}
            $
    \InputIfFileExists{zx/onb1_left.tikz}{}{\input{./tikz/zx/onb1_left.tikz}}
 \axeq{\text{id}} 
    \InputIfFileExists{zx/onb1_right.tikz}{}{\input{./tikz/zx/onb1_right.tikz}}
$ & $\quad$ & $
    \InputIfFileExists{zx/onb2_left.tikz}{}{\input{./tikz/zx/onb2_left.tikz}}
 \axeq{\emptyset} 
    \InputIfFileExists{zx/onb2_right.tikz}{}{\input{./tikz/zx/onb2_right.tikz}}
$
        \end{tabular}
    \end{center}
    The derivation below illustrates the behaviour of $CU$ when the control qubit is in state $\qState{0}$.\begin{align}\label{ex:contextual}
        \begin{split}
            
    \InputIfFileExists{zx/contextual_proof/step1.tikz}{}{\input{./tikz/zx/contextual_proof/step1.tikz}}
 &\axeq{\refeq{ax:diagnat}} 
    \InputIfFileExists{zx/contextual_proof/step2.tikz}{}{\input{./tikz/zx/contextual_proof/step2.tikz}}
 \axeq{\ast_1}  
    \InputIfFileExists{zx/contextual_proof/step3.tikz}{}{\input{./tikz/zx/contextual_proof/step3.tikz}}
 \axeq{\ast_2}  
    \InputIfFileExists{zx/contextual_proof/step4.tikz}{}{\input{./tikz/zx/contextual_proof/step4.tikz}}
 \\
            &\axeq{\refeq{ax:bangnat}} 
    \InputIfFileExists{zx/contextual_proof/step5.tikz}{}{\input{./tikz/zx/contextual_proof/step5.tikz}}
 \axeq{\refeq{ax:diagun}, \refeq{ax:codiagun}} 
    \InputIfFileExists{zx/contextual_proof/step6.tikz}{}{\input{./tikz/zx/contextual_proof/step6.tikz}}

        \end{split}
    \end{align}
    To prove steps $(\ast_1)$ and $(\ast_2)$ it is necessary to decompose the diagram via $\per$. For example, $(\ast_2)$ is 
    \begin{align*}
        
    \InputIfFileExists{zx/contextual_deriv/step1.tikz}{}{\input{./tikz/zx/contextual_deriv/step1.tikz}}

        \axeq{\text{def. }\per}
        
    \InputIfFileExists{zx/contextual_deriv/step2_left.tikz}{}{\input{./tikz/zx/contextual_deriv/step2_left.tikz}}

        \per    
        
    \InputIfFileExists{zx/contextual_deriv/step2_right.tikz}{}{\input{./tikz/zx/contextual_deriv/step2_right.tikz}}

        \stackrel{\eqref{ax:onb2}}{=_{\basicR}}
        
    \InputIfFileExists{zx/contextual_deriv/step3_left.tikz}{}{\input{./tikz/zx/contextual_deriv/step3_left.tikz}}

        \per    
        
    \InputIfFileExists{zx/contextual_deriv/step3_right.tikz}{}{\input{./tikz/zx/contextual_deriv/step3_right.tikz}}

        \axeq{\text{def. }\per}
        
    \InputIfFileExists{zx/contextual_deriv/step4.tikz}{}{\input{./tikz/zx/contextual_deriv/step4.tikz}}

    \end{align*}
\end{example}

The proof above is not entirely graphical because of the decomposition via $\per$. In the following we show that one can easily avoid this inconvenience by taking the right whiskering, for all monomials $U$, of each of the axioms in $\basicR$. 
In other words, rather than $\basicR$, we consider the following set of axioms
\begin{equation}\label{eq:WA}\wiskbasicR = \{(\RW{U}{\t_1}, \RW{U}{\t_2}) \mid (\t_1,\t_2)\in \basicR \text{ and } U\in \sort^\star\}\end{equation}
and we write $\WprecongBA$ for the smallest precongruence (w.r.t. $\piu$ and $;$) generated by $\wiskbasicR$, i.e.\ the relation inductively defined as 
\[
\begin{array}{ccccc}
\inferrule*[right=($\wiskbasicR$)]{\t_1 \wiskbasicR \t_2}{\t_1 \WprecongBA \t_2}
&
\inferrule*[right=($r$)]{-}{\t \WprecongBA \t}
&
\inferrule*[right=($t$)]{\t_1 \WprecongBA \t_2 \quad \t_2 \WprecongBA \t_3}{\t_1 \WprecongBA \t_3}
&
\inferrule*[right=($\piu$)]{\t_1 \WprecongBA \t_2 \quad \s_1 \WprecongBA \s_2}{\t_1\piu\s_1 \WprecongBA \t_2 \piu \s_2}
&
\inferrule*[right=($;$)]{\t_1 \WprecongBA \t_2 \quad \s_1 \WprecongBA \s_2}{\t_1;\s_1 \WprecongBA \t_2;\s_2}
\end{array}
\]
Observe that in the above definition we do not close $\WprecongBA$ by $\per$. Yet, as stated by the following theorem, $\WprecongBA$ coincides with $\precongB$.

\begin{theorem}\label{thm:contextual}
For all tapes $\t_1,\t_2$, $\t_1  \precongB \t_2$ if and only if $\t_1  \WprecongBA \t_2$.
\end{theorem}
\begin{proof}
	To prove that $\precongB \subseteq  \WprecongBA $, first observe that $\basicR \subseteq \wiskbasicR$ by \eqref{eq:whisk:uno}. Thus, to conclude it is enough to show that $\WprecongBA$ is closed under $\per$, i.e.\ that if   $\t_1 \WprecongBA \t_2$ and  $\s_1 \WprecongBA \s_2$, then  $\t_1\per \s_1 \WprecongBA \t_2 \per \s_2$. But this is easy using the definition of $\per$ and the algebra of whiskering.
	
	To prove that $ \WprecongBA \subseteq  \precongB $, it is enough to show that $\wiskbasicR \subseteq \precongB$: if $\t_1 \wiskbasicR \t_2$, there exists $(\t_1', \t_2') \in \basicR$ such that $\RW{U}{\t_i'} = \t_i$ for some monomial $U$. Now, it is easy to see that $\RW{U}{\t_i} = \t_i \per \id{U}$, thus by rules ($r$) and ($\per$), $\t_1 \precongB \t_2$.
\end{proof}

The proof in Example~\ref{ex:contextual} can then be carried out completely diagrammatically by virtue of Theorem~\ref{thm:contextual}. 
Indeed, steps $(\ast_1)$ and $(\ast_2)$ become trivial by means of the whiskered axioms 
\[ 
    \InputIfFileExists{zx/w_onb1_left.tikz}{}{\input{./tikz/zx/w_onb1_left.tikz}}
 \axeq{\text{Wid}} 
    \InputIfFileExists{zx/w_onb1_right.tikz}{}{\input{./tikz/zx/w_onb1_right.tikz}}
 \qquad 
    \InputIfFileExists{zx/w_onb2_left.tikz}{}{\input{./tikz/zx/w_onb2_left.tikz}}
 \axeq{\text{W}\emptyset} 
    \InputIfFileExists{zx/w_onb2_right.tikz}{}{\input{./tikz/zx/w_onb2_right.tikz}}
 \]

\begin{remark}\label{rem:tapes universal language}
Tape diagrams for finite coproduct rig categories, namely rig categories where $\piu$ is a coproduct and $\zero$ is an initial object, can be obtained by discarding from the definition of $\CatTape$ the comonoids $(\diag{U},\bang{U})$. Similarly, by ignoring the monoids $(\codiag{U},\cobang{U})$, one obtains tape diagrams for finite product rig categories. All the results in the current section hold verbatim for these constructions. However, our focus on finite biproduct categories is justified by the fact that, thanks to Theorem \ref{thm:equivalentsignature} which only holds when $\piu$ is a biproduct, tape diagrams provide a universal language for fb rig categories: all \emph{rig} signatures generate fb rig categories whose arrows are tapes. In the next section, we illustrate several results that rely on the finite biproduct structure of $\CatTape$.
\end{remark}

 \section{Tapes as Matrices}\label{sc:matrix}

Like any category with  finite biproducts, $\CatTape$ is enriched over $\CMon$, the category of commutative monoids. For all polynomials $P,Q$, the homset $\CatTape[P,Q]$ carries a commutative monoid defined as
\begin{equation}\label{eq:+zero}
\begin{array}{rcccr}
\t_1 + \t_2 &\defeq& \Tconv{\t_1}{\t_2}{P}{Q} &\qquad \qquad \qquad& \text{(i.e. $\t_1 + \t_2 \defeq \diag P ; (\t_1 \piu \t_2) ; \codiag Q$)} \\
\zerotape_{P,Q} &\defeq& \TPolycounitunit{P}{Q} &\qquad \qquad \qquad& \text{(i.e. $\zerotape_{P,Q} \defeq \bang P ; \cobang Q$)}
\end{array}\end{equation}
for all $\t_1,\t_2 \colon P \to Q$.
This structure distributes not only over $;$
but also with respect to $\per$. \begin{proposition}\label{prop:dist+}
	Let $\t_1, \t_2\colon P \to Q$ and $\s$ of the appropriate type. It holds that
	\smallskip
	
	\begin{minipage}{0.5\linewidth}
		\begin{enumerate}
			\item $(\t_1 + \t_2) ; \s = (\t_1 ; \s) + (\t_2 ; \s)$
			\item $ \s ; (\t_1 + \t_2) = (\s ; \t_1) + (\s ; \t_2) $
			\item $\zerotape ; \s = \zerotape = \s ; \zerotape $
		\end{enumerate}
	\end{minipage}
	\begin{minipage}{0.5\linewidth}
		\begin{enumerate}  \setcounter{enumi}{3}
			\item $(\t_1 + \t_2) \per \s= (\t_1 \per \s) + (\t_2 \per \s)$
			\item $ \s \per (\t_1 + \t_2) = (\s \per \t_1) + (\s \per \t_2 )$
			\item $\zerotape \per \s= \zerotape = \s \per \zerotape$
		\end{enumerate}
	\end{minipage}
\end{proposition}
In this section, we illustrate how such  enrichment can be exploited to define a matrix calculus of tapes. First of all we need to identify the entries of these matrices. 
Let $\monomial$ be the full subcategory of  $\CatTape$ whose objects are just monomials (i.e.\ unary sums). It is immediate to see that  $\monomial$ is 
a ssm category (w.r.t. $\per$), because if $U$ and $V$ are two monomials, then so is $U \per V$ by definition of $\per$. It is also clearly enriched over $\CMon$. 
Here are two examples of morphisms in $\monomial$. \\
\begin{minipage}{0.38\textwidth}
	\begin{equation}\label{ex:mnm1}
		
    \InputIfFileExists{tapes/examples/monomial.tikz}{}{\input{./tikz/tapes/examples/monomial.tikz}}

	\end{equation}
\end{minipage}
\hfill
\begin{minipage}{0.55\textwidth}
	\begin{equation}\label{ex:mnm2}
		
    \InputIfFileExists{tapes/examples/monomial2.tikz}{}{\input{./tikz/tapes/examples/monomial2.tikz}}

	\end{equation}
\end{minipage} \\
Notice that \eqref{ex:mnm1} is in fact $\tape c + \tape c + \tape d$, with $c,\, d \in \CatString$. Now, by definition, a morphism in $\monomial$ is a tape of $\CatTape$ with only one `input' and one `output', but in between these two it can be arbitrarily complicated, like in~\eqref{ex:mnm2} above. However, it turns out that every tape in $\monomial$ can be written as the diagram in~\eqref{ex:mnm1}, that is, as a local sum $\sum_i \tape{c_i}$. This is a consequence of the fact that, as we will see in Corollary~\ref{cor:monomials in tape are isomorphic to multisets of string diagrams},  $\monomial$ is isomorphic to $\fCMon{\CatString}$: the free $\CMon$-enriched category generated by $\CatString$.

\begin{definition}
	Let $\Cat C$ be any category. The \emph{free $\CMon$-enriched category generated by $\Cat C$}, denoted as $\fCMon {\Cat C}$, is the category whose objects are those of $\Cat C$, while $\fCMon{\Cat C}[A,B]$ is the free commutative monoid generated by $\Cat C[A,B]$:
a morphism in $\fCMon{\Cat C}[A,B]$ is a finite multiset of morphisms in $\Cat C[A,B]$. We write multisets as $\multiset{a_1,\dots,a_n}$ where the $a_i$ are not necessarily distinct.
The identity for $A$ in $\fCMon{\Cat C}$ is $\multiset{\id{A}^{\Cat C}}$, while if $f \!\colon\! A \!\to\! B$ and $g \!\colon\! B \!\to\! C$ are morphisms of $\fCMon{\Cat C}$, then $f ; g \defeq \multiset{ a ; b \mid a \in f,\, b \in g}$ ($a ; b$ has multiplicity equal to the product of the multiplicities of $a$ in the multiset $f$ and $b$ in the multiset $g$). Addition of multisets is given by union, with the empty multiset being the neutral element.
\end{definition}

The isomorphism between  $\monomial$ and $\fCMon{\CatString}$ assigns to each arrow $\m \colon U \to V$ in $\monomial$, hereafter referred to as \emph{monomial tapes}, a multiset whose elements are the morphisms of $\CatString$ appearing in a path in the diagrammatic representation of $\m$. For instance, the monomial tape in~\eqref{ex:mnm1} corresponds to the multiset $\multiset{c,c,d}$, while the one in~\eqref{ex:mnm2} to $\multiset{{c;e},\,{c;f},\,{c;g},\,{c;e},\,{c;f},\,{c;g},\, {d;e},\,{d;f},\,{d;g}}$.
Vice versa, every multiset $\multiset{c_1,\dots,c_n}$ of arrows in $\CatString$ corresponds to the monomial tape $\m = \sum_{i=1}^n \tape{c_i}$.

We can now consider an arbitrary tape $\t \colon \Piu[i=1][n]{U_i} \to \Piu[j=1][m]{V_j}$. One can represent $\t$ as a $m \times n$ matrix $\matr \t$ whose $(j,i)$ entry (row $j$, column $i$), is the monomial tape
\begin{equation}\label{eq:tji}
\t_{ji} \defeq \Bigl( 
\begin{tikzcd}
U_i \ar[r,"\inj i"] & \PiuL[k=1][n]{U_k} \ar[r,"\t"] & \PiuL[k=1][m]{V_k} \ar[r,"\proj j"] & V_j
\end{tikzcd}
\Bigr)
\end{equation}
where $\inj i \defeq  \Piu[k=1][i-1]{\cobang{U_k}} \piu \id{U_i} \piu \Piu[k=i+1][n]{\cobang{U_k}} $ and $\proj j$ is defined dually.
Hence we can associate to $\t$ a $m \times n$ matrix $\FF(\t)$ whose entries are the multisets corresponding to the monomial tapes $\t_{ji}$. 

\begin{example}\label{ex:tapetomat}
	Consider again $\t$ and $\s$ from Example~\ref{ex:diagWhisk}, then \[
	\FF(\t) = 
	\begin{pNiceMatrix}[first-col,first-row]
	\rotatebox{90}{$\Lsh$} & U & V \\
	W & \multiset{\EntryD{d}{e}} & \multiset{\EntryT{c}{d}{e}} \\
	Z & \multiset{\EntryU{d}} & \multiset{\EntryD{c}{d}}
	\end{pNiceMatrix}
	\quad
	\text{and}
	\quad
	\FF(\s) = 
	\begin{pNiceMatrix}[first-col,first-row]
	\rotatebox{90}{$\Lsh$} & U' & V' \\
	W' & \multiset{\EntryD{c'}{d'}} & \multiset{\EntryU{d'}} \\
	Z' & \multiset{\EntryT{c'}{d'}{e'}} & \multiset{\EntryD{d'}{e'}}
	\end{pNiceMatrix}
	\]
\end{example}

This correspondence forms an isomorphism that we now illustrate in detail. Given a $\CMon$-enriched category $\Cat S$, one can form the \emph{biproduct completion}~\cite{coecke2017two,mac_lane_categories_1978} of $\Cat S$, denoted as $\Mat{\Cat S}$. Its objects are formal $\piu$'s of objects of $\Cat S$, while a morphism $M \colon \Piu[k=1][n]{A_k} \to \Piu[k=1][m]{B_k}$ is a $m \times n$ matrix where $M_{ji} \in \Cat S[A_i,B_j]$. Composition is given by matrix multiplication, with the addition being the plus operation on the homsets (provided by the enrichment) and multiplication being composition. The identity morphism of $\Piu[k=1][n]{A_k}$ is given by the $n \times n$ matrix $(\delta_{ji})$, where $\delta_{ji} = \id{A_j}$ if $i=j$, while if $i \neq j$, then $\delta_{ji}$ is the zero morphism of $\Cat S[A_i,A_j]$.

\begin{theorem}\label{thm:F_2(C) isomorphic to biproduct completion}
	Let $\Cat C$ be any category. Then $F_2(\Cat C) \cong \Mat{\fCMon{\Cat C}}$ as fb categories.
\end{theorem}
\begin{proof}
	By~\cite[Exercises VIII.2.5-6]{mac_lane_categories_1978}, we have a pair of adjunctions
	\[
	\begin{tikzcd}
		\CAT \ar[r,bend left,"{\fCMon{(-)}}"name=F]   & \CMonCat \ar[l,bend left,"{}"name=G] \ar[r,bend left,"{\Mat{-}}"name=H] & \fbCat \ar[l,bend left,""name=K]
		\arrow[from=F,to=G,draw=none,"\perp"description]
		\arrow[from=H,to=K,draw=none,"\perp"description]
	\end{tikzcd}
	\]
	where the right adjoints are forgetful functors. Since adjunctions compose, the composite functor $\Mat{\fCMon{(-)}}$ is left adjoint to the forgetful functor $U \colon \fbCat \to \CAT$, as is $F_2$: hence they are naturally isomorphic.
This makes $F_2(\Cat C)$ and $\Mat{\fCMon{\Cat C}}$ isomorphic as fb categories. Explicitly, the functor $\FF \colon F_2(\Cat C) \to \Mat{\fCMon{\Cat C}} $
	is the identity on objects and it is defined on morphisms by induction:
	\begin{itemize}
		\item\label{itemize:definition of F} $\FF(\tapeFunct{c}) = 
		\begin{pmatrix}
			\multiset{c}
		\end{pmatrix}$ (the $1 \times 1$ matrix of the multiset consisting of one copy of $c$)
		\item $\FF(\id{})=\id{}$, $\FF(\t_2 \circ \t_1) = \FF(\t_2) \circ \FF(\t_1)$
		\item Given $A,B \in \Cat C$, $\FF(\symm{A}{B})=
		\begin{pmatrix}
			\emptyset & \multiset{\id B} \\
			\multiset{\id A} & \emptyset
		\end{pmatrix}$
		of size $2 \times 2$,
		\begin{align*}
			\FF(\diag A) &= 
			\begin{pmatrix}
				\multiset{\id A} \\ \multiset{\id A}
			\end{pmatrix}
			& \FF(\codiag A) &=
			\begin{pmatrix}
				\multiset{\id A} & \multiset{\id A}
			\end{pmatrix} \\
			\FF(\bang A) &= \text{empty matrix of size $0 \times 1$} & \FF(\cobang A) &=  \text{empty matrix of size $1 \times 0$}
		\end{align*}
		\item For $f_1 \colon \PiuL[i=1][n]{A_i} \to \PiuL[i=1][n']{A_i'}$ and $f_2 \colon \PiuL[i=1][m]{B_j} \to \PiuL[i=1][m']{B_j'}$, 
		\[
		\FF(f_1 \piu f_2) = 
		\begin{pmatrix}
			\FF(f_1) & \emptyset_{n' \times m} \\
			\emptyset_{m' \times n} & \FF(f_2)
		\end{pmatrix}\qedhere
		\]
	\end{itemize}
\end{proof}

\subsection{Kronecker product in $\Mat{\fCMon{\CatString}}$}
If  $(\Cat C, \per_{\Cat{C}}, \uno_{\Cat{C}})$ is a ssm category, then we can define a monoidal product on  $\fCMon{\Cat C}$:
\begin{equation}\label{eq:monoidalproductinside}
f \per_{\fCMon{\Cat C}} g \defeq \multiset{ a \per_{\Cat C} b \mid a \in f,\, b \in g } \colon A \per_{\Cat C} B \to A' \per_{\Cat C} B'.
\end{equation}
In turn, this allows us to define a monoidal product in $\Mat{\fCMon{\Cat C}}$ \emph{à la} Kronecker. We will denote it as $\kron$. On objects it is given as
$\bigl(\Piu[i=1][n]{A_i} \bigr) \kron \bigl( \Piu[j=1][m]{B_j} \bigr) \defeq \Piu[i=1][n]{\Piu[j=1][m]{A_iB_j}}$.
If $M \colon \Piu[i=1][n]{A_i} \to \Piu[i'=1][n']{A_{i'}'}$ and $N \colon \Piu[j=1][m]{B_j} \to \Piu[j'=1][m']{B_{j'}'}$, then $M \kron N$ is the matrix of size $n' m' \times n  m$ defined as in the usual Kronecker product of matrices (with $\per_{\fCMon{\Cat C}}$ playing the role of multiplication).

\begin{theorem}\label{thm:tensor of matrices is Kronecker}
	Let $\Cat C$ be an ssm category. Then $\Mat{\fCMon{\Cat C}}$ is a fb rig category, with tensor given by $\kron$, and  $F_2(\Cat C) \cong \Mat{\fCMon{\Cat C}}$ as fb rig categories.
\end{theorem}
\begin{proof}
	One can define on $F_2(\Cat C)$ a monoidal product $\otimes$ distributing on $\piu$ in a completely analogous way of $\CatTape$, making it a fb rig category. This means that $\Mat{\fCMon{\Cat C}}$ inherits the structure of a strict fb rig category from $F_2(\Cat C)$ via the isomorphism $\FF$ of Theorem~\ref{thm:F_2(C) isomorphic to biproduct completion}, by simply setting $M {\otimes} N \defeq \FF(\FF^{-1}(M) \perT \FF^{-1}(N))$ and similarly for the rest of the rig structure. This immediately makes $\FF$ an isomorphism of fb rig categories. Moreover, one can prove that $\otimes = \kron$ by means of direct calculations. 
\end{proof}

We conclude by applying our general result to $\Cat C = \CatString$.

\begin{corollary}\label{thm:tapes as matrices}$\CatTape \cong \Mat{\fCMon\CatString}$ as fb rig categories.
\end{corollary}
\begin{corollary}\label{cor:monomials in tape are isomorphic to multisets of string diagrams}
	$\monomial \cong \fCMon\CatString$ as ssm categories. \end{corollary}

\begin{example} Recall from Example~\ref{ex:diagPer} the tape diagram $\t \per \s$. The corresponding matrix, which can be computed as the Kronecker product of $\FF(\t)$ and $\FF(\s)$ in Example~\ref{ex:tapetomat}, is illustrated below. 
	\[
	\FF(\t \per \s) = 
	\begin{pNiceMatrix}[first-col,first-row]
	\rotatebox{90}{$\Lsh$} & UU' & UV' & VU' & VV' \\
	WW' & \multiset{\EntryDxD{d}{e}{c'}{d'}} & \multiset{\EntryDxU{d}{e}{d'}} & \multiset{\EntryTxD{c}{d}{e}{c'}{d'}} & \multiset{\EntryTxU{c}{d}{e}{d'}} \\
	WZ' & \multiset{\EntryDxT{d}{e}{c'}{d'}{e'}} & \multiset{\EntryDxD{d}{e}{d'}{e'}} & \multiset{\EntryTxT{c}{d}{e}{c'}{d'}{e'}} & \multiset{\EntryTxD{c}{d}{e}{d'}{e'}} \\
	ZW' & \multiset{\EntryUxD{d}{c'}{d'}} & \multiset{\EntryUxU{d}{d'}} & \multiset{\EntryDxD{c}{d}{c'}{d'}} & \multiset{\EntryDxU{c}{d}{d'}} \\
	ZZ' & \multiset{\EntryUxT{d}{c'}{d'}{e'}} & \multiset{\EntryUxD{d}{d'}{e'}} & \multiset{\EntryDxT{c}{d}{c'}{d'}{e'}} & \multiset{\EntryDxD{c}{d}{d'}{e'}} \\
	\end{pNiceMatrix}
	\]
\end{example}

\subsection{Poset Enrichment}\label{sec:preorderenrichment}
We saw that $\CatTape$ is isomorphic to a category of matrices in Corollary~\ref{thm:tapes as matrices}. Here we want to extend the isomorphism to the case where the category of tapes is equipped with a poset generated by a tape theory. We shall focus on a particular case of tape theory, namely the one arising from a monoidal theory $(\sign,\basicR)$ by simply adding the four tape axioms in Figure~\ref{fig:rel axioms}. These axioms force $\codiag{U}$ to be left adjoint to $\diag{U}$ and $\cobang{U}$ to $\bang{U}$. Moreover, they make $+$, as defined in~\eqref{eq:+zero}, idempotent, thus a join $\sqcup$, with $\zerotape$ being its bottom element $\bot$ (see Figure ~\ref{fig:derivationUPROPS}).

For a monoidal theory $(\sign,\basicR)$, we call \emph{the generated tape theory} $(\sign,\tilde \basicR)$ the tape theory where $\tilde\basicR$ consists of the four inequalities in Figure~\ref{fig:rel axioms} together with all those pairs $(\tape c, \tape d)$ whenever $(c,d) \in \basicR$. We can then form the category $\CatTapeI{\tilde\basicR}$ as explained in Section~\ref{sc:contextual}. 
It turns out that $\CatTapeI{\tilde\basicR}$ is isomorphic to a certain category of matrices that we are going to illustrate next.

Let $(X,\le)$ be a preordered set and $S \subseteq X$. The \emph{downward closure} of $S$ is defined to be the set $\downward S = \{ x \in X \mid \exists s \in S \ldotp x \le s \}$. $S$ is said to be \emph{downward closed} if $S = \downward S$; a downward closed set $A \subseteq X$ is said to be \emph{finitely generated} if there exists a finite set $B\subseteq X$ such that $A=\downward B$. Consider now  an arbitrary category $\Cat C$ enriched over a preorder $\le_{\Cat C}$. We can form a new category $\downsetification{\Cat C}$ whose objects are $Ob(\Cat C)$ while a morphism of type  $ X \to Y$ in $\downsetification{\Cat C}$ is a finitely generated downward closed subset of $\Cat C[X,Y]$. $\downsetification{\Cat C}[X,Y]$ is partially ordered by inclusion, has all finite joins given by unions, composition is defined using the composition in $\Cat C$ and this makes $\downsetification{\Cat C}$ a finite join-semilattice enriched category. For a monoidal theory $(\sign,\basicR)$, one can construct the poset enriched monoidal category $\posetification{\Cat{C}_{\sign,\basicR}}$ as in Section~\ref{sc:contextual} and, since $\downsetification{\posetification{\Cat{C}_{\sign,\basicR}}}$ is enriched over finite join semilattices (and thus commutative monoids), one can take its biproduct completion. 
\begin{theorem}\label{thm:tapes as matrices poset version}
	$\CatTapeI{\tilde\basicR} \cong \Mat{\downsetification{\posetification{\Cat{C}_{\sign,\basicR}}} }$ as poset-enriched fb rig categories.
\end{theorem}
\begin{proof}
	Let $\Cat C$ be preorder-enriched: then we can equip $\fCMon{\Cat C}[X,Y]$ with the Egli-Milner preorder 
	\[
	\multiset{c_1,\dots,c_n} \le^{EM} \multiset{d_1,\dots,d_m} \iff \forall i \ldotp \exists j \ldotp c_i \le_{\Cat C} d_j,
	\]
	making  $\fCMon{\Cat C}$, and in turn $\Mat{\fCMon{\Cat{C}}}$, preorder-enriched.
We can also enrich $F_2(\Cat C)$ in a way that the isomorphism $\FF$ of Theorem~\ref{thm:F_2(C) isomorphic to biproduct completion} becomes  preorder-enriched. To do so we add on the homsets of $F_2(\Cat C)$ the precongruence, with respect to composition, $\piu$ and $\per$, generated by the following axioms:
	\[
	f \preceq_{\Cat C} g \Rightarrow \tape{f} \preceq \tape{g} \qquad 
	\diag A ; \codiag A \preceq \id A \qquad 
	\id{A \piu A} \preceq {\codiag{A} ; \diag A} \qquad 
	\bang A ; \cobang A \preceq \id A \qquad
	\id\zero \preceq {\cobang{A} ; \bang{A}}
	\]
(The last four axioms are the generalised version of Figure~\ref{fig:rel axioms} for $F_2(\Cat C)$.) 
	One can then prove that $\FF$ preserves and reflects the inequalities of $F_2(\Cat C)$. Taking now $\posetification{F_2(\Cat C)}$ and $\posetification{\Mat{\fCMon{\Cat C} } } $, we obtain a \emph{poset}-enriched isomorphism. Observe that $\posetification{\Mat{\fCMon{\Cat C}}} \cong \Mat{\posetification{ (\fCMon{\Cat C}) } }$, because the order on matrices is entry by entry, and that $\posetification{(\fCMon{\Cat C})} \cong \downsetification{\Cat C}$, because for all $S,T \in \fCMon{\Cat C}[X,Y]$, $S \le_{EM} T$ iff $\downward{S} \subseteq \downward{T}$.
	
	If we apply this to $\Cat C={\Cat C}_{\sign,\basicR}$, then we have that  $\posetification{F_2(\Cat C)} = \CatTapeI{\tilde{\basicR}}$ and we obtain
	$\CatTapeI{\tilde{\basicR}} \cong \Mat{\downsetification{{\Cat C}_{\sign,\basicR}}}$. We conclude by simply observing that $\downsetification{{\Cat C}_{\sign,\basicR}} \cong \downsetification{\posetification{\Cat{C}_{\sign,\basicR}}}$.
\end{proof}

A consequence of Theorem~\ref{thm:tapes as matrices poset version} is that we can characterise the partial order of $\CatTapeI{\tilde\basicR}[P,Q]$ as follows:
for arbitrary monomials $\sum_{h=1}^n \tape{c_h}$ and $\sum_{k=1}^m \tape{d_k}$ we define $\leq_{\basicR}^{EM} $ as
\begin{equation*}
\sum_{h=1}^n \tape{c_h} \leq_{\basicR}^{EM} \sum_{k=1}^m \tape{d_k} \; \iff \; \forall h \ldotp \exists k \ldotp c_h \leq_{\basicR} d_k
\end{equation*}
and we extend it to arbitrary tapes $\t,\s \colon \Piu[i]{U_i} \to \Piu[j]{V_j}$ as
\begin{equation*}
\t \leq_{\basicR}^{EM} \s \; \iff \; \forall i,j \ldotp \t_{ji} \leq_{\basicR}^{EM} \s_{ji} 
\end{equation*}
where $\t_{ji}$ and $\s_{ji}$ are the monomials in~\eqref{eq:tji}.

\begin{corollary}\label{cor:poset}
$\t \le_{\tilde{\basicR}} \s $ iff $\t \leq_{\basicR}^{EM} \s$.
\end{corollary}

\begin{remark}\label{remark I empty}
	Consider the case of $\mathbb I$ being $\zero$, the empty set. Then Theorem~\ref{thm:tapes as matrices poset version} asserts that $\CatTapeI{\tilde\zero}$, which is simply the poset enriched version of $\CatTape$ generated by the inequalities of Figure~\ref{fig:rel axioms}, is isomorphic to the biproduct completion of the free \emph{join-semilattice enriched} category generated by $\CatString$. The latter is the same as $\fCMon\CatString$ except that its homsets are sets, rather than multisets, of arrows of $\CatString$. Indeed, since the preorder on  $\Cat{C}_{\sign,\zero}[X,Y]$ is discrete, we have that the finitely generated downward closed subsets of $\Cat{C}_{\sign,\zero}[X,Y]$ are exactly finite subsets of $\Cat{C}_{\sign,\zero}[X,Y]$. \end{remark}

\begin{example}[$\CatTapeCB$]\label{ex:TAPECB}\label{ex:poTCB}We recall from \cite{GCQ} $\CB$: the  cartesian bicategory freely generated by a single sorted, i.e. $\sort=\{A\}$, monoidal signature $\sign$. In a nutshell, $\CB$ can be described as $\posetification{\Cat{C}_{\Gamma,\basicR}}$ where the monoidal signature $\Gamma$ is
$\Sigma \cup \{ \cocopier{A} \colon A \perG A \to A, \;\; \codischarger{A} \colon \unoG \to A, \;\; \copier{A} \colon A \to A \perG A, \;\; \discharger{A} \colon A \to \unoG\}$ and the set of axioms $\basicR$ consists of those of cartesian bicategories, see Figure~\ref{fig:cb axioms}.
There we draw, for all $U\in \sort^\star$\footnote{The coherence axioms of (co)monoids (in Figure~\ref{fig:fbcoherence}) provide a recipe to define inductively, $\cocopier{U}$, $\codischarger{U}$, $\copier{U}$, $\discharger{U}$, for all $U\in \sort^\star$.}, $\cocopier{U}$, $\codischarger{U}$, $\copier{U}$, $\discharger{U}$ as
\[\CBcocopier{U} \qquad \CBcodischarger{U} \qquad \CBcopier{U} \qquad \CBdischarger{U}\]
In the generated tape theory $(\sign, \tilde\basicR)$, the set of axioms $\tilde\basicR$ consists of those in Figures~\ref{fig:rel axioms} and~\ref{fig:cb axioms}.
In the next section we will show that $\CatTapeIGamma{\tilde\basicR}$ provides a complete calculus for relations. An essential ingredient is Corollary~\ref{cor:poset} above. Hereafter we will write $\CatTapeCB$ for $\CatTapeIGamma{\tilde\basicR}$ and denote the poset of $\CB$ as $\minorinCB$ while the one in $\CatTapeCB$ as $\minorinTCB$.
\end{example}

\begin{example}[$\sqcup$-props]\label{ex:mu-props} By exploiting a mixture of algebraic and diagrammatic syntax, the authors of \cite{DBLP:conf/fossacs/BoisseauP22} introduced $\sqcup$-props to model piecewise linear systems (e.g.\ diodes). Tape diagrams, instead, provide a purely graphical calculus for $\sqcup$-props.

Consider a single sorted, i.e. $\sort= \{A\}$, monoidal signature $\sign$ and take $\basicR$ to be the empty set $\zero$ (no axioms). The objects of $\CatTapeI{\tilde\zero}$ can easily be seen to be in one-to-one correspondence with words of natural numbers: $\Piu[i][n]{A^{m_i}}\mapsto m_1m_2 \dots m_n$. In particular monomials are natural numbers and $\per$ on monomials is just addition. Thus, by taking $\monomial$, the full subcategory of $\CatTapeI{\tilde\zero}$ where objects are monomials, one obtains a \emph{prop} \cite{Lack2004a, MacLane1965}. It is immediate to see that $\monomial$ is a \emph{$\sqcup$-prop}: every homset carries a join semilattice with bottom (as in \eqref{eq:+zero}), which is preserved by composition and monoidal product (Proposition~\ref{prop:dist+}). 
Most importantly, $\monomial$ is the $\sqcup$-prop freely generated by $\sign$, as defined in \cite{DBLP:conf/fossacs/BoisseauP22}. One can readily see this by means of Theorem~\ref{thm:tapes as matrices poset version} and Remark~\ref{remark I empty}.
Therefore, whenever one is interested in the $\sqcup$-prop freely generated by some monoidal signature $\sign$, one can rather embed it into $\CatTapeI{\tilde\zero}$ and exploit the graphical calculus of tapes (see Figure~\ref{fig:uprop} for an example). Notice that there is a little mismatch with the definition in \cite{DBLP:conf/fossacs/BoisseauP22}, where the freely generated $\sqcup$-prop has as arrows only
\emph{non-empty} finite sets. The empty set is instead denoted in $\monomial$ by $\zerotape$: the presence of $\zerotape$ seems however a feature rather than an issue as illustrated, for instance, by the axiom \eqref{ax:onb2}.
\end{example}

\section{Back to relations}\label{sec:CBPOPL}

The tape axioms introduced in Section \ref{sec:preorderenrichment} forcing  $(\codiag{X}, \cobang{X})$ to be left adjoint to $(\diag{X},\bang{X})$ (Figure~\ref{fig:rel axioms}) suggest a general categorical notion reconciling the two monoidal structures of $\Rel$ (Section \ref{sec:2monREL}).

\begin{definition}\label{def:fbcbrig}
A poset enriched rig category $\Cat{C}$ is said to be a \emph{fb-cb rig category} (or simply fb-cb category for short) if 
\begin{enumerate}
\item $(\Cat{C}, \piu, \zero)$ is a finite biproduct category;
\item $(\Cat{C}, \per, \uno)$ is a cartesian bicategory;
\item the monoid  $(\codiag{X}, \cobang{X})$ of $\piu$ is left adjoint to the  comonoid $(\diag{X},\bang{X})$, i.e.\
\[
\id{\zero} \leq \cobang{X} ; \bang{X} \quad \id{X}  \piu \id X \leq  \codiag{X};\diag{X} \quad \bang{X} ; \cobang{X} \leq \id X \quad \text{and} \quad \diag{X};\codiag{X} \leq \id{X}
\]
\item\label{fbcb coherence conditions} the (co)monoids of both  $(\Cat{C}, \piu, \zero)$ and $(\Cat{C}, \per, \uno)$ satisfy the following coherence axioms:
\[
	\begin{array}{cc}
		\copier{X \piu Y} = (\copier{X} \piu \cobang{XY} \piu \cobang{YX} \piu \copier{Y}) ; (\Idl{X}{X}{Y} \piu \Idl{Y}{X}{Y}) ; (\Idr{X}{Y}{X \piu Y}) & \discharger{X \piu Y} = (\discharger{X} \piu \discharger{Y}) ; \codiag{\uno} \\
		\cocopier{X \piu Y} = (\dr{X}{Y}{X \piu Y}) ; (\dl{X}{X}{Y} \piu \dl{Y}{X}{Y}) ; (\cocopier{X} \piu \bang{XY} \piu \bang{YX} \piu \cocopier{Y}) & \codischarger{X \piu Y} = \diag{\uno} ; (\codischarger{X} \piu \codischarger{Y})
	\end{array}
\]
\end{enumerate}
A \emph{morphism of fb-cb categories} is a poset enriched rig functor that is both a morphism of finite biproduct categories and a morphism of cartesian bicategories.
\end{definition}
We have already seen in Section~\ref{sec:2monREL} that $(\Rel,\piu, \zero)$ is an fb-category and $(\Rel,\per, \uno)$ is a cartesian bicategory. To conclude that $\Rel$ is an fb-cb category is enough to check that conditions (3) and (4) above are satisfied: this is trivial by using the definitions of the two (co)monoids of $\Rel$ in~\eqref{eq:comonoidsREL}.
\begin{remark}\label{rmk:relsem}
By exploiting the structure of fb-cb category, one can express finite unions and intersections in $\Rel$. The reader can easily check that for all relations $R,S\colon X \to Y$, it holds that
\[R\cup S = \diag{X} ; (R \piu S); \codiag{Y} \qquad \{\} = \bang{X}; \cobang{Y} \qquad R\cap S = \copier{X} ; (R \per S) ; \cocopier{Y} \qquad X\per Y = \discharger{X}; \codischarger{Y}\]
\end{remark}

Another example of fb-rig category is $\CatTapeCB$ from Example \ref{ex:TAPECB}. 

\begin{theorem}\label{thm:Tape Cartesian}
$\CatTapeCB$ is an fb-cb category.
\end{theorem}
\begin{proof} 
By construction $\CatTapeCB$ is a poset enriched fb rig category with  $(\codiag{X}, \cobang{X})$ left adjoint to $(\diag{X},\bang{X})$. To prove that $(\CatTapeCB, \per, \uno)$ is a cartesian bicategory, we need to define 
$\copier{P} \colon P \to P\per P$, $\discharger{P} \colon P \to \uno$, $\cocopier{P} \colon P \to P \per P$ and  $\codischarger{P} \colon \uno \to P$ for all polynomials $P$.  The coherence axioms in Definition~\ref{def:fbcbrig} provide us the recipe: \begin{equation}\label{eq:copierind}
        \begin{array}{rcl C{0.5cm} rcl}
            \copier{\zero} &\defeq& \id{\zero} && \discharger{\zero} &\defeq& \cobang{\uno}  \\
            \copier{U \piu P'} &\defeq& \Tcopier{U} \piu \cobang{UP'} \piu ((\cobang{P'U} \piu \copier{P'}) ; \Idl{P'}{U}{P'}) && \discharger{U \piu P'} &\defeq& (\Tdischarger{U} \piu \discharger{P'}) ; \codiag{\uno}
        \end{array}
    \end{equation}
    \begin{equation}\label{eq:cocopierind}
        \begin{array}{rcl C{0.5cm} rcl}
            \cocopier{\zero} &\defeq& \id{\zero} && \codischarger{\zero} &\defeq& \bang{\uno}  \\
            \cocopier{U \piu P'} &\defeq& \Tcocopier{U} \piu \bang{UP'} \piu (\dl{P'}{U}{P'} ; (\bang{P'U} \piu \cocopier{P'})) && \codischarger{U \piu P'} &\defeq& \diag{\uno} ; (\Tcodischarger{U} \piu \codischarger{P'})
        \end{array}
    \end{equation}
 For instance, $\copier{A \piu B} \colon A \piu B \to (A \piu B)\per(A \piu B) = AA \piu AB \piu BA \piu BB$ and $\discharger{A \piu B} \colon A \piu B \to \uno$ are
    \[\copier{A \piu B} = 
    \InputIfFileExists{cb/examples/copierApB.tikz}{}{\input{./tikz/cb/examples/copierApB.tikz}}
 \qquad \discharger{A \piu B} = 
    \InputIfFileExists{cb/examples/dischargerApB.tikz}{}{\input{./tikz/cb/examples/dischargerApB.tikz}}
 \qedhere \]
\end{proof}

We now focus on fb-cb morphisms from $\CatTapeCB$ to $\Rel$.
In \cite{GCQ} it is shown that every interpretation $\interpretation$ of the monoidal signature $\Sigma$ in $\Rel$ gives rise to a morphism of cartesian bicategories $\CBdsem{\cdot}_{\interpretation}\colon \CB \to \sRel$. This extends in turn to a morphism of fb-cb categories $\TCBdsem{\cdot}_{\interpretation} \colon \CatTapeCB \to \sRel$ as illustrated in Table~\ref{tabSemantic}. More generally, the following holds.

\begin{table}
\tiny{
\begin{center}
\begin{tabular}{ccccc}
	\toprule
$\CBdsem{\CBRel{R}{U}{V}}_{\interpretation} = \alpha_{\sign}(R) $&$ \!\!\!\CBdsem{\CBcopier{A}}_{\interpretation} = \copier{X}  $&$ \!\!\! \CBdsem{\CBdischarger{A}}_{\interpretation} = \discharger{X}  $&$\!\!\! \CBdsem{\CBcocopier{A}}_{\interpretation} = \cocopier{X}  $&$\!\!\! \CBdsem{\CBcodischarger{A}}_{\interpretation} = \codischarger{X}$\\
$\CBdsem{\id{A}}_{\interpretation}= \id{X} $&$\!\!\! \CBdsem{\id{1}}_{\interpretation}= \id{1} $&$\!\!\!\!\!\! \CBdsem{\symmt{A}{A}}_{\interpretation} = \symmt{X}{X}  $&$\!\!\! \CBdsem{c_1;c_2}_{\interpretation} = \CBdsem{c_1}_{\interpretation}; \CBdsem{c_2}_{\interpretation}  $&$\!\!\! \CBdsem{c_1\per c_2}_{\interpretation} = \CBdsem{c_1}_{\interpretation} \per \CBdsem{c_2}_{\interpretation}$\\
\midrule
$\TCBdsem{\Tcirc{c}{U}{V}}_{\interpretation}=\CBdsem{c}_{\interpretation} $&$\!\!\! \TCBdsem{\Tcomonoid{U}}_{\interpretation}=\diag{W} $&$\!\!\! \TCBdsem{\Tcounit{U}}_{\interpretation} = \bang{W} $&$\!\!\!\TCBdsem{\Tmonoid{U}}_{\interpretation}= \codiag{W} $&$\!\!\! \TCBdsem{\Tunit{U}}_{\interpretation} = \cobang{W}$ \\
$\TCBdsem{\id{U}}_{\interpretation}=\id{W} $&$\!\!\! \TCBdsem{\id{\zero}}_{\interpretation} = \id{\zero}  $&$\!\!\! \TCBdsem{\symmp{U}{U'}}_{\interpretation}= \symmp{W}{W'} $&$\!\!\! \TCBdsem{\s;\t}_{\interpretation} = \TCBdsem{\s}_{\interpretation} ; \TCBdsem{\t}_{\interpretation} $&$\!\!\! \TCBdsem{\s \piu \t}_{\interpretation} = \TCBdsem{\s}_{\interpretation} \piu \TCBdsem{\t}_{\interpretation} $ \\
\bottomrule
\end{tabular}
\end{center}}
\caption{The morphisms $\CBdsem{\cdot}_{\interpretation} \colon \CB \to \Rel$ and $\TCBdsem{\cdot}_{\interpretation} \colon \CatTapeCB \to \Rel$ extending an interpretation $\interpretation=(\alpha_{\sort}, \alpha_{\sign})$ of a monoidal signature $\sign$ in $\Rel$. We write $X,W,W'$ for $\alpha_\sort(A)$, $\alpha^\sharp_\sort(U)$ and $\alpha^\sharp_\sort(U')$ respectively, and $\cocopier{X}$, $\codischarger{X}$, $\copier{X}$, $\discharger{X}$, $\codiag{W}$ $\cobang{W}$, $\diag{W}$, $\bang{W}$ for the (co)monoids in $\Rel$ defined as in~\eqref{eq:comonoidsREL}.}
\label{tabSemantic}
\end{table}

\begin{proposition}\label{prop:bijective}
The following are (pairwise) in bijective correspondence:
\begin{enumerate}
\item interpretations of the monoidal signature $\sign$ in $\Rel$;\label{item1}
\item morphisms of cartesian bicategories $\CB \to \sRel$;\label{item2}
\item morphisms of sesquistrict fb-cb categories $\CatTapeCB \to \sRel$.\label{item3}
\end{enumerate}
\end{proposition}

Arrows in $\CatTapeCB$ can be thought of as expressions and $\TCBdsem{\cdot}_{\interpretation}  \colon \CatTapeCB \to \sRel$ as a semantic map assigning to each expression $\t$ the denoted relation. For instance 
by using the definition in Table~\ref{tabSemantic} and Remark~\ref{rmk:relsem}, the reader can easily check that
\begin{align}\label{eq:examplesem}
    \TCBdsem{
    \InputIfFileExists{relations/RoSaT.tikz}{}{\input{./tikz/relations/RoSaT.tikz}}
}_{\interpretation} &=\alpha_{\sign}(R) \cup (\alpha_{\sign}(S) \cap \alpha_{\sign}(T)).
\end{align}

It turns out that if $\TCBdsem{\s}_{\interpretation} \subseteq \TCBdsem{\t}_{\interpretation}$ for all interpretations $\interpretation$ of $\sign$ in $\Rel$, then one can derive that $\s \minorinTCB \t$ in $\CatTapeCB$. In other words, the axioms of $\CatTapeCB$ are \emph{complete} with respect to relational inclusion.

\begin{theorem}[Completeness]\label{thm:completeness} Let $\s, \t$ be arrows in $\CatTapeCB$. Then 
$\s \minorinTCB \t$ if and only if $\mathcal{M}(\s)\leq \mathcal{M}(\t)$  for all morphisms of sesquistrict fb-cb categories $\mathcal{M} \colon \CatTapeCB \to \Rel$.
\end{theorem}
\begin{proof}
	Soundness is trivial: since the $\mathcal{M}$'s by definition preserve the ordering, then $\s \minorinTCB \t$ entails $\mathcal{M}(\s)\leq \mathcal{M}(\t)$ for all $\mathcal M$.
	For completeness we will make use of the following facts, where every $\mathcal M$ appearing below is intended to be a morphism of fb-cb  categories:
	\begin{enumerate}
		\item 	For all  $\mathcal{M}\colon \CatTapeCB \to \Rel$ and $c,d\in \CB[U,V]$, $\mathcal{M}(\tape{c} + \tape{d}) =  \mathcal{M}(\tape{c}) \cup \mathcal{M}(\tape{d})$.
		\item Let $c,d_1, \dots, d_m\in \CB[U,V]$.
		If for all $\mathcal{M}\colon \CatTapeCB \to \Rel$ we have $\mathcal{M}(\tape{c}) \leq \mathcal{M}(\sum_{j=1}^{m} \tape{d_j})$, then there exists a $j$ such that $c \minorinCB d_j$. 
		\item Let $c_1,\dots c_n,d_1, \dots, d_m\in \CB[U,V]$.
		If for all $\mathcal{M}\colon \CatTapeCB \to \Rel$ we have $\mathcal{M}(\sum_{i=1}^{n} \tape{c_i}) \leq \mathcal{M}(\sum_{j=1}^{m} \tape{d_j})$, then for all $i$ there exists a $j$ such that $c_i \minorinCB d_j$.
		\item Let $\s, \t$ be arrows in $\monomial$. If for all  $\mathcal{M} \colon \CatTapeCB \to \Rel$ we have $\mathcal{M}(\s)\leq \mathcal{M}(\t)$, then
		$\s \minorinTCB \t$.
		\item Let $\Cat{C}$ be a fb category monoidally enriched over a preorder $\leq$.
		Let $f,g\colon \Piu[i=1][n] {A_i} \to \Piu[j=1][m] {B_j}$ be arrows in $\Cat{C}$. Let $f_{ji}$ and $g_{ji}$ be, respectively, $\mu_i ; f; \pi_j$ and $\mu_i ; g; \pi_j$. Then $f\leq g$ if and only if $f_{ji} \leq g_{ji}$ for all $i,j$.
	\end{enumerate}
(1) and (5) are easy to prove. (2) follows from the proof of Theorem 17 in~\cite{GCQ}: for all $c\in \CB[U,V]$, there exist a morphism $\mathcal{U}_c \!\colon\! \CB \!\to\! \Rel$ and an element $(\iota,\omega) \!\in\! \mathcal{U}_c(c)$ such that
for all $d \!\in\! \CB[U,V]$, if $(\iota,\omega) \!\in\! \mathcal{U}_c(d)$ then $c \minorinCB d$. (3) follows from (1) and (2). (4) follows from (3) and Corollary~\ref{cor:poset}. 
	
By recalling that, for arbitrary tape diagrams $\s,\t\colon \Piu {U_i} \to \Piu {V_j}$, $\s_{ji}$ and $\t_{ji}$ are in $\monomial$, we can easily extend (4):
if $\mathcal{M}(\s)\leq \mathcal{M}(\t)$ for all morphisms of fb-cb categories $\mathcal{M} \colon \CatTapeCB \to \Rel$, 
then, by applying (5) to $\Rel$, $\mathcal{M}(\s)_{ji} \leq \mathcal{M}(\t)_{ji}$. Hence $\mathcal{M}(\s_{ji}) = \mathcal{M}(\s)_{ji} \leq \mathcal{M}(\t)_{ji} = \mathcal{M}(\t_{ji})$ for all fb-cb morphisms $\mathcal M$.
By (4) we have $\s_{ji} \minorinTCB \t_{ji}$ and thus, again by (5), $\s \minorinTCB \t$.
\end{proof}

\begin{remark}
	Point (2) in the above proof already appears in different flavours in the literature: for instance it closely resembles Lemma 13 in~\cite{chandra1977optimal} and Lemma 3 in~\cite{andreka_equational_1995}. In this sense, the proof above is analogous to the one in~\cite{sagiv1980equivalences} that provides an algorithm for checking inclusion of disjunctive-conjunctive queries by relying on the algorithm in~\cite{chandra1977optimal} for conjunctive queries.
\end{remark}

\paragraph{Deconstructing the calculus of  relations with tapes}\begin{table}
	\[
	\begin{array}{rclcrcl}
		\toprule
		\encoding{R} &=& 
    \InputIfFileExists{relations/R.tikz}{}{\input{./tikz/relations/R.tikz}}
 & &\encoding{\op{E}} &=& 
    \InputIfFileExists{relations/op.tikz}{}{\input{./tikz/relations/op.tikz}}
\\[1.5ex]
		\encoding{\bot} &=& 
    \InputIfFileExists{relations/bot.tikz}{}{\input{./tikz/relations/bot.tikz}}
& & \encoding{E_1 \cup E_2} &=& \Tcomonoid{} \; ; \; (\encoding{E_1} \piu \encoding{E_2}) \; ; \; \Tmonoid{} \\[3ex]
		\encoding{\top} &=& 
    \InputIfFileExists{relations/top.tikz}{}{\input{./tikz/relations/top.tikz}}
 && \encoding{E_1 \cap E_2} &=& \TapedComonoid \;  ; \; (\encoding{E_1} \per \encoding{E_2}) \; ; \; \TapedMonoid \\[1.5ex]
		\encoding{1} &=& 
    \InputIfFileExists{relations/id.tikz}{}{\input{./tikz/relations/id.tikz}}
 & &\encoding{E_1 ; E_2} &=& \encoding{E_1} ; \encoding{E_2} \\
		\bottomrule
	\end{array}
	\]
	\caption{The encoding $\encoding{-}$$\colon \CRS \to \CatTapeCB[A,A]$}\label{tab:encoding}
\end{table}

We conclude by showing how tapes can help in dealing with $\minorExpression$, i.e.\ the semantic inclusion of the calculus of binary relations (Section \ref{sc:leading example}).

Recall from Example~\ref{ex:signature} that the set $\sign$ in $\CRS$ can be regarded as a monoidal signature $\sign$ with set of sorts $\sort=\{A\}$. From this signature one constructs $\CatTapeCB$ as prescribed in Example~\ref{ex:TAPECB} and encodes expressions of $\CRS$ into tapes of $\CatTapeCB$.
The encoding $\encoding{\cdot}$ is illustrated in Table~\ref{tab:encoding} where, to make the notation lighter, we avoided labelling  all the wires with $A$. It is easy to see that all expressions are mapped into tapes of type $A \to A$. 
For instance,
\begin{align*}
    \encoding{R \cup (S \cap T)} &= \Tcomonoid{} \; ; \; (\encoding{R} \piu \encoding{S \cap T}) \; ; \; \Tmonoid{} \\\displaybreak[0]
    &= \Tcomonoid{} \; ; \; (\Trel{R} \piu (\TapedComonoid \;  ; \; (\encoding{S} \per \encoding{T}) \; ; \; \TapedMonoid)) \; ; \; \Tmonoid{} \\\displaybreak[0]
    &= \Tcomonoid{} \; ; \; (\Trel{R} \piu (\TapedComonoid \;  ; \; (\Trel{S} \per \Trel{T}) \; ; \; \TapedMonoid)) \; ; \; \Tmonoid{} \\\displaybreak[0]
    &= 
    \InputIfFileExists{relations/RoSaT.tikz}{}{\input{./tikz/relations/RoSaT.tikz}}

\end{align*}
Observe that $\TCBdsem{\encoding{R \cup (S \cap T)}}$, illustrated in~\eqref{eq:examplesem}, coincides with $\dsemRel{R \cup (S \cap T)}$ defined in~\eqref{eq:sematicsExpr}.
More generally, a simple inductive argument confirms that $\encoding{\cdot}$ preserves the semantics.
\begin{proposition}\label{prop:encodingsound}
For all expressions $E\in \CRS$ and interpretations $\interpretation$, $\dsemRel{E} = \TCBdsem{\encoding{E}}_{\interpretation}$.
\end{proposition}

By Propositions~\ref{prop:bijective},~\ref{prop:encodingsound} and the completeness theorem the next result follows immediately.
\begin{corollary}\label{crlFinal}
For all $E_1,E_2 \in \CRS$, $E_1 \minorExpression E_2$ if and only if $\encoding{E_1} \minorinTCB \encoding{E_2}$.
\end{corollary}

\begin{remark}
	The reader may wonder whether it is possible to encode tape diagrams into the calculus of relations. This is not the case, even when considering only tapes in $\CatTapeCB[A,A]$. To prove this it is enough to observe that a tape of type $A \to A$ can express the graph (8) in~\cite{DBLP:conf/stacs/Pous18} which is known to be not expressible in $\CRS$.  More generally, $\CatTapeCB$ can express coherent logic, i.e.\ the fragment of first order logic consisting of $\exists$, $\wedge$, $\top$, $\vee$ and $\bot$, while the positive fragment of the calculus of relations expresses the restriction of coherent logic to formulae using at most three variables: indeed one needs four variables to express the graph (8) in~\cite{DBLP:conf/stacs/Pous18}. The curious reader may try to extend the encoding in Figure 4 of \cite{GCQ} with some tapes expressing $\vee$ and $\bot$.
\end{remark}

\section{Conclusion}

\paragraph{Related work} The idea of inserting string diagrams into superstructures, like we do when passing from $\CatString$ to $\CatTape$, is not new. Here we mention a few examples, but this list is by no means exhaustive. Functorial boxes~\cite{mellies_functorial_2006}, for instance, are a graphical expedient that allows one to draw in a simple, intuitive manner monoidal functors $F \colon \Cat C \to \Cat D$: the image along $F$ of a string diagram $c$ in $\Cat C$ can be depicted as a box, labelled by $F$, drawn around $c$. A clear analogy to our tape diagrams, however, as a diagrammatic calculus for finite biproduct rig categories, does not seem to be evident. A more apparent resemblance with our drawings are \emph{twisting polygraphs} used in~\cite{acclavio_proof_2019} to give a formal syntax for the representation of proof nets in Multiplicative Linear Logic. In there the author develops a string diagrammatic calculus, where diagrams look like tree-shaped ``tapes'', whose leaves are the axioms appearing in the proof. However, there is no mention of rig categories in~\cite{acclavio_proof_2019}. 
Also in~\cite{bartlett_modular_2015} the authors use a diagrammatic language consisting of certain surfaces that host internal string diagrams within them but these are, like sheet diagrams in~\cite{comfort2020sheet}, inherently three dimensional.

All in all, sheet diagrams are the closest related structures to tape diagrams. Similarities include the fact that the objects are polynomials in both cases and indeed also the category of sheet diagrams turns out to be a sesquistrict rig category. Moreover, the definition of $\per$ in~\cite{comfort2020sheet} exploits a notion of whiskering given in terms of diagrammatic manipulations, while in our approach it is defined inductively.
The key difference between the two languages is that sheet diagrams may have nodes (morphisms) in the intersection of two different surfaces. In the presence of a biproduct, by virtue of our Theorem~\ref{thm:equivalentsignature}, we can reduce a rig signature into a monoidal one. In this case, all the generators will appear as nodes in a single sheet and never in an intersection of two or more. The only morphisms that will appear in an intersection are (co)diagonals whose existence is guaranteed by the biproduct structure of $\piu$.
\begin{equation*}
    \begin{array}{ccc}
        \includegraphics[scale=.26]{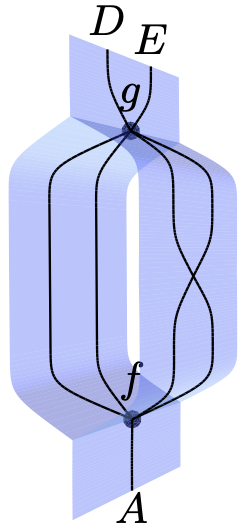}
        &
        \qquad\qquad\qquad\qquad
        &
        \begin{tikzpicture}
            \begin{pgfonlayer}{nodelayer}
                \node [style=none] (0) at (-4.5, 3.75) {};
                \node [style=none] (13) at (-4.5, 3.1) {};
                \node [style=none] (14) at (-4.5, 4.4) {};
                \node [style=none] (24) at (-3, 3.75) {};
                \node [style=box, scale=0.7] (25) at (-1.25, 5) {$f_1$};
                \node [style=box, scale=0.7] (26) at (-1.25, 2.5) {$f_2$};
                \node [style=none] (30) at (-1.75, 1.85) {};
                \node [style=none] (31) at (-1.75, 5.65) {};
                \node [style=none] (32) at (-1.75, 4.35) {};
                \node [style=none] (33) at (-2.4, 3.75) {};
                \node [style=none] (34) at (-1.75, 3.15) {};
                \node [style=none] (35) at (-3.5, 3.1) {};
                \node [style=none] (36) at (-3.5, 4.4) {};
                \node [style=none] (65) at (-0.85, 2.775) {};
                \node [style=none] (75) at (-0.85, 5.275) {};
                \node [style=label] (77) at (-5, 3.75) {$A$};
                \node [style=none] (80) at (-0.85, 2.225) {};
                \node [style=none] (81) at (-0.85, 4.725) {};
                \node [style=none] (86) at (-1.75, 5) {};
                \node [style=none] (87) at (-1.75, 2.5) {};
                \node [style=none] (88) at (4.65, 4.025) {};
                \node [style=none] (89) at (4.65, 3.1) {};
                \node [style=none] (90) at (4.65, 4.4) {};
                \node [style=none] (91) at (3.175, 4.025) {};
                \node [style=box, scale=0.8] (92) at (1.4, 5) {$g_1$};
                \node [style=box, scale=0.8] (93) at (1.4, 2.5) {$g_2$};
                \node [style=none] (95) at (1.9, 1.85) {};
                \node [style=none] (96) at (1.9, 5.65) {};
                \node [style=none] (97) at (1.9, 4.35) {};
                \node [style=none] (98) at (2.55, 3.75) {};
                \node [style=none] (99) at (1.9, 3.15) {};
                \node [style=none] (100) at (3.65, 3.1) {};
                \node [style=none] (101) at (3.65, 4.4) {};
                \node [style=none] (104) at (1, 2.775) {};
                \node [style=none] (107) at (1, 5.275) {};
                \node [style=label] (109) at (5.15, 4.025) {$D$};
                \node [style=none] (112) at (1, 2.225) {};
                \node [style=none] (113) at (1, 4.725) {};
                \node [style=none] (118) at (1.9, 5.25) {};
                \node [style=none] (119) at (1.9, 2.75) {};
                \node [style=none] (120) at (4.65, 3.475) {};
                \node [style=none] (121) at (3.175, 3.475) {};
                \node [style=none] (122) at (1.9, 4.75) {};
                \node [style=none] (123) at (1.9, 2.25) {};
                \node [style=label] (124) at (5.15, 3.45) {$E$};
            \end{pgfonlayer}
            \begin{pgfonlayer}{edgelayer}
                \draw [style=tape] (101.center)
                     to (90.center)
                     to (89.center)
                     to (100.center)
                     to [bend left] (95.center)
                     to (30.center)
                     to [bend left] (35.center)
                     to (13.center)
                     to (14.center)
                     to (36.center)
                     to [bend left] (31.center)
                     to (96.center)
                     to [bend left] cycle;
                \draw [style=tapeNoFill, fill=white] (33.center)
                     to [bend right=45] (34.center)
                     to (99.center)
                     to [bend right=45] (98.center)
                     to [bend right=45] (97.center)
                     to (32.center)
                     to [bend right=45] cycle;
                \draw (24.center) to (0.center);
                \draw [bend left=45] (24.center) to (86.center);
                \draw [bend right=45] (24.center) to (87.center);
                \draw (86.center) to (25);
                \draw (87.center) to (26);
                \draw (91.center) to (88.center);
                \draw [bend right=45] (91.center) to (118.center);
                \draw [bend left=45] (91.center) to (119.center);
                \draw (121.center) to (120.center);
                \draw [bend right=45] (121.center) to (122.center);
                \draw [bend left=45] (121.center) to (123.center);
                \draw (113.center) to (81.center);
                \draw (75.center) to (107.center);
                \draw [in=-180, out=0] (80.center) to (104.center);
                \draw [in=0, out=180] (112.center) to (65.center);
            \end{pgfonlayer}
        \end{tikzpicture}
    \end{array}
\end{equation*}
For example, consider the sheet diagram above (on the left) as a representation of a $\Cat{Vect}$ morphism. Since $\Cat{Vect}$ has biproducts the generator $f \colon A \to BC \piu BC$ is decomposed as $\diag{A} ; (f_1 \piu f_2)$ where $f_1, f_2 \colon A \to BC$ are now generators of the novel monoidal signature. Applying the same procedure to $g \colon BC \piu CB \to DE$ yields the tape diagram on the right.
When $\piu$ is not a biproduct, Theorem~\refeq{thm:equivalentsignature} does not hold and thus the procedure described above may fail. One example is $\Cat{Set}$, where $+$ is not a biproduct and thus a function $f \colon A \times B \to X + Y$ cannot be canonically decomposed in two functions $f_1 \colon A \times B \to X$ and $f_2 \colon A \times B \to Y$. Therefore $f \colon A \times B \to X + Y$ can be drawn using sheet diagrams, crucially in 3 dimensions, but not using tape diagrams.

\paragraph{Conclusion and future work} Like string diagrams provide, by virtue of Theorem 2.3 in \cite{joyal1991geometry}, a graphical language for symmetric monoidal categories,
tape diagrams are, by Theorems~\ref{thm:equivalentsignature} and~\ref{thm:Tapes is free sesquistrict generated by sigma}, a universal language for rig categories with finite biproducts.
In particular, thanks to Corollary~\ref{co:rig semantics}, whenever the semantic domain of a string diagrammatic language carries the structure of a finite biproduct rig category (or even a finite (co)product, see Remark \ref{rem:tapes universal language}), one can wrap string diagrams into tapes and obtain a meaningful language. 
By applying this approach to the ZX-calculus, we are able to easily specify a primitive form of quantum control (Examples~\ref{ex:zxcalculus} and~\ref{ex:zxcontext}). Other relevant instances of tape diagrams are $\sqcup$-props from~\cite{DBLP:conf/fossacs/BoisseauP22} (Example~\ref{ex:mu-props}).

The leading example  investigated in this paper is $\CR$, the positive fragment of the calculus of relations~\cite{tarski1941calculus}. The tape diagrams axioms in Figures~\ref{fig:tapesax},~\ref{fig:rel axioms} and~\ref{fig:cb axioms} allow us to prove all and only the valid equivalences between expressions of $\CR$ (Corollary~\ref{crlFinal}). This result follows easily from the completeness theorem (Theorem~\ref{thm:completeness}) that is close in spirit to those in~\cite{hasegawa2008finite} and~\cite{selinger2012finite}. Note that Corollary~\ref{crlFinal} does not contradict the impossibility of a finite axiomatisation proved in~\cite{hodkinson2000axiomatizability}, as our syntax is a radical departure from $\CR$. However, the fact that a simple axiomatisation---consisting of several well known algebraic structures---is possible with tape diagrams seems to suggest that the tape notation is more suitable than the traditional one.

The obvious next step consists in considering the extension of $\CR$ with Kleene star (reflexive and transitive closure), see e.g.~\cite{brunet2015petri,DBLP:conf/stacs/Pous18}. In terms of tapes this can be obtained by adding a trace to the monoidal structure given by $\piu$. Note that $\CatTapeCB$ is actually traced on $\per$, but this trace corresponds to feedbacks and not to Kleene star (iteration). The relationship between the two traced monoidal structures of $\Rel$ has been studied in~\cite{selinger1998note}, rephrasing an early work by Bainbridge~\cite{bainbridge1976feedback}, who was the first to observe the duality between \emph{data flow} and \emph{control flow} provided by $(\Rel, \per, \uno)$ and $(\Rel, \piu, \zero)$. Indeed, the language resulting from adding a trace to $\CatTapeCB$ would represent data flow at the level of circuits and control flow at the level of tapes. Such language would be similar in spirit to a diagrammatic version of Hoare Logic~\cite{hoare1969axiomatic} consisting of imperative programs, predicates on them and, in place of a proof system, diagrammatic laws. The ubiquity of our axioms gives us the hope that this language could be easily adjusted to deal with concurrent and quantum computations. Preliminary inspirations come, respectively, from ~\cite{hoare2011concurrent,kappe2018concurrent,DBLP:journals/acta/BaldanG19} and~\cite{harding2008orthomodularity}.

\begin{acks}
The authors would like to thank Dusko Pavlovic and Pawe{\l} Soboci\'{n}ski for the early discussions that eventually lead to tape diagrams. While writing the paper, we have received useful comments by Farzad Kianvash, Aleks Kissinger, Samuel Mimram, Chad Nester, Robin Piedeleu and Donald Yau.
The authors would also like to thank the anonymous referees for
their valuable comments and helpful suggestions. This work is
supported by the \grantsponsor{MUR}{Ministero dell'Università e della Ricerca of Italy}{www.mur.gov.it/} under Grant
No.~\grantnum{MUR}{201784YSZ5, PRIN2017 – ASPRA (Analysis of Program Analyses)} and partially supported by the \grantsponsor{UniPi}{Università di Pisa}{https://www.unipi.it} under Grant No.~\grantnum{UniPi}{PRA-2022-2023 FM4HD}.
\end{acks}

\bibliographystyle{ACM-Reference-Format}
\bibliography{references}

\appendix

 \newpage

\section{Coherence Axioms}\label{app:coherence axioms}

In this Appendix we collect together various Figures, listing the coherence axioms required by the definition of the algebraic structures we consider in the article.

 \begin{figure}[H]
     \begin{equation} \label{ax:monoidaltriangle}
         \begin{tikzcd}[column sep=tiny]
    (X \perG \unoG) \perG Y \arrow[rr, "{\alpha_{X, \unoG, Y}}"] \arrow[rd, "\rho_X \perG id_Y"'] &             & X \perG (1 \perG Y) \arrow[ld, "id_X \perG \lambda_Y"] \\
                                                                                               & X \perG Y &                                                            
\end{tikzcd}  \tag{M1}
     \end{equation}
     \begin{equation}\label{ax:monoidalpentagone}
         \begin{tikzcd}[column sep=tiny]
                                                                                                                           & (X \perG Y) \perG (Z \perG W) \arrow[rd, "{\alpha_{X, Y, Z \perG W}}"] &                                                                                 \\
((X \perG Y) \perG Z) \perG W \arrow[ru, "{\alpha_{X \perG Y, Z, W}}"] \arrow[d, "{\alpha_{X,Y,Z} \perG id_W}"'] &                                                                                & X \perG (Y \perG (Z \perG W))                                             \\
(X \perG (Y \perG Z)) \perG W \arrow[rr, "{\alpha_{X, Y \perG Z, W}}"']                                            &                                                                                & X \perG ((Y \perG Z) \perG W) \arrow[u, "{id_X \perG \alpha_{Y,Z,W}}"']
\end{tikzcd}  \tag{M2}
     \end{equation}
     \caption{Coherence axioms of monoidal categories}
     \label{fig:moncatax}
 \end{figure}

\begin{figure}[H]
   \begin{minipage}[t]{0.48\textwidth}
       \begin{equation}\label{eq:symmax1}
           \begin{tikzcd}[row sep=normal]
X \perG Y \arrow[r, "{\sigma_{X,Y}}"] \arrow[rd, Rightarrow, no head, shift right] & Y \perG X \arrow[d, "{\sigma_{Y,X}}"] \\
                                                                        & X \perG Y                            
\end{tikzcd}  \tag{S1}
       \end{equation}
   \end{minipage}
   \hfill
   \begin{minipage}[t]{0.48\textwidth}
        \begin{equation}\label{eq:symmax2}
            \begin{tikzcd}[row sep=normal]
X \perG \unoG \arrow[r, "{\sigma_{X,\unoG}}"] \arrow[d, "\rho_X"'] & \unoG \perG X \arrow[d, "\lambda_X"] \\
X \arrow[r, Rightarrow, no head]                                & X                                    
\end{tikzcd}  \tag{S2}
        \end{equation}    
   \end{minipage}
    \begin{equation}\label{eq:symmax3}
        \begin{tikzcd}
                                                                                                  & X \perG (Y \perG Z) \arrow[r, "{id_X \perG \sigma_{Y,Z}}"] & X \perG (Z \perG Y) \arrow[rd, "{\alpha^-_{X, Z, Y}}"]       &                         \\
(X \perG Y) \perG Z \arrow[ru, "{\alpha_{X, Y, Z}}"] \arrow[rd, "{\sigma_{X \perG Y, Z}}"'] &                                                                  &                                                                     & (X \perG Z) \perG Y \\
                                                                                                  & Z \perG (X \perG Y) \arrow[r, "{\alpha^-_{Z, X, Y}}"']    & (Z \perG X) \perG Y \arrow[ru, "{\sigma_{Z, X} \perG id_Y}"'] &                        
\end{tikzcd}  \tag{S3}
    \end{equation}
    \caption{Coherence axioms of symmetric monoidal categories}
    \label{fig:symmmoncatax}
\end{figure}

\begin{figure}[H]
    \begin{equation}\label{eq:coherence diag}\tag{FP1}
        \begin{tikzcd}[column sep=4.5em,baseline=(current  bounding  box.center)]
            X \perG Y \ar[r,"\diag{X \perG Y}"] \ar[d,"\diag X \perG \diag Y"'] & (X \perG Y) \perG (X \perG Y) \\
            (X \perG Y) \perG (Y \perG Y) \ar[d,"\assoc X X {Y \perG Y}"'] \\
            X \perG (X \perG (Y \perG Y)) \ar[d,"\id X \perG \Iassoc X Y Y"'] & X \perG (Y \perG (X \perG Y)) \ar[uu,"\Iassoc X Y {X \perG Y}"'] \\
            X \perG ((X \perG Y) \perG Y) \ar[r,"\id X \perG (\symm{X}{Y} \perG \id Y)"] & X \perG ((Y \perG X) \perG Y) \ar[u,"\id X \perG \assoc Y X Y"'] 
        \end{tikzcd}
    \end{equation}
    \\
	\begin{minipage}[b]{0.33\textwidth}
	 	\begin{equation}\label{eq:coherence bang}\tag{FP2}
            \begin{tikzcd}[baseline=(current  bounding  box.center)]
                X \perG Y \ar[r,"\bang{X \perG Y}"] \ar[d,"\bang X \perG \bang Y"'] & \unoG \\
                \unoG \perG \unoG \ar[ur,"\lunit \unoG"']
            \end{tikzcd}
	    \end{equation} 
	\end{minipage}
	\hfill
	\begin{minipage}[b]{0.26\textwidth}
		\begin{equation}\tag{FP3}
		\begin{tikzcd}
		I \ar[r,shift left=2,"\diag I"] \ar[r,shift right=2,"\Ilunit I"'] & I \perG I
		\end{tikzcd}
		\end{equation}
		\end{minipage}
	\hfill
	\begin{minipage}[b]{0.26\textwidth}
		\begin{equation}\label{eq:bang I = id I}\tag{FP4}
            \begin{tikzcd}
            I \ar[r,shift left=2,"\bang I"] \ar[r,shift right=2,"\id I"'] & I
            \end{tikzcd}
		\end{equation}
	\end{minipage}

\begin{equation}\label{eq:coherence codiag}\tag{FC1}
        \begin{tikzcd}[column sep=4.5em,baseline=(current  bounding  box.center)]
        (X \perG Y) \perG (X \perG Y) \ar[dd,"\assoc X Y {X \perG Y}"'] \ar[r,"\codiag{X \perG Y}"] & X \perG Y \\
        & (X \perG X) \perG (Y \perG Y) \ar[u,"\codiag X \perG \codiag Y"']\\
            X \perG (Y \perG (X \perG Y)) \ar[d,"\id X \perG \Iassoc Y X Y"'] & X \perG (X \perG (Y \perG Y)) \ar[u,"\Iassoc X X {Y \perG Y}"']  \\
        X \perG ((Y \perG X) \perG Y) \ar[r,"\id X \perG ( \symm{Y}{X} \perG \id Y)"] & X \perG (( X \perG Y) \perG Y) \ar[u,"\id X \perG \assoc X Y Y"']
        \end{tikzcd}
    \end{equation}
    \\
\begin{minipage}[b]{0.33\textwidth}
	\begin{equation}\label{eq:coherence cobang}\tag{FC2}
        \begin{tikzcd}[baseline=(current  bounding  box.center)]
        \unoG \ar[r,"\cobang{X \perG Y}"] \ar[d,"\Ilunit \unoG"']  & X \perG Y \\
        \unoG \perG \unoG \ar[ur,"\cobang X \perG \cobang Y"']  
        \end{tikzcd}
	\end{equation}
\end{minipage}
\hfill
\begin{minipage}[b]{0.26\textwidth}
	\begin{equation}\tag{FC3}
        \begin{tikzcd}
        I \perG I \ar[r,shift left=2,"\codiag I"] \ar[r,shift right=2,"\lunit I"'] &  I
        \end{tikzcd}
	\end{equation}
\end{minipage}
\hfill
\begin{minipage}[b]{0.26\textwidth}
	\begin{equation}\label{eq:cobang I = id I}\tag{FC4}
        \begin{tikzcd}
        I \ar[r,shift left=2,"\cobang I"] \ar[r,shift right=2,"\id I"'] & I
        \end{tikzcd}
	\end{equation}
\end{minipage}
\caption{Coherence axioms for (co)commutative (co)monoids}
\label{fig:fbcoherence}
\end{figure}

\begin{figure}[H]
    \begin{minipage}[t]{0.45\textwidth}
        \begin{equation}
            \label{eq:rigax1}\tag{R1}
            \adjustbox{scale=0.9}{
                \begin{tikzcd}[row sep=normal]
                    {(X\piu Y)Z} & {XZ \piu YZ} \\
                    {Z(X \piu Y)} & {ZX \piu ZY}
                    \arrow["{\dr{X}{Y}{Z}}", from=1-1, to=1-2]
                    \arrow["{\symmt{X \piu Y}{Z}}"', from=1-1, to=2-1]
                    \arrow["{\dl{Z}{X}{Y}}"', from=2-1, to=2-2]
                    \arrow["{\symmt{X}{Z} \piu \symmt{Y}{Z}}", from=1-2, to=2-2]
                \end{tikzcd}
            }
        \end{equation}    
    \end{minipage}
    \hfill
    \begin{minipage}[t]{0.45\textwidth}
        \begin{equation}
            \label{eq:rigax2}\tag{R2}
            \adjustbox{scale=0.9}{\begin{tikzcd}
	{(X \piu Y)Z} & {XZ \piu YZ} \\
	{(Y \piu X)Z} & {YZ \piu XZ}
	\arrow["{\dr{X}{Y}{Z}}", from=1-1, to=1-2]
	\arrow["{\symmp{X}{Y}\per \id Z}"', from=1-1, to=2-1]
	\arrow["{\dr{Y}{X}{Z}}"', from=2-1, to=2-2]
	\arrow["{\symmp{XZ}{YZ}}", from=1-2, to=2-2]
\end{tikzcd}}
        \end{equation}
    \end{minipage}
    
    \begin{equation}
        \label{eq:rigax3}\tag{R3}
        \adjustbox{scale=0.9}{\begin{tikzcd}[column sep=52pt]
	{((X \piu Y)\piu Z)W} & {(X \piu Y)W \piu ZW} & {(XW \piu YW) \piu ZW} \\
	{(X \piu (Y \piu Z))W} & {XW \piu (Y \piu Z)W} & {XW \piu (YW \piu ZW)}
	\arrow["{\dr{X \piu Y}{Z}{W}}", from=1-1, to=1-2]
	\arrow["{\assocp{X}{Y}{Z} \per id_W}"', from=1-1, to=2-1]
	\arrow["{\dr{X}{Y \piu Z}{W}}"', from=2-1, to=2-2]
	\arrow["{\dr{X}{Y}{W} \per id_{ZW}}", from=1-2, to=1-3]
	\arrow["{id_{XW} \per \dr{Y}{Z}{W}}"', from=2-2, to=2-3]
	\arrow["{\assocp{XW}{YW}{ZW}}", from=1-3, to=2-3]
\end{tikzcd}}
    \end{equation}
    \begin{equation}
        \label{eq:rigax4}\tag{R4}
        \adjustbox{scale=0.9}{\begin{tikzcd}[column sep=56pt]
	{((X \piu Y)Z)W} & {(XZ \piu YZ)W} & {(XZ)W \piu (YZ)W} \\
	{(X \piu Y)(ZW)} && {X(ZW) \piu Y(ZW)}
	\arrow["{\assoct{X\piu Y}{Z}{W}}"', from=1-1, to=2-1]
	\arrow["{\dr{X}{Y}{Z} \per \id W}", from=1-1, to=1-2]
	\arrow["{\dr{XZ}{YZ}{W}}", from=1-2, to=1-3]
	\arrow["{\assoct{X}{Z}{W} \piu \assoct{Y}{Z}{W}}", from=1-3, to=2-3]
	\arrow["{\dr{X}{Y}{ZW}}"', from=2-1, to=2-3]
\end{tikzcd}}
    \end{equation}
    \begin{equation}
        \label{eq:rigax5}\tag{R5}
        \adjustbox{scale=0.9}{\begin{tikzcd}[column sep=normal]
	{(X \piu Y)(Z \piu W)} & {X(Z \piu W) \piu Y(Z \piu W)} & {} \\
	{(X \piu Y)Z \piu (X \piu Y)W} & {(XZ \piu XW) \piu (YZ \piu YW)} \\
	{(XZ \piu YZ) \piu (XW \piu YW)} & {XZ \piu (XW \piu (YZ \piu YW))} \\
	{XZ \piu (YZ \piu (XW \piu YW))} & {XZ \piu ((XW \piu YZ) \piu YW)} \\
	{XZ \piu ((YZ \piu XW) \piu YW)} & {XZ \piu ((YZ \piu XW) \piu YW)}
	\arrow["{\dr{X}{Y}{Z \piu W}}", from=1-1, to=1-2]
	\arrow[Rightarrow, no head, from=5-1, to=5-2]
	\arrow["{\dl{X \piu Y}{Z}{W}}"', from=1-1, to=2-1]
	\arrow["{\dr{X}{Y}{Z} \piu \dr{X}{Y}{W}}"', from=2-1, to=3-1]
	\arrow["{\assocp{XZ}{YZ}{XW \piu YW}}"', from=3-1, to=4-1]
	\arrow["{id_{XZ} \piu \Iassocp{YZ}{XW}{YW}}"', from=4-1, to=5-1]
	\arrow["{\dl{X}{Z}{W} \piu \dl{Y}{Z}{W}}", from=1-2, to=2-2]
	\arrow["{\assocp{XZ}{XW}{YZ \piu YW}}", from=2-2, to=3-2]
	\arrow["{id_{XZ} \piu \Iassocp{XW}{YZ}{YW}}", from=3-2, to=4-2]
	\arrow["{id_{XZ} \piu (\symmp{XW}{YZ} \piu id_{YW})}", from=4-2, to=5-2]
\end{tikzcd}}
    \end{equation}

    \begin{minipage}[t]{0.25\textwidth}
        \begin{equation}
            \label{eq:rigax6}\tag{R6}
            \adjustbox{scale=0.9}{\begin{tikzcd}
	{\zero \per \zero} & \zero
	\arrow["{\annl \zero}", shift left=2, from=1-1, to=1-2]
	\arrow["{\annr \zero}"', shift right=2, from=1-1, to=1-2]
\end{tikzcd}}
        \end{equation}
    \end{minipage}
    \hfill
    \begin{minipage}[t]{0.45\textwidth}
        \begin{equation}
            \label{eq:rigax7}\tag{R7}
            \adjustbox{scale=0.9}{\begin{tikzcd}
	{(X \piu Y)\zero} & {X\zero \piu Y\zero} \\
	\zero & {\zero \piu \zero}
	\arrow["{\lunitp{\zero}}", from=2-2, to=2-1]
	\arrow["{\annr{X \piu Y}}"', from=1-1, to=2-1]
	\arrow["{\annr X \piu \annr Y}", from=1-2, to=2-2]
	\arrow["{\dr{X}{Y}{\zero}}", from=1-1, to=1-2]
\end{tikzcd}}
        \end{equation}
    \end{minipage}
    \hfill
    \begin{minipage}[t]{0.25\textwidth}
        \begin{equation}
            \label{eq:rigax8}\tag{R8}
            \adjustbox{scale=0.9}{\begin{tikzcd}
	{\zero \per \uno} & \zero
	\arrow["{\annl \uno}", shift left=2, from=1-1, to=1-2]
	\arrow["{\runitt \zero}"', shift right=2, from=1-1, to=1-2]
\end{tikzcd}}
        \end{equation}
    \end{minipage}
    \\
    \begin{minipage}[t]{0.48\textwidth}
        \begin{equation}
            \label{eq:rigax9}\tag{R9}
            \adjustbox{scale=0.9}{\begin{tikzcd}[column sep=15pt]
	{X \per \zero} && {\zero \per X} \\
	& \zero
	\arrow["{\annl X}", from=1-3, to=2-2]
	\arrow["{\symmt{X}{\zero}}", from=1-1, to=1-3]
	\arrow["{\annr X}"', from=1-1, to=2-2]
\end{tikzcd}}
        \end{equation}    
    \end{minipage}
    \hfill
    \begin{minipage}[t]{0.48\textwidth}
        \begin{equation}
            \label{eq:rigax10}\tag{R10}
            \adjustbox{scale=0.9}{\begin{tikzcd}
	{(XY)0} && {X(Y\zero)} \\
	\zero && X\zero
	\arrow["{\assoct{X}{Y}{\zero}}", from=1-1, to=1-3]
	\arrow["{id_X \per \annr Y}", from=1-3, to=2-3]
	\arrow["{\annr{XY}}"', from=1-1, to=2-1]
	\arrow["{\annr X}", from=2-3, to=2-1]
\end{tikzcd}}
        \end{equation}
    \end{minipage}
    \\
    \begin{minipage}[t]{0.50\textwidth}
        \begin{equation}
            \label{eq:rigax11}\tag{R11}
            \adjustbox{scale=0.9}{\begin{tikzcd}
	{(\zero \piu X)Y} && {\zero Y \piu XY} \\
	XY && {\zero \piu XY}
	\arrow["{\dr{\zero}{X}{Y}}", from=1-1, to=1-3]
	\arrow["{\annl Y \piu id_{XY}}", from=1-3, to=2-3]
	\arrow["{\lunitp{X} \per \id Y}"', from=1-1, to=2-1]
	\arrow["{\lunitp{XY}}", from=2-3, to=2-1]
\end{tikzcd}}
        \end{equation}    
    \end{minipage}
    \hfill
    \begin{minipage}[t]{0.46\textwidth}
        \begin{equation}
            \label{eq:rigax12}\tag{R12}
            \adjustbox{scale=0.9}{\begin{tikzcd}[column sep=tiny]
	{(X \piu Y)\uno} && {X\uno \piu Y\uno } \\
	& {X \piu Y}
	\arrow["{\dr{X}{Y}{\uno}}", from=1-1, to=1-3]
	\arrow["{\runitt X \piu \runitt Y}", from=1-3, to=2-2]
	\arrow["{\runitt{X \piu Y}}"', from=1-1, to=2-2]
\end{tikzcd}}
        \end{equation}
    \end{minipage}
    \caption{Coherence Axioms of symmetric rig categories}
    \label{fig:rigax}
\end{figure}

\begin{figure}[H]
    \begin{equation}
        \label{eq:dl1}
        \begin{tikzcd}[row sep=normal]
	{X(Y\piu Z)} & {XY \piu XZ} \\
	{X(Z\piu Y)} & {XZ \piu XY}
	\arrow["{\dl{X}{Y}{Z}}", from=1-1, to=1-2]
	\arrow["{id_X \per \symmp{Y}{Z}}"', from=1-1, to=2-1]
	\arrow["{\dl{X}{Z}{Y}}"', from=2-1, to=2-2]
	\arrow["{\symmp{XY}{XZ}}", from=1-2, to=2-2]
\end{tikzcd}     \end{equation}
    \begin{equation}
        \label{eq:dl2}
        \begin{tikzcd}[column sep=52pt]
	X[(Y \piu Z) \piu W] \ar[r,"\dl{X}{Y \piu Z}{W}"] \ar[d,"\id{X} \per \assocp Y Z W"'] & X(Y \piu Z) \piu XW \ar[r,"\dl{X}{Y}{Z} \piu \id{XW}"] & (XY \piu XZ) \piu XW \ar[d,"\assocp {XY}{XZ}{XW}"] \\
	X[Y\piu(Z \piu W)] \ar[r,"\dl{X}{Y}{Z \piu W}"] & XY \piu X(Z \piu W) \ar[r,"\id{XY} \piu \dl{X}{Z}{W}"] & XY \piu (XZ \piu XW)
\end{tikzcd}     \end{equation}
    \begin{equation}
        \label{eq:dl3}
        \begin{tikzcd}[column sep=48pt]
	{(XY)(Z \piu W)} && {(XY)Z \piu (XY)W} \\
	{X(Y(Z \piu W))} & {X(YZ \piu YW)} & {X(YZ) \piu X(YW)}
	\arrow["{\assoct{X}{Y}{Z \piu W}}"', from=1-1, to=2-1]
	\arrow["{id_X \per \dl{Y}{Z}{W}}"', from=2-1, to=2-2]
	\arrow["{\dl{X}{YZ}{YW}}"', from=2-2, to=2-3]
	\arrow["{\assoct{X}{Y}{Z} \piu \assoct{X}{Y}{W}}", from=1-3, to=2-3]
	\arrow["{\dl{XY}{Z}{W}}", from=1-1, to=1-3]
\end{tikzcd}     \end{equation}
    \begin{minipage}[t]{0.44\textwidth}
    \begin{equation}
        \label{eq:dl4}
        \begin{tikzcd}
	{\zero(X \piu Y)} & {\zero X \piu \zero Y} \\
	\zero & {\zero \piu \zero}
	\arrow["{\lunitp{\zero}}", from=2-2, to=2-1]
	\arrow["{\annl{X \piu Y}}"', from=1-1, to=2-1]
	\arrow["{\annl X \piu \annl Y}", from=1-2, to=2-2]
	\arrow["{\dl{\zero}{X}{Y}}", from=1-1, to=1-2]
\end{tikzcd}     \end{equation}
    \end{minipage}
    \hfill
    \begin{minipage}[t]{0.54\textwidth}
    \begin{equation}
        \label{eq:dl5}
        \begin{tikzcd}
	{X(\zero \piu Y)} && {X\zero \piu XY} \\
	XY && {\zero \piu XY}
	\arrow["{\dl{X}{\zero}{Y}}", from=1-1, to=1-3]
	\arrow["{\annr X \piu id_{XY}}", from=1-3, to=2-3]
	\arrow["{id_X \per \lunitp{Y}}"', from=1-1, to=2-1]
	\arrow["{\lunitp{XY}}", from=2-3, to=2-1]
\end{tikzcd}     \end{equation}
    \end{minipage}
    \\
    \begin{minipage}[t]{0.48\textwidth}
        \begin{equation}
            \label{eq:dl7}
            \begin{tikzcd}
	{X(Y\piu \zero)} && {XY \piu X\zero} \\
	XY && {XY \piu \zero}
	\arrow["{\dl{X}{Y}{\zero}}", from=1-1, to=1-3]
	\arrow["{ id_{XY} \piu \annr X}", from=1-3, to=2-3]
	\arrow["{id_X \per \runitp{Y}}"', from=1-1, to=2-1]
	\arrow["{\runitp{XY}}", from=2-3, to=2-1]
\end{tikzcd}         \end{equation}
        \end{minipage}
        \hfill
        \begin{minipage}[t]{0.50\textwidth}
        \begin{equation}
            \label{eq:dl8}
            \begin{tikzcd}
	{(X \piu \zero)Y} && {XY \piu \zero Y} \\
	XY && {XY \piu \zero}
	\arrow["{\dr{X}{\zero}{Y}}", from=1-1, to=1-3]
	\arrow["{\id{XY} \piu \annl{Y}}", from=1-3, to=2-3]
	\arrow["{\runitp{X} \per \id Y}"', from=1-1, to=2-1]
	\arrow["{\runitp{XY}}", from=2-3, to=2-1]
\end{tikzcd}         \end{equation}
    \end{minipage}
    \begin{equation}
        \label{eq:dl6}
        \begin{tikzcd}[column sep=tiny]
	{\uno(X \piu Y)} && {\uno X \piu \uno Y } \\
	& {X \piu Y}
	\arrow["{\dl{\uno}{X}{Y}}", from=1-1, to=1-3]
	\arrow["{\lunitp X \piu \lunitp Y}", from=1-3, to=2-2]
	\arrow["{\lunitp{X \piu Y}}"', from=1-1, to=2-2]
\end{tikzcd}     \end{equation}
    \caption{Derived laws of symmetric rig categories}
    \label{fig:dlaw}
\end{figure}  \section{Additional figures}\label{app:morefigures}

In this Appendix we provide some additional figures representing useful/interesting axioms and derivations.

\begin{figure}
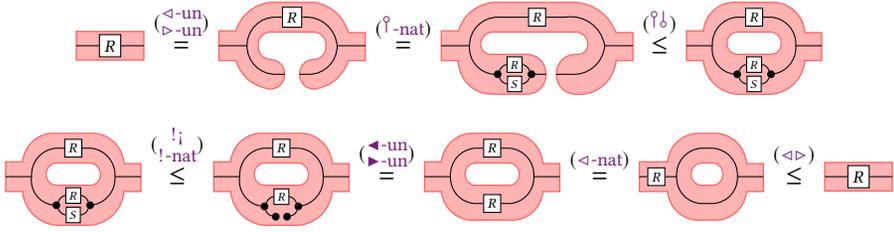

    \[ 
        
    \InputIfFileExists{topos/deriv1/step1.tikz}{}{\input{./tikz/topos/deriv1/step1.tikz}}
 \axeq{\substack{\refeq{ax:diagun} \\ \refeq{ax:codiagun}}} 
    \InputIfFileExists{topos/deriv1/step2.tikz}{}{\input{./tikz/topos/deriv1/step2.tikz}}
 \axeq{\refeq{ax:bangnat}} 
    \InputIfFileExists{topos/deriv1/step3.tikz}{}{\input{./tikz/topos/deriv1/step3.tikz}}
 \axsubeq{\refeq{ax:relBangCobang}} 
    \InputIfFileExists{topos/deriv1/step4.tikz}{}{\input{./tikz/topos/deriv1/step4.tikz}}
 
    \]
    
    \[
        
    \InputIfFileExists{topos/deriv1/step4.tikz}{}{\input{./tikz/topos/deriv1/step4.tikz}}
 \axsubeq{\substack{\refeq{ax:dischargeradj2} \\ \refeq{ax:dischargernat}}} 
    \InputIfFileExists{topos/deriv2/step3.tikz}{}{\input{./tikz/topos/deriv2/step3.tikz}}
 \axeq{\substack{\refeq{ax:copierun} \\ \refeq{ax:cocopierun}}} 
    \InputIfFileExists{topos/deriv2/step6.tikz}{}{\input{./tikz/topos/deriv2/step6.tikz}}
 \axeq{\refeq{ax:diagnat}} 
    \InputIfFileExists{topos/deriv2/step7.tikz}{}{\input{./tikz/topos/deriv2/step7.tikz}}
 \axsubeq{\refeq{ax:relDiagCodiag}} 
    \InputIfFileExists{topos/deriv1/step1.tikz}{}{\input{./tikz/topos/deriv1/step1.tikz}}

    \]    
    \caption{Graphical proof for $R \cup (R \cap S) \equiv_{\CR} R$}
    \label{fig:R or R and S}
\end{figure}

\begin{figure}[H]
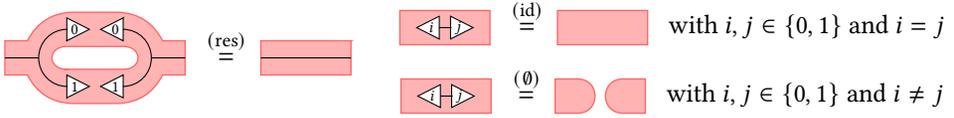

    \begin{center}
        \begin{tabular}{cc}
            \multirow{2}{*}{
    \InputIfFileExists{zx/idemp_left.tikz}{}{\input{./tikz/zx/idemp_left.tikz}}
 $\axeq{\text{res}}$ 
    \InputIfFileExists{zx/idemp_right.tikz}{}{\input{./tikz/zx/idemp_right.tikz}}
}& \sr {
    \InputIfFileExists{zx/inner_product.tikz}{}{\input{./tikz/zx/inner_product.tikz}}
 $\axeq{\text{id}}$ 
    \InputIfFileExists{zx/onb1_right.tikz}{}{\input{./tikz/zx/onb1_right.tikz}}
} \text{with $i,j \in \{0, 1\}$ and $i = j$}\\
            & \sr {
    \InputIfFileExists{zx/inner_product.tikz}{}{\input{./tikz/zx/inner_product.tikz}}
 $\axeq{\emptyset}$ 
    \InputIfFileExists{zx/onb2_right.tikz}{}{\input{./tikz/zx/onb2_right.tikz}}
 \text{with $i,j \in \{0, 1\}$ and $i\neq j$}}\\
        \end{tabular}
    \end{center}
\caption{The axioms above are a graphical rendering of the necessary and sufficient conditions for $\ket{0}$ and $\ket{1}$
to form an orthonormal basis (see~\cite[Theorem 5.32]{Coecke2017}). In particular, axiom $(\text{id})$ asserts that the basis vectors have norm $1$, while $(\emptyset)$ says that they are orthogonal to each other.}
\label{fig:orthonormal}
\end{figure}

\begin{figure}[H]
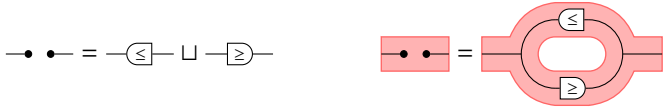

		\[ \Ctotrel{\uno} = \Ccoreg{\leq}{\uno} \sqcup \Creg{\geq}{\uno} \qquad \qquad 
    \InputIfFileExists{tapes/totrel.tikz}{}{\input{./tikz/tapes/totrel.tikz}}
 = 
    \InputIfFileExists{tapes/leq_geq.tikz}{}{\input{./tikz/tapes/leq_geq.tikz}}
 \]
		\caption{The axiom of GPLA~\cite{DBLP:conf/fossacs/BoisseauP22} on the left expresses the totality of an ordering relation $\leq$. Within tape diagrams the same axiom can be rendered completely graphically as shown on the right.}
		\label{fig:uprop}
\end{figure}

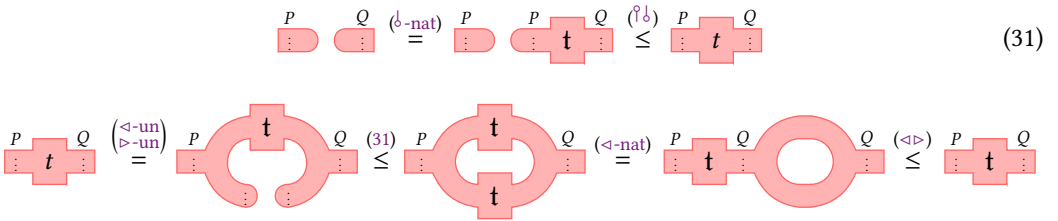
\begin{figure}[H]
	\begin{equation}\label{eq:bangcobang}
		\TPolycounitunit{P}{Q} 
		\stackrel{\eqref{ax:cobangnat}}{=}
		\TPolyCounit{P} \; \begin{tikzpicture}
			\begin{pgfonlayer}{nodelayer}
				\node [style=none] (35) at (-2, -0.5) {};
				\node [style=none] (36) at (-2, 0.5) {};
				\node [style=dots] (90) at (1.5, 0.025) {$\vdots$};
				\node [style=none] (91) at (2, -0.5) {};
				\node [style=none] (92) at (2, 0.5) {};
				\node [style=label] (94) at (-1.5, 1) {$P$};
				\node [style=label] (95) at (1.5, 1) {$Q$};
				\node [style=none] (96) at (-0.75, -1) {};
				\node [style=none] (97) at (0.75, -1) {};
				\node [style=none] (98) at (0.75, 1) {};
				\node [style=none] (99) at (-0.75, 1) {};
				\node [style=none] (104) at (0, 0) {$\t$};
				\node [style=dots] (105) at (-1.5, 0.025) {$\vdots$};
				\node [style=none] (106) at (-0.75, -0.5) {};
				\node [style=none] (107) at (0.75, -0.5) {};
				\node [style=none] (108) at (0.75, 0.5) {};
				\node [style=none] (109) at (-0.75, 0.5) {};
			\end{pgfonlayer}
			\begin{pgfonlayer}{edgelayer}
				\draw [style=tape] (97.center)
					 to (96.center)
					 to (106.center)
					 to (35.center)
					 to [bend left=90, looseness=1.75] (36.center)
					 to (109.center)
					 to (99.center)
					 to (98.center)
					 to (108.center)
					 to (92.center)
					 to (91.center)
					 to (107.center)
					 to cycle;
			\end{pgfonlayer}
		\end{tikzpicture}
		\stackrel{\eqref{ax:relBangCobang}}{\leq}
		\Tbox{t}{P}{Q}
	\end{equation}
	\\
	\begin{equation*}
		\Tbox{t}{P}{Q} 
		\stackrel{\left( \substack{\refeq{ax:diagun} \\ \refeq{ax:codiagun}} \right)}{=}
		\begin{tikzpicture}
			\begin{pgfonlayer}{nodelayer}
				\node [style=none] (8) at (-0.95, -2) {};
				\node [style=none] (9) at (-0.75, 2) {};
				\node [style=none] (10) at (-0.75, 1) {};
				\node [style=none] (11) at (-1.75, 0) {};
				\node [style=none] (12) at (-0.95, -1) {};
				\node [style=none] (13) at (-2.75, -0.5) {};
				\node [style=none] (14) at (-2.75, 0.5) {};
				\node [style=none] (24) at (-4, 0) {};
				\node [style=none] (35) at (-4, -0.5) {};
				\node [style=none] (36) at (-4, 0.5) {};
				\node [style=none] (83) at (0.95, -2) {};
				\node [style=none] (84) at (0.75, 2) {};
				\node [style=none] (85) at (0.75, 1) {};
				\node [style=none] (86) at (1.75, 0) {};
				\node [style=none] (87) at (0.95, -1) {};
				\node [style=none] (88) at (2.75, -0.5) {};
				\node [style=none] (89) at (2.75, 0.5) {};
				\node [style=none] (90) at (4, 0) {};
				\node [style=none] (91) at (4, -0.5) {};
				\node [style=none] (92) at (4, 0.5) {};
				\node [style=label] (94) at (-3.25, 1) {$P$};
				\node [style=label] (95) at (3.25, 1) {$Q$};
				\node [style=none] (96) at (-0.75, 0.5) {};
				\node [style=none] (97) at (0.75, 0.5) {};
				\node [style=none] (98) at (0.75, 2.5) {};
				\node [style=none] (99) at (-0.75, 2.5) {};
				\node [style=none] (104) at (0, 1.5) {$\t$};
				\node [style=dots] (106) at (3.25, 0.025) {$\vdots$};
				\node [style=dots] (107) at (-3.25, 0.025) {$\vdots$};
				\node [style=dots] (108) at (0.95, -1.475) {$\vdots$};
				\node [style=dots] (109) at (-0.95, -1.475) {$\vdots$};
			\end{pgfonlayer}
			\begin{pgfonlayer}{edgelayer}
				\draw [style=tape] (11.center)
					 to [bend left=45] (10.center)
					 to (96.center)
					 to (97.center)
					 to (85.center)
					 to [bend left=45] (86.center)
					 to [bend left=45] (87.center)
					 to [bend right=90, looseness=2.00] (83.center)
					 to [bend right] (88.center)
					 to (91.center)
					 to (92.center)
					 to (89.center)
					 to [bend right] (84.center)
					 to (98.center)
					 to (99.center)
					 to (9.center)
					 to [bend right] (14.center)
					 to (36.center)
					 to (35.center)
					 to (13.center)
					 to [bend right] (8.center)
					 to [bend right=90, looseness=1.75] (12.center)
					 to [bend left=45] cycle;
			\end{pgfonlayer}
		\end{tikzpicture}	
		\stackrel{\eqref{eq:bangcobang}}{\leq}
		\Tconv{\t}{\t}{P}{Q}
		\stackrel{\eqref{ax:diagnat}}{=} 
		\Ttconv{\t}{P}{Q}
		\stackrel{\eqref{ax:relDiagCodiag}}{\leq}
		\Tbox{\t}{P}{Q}
	\end{equation*}
	\caption{Graphical derivation proving that with the axioms in Figure \ref{fig:rel axioms} $+$ becomes idempotent with $\zerotape$ as bottom element}
	\label{fig:derivationUPROPS}
\end{figure}

\begin{figure}[H]
    \mylabel{ax:stringsymmtinv}{$\sigma^{\per}$-inv}
    \mylabel{ax:stringsymmtnat}{$\sigma^{\per}$-nat}
    \mylabel{ax:stringcopieras}{$\copier{}$-as}
    \mylabel{ax:stringcopierun}{$\copier{}$-un}
    \mylabel{ax:stringcopierco}{$\copier{}$-co}
    \mylabel{ax:stringcocopieras}{$\cocopier{}$-as}
    \mylabel{ax:stringcocopierun}{$\cocopier{}$-un}
    \mylabel{ax:stringcocopierco}{$\cocopier{}$-co}
    \mylabel{ax:stringspecfrob}{S}
    \mylabel{ax:stringfrob}{F}
    \mylabel{ax:stringcopiernat}{$\copier{}$-nat}
    \mylabel{ax:stringdischargernat}{$\discharger{}$-nat}
    \mylabel{ax:stringdischargeradj1}{\codischarger{}\discharger{}}
    \mylabel{ax:stringdischargeradj2}{\discharger{}\codischarger{}}
    \mylabel{ax:stringcopieradj1}{\cocopier{}\copier{}}
    \mylabel{ax:stringcopieradj2}{\copier{}\cocopier{}}
    \scalebox{0.9}{
    \begin{tabular}{rcl rcl}
        
    \InputIfFileExists{cbString/symm_inv_left.tikz}{}{\input{./tikz/cbString/symm_inv_left.tikz}}
 & $\axeq{\sigma^{\per}\text{-inv}}$ & 
    \InputIfFileExists{cbString/symm_inv_right.tikz}{}{\input{./tikz/cbString/symm_inv_right.tikz}}
 & 
    \InputIfFileExists{cbString/symm_nat_left.tikz}{}{\input{./tikz/cbString/symm_nat_left.tikz}}
 & $\axeq{\sigma^{\per}\text{-nat}}$ & 
    \InputIfFileExists{cbString/symm_nat_right.tikz}{}{\input{./tikz/cbString/symm_nat_right.tikz}}
 \\
        \hline
        
    \InputIfFileExists{cbString/monoid_assoc_left.tikz}{}{\input{./tikz/cbString/monoid_assoc_left.tikz}}
 & $\axeq{\copier{}\text{-as}}$ & 
    \InputIfFileExists{cbString/monoid_assoc_right.tikz}{}{\input{./tikz/cbString/monoid_assoc_right.tikz}}
 & 
    \InputIfFileExists{cbString/comonoid_assoc_left.tikz}{}{\input{./tikz/cbString/comonoid_assoc_left.tikz}}
 & $\axeq{\cocopier{}\text{-as}}$ & 
    \InputIfFileExists{cbString/comonoid_assoc_right.tikz}{}{\input{./tikz/cbString/comonoid_assoc_right.tikz}}
 \\
        
    \InputIfFileExists{cbString/monoid_unit_left.tikz}{}{\input{./tikz/cbString/monoid_unit_left.tikz}}
 & $\axeq{\copier{}\text{-un}}$ & 
    \InputIfFileExists{cbString/monoid_unit_right.tikz}{}{\input{./tikz/cbString/monoid_unit_right.tikz}}
 & 
    \InputIfFileExists{cbString/comonoid_unit_left.tikz}{}{\input{./tikz/cbString/comonoid_unit_left.tikz}}
 & $\axeq{\cocopier{}\text{-un}}$ & 
    \InputIfFileExists{cbString/comonoid_unit_right.tikz}{}{\input{./tikz/cbString/comonoid_unit_right.tikz}}
 \\
        
    \InputIfFileExists{cbString/monoid_comm_left.tikz}{}{\input{./tikz/cbString/monoid_comm_left.tikz}}
 & $\axeq{\copier{}\text{-co}}$ & 
    \InputIfFileExists{cbString/monoid_comm_right.tikz}{}{\input{./tikz/cbString/monoid_comm_right.tikz}}
 & 
    \InputIfFileExists{cbString/comonoid_comm_left.tikz}{}{\input{./tikz/cbString/comonoid_comm_left.tikz}}
 & $\axeq{\cocopier{}\text{-co}}$ & 
    \InputIfFileExists{cbString/comonoid_comm_right.tikz}{}{\input{./tikz/cbString/comonoid_comm_right.tikz}}
 \\
        \hline
        
    \InputIfFileExists{cbString/spec_frob_left.tikz}{}{\input{./tikz/cbString/spec_frob_left.tikz}}
 & $\axeq{S}$ & 
    \InputIfFileExists{cbString/monoid_unit_right.tikz}{}{\input{./tikz/cbString/monoid_unit_right.tikz}}
 & 
    \InputIfFileExists{cbString/frob_left.tikz}{}{\input{./tikz/cbString/frob_left.tikz}}
 & $\axeq{F}$ & 
    \InputIfFileExists{cbString/frob_center.tikz}{}{\input{./tikz/cbString/frob_center.tikz}}
 \\
        \hline
        
    \InputIfFileExists{cbString/copier_nat_left.tikz}{}{\input{./tikz/cbString/copier_nat_left.tikz}}
 & $\axsubeq{\copier{}\text{-nat}}$ & 
    \InputIfFileExists{cbString/copier_nat_right.tikz}{}{\input{./tikz/cbString/copier_nat_right.tikz}}
 & 
    \InputIfFileExists{cbString/discharger_nat_left.tikz}{}{\input{./tikz/cbString/discharger_nat_left.tikz}}
 & $\axsubeq{\discharger{}\text{-nat}}$ & 
    \InputIfFileExists{cbString/discharger_nat_right.tikz}{}{\input{./tikz/cbString/discharger_nat_right.tikz}}
\\
        \hline
        
    \InputIfFileExists{cbString/adjoint1_bangs_left.tikz}{}{\input{./tikz/cbString/adjoint1_bangs_left.tikz}}
 & $\axsubeq{\codischarger{}\discharger{}}$ & 
    \InputIfFileExists{cbString/empty.tikz}{}{\input{./tikz/cbString/empty.tikz}}
 & 
    \InputIfFileExists{cbString/monoid_unit_right.tikz}{}{\input{./tikz/cbString/monoid_unit_right.tikz}}
 & $\axsubeq{\discharger{}\codischarger{}}$ & 
    \InputIfFileExists{cbString/adjoint2_bangs_right.tikz}{}{\input{./tikz/cbString/adjoint2_bangs_right.tikz}}
 \\
        
    \InputIfFileExists{cbString/frob_center.tikz}{}{\input{./tikz/cbString/frob_center.tikz}}
 & $\axsubeq{\cocopier{}\copier{}}$ & 
    \InputIfFileExists{cbString/adjoint1_diags_right.tikz}{}{\input{./tikz/cbString/adjoint1_diags_right.tikz}}
 & 
    \InputIfFileExists{cbString/monoid_unit_right.tikz}{}{\input{./tikz/cbString/monoid_unit_right.tikz}}
 & $\axsubeq{\copier{}\cocopier{}}$ & 
    \InputIfFileExists{cbString/spec_frob_left.tikz}{}{\input{./tikz/cbString/spec_frob_left.tikz}}

    \end{tabular}
    }
    \caption{Axioms of $\Cat{CB}_\Sigma$}
    \label{fig:cb sigma ax}
\end{figure}  \section{Appendix to Section~\ref{sc:monoidal}}

In this Appendix we show how Definition~\ref{def:biproduct category} is equivalent to the more standard definition of a category with finite biproducts.

Traditionally a category $\Cat C$ with a zero object $\zero$ is said to have finite biproducts if for every pair of objects $X_1$ and $X_2$ there exists a third object $X_1 \piu X_2$ and morphisms $\proj i \colon  X_1 \piu X_2 \to X_i$, $\inj i \colon X_i \to X_1 \piu X_2$ such that $(X_1 \piu X_2, \proj 1, \proj 2)$ is a product, $(X_1 \piu X_2, \inj 1, \inj 2)$ is a coproduct and
\[\inj i ; \proj j = \delta_{i,j}\]
where $\delta_{i,j} = \id{X_i}$ if $i = j$, otherwise $\delta_{i,j} = \zero_{X_i,X_j}$ (i.e.\ $\bang{X_i}; \cobang{X_j}$). This is consistent with Definition~\ref{def:biproduct category}, as we show now.

For any two objects $X_1$, $X_2$ of a  fb-category $\Cat C$, $X_1 \perG X_2$ is the categorical product $X_1 \times X_2$: the projections $\pi_1\colon X_1 \perG X_2 \to X_1$ and $\pi_2\colon X_1 \perG X_2 \to X_2$ are  
\[\begin{tikzcd}[column sep=3em]
	X_1 \perG X_2 \ar[r,"\id{X_1} \perG \, \bang{X_2}"] &  X_1 \perG \unoG \ar[r,"\runit{X_1}"]  & X_1
\end{tikzcd}
\quad \text{and} \quad
\begin{tikzcd}[column sep=3em]
	X_1 \perG X_2 \ar[r," \bang{X_1} \perG \id{X_2} "] &  \unoG \perG X_2 \ar[r,"\lunit{X_2}"]  & X_2.
\end{tikzcd}	
\]
The unit $\unoG$ is the terminal object and $\bang X$ the unique morphism of type $X \to \unoG$. 
For $f_1\colon Y\to X_1$, $f_2 \colon Y \to X_2$, their pairing $\pairing{f_1,f_2}\colon Y \to X_1 \perG X_2$ is given by \[\begin{tikzcd}
	Y  \ar[r,"\diag{Y}"] & Y \perG Y \ar[r,"f_1 \perG f_2"] &  X_1 \perG X_2.
\end{tikzcd}\]

$X_1 \perG X_2$ is also a categorical coproduct, with injections $\inj i \colon X_i \to X_1 \perG X_2$ and copairing $\copairing{f_1,f_2} \colon X_1 \perG X_2 \to Y$, for $f_1 \colon X_1 \to Y$ and $f_2 \colon X_2 \to Y$, in the dual way. Also dually $\unoG$ is the initial object and $\cobang X$ the unique morphism of type $\unoG \to X$.

Hence  $\unoG$ is a zero object and $\bang X ; \cobang Y \colon X \to Y$ is a zero morphism. Also the third requirement $\inj i ; \proj j = \delta_{i,j}$ holds. Indeed,

\[
\begin{tikzcd}[column sep=3em]
	X_1 \ar[r,"\Irunit {X_1}"] \ar[ddrr,bend right,"\id{X_1}"'] \ar[rr,start anchor=north east,bend left,"\inj 1"] & X_1 \piu \zero \ar[r,"{\id{X_1} \piu \cobang{X_2} }"] \ar[dr,"\id{X_1} \piu \cobang{X_2};\bang{X_2}"description] \ar[dr,bend right=45,"\id{X_1} \piu \id\zero"'] & X_1 \piu X_2 \ar[d,"\id{X_1} \piu \, \bang{X_2}"] \ar[dd,"\proj 1",rounded corners,to path={[pos=0.75] -| 
		([xshift=12em]\tikztotarget.east)	\tikztonodes --  (\tikztotarget)  }]
	\\
	& & X_1 \piu \zero \ar[d,"\runit{X_1}"] \\
	& & X_1
\end{tikzcd}
\qquad
\begin{tikzcd}
	X_2 \ar[r,"\Ilunit {X_2}"] \ar[ddrr,bend right,"\id{X_2}"'] \ar[rr,start anchor=north east,bend left,"\inj 2"] & \zero \piu X_2 \ar[r,"{ \cobang{X_1} } \piu \id{X_2}"] \ar[dr,"\cobang{X_1};\bang{X_1} \piu \id{X_2}"description] \ar[dr,bend right=45,"\id\zero \piu \id{X_2}"'] & X_1 \piu X_2 \ar[d," \bang{X_1} \piu \id{X_2}"] \ar[dd,"\proj 2",rounded corners,to path={[pos=0.75] -| 
		([xshift=12em]\tikztotarget.east)	\tikztonodes --  (\tikztotarget)  }] \\
	& & \zero \piu X_2 \ar[d,"\lunit{X_2}"] \\
	& & X_2
\end{tikzcd}
\]
and
\[
\begin{tikzcd}
	X_1 \ar[r,"\Irunit {X_1}"] \ar[d,"\bang{X_1}"'] \ar[rr,start anchor=north east,bend left,"\inj 1"] & X_1 \piu \zero \ar[r,"{\id{X_1} \piu \cobang{X_2} }"] \ar[d,"\bang{X_1} \piu \id\zero"]  & X_1 \piu X_2 \ar[d," \bang{X_1} \piu \id{X_2}"] \ar[dd,"\proj 2",rounded corners,to path={[pos=0.75] -| 
		([xshift=12em]\tikztotarget.east)	\tikztonodes --  (\tikztotarget)  }] \\
	\zero \ar[r,"\Irunit\zero"] \ar[dr,"\id{\zero}"'] & \zero \piu \zero \ar[r,"\id\zero \piu \cobang{X_2}"] \ar[d,"\lunit\zero"] & \zero \piu X_2 \ar[d,"\lunit{X_2}"] \\
	& \zero \ar[r,"\cobang{X_2}"] & X_2
\end{tikzcd}
\qquad
\begin{tikzcd}
	X_2 \ar[r,"\Ilunit {X_2}"] \ar[d,"\bang{X_2}"']  \ar[rr,start anchor=north east,bend left,"\inj 2"] & \zero \piu X_2 \ar[r,"\cobang{X_1} \piu \id{X_2}"] \ar[d,"\id\zero \piu \bang{X_2}"] & X_1 \piu X_2 \ar[d,"\id{X_1} \piu \bang{X_2}"] \ar[dd,"\proj 1",rounded corners,to path={[pos=0.75] -| 
		([xshift=12em]\tikztotarget.east)	\tikztonodes --  (\tikztotarget)  }] \\
	\zero \ar[r,"\Ilunit\zero"] \ar[dr,"\id{\zero}"'] & \zero \piu \zero \ar[r,"\cobang{X_1} \piu \id\zero"] \ar[d,"\runit\zero"] & X_1 \piu \zero \ar[d,"\runit{X_1}"] \\
	& \zero \ar[r,"\cobang{X_1}"] & X_1
\end{tikzcd}
\]
commute, using naturality of $\lunit{}$, $\runit{}$ and the fact that $\lunit\zero = \runit\zero$.  \section{Appendix to Section~\ref{sc:rig}}\label{appendix:rig}

In this Appendix we state and prove some additional facts for fb-rig categories (Proposition~\ref{prop:fbrig}) and for sesquistrict rig categories (Lemma~\ref{lm:sesqui:symmt}), which combined together yield Lemma~\ref{lm:per biprod} for sesquistrict fb rig categories. We also introduce the notion of \emph{monomial} and \emph{polynomial} for arbitrary sesquistrict rig categories (Definition~\ref{def:monomials and polynomials in arbitrary sesquistrict}). Finally, we give the proofs of the results claimed in Section~\ref{sc:rig}, namely Theorem~\ref{thm:strict(C) is sesquistrict}, Lemma~\ref{lemma:every functor in strict(D) is well behaved} and Theorem~\ref{thm:equivalentsignature}.

\begin{proposition}\label{prop:fbrig}
	In any fb-rig category the following diagrams commute:
	\[
	\begin{tikzcd}
		X \per (Y \piu Y) \ar[dr,"{\dl X Y Y}"] \\
		X \per Y \ar[u,"\id X \per \diag Y"] \ar[r,"\diag{X \per Y}"] \ar[d,"\diag X \per \id Y"'] & (X \per Y) \piu (X \per Y) \\
		(X \piu X) \per Y \ar[ur,"{\dr X X Y}"'] 
	\end{tikzcd}
	\qquad
	\begin{tikzcd}[column sep=4em]
		X \per \zero \ar[dr,"\annr X"] \\
		X \per Y \ar[u,"\id X \per \bang Y"] \ar[r,"\bang{X \per Y}"] \ar[d,"\bang X \per \id Y"'] & \zero \\
		\zero \per Y \ar[ur,"\annl Y"']
	\end{tikzcd}
	\]
	Dually, also the following diagrams commute:
	\[
	\begin{tikzcd}[column sep=4em]
		X \per (Y \piu Y) \ar[dr,"\id X \per \codiag Y"] \\
		(X \per Y) \piu (X \per Y) \ar[u,"{\Idl X Y Y}"] \ar[r,"\codiag{X \per Y}"] \ar[d,"{\Idr X Y Y}"'] & X \per Y \\
		(X \piu X) \per Y \ar[ur,"\codiag X \per \id Y"']
	\end{tikzcd}
	\qquad
	\begin{tikzcd}
		X \per \zero \ar[dr,"\id X \per \cobang Y"] \\
		\zero \ar[u,"\Iannr X"] \ar[r,"\cobang{X \per Y}"] \ar[d,"\Iannl Y"'] & X \per Y \\
		\zero \per Y \ar[ur,"\cobang X \per \id Y"']
	\end{tikzcd}
	\]
\end{proposition}
\begin{proof}
	The diagrams involving $\zero$ are trivially commuting due to the fact that it is a zero object.
	
	We show that $\diag{X \per Y} = (\diag X \per \id Y) ; \dr X X Y$. To do that, we prove that the morphism $(\diag X \per \id Y) ; \dr X X Y$ satisfies the same universal property of $\diag{X \per Y} = \pairing{\id{X \per Y}, \id{X \per Y}}$; in other words we show that
	\begin{enumerate}
		\item $(\diag X \per \id Y) ; \dr X X Y ; \proj 1 = \id{X \per Y}$, 
		\item $(\diag X \per \id Y) ; \dr X X Y ; \proj 2 = \id{X \per Y}$
	\end{enumerate}
	where $\proj 1, \proj 2 \colon (X \per Y) \piu (X \per Y) \to X \per Y$. Recall that $\proj 1 = (\id{X \per Y} \piu \bang{X \per Y}) ; \runitp{X \per Y}$ and that $\proj 2 = (\bang{X \per Y} \piu \id{X \per Y}) ; \lunitp{X \per Y}$. We have that the following diagram commutes:
	\[
	\begin{tikzcd}[column sep={3.5em,},bend angle=10,row sep=3em]
		& (X \piu X) \per Y \ar[r,"{\dr X X Y}"] \ar[d,"(\id X \piu \bang X) \per \id Y"{description,name=f}] & (X \per Y) \piu (X \per Y)  \ar[r,"\id{X \per Y} \piu \bang{X \per Y}"] \ar[d,"(\id X \per \id Y) \piu (\bang X \per \id Y)"{description,name=g}] & (X \per Y) \piu \zero \ar[dr,"\runitp{X \per Y}"]\\
		X \per Y \ar[ur,"\diag X \per \id Y"] 
		\arrow[rrrr,"\id{X \per Y}"',rounded corners,to path={ -- ([yshift=-5cm]\tikztostart.south) -- ([yshift=-5cm]\tikztotarget.south) \tikztonodes -| ([yshift=-2cm]\tikztotarget.south) -- (\tikztotarget)}]
		& (X \piu \zero) \per Y \ar[r,"{\dr X \zero Y}"] \ar[rrr,bend right,"\runitp X \per \id Y"'] & (X \per Y) \piu (\zero \per X) \ar[ur,swap,"\id{X \per Y} \piu \annl X"name=h] \ar[rr,draw=none,"\ast_3"description] & & X \per Y \\[-1em]
		& & \ast_4
		\arrow[from=f,to=g,draw=none,"\ast_1"description]
		\arrow[from=1-3,to=h,draw=none,"\ast_2"description]
	\end{tikzcd}
	\]
	because of the naturality of $\delta^r$ ($\ast_1$), terminality of $\zero$ ($\ast_2$), the derived law \eqref{eq:dl8} ($\ast_3$) and the fact that $\diag X ; \proj 1 = \id X$ ($\ast_4$). This proves the first equality. As per the second, for similar reasons the following diagram commutes:
	\[
	\begin{tikzcd}[column sep={3.5em,},bend angle=10,row sep=3em]
		& (X \piu X) \per Y \ar[r,"{\dr X X Y}"] \ar[d,"(\bang X \piu \id X) \per \id Y"{description,name=f}] & (X \per Y) \piu (X \per Y)  \ar[r,"\bang{X \per Y} \piu \id{X \per Y}"] \ar[d,"(\bang X \per \id Y) \piu (\id X \per \id Y)"{description,name=g}] & \zero \piu (X \per Y)  \ar[dr,"\lunitp{X \per Y}"]\\
		X \per Y \ar[ur,"\diag X \per \id Y"] 
		\arrow[rrrr,"\id{X \per Y}"',rounded corners,to path={ -- ([yshift=-5cm]\tikztostart.south) -- ([yshift=-5cm]\tikztotarget.south) \tikztonodes -| ([yshift=-2cm]\tikztotarget.south) -- (\tikztotarget)}]
		& (\zero \piu X) \per Y \ar[r,"{\dr \zero X Y}"] \ar[rrr,bend right,"\lunitp X \per \id Y"'] & (\zero \per Y) \piu (X \per Y) \ar[ur,swap,"\annl Y \piu \id{X \per Y}"name=h] \ar[rr,draw=none,"\ast_5"description] & & X \per Y \\
	\end{tikzcd}
	\]
	where now $\ast_5$ commutes because of Axiom~\eqref{eq:rigax11}. 
	
	The fact that $\codiag{X \per Y} = \Idr X X Y ; (\codiag X \per \id Y)$ can be proved in a similar way using the derived laws~\eqref{eq:dl5} and~\eqref{eq:dl7}.
\end{proof} 

\subsection{Additional facts for sesquistrict rig categories}

\begin{definition}\label{def:monomials and polynomials in arbitrary sesquistrict}
	Let $H \colon \Cat M \to \Cat D$ be a sesquistrict rig category. A \emph{monomial} in $\Cat D$ is an object of the form $\Per[j=1][m]{H(A_j)}$ where $A_j \in \Cat M$. A \emph{polynomial} in $\Cat D$ is an object of the form $\Piu[i=1][n] U_i$ where $U_i$ is a monomial for all $i=1\dots n$.
\end{definition}

\begin{lemma}\label{lm:sesqui:symmt}
	Let $H \colon \Cat M \to \Cat D$ be a sesquistrict rig category. For all monomials $U,V$ and polynomials $Y,Z$, the following hold:
	\begin{enumerate}
		\item $\dl{U}{Y}{Z}=\id{UY \piu UZ}$.
		\item If  $Y=\Piu[i]{U_i}$, then $\symmt{Y}{V} = \Piu[i]{\symmt{U_i}{V}}$. 
	\end{enumerate}
\end{lemma}
\begin{proof}
	We prove the first item by induction on $U$. 
	
	\textbf{Case $U = \uno$:} $\dl{1}{Y}{Z} \axeq{\ref{eq:dl6}}\id{Y\piu Z}$.
	
	\textbf{Case $U = H(A)U'$:}
	\begin{align*}
		\dl{H(A)U'}{Y}{Z} &= \tag{\ref{eq:dl3}} (\id{H(A)} \per \dl{U'}{Y}{Z}) ; \dl{H(A)}{U'Y}{U'Z} \\
		&= \tag{Ind. hp., sesquistrictness} (\id{H(A)} \per \id{U'Y\piu U'Z}) ; \id{H(A)U'Y \piu H(A)U'Z } \\
		&= \id{H(A)U'Y \piu H(A)U'Z } 
	\end{align*}
	
	We prove the second point by induction on $Y$.

	\textbf{Case $Y = \zero$:} it follows from axiom~\eqref{eq:rigax9}.

	\textbf{Case $Y = U \piu Y'$:} Let $Y' = \Piu[i']{U'_{i'}}$
	\begin{align*}
		\symmt{U \piu Y'}{V} &= \tag{\ref{eq:rigax1}} (\symmt{U}{V} \piu \symmt{Y'}{V}) ; \Idl{V}{U}{Y'} \\
		&= \tag{Lemma \ref{lm:sesqui:symmt}.1} \symmt{U}{V} \piu \symmt{Y'}{V} \\
		&= \tag{Ind. hp.} \symmt{U}{V} \piu \Piu[i']{\symmt{U'_{i'}}{V}} 
	\end{align*}
\end{proof}

The following useful result is an immediate consequence of Proposition~\ref{prop:fbrig} and Lemma~\ref{lm:sesqui:symmt}.
\begin{lemma}\label{lm:per biprod}
	Let $H \colon \Cat M \to \Cat D$ be a sesquistrict fb rig category, $U$ a monomial and $Y$ a polynomial. Then:
	\begin{enumerate}
		\item $\id{U} \per \diag{Y} = \diag{UY} = \diag{U} \per \id{Y}$.
		\item $\id{U} \per \bang{Y} = \bang{UY} = \bang{U} \per \id{Y}$.
		\item $\id{U} \per \codiag{Y} = \codiag{UY} = \codiag{U} \per \id{Y}$.
		\item $\id{U} \per \cobang{Y} = \cobang{UY} = \cobang{U} \per \id{Y}$.
	\end{enumerate}
\end{lemma}

\subsection{Proofs of Section~\ref{sc:rig}}

\begin{proof}[Proof of Theorem~\ref{thm:strict(C) is sesquistrict}]	
	We recall from~\cite{johnson2021bimonoidal} some facts regarding the strictification $\sCatT C$ of $\Cat C$.
	
	Let $P=\Piu[i=1][r]{U_i}$ and $Q=\Piu[j=1][s]{V_j}$ be objects of $\sCatT C$ (the $U_i$'s and $V_j$'s are words of objects of $\Cat C$). Then $P \per Q$ is given as in~\eqref{def:productPolynomials} and it is immediate to see that $A \per (P \piu Q) = (A \per P) \piu (A \per Q)$ when $A \in Ob(\Cat C)$. We need to show that $\id{A \per (P \piu Q)} = \dl{A}{P}{Q}$ in $\sCatT C$.

	In $\sCatT C$, the left distributor $\dl S P Q \colon S \per (P \piu Q) \to (S \per P) \piu (S \per Q)$ is defined as
	\[
	\begin{tikzcd}
		S \per (P \piu Q) \ar[r,"\dl S P Q"] \ar[d,"\symmt{S}{P \piu Q}"'] & (S \per P) \piu (S \per Q) \\
 		(P \piu Q) \per S \ar[r,"\id{}"] & (P \per S) \piu (Q \per S) \ar[u,"{\symmt{P}{S} \piu \symmt{Q}{S}}"']
	\end{tikzcd} \qquad \text{(in $\sCatT C$)}
	\]
	where the symmetry $\symmt{R}{T} \colon R \per T \to T \per R$ in $\sCatT C$ is defined as the following composite in $\Cat C$:
	\[
	\begin{tikzcd}
		\pi(R \per T) \ar[r,"{\symmt{R}{T}}"] \ar[d,"{\cong_{Lap}}^{-1}"'] & \pi(T \per R) \\
		\pi(R) \per \pi(T) \ar[r,"{\symmt{\pi(R)}{\pi(T)}}"] & \pi(T) \per \pi(R) \ar[u,"\cong_{Lap}"']
	\end{tikzcd}
	\]
	The arrow $\cong_{Lap}$ is the isomorphism built up using the structure of rig category whose existence and uniqueness is guaranteed by the coherence theorem for rig categories~\cite{laplaza_coherence_1972,johnson2021bimonoidal}.
	
In order to prove that $\id{A \per (P \piu Q)} = \dl{A}{P}{Q}$ in $\sCatT C$, we will use the coherence theorem for rig categories as formulated in~\cite{johnson2021bimonoidal}. Suppose that $U_i=B^i_1\dots B^i_{n_i}$ and $V_j = C^j_1\dots C^j_{m_j}$. Now, we have that $\pi(A \per (P \piu Q))$ is the right bracketing of the following object of $\Cat C$:
	 \[
	 \Piu[i=1][r]{\Per[k=1][n_i]{(A \per B^i_k)}} \piu \Piu[j=1][m]{\Per[k=1][m_j]{(A \per C^j_k)}}
	 \]
	 This object, as a term of the free $\{\piu,\per\}$-algebra generated by the set
	 \[
	 X = \{A,B^1_1,\dots,B^1_{n_1},\dots,B^r_1,\dots,B^r_{n_r}, C^1_1,\dots,C^1_{m_1},\dots,C^s_1,\dots,C^s_{n_s},\zero^X,\uno^X\},
	 \] is \emph{regular} in the sense of~\cite[Definition 3.1.25]{johnson2021bimonoidal}. Indeed, for every $i$ the factors $A, B^i_1,\dots,B^i_{n_i}$ are all distinct elements of $X$, and similarly for every $j$ the factors $A, C^j_1,\dots,C^j_{m_j}$ are all distinct elements of $X$; moreover, all the $r + s$ addends $\Per[k=1][n_i]{(A \per B^i_k)}$ and $\Per[k=1][m_j]{(A \per C^j_k)}$ are all pairwise distinct elements of the free $\{\piu,\per\}$-algebra on $X$. Hence, by the coherence theorem for rig categories, as stated in Theorem~3.9.1 of~\cite{johnson2021bimonoidal}, there exists a unique isomorphism, built up using instances of the natural isomorphisms given by the structure of rig category and involving only the objects in $X$, from $\pi(A \per (P \piu Q))$ to itself. Since $\id{A \per (P \piu Q)}$ and $\dl{A}{P}{Q}$ are two such structural isomorphisms, they must coincide. 
\end{proof}

\begin{proof}[Proof of Lemma~\ref{lemma:every functor in strict(D) is well behaved}]Simply postcompose $F$ with the two functors $\sCatT D \to \Cat D \to \sCatT D$ that make the equivalence $\Cat D \simeq \sCatT D$.
\end{proof}

\begin{proof}[Proof of Theorem~\ref{thm:equivalentsignature}]
	Let $\sign$ be a rig signature and let $\gen$ be a generator in $\sign$, say \[\gen \colon \Piu[i=1][n_\gen]{U_i} \to \Piu[j=1][m_\gen]{V_j}\]
	where $U_i$, $V_j$ are monomials for $i=1 \dots n_\gen$ and $j=1 \dots m_\gen$. For each $(i,j)$, we define a new generator
	\[
	\gen_{j,i} \colon U_i \to V_j
	\]
	having as arity the monomial $U_i$ and as coarity the monomial $V_j$. These $\gen_{j,i}$'s are now formal symbols that we use to define a signature $\sign_M$, obtained from $\sign$ by replacing each $\gen \in \sign$ with all the $\gen_{j,i}$'s:
\[
	\sign_M \defeq \{ \gen_{j,i} \mid \gen\in \sign,\, j = 1\dots m_\gen,\, i=1\dots n_\gen \}\text{.}
	\]
	Observe that $\sign_M$ is a monoidal signature since it only contains symbols having as arity and coarity some monomials.
	
	For example, assume that $\sign$ consists of a single generator $\gen \colon AB \piu C \to A \piu B \piu C$ depicted as follows:
	\[
    \InputIfFileExists{tapes/examples/rigGen.tikz}{}{\input{./tikz/tapes/examples/rigGen.tikz}}
\]
	Then $\sign_M$ is the signature consisting of the following six generators:
	\[ \monGenDue{\gen_{1,1}}{A}{B}{A} \quad \monGenDue{\gen_{2,1}}{A}{B}{B} \quad  \monGenDue{\gen_{3,1}}{A}{B}{C}\]
	\[ \monGen{\gen_{1,2}}{C}{A} \quad \monGen{\gen_{2,2}}{C}{B} \quad \monGen{\gen_{3,2}}{C}{C}\]
	
	Now, call $\fssbRig$ and $\fssbRigM$ the free sesquistrict fb rig categories generated, respectively, by $\sign$ and $\sign_M$. We show that there is an isomorphism of rig categories 
	\[\begin{tikzcd} \fssbRig \ar[r,bend left,"F"above]  &\fssbRigM \ar[l,bend left,"G"below]
	\end{tikzcd}\]
	The functor $F$ is the unique strict fb-rig functor mapping every $\gen \in \sign$ with arity $\Piu[i][n]{U_i} $ and coarity $\Piu[j][m]{V_j}$ into
	\[\begin{tikzcd}[ampersand replacement=\&,column sep=3em]
		\PiuL[i=1][n]{U_i} \ar[r,"{\PiuL[i=1][n]{\diagg{U_i} m}}"] \& \PiuL[i=1][n]{\PiuL[j=1][m]{U_i}} \ar[r,"{\PiuL[i=1][n]{\PiuL[j=1][m]{\gen_{ji}}}}"] \& \PiuL[i=1][n]{\PiuL[j=1][m]{V_j}} \ar[r,"\codiagg{\Piu[j=1][m]{V_j}}{n}"] \& \PiuL[j=1][m]{V_j}
	\end{tikzcd}
	\]
	Similarly, the functor $G$ is the unique strict fb-rig functor mapping every $\gen_{ji} \in \sign_M$ into 
	\[
	\begin{tikzcd}[ampersand replacement=\&,column sep=3em]
		{U_i} \ar[r,"\inj{i}"] \& \PiuL[i=1][n]{U_i} \ar[r,"s"] \& \PiuL[j=1][m]{V_j} \ar[r,"\proj{j}"] \& {V_j}
	\end{tikzcd}
	\]
	To prove that $GF=Id_{\fssbRig}$, it is enough to check that $GF(\gen)=\gen$. Since $G$ preserves the finite biproduct rig structure, it holds that $GF(\gen)$ is
	\[\begin{tikzcd}[ampersand replacement=\&,column sep=4em]
		\PiuL[i=1][n]{U_i} \ar[r,"{\PiuL[i=1][n]{\diagg{U_i} m}}"] \& \PiuL[i=1][n]{\PiuL[j=1][m]{U_i}} \ar[r,"{\PiuL[i=1][n]{\PiuL[j=1][m]{G(\gen_{ji})}}}"] \& \PiuL[i=1][n]{\PiuL[j=1][m]{V_j}} \ar[r,"\codiagg{\Piu[j=1][m]{V_j}}{n}"] \& \PiuL[j=1][m]{V_j}
	\end{tikzcd}
	\]
	which, by Proposition~\ref{prop:f as composite of diagonals-fji-codiagonals}, is exactly $\gen$.

	To prove that  $FG(\gen_{ji})=\gen_{ji}$, just observe that $FG(\gen_{ji})$ is the upper leg of the following diagram:
\[
		\begin{tikzcd}[row sep={3em}]& & \PPiuL h n k m {U_h} \ar[r,"{\PPiuL h n k m {s_{kh}}}"] & \PPiuL h n k m {V_k} \ar[dr,"\codiagg{\Piu[k]{V_k}} n"]\\
			& \PiuL[h=1][n]{U_h} \ar[ur,"{\PiuL[h=1][n]{\diagg{U_h} m}}"]  & \PiuL[k=1][m]{U_i} \ar[u,"\inj i"description] \ar[r,"{\PiuL[k=1][m]{s_{ki}}}"] \ar[d,"\proj j"description] & \PiuL[k=1][m]{V_k} \ar[r,"\id{}"] \ar[u,"\inj i"description] \ar[d,"\proj j"description] & \PiuL[k=1][m]{V_k} \ar[dr,"\proj j"]\\
			U_i \ar[ur,"\inj i"] \ar[urr,"\diagg{U_i} m "description] \ar[rr,"\id{}"] & & U_i  \ar[r,"s_{ji}"] & V_j \ar[rr,"\id{}"] & & V_j
		\end{tikzcd}
		\]
		which commutes because of the naturality of the injections and the projections for the biproduct $\piu$, as well as the universal property of $\codiagg{\Piu[k]{V_k}} n$ as the copairing of n copies of $\id{\Piu[k]{V_k}}$ (in the top-right triangle).	
	\end{proof}  \section{Appendix to Section~\ref{sc:tape}}\label{sc:appendixTapes}

In this Appendix we study, in detail, the properties of $\per_{\CatTape}$. It is divided in four subsections. In Section~\ref{appTape1} we state and prove several properties of the distributors and the symmetries (Lemma~\ref{lemma:tapesymdis} and~\ref{lm:tape:dl aux}). In Section~\ref{appTape2} we present some additional lemmas for monomial whiskerings (Lemma~\ref{lm:whisk} and~\ref{lm:tape:perC}), which are needed to prove, in Section~\ref{app:whiskering}, the 17 equations stated in Table~\ref{table:whisk}: the algebra of whiskerings. Finally, in Section~\ref{appTape4} we prove the remaining results that were claimed in Section~\ref{sc:tape}, namely Theorems~\ref{thm:taperig},~\ref{thm:Tapes is free sesquistrict generated by sigma},~\ref{thm:contextual} and Corollary~\ref{co:rig semantics}.

\subsection{Properties of distributors and symmetries in $\CatTape$}\label{appTape1}

The following lemma provides a sanity check for our inductive definitions.
\begin{lemma}\label{lemma:tapesymdis}
For all $P,P',Q,R$, monomials $U,V$ the followings hold:
\begin{enumerate}
    \item $\dl{U}{Q}{R} = \id{U(Q\piu R)}$\label{lm:tape:dl U}
    \item $\dl{P \piu P'}{Q}{R} = (\dl{P}{Q}{R} \piu \dl{P'}{Q}{R});(\id{PQ} \piu \symmp{PR}{P'Q} \piu \id{P'R})$\label{lm:tape:dlG}
    \item $\dl{P}{\zero}{R} = \id{PR}$\label{lm:tape:dl zero}
    \item $\dl{P}{Q}{\zero} = \id{PQ}$\label{lm:tape:dl zero ultimo}
    \item $\symmt{P}{Q} ; \symmt{Q}{P} = \id{P  Q}$\label{lm:tape:symmtInv}
    \item $\symmt{P}{Q \piu R} = \dl{P}{Q}{R} ; (\symmt{P}{Q} \piu \symmt{P}{R})$\label{lm:tape:symmt dl}
    \item $\symmt{P}{\uno} = \id{P}$\label{lm:tape:symmt uno}
    \item $\symmt{P}{V} = \Piu[i]{\tapesymm{U_i}{V}}$ for $P = \Piu[i]{U_i}$. In particular $\symmt{U}{V}=\tape{\symm{U}{V}}$. \label{lm:tape:symmt V}
\end{enumerate}
\end{lemma}

\begin{proof}[Proof of Lemma \ref{lemma:tapesymdis}.\ref{lm:tape:dl U}] It holds by definition of $\delta^l$.
\end{proof}

\begin{proof}[Proof of Lemma \ref{lemma:tapesymdis}.\ref{lm:tape:dlG}] We prove that $\dl{P \piu P'}{Q}{R} = (\dl{P}{Q}{R} \piu \dl{P'}{Q}{R});(\id{PQ} \piu \symmp{PR}{P'Q} \piu \id{P'R})$ by induction on $P$.

    \textbf{Case $P = \zero$:} it holds by definition of $\delta^l$.

\textbf{Case $P = U \piu P''$:} 
    \begingroup
    \allowdisplaybreaks
    \begin{align*}
        \dl{U \piu P'' \piu P'}{Q}{R} &= \dl{U \piu (P'' \piu P')}{Q}{R} \\
        &=\tag{Def. $\delta^l$} (\id{U(Q \piu R)} \piu \dl{P'' \piu P'}{Q}{R});(\id{UQ} \piu \symmp{UR}{P''Q \piu P'Q} \piu \id{P''R \piu P'R})\\
        &=\tag{Rem.~\ref{rem:per tape is right distributive}} 
    \InputIfFileExists{lemmaDlG/step1.tikz}{}{\input{./tikz/lemmaDlG/step1.tikz}}
 \\
        &=\tag{Ind. hp.} 
    \InputIfFileExists{lemmaDlG/step2.tikz}{}{\input{./tikz/lemmaDlG/step2.tikz}}
\\
        &=\tag{Funct. $\piu$} 
    \InputIfFileExists{lemmaDlG/step4.tikz}{}{\input{./tikz/lemmaDlG/step4.tikz}}
\\
&=\begin{multlined}[t]
            (((\id{U(Q\piu R)} \piu \dl{P''}{Q}{R});(\id{UQ} \piu \symmp{UR}{P''Q} \piu \id{P''R})) \piu \dl{P'}{Q}{R}); \\ (\id{UQ \piu P''Q} \piu \symmp{UR \piu P''R}{P'Q} \piu \id{P'R})
        \end{multlined} \\
        &=\tag{Def. $\delta^l$}(\dl{U \piu P''}{Q}{R} \piu \dl{P'}{Q}{R});(\id{UQ \piu P''Q} \piu \symmp{UR \piu P''R}{P'Q} \piu \id{P'R})
    \end{align*}
    \endgroup
\end{proof}

\begin{proof}[Proof of Lemma \ref{lemma:tapesymdis}.\ref{lm:tape:dl zero}] We prove that $\dl{P}{\zero}{R} = \id{PR}$ by induction on $P$.

    \textbf{Case $P = \zero$:} it holds by definition of $\delta^l$.

    \textbf{Case $P = U \piu P'$:} 
    \begin{align*}
        \dl{U \piu P'}{\zero}{R} &=\tag{Def. $\delta^l$} (\id{U(\zero \piu R)} \piu \dl{P'}{\zero}{R});(\id{U\zero} \piu \symmp{UR}{P'\zero} \piu \id{P'R}) \\
        &=\tag{Ind. hp.} (\id{UR} \piu \id{P'R});(\id{\zero} \piu \symmp{UR}{\zero} \piu \id{P'R}) \\
        &=\tag{\ref{eq:symmax2}} (\id{UR} \piu \id{P'R});(\id{UR} \piu \id{P'R}) \\
        &=\tag{Funct. $\piu$} \id{UR \piu P'R} \\
        &=\tag{Rem.~\ref{rem:per tape is right distributive}} \id{(U \piu P')R} \\
    \end{align*}
\end{proof}

\begin{proof}[Proof of Lemma \ref{lemma:tapesymdis}.\ref{lm:tape:dl zero ultimo}] Analogous to the proof of Lemma~\ref{lemma:tapesymdis}.\ref{lm:tape:dl zero} above.
\end{proof}

\begin{proof}[Proof of Lemma \ref{lemma:tapesymdis}.\ref{lm:tape:symmtInv}] 
We want to prove that $\symmt{P}{Q} ; \symmt{Q}{P} = \id{P  Q}$. 
    First we prove the following three equations:
    \begin{align*}
        \symmt{\zero}{Q} &= \id{\zero} \tag{$\ast_1$} \\
        \symmt{U \piu P'}{V} &= (\symmt{U}{V} \piu \symmt{P'}{V}) ; \Idl{V}{U}{P'} \tag{$\ast_{2.1}$} \\
        \symmt{U \piu P'}{Q} &= (\Piu[j]{\tapesymm{U}{V_j}} \piu \symmt{P'}{Q}) ; \Idl{Q}{U}{P'} \text{ with $Q = \Piu[j]{V_j}$} \tag{$\ast_2$}
    \end{align*} 
    For $(\ast_1)$ we proceed by induction on $Q$:
    
    \textbf{Case $Q = \zero$:} it holds by definition of $\symmt$.
    
    \textbf{Case $Q = V \piu Q'$:}
    \begin{align*}
        \symmt{\zero}{V \piu Q'} &=\tag{Def. $\symmt$} \dl{\zero}{V}{Q'} ; (\Piu[i=1][0]{\tapesymm{U_i}{V}} \piu \symmt{\zero}{Q'}) \\
        &=\tag{Def. $\delta^l$, ind. hp.} \id\zero
    \end{align*}

    For $(\ast_{2.1})$, suppose that $P' = \Piu[i']{U_{i'}'}$ and observe that the following holds:
    \begin{align*}
        \symmt{U \piu P'}{V} &=\tag{Def. $\symmt$} \dl{U \piu P'}{V}{\zero} ; (\tapesymm{U}{V} \piu \Piu[i']{\tapesymm{U'_{i'}}{V}} \piu \symmt{U}{\zero}) \\
        &=\tag{Def. $\symmt$} \dl{U \piu P'}{V}{\zero} ; (\symmt{U}{V} \piu \symmt{P'}{V}) \\
        &=\tag{Lemma~\ref{lemma:tapesymdis}.\ref{lm:tape:dl zero}} (\symmt{U}{V} \piu \symmt{P'}{V}) \\
        &=\tag{Lemma~\ref{lemma:tapesymdis}.\ref{lm:tape:dl U}} (\symmt{U}{V} \piu \symmt{P'}{V}) ; \Idl{V}{U}{P'}
    \end{align*}

    For $(\ast_2)$ we proceed by induction on $Q:$
    
    \textbf{Case $Q = \zero$:}
    \begin{align*}
        \symmt{U \piu P'}{\zero} &=\tag{Def. $\symmt$} \id\zero \\
        &=\tag{Def. $\delta^l$} (\Piu[i=1][0]{\tapesymm{U_i}{V}} \piu \symmt{\zero}{Q'}) ; \dl{\zero}{V}{Q'}\\
    \end{align*}
    
    \textbf{Case $Q = V \piu Q'$:}
    \begin{align*}
        \symmt{U \piu P'}{V \piu Q'} &=\tag{Def. $\symmt$} \dl{U \piu P'}{V}{Q'} ; (\tapesymm{U}{V} \piu \Piu[i']{\tapesymm{U'_{i'}}{V}} \piu \symmt{U \piu P'}{Q'}) \\
        &=\tag{Lemma~\ref{lemma:tapesymdis}.\ref{lm:tape:symmt V}} \dl{U \piu P'}{V}{Q'} ; (\symmt{U \piu P'}{V} \piu \symmt{U \piu P'}{Q'}) \\
        &=\tag{$\ast_{2.1}$, ind. hp.} \dl{U \piu P'}{V}{Q'} ; (((\symmt{U}{V} \piu \symmt{P'}{V}) ; \Idl{V}{U}{P'}) \piu ((\symmt{U}{Q'} \piu \symmt{P'}{Q'}) ; \Idl{Q'}{U}{P'})) \\
        &=\tag{Def.\ $\delta^l$}\begin{multlined}[t]
            (\dl{U}{V}{Q'} \piu \dl{P'}{V}{Q'}) ; (\id{UV} \piu \symmp{UQ'}{P'V} \piu \id{P'Q'}) ; \\
            (((\symmt{U}{V} \piu \symmt{P'}{V}) ; \Idl{V}{U}{P'}) \piu ((\symmt{U}{Q'} \piu \symmt{P'}{Q'}) ; \Idl{Q}{U}{P'}))
        \end{multlined} \\
        &=\tag{Funct.\ $\piu$}\begin{multlined}[t]
            (\dl{U}{V}{Q'} \piu \dl{P'}{V}{Q'}) ; (\id{UV} \piu \symmp{UQ'}{P'V} \piu \id{P'Q'}) ; \\
            (\symmt{U}{V} \piu \symmt{P'}{V} \piu \symmt{U}{Q'} \piu \symmt{P'}{Q'}) ; (\Idl{V}{U}{P'} \piu \Idl{Q'}{U}{P'})
        \end{multlined} \\
        &=\tag{\ref{ax:symmpnat}}\begin{multlined}[t]
            (\dl{U}{V}{Q'} \piu \dl{P'}{V}{Q'}) ; (\symmt{U}{V} \piu \symmt{U}{Q'} \piu \symmt{P'}{V} \piu \symmt{P'}{Q'}) ; \\
            (\id{VU} \piu \symmp{Q'U}{VP'} \piu \id{Q'P}) ; (\Idl{V}{U}{P'} \piu \Idl{Q'}{U}{P'})
        \end{multlined} \\
        &=\tag{Def.\ $\delta^l$} (\dl{U}{V}{Q'} \piu \dl{P'}{V}{Q'}) ; (\symmt{U}{V} \piu \symmt{U}{Q'} \piu \symmt{P'}{V} \piu \symmt{P'}{Q'}) ; \Idl{V \piu Q'}{U}{P'} \\
        &=\tag{Funct. $\piu$} (\dl{U}{V}{Q'} ; (\symmt{U}{V} \piu \symmt{U}{Q'})) \piu (\dl{P'}{V}{Q'} ; (\symmt{P'}{V} \piu \symmt{P'}{Q'})) ; \Idl{V \piu Q'}{U}{P'} \\
        &=\tag{Def.\ $\symmt$, Lemma~\ref{lemma:tapesymdis}.\ref{lm:tape:symmt V} } (\symmt{U}{V \piu Q'} \piu \symmt{P'}{V \piu Q'}) ; \Idl{V \piu Q'}{U}{P'}
    \end{align*}
    
    The rest of the proof is by induction on $Q$:

    \textbf{Case $Q = \zero$} \begin{align*}
        \symmt{P}{\zero} ; \symmt{\zero}{P} &=\tag{Def. $\symmt$} \id\zero ; \symmt{\zero}{P} \\
        &=\tag{$\ast_1$} \id\zero ; \id\zero \\
        &= \id\zero \\
    \end{align*}

    \textbf{Case $Q = V \piu Q'$} \begin{align*}
        \symmt{P}{V \piu Q'} ; \symmt{V \piu Q'}{P} &=\tag{Def. $\symmt$, $\ast_2$} \dl{P}{V}{Q'} ; (\Piu[i]{\tapesymm{U_i}{V}} \piu \symmt{P}{Q'}) ; (\Piu[i]{\tapesymm{V}{U_i}} \piu \symmt{Q'}{P}) ; \Idl{P}{V}{Q'} \\
        &=\tag{Funct. $\piu$} \dl{P}{V}{Q'} ; ((\Piu[i]{\tapesymm{U_i}{V}} ; \Piu[i]{\tapesymm{V}{U_i}}) \piu (\symmt{P}{Q'} ; \symmt{Q'}{P})) ; \Idl{P}{V}{Q'} \\
        &=\tag{\ref{eq:symmax1}, ind. hp.} \dl{P}{V}{Q'} ; (\id{PV} \piu \id{PQ'}) ; \Idl{P}{V}{Q'} \\  
        &=\tag{Iso} \id{P(V \piu Q')}
    \end{align*}
\end{proof}

\begin{proof}[Proof of Lemma \ref{lemma:tapesymdis}.\ref{lm:tape:symmt V}] It holds by definition of $\symmt$.
\end{proof}

\begin{lemma}\label{lm:tape:dl aux}
    The following holds for all monomials $V$ and polynomials $P,Q,R$:
    \[ \dl{P}{V \piu Q}{R} ; (\dl{P}{V}{Q} \piu \id{PR}) = \dl{P}{V}{Q \piu R} ; (\id{PV} \piu \dl{P}{Q}{R}) \]
\end{lemma}
\begin{proof} By induction on $P$.

    \textbf{Case $P = \zero$:} it holds since both sides of the equation are $\id\zero$ by definition of $\delta^l$.

    \textbf{Case $P = U \piu P'$:} For the inductive case we will mix the classical syntax with tape diagrams to ease the reading.
    \begingroup
    \allowdisplaybreaks
    \begin{align*}
        &\dl{U \piu P'}{V \piu Q}{R} ; (\dl{U \piu P'}{V}{Q} \piu \id{UR \piu P'R}) \\
        &=\tag{Def. $\delta^l$} \begin{multlined}[t]
(\id{U((V \piu Q) \piu R)} \piu \dl{P'}{V \piu Q}{R}) ; (\id{UV \piu UQ} \piu \symmp{UR}{P'(V \piu Q)} \piu \id{P'R}) ; \\
            (\dl{U \piu P'}{V}{Q} \piu \id{UR \piu P'R})
        \end{multlined}\\
        &=\tag{Def. $\delta^l$} \begin{multlined}[t]
(\id{U((V \piu Q) \piu R)} \piu \dl{P'}{V \piu Q}{R}) ; (\id{UV \piu UQ} \piu \symmp{UR}{P'(V \piu Q)} \piu \id{P'R}) ; \\
            (((\id{U(V \piu Q)} \piu \dl{P'}{V}{Q});(\id{UV} \piu \symmp{UQ}{P'V} \piu \id{P'Q})) \piu \id{UR \piu P'R})
        \end{multlined}\\
        &= 
    \InputIfFileExists{lemmaDl/step1.tikz}{}{\input{./tikz/lemmaDl/step1.tikz}}
 \\
        &=\tag{\ref{ax:symmpnat}} 
    \InputIfFileExists{lemmaDl/step2.tikz}{}{\input{./tikz/lemmaDl/step2.tikz}}
 \\
        &=\tag{Ind. hp.} 
    \InputIfFileExists{lemmaDl/step3.tikz}{}{\input{./tikz/lemmaDl/step3.tikz}}
 \\
&= \tag{Funct. $\piu$}  
    \InputIfFileExists{lemmaDl/step5.tikz}{}{\input{./tikz/lemmaDl/step5.tikz}}
 \\
&= \begin{multlined}[t]
            (\id{U(V\piu (Q \piu R))} \piu \dl{P'}{V}{Q \piu R}) ; (\id{UV} \piu \symmp{U(Q \piu R)}{P'V} \piu \id{P'(Q \piu R)}) ; \\
            (\id{UV \piu P'V} \piu ((\id{U(Q \piu R)} \piu \dl{P'}{Q}{R});(\id{UQ} \piu \symmp{UR}{P'Q} \piu \id{P'R})))
        \end{multlined}\\
        &=\tag{Def. $\delta^l$} \dl{U \piu P'}{V}{Q \piu R} ; (\id{UV \piu P'V} \piu \dl{U \piu P'}{Q}{R}) \\
    \end{align*}
    \endgroup
\end{proof}

\begin{proof}[Proof of Lemma \ref{lemma:tapesymdis}.\ref{lm:tape:symmt dl}]
    We want to prove that $\symmt{P}{Q \piu R} = \dl{P}{Q}{R} ; (\symmt{P}{Q} \piu \symmt{P}{R})$ by induction on $Q$.
    
    \textbf{Case $Q=\zero$:} By Lemma~\ref{lemma:tapesymdis}.\ref{lm:tape:dl zero}, $\dl {P}{\zero}{R}=\id R$ and by definition of $\symmt$, $\symmt{P}{\zero}=\id{\zero}$. The right-hand side reduces therefore to $\symmp{P}{R}$, as desired.

    \textbf{Case $Q=V \piu Q'$:}
    \begin{align*}
        \symmt{P}{V \piu Q' \piu R} &=\tag{Def. $\symmt$} \dl{P}{V}{Q' \piu R} ; (\Piu[i]{\tapesymm{U_i}{V}} \piu \symmt{P}{Q' \piu R}) \\
        &=\tag{Ind. hp.} \dl{P}{V}{Q' \piu R} ; (\Piu[i]{\tapesymm{U_i}{V}} \piu (\dl{P}{Q'}{R};(\symmt{P}{Q'} \piu \symmt{P}{R}))) \\
        &=\tag{Funct. $\piu$} \dl{P}{V}{Q' \piu R} ; (\id{PV} \piu \dl{P}{Q'}{R}) ; (\Piu[i]{\tapesymm{U_i}{V}} \piu \symmt{P}{Q'} \piu \symmt{P}{R}) \\
        &=\tag{Lemma~\ref{lm:tape:dl aux}} \dl{P}{V \piu Q'}{R} ; (\dl{P}{V}{Q'} \piu \id{PR}) ; (\Piu[i]{\tapesymm{U_i}{V}} \piu \symmt{P}{Q'} \piu \symmt{P}{R}) \\
        &=\tag{Funct. $\piu$} \dl{P}{V \piu Q'}{R} ; ((\dl{P}{V}{Q'};(\Piu[i]{\tapesymm{U_i}{V}} \piu \symmt{P}{Q'})) \piu \symmt{P}{R}) \\
        &=\tag{Def. $\symmt$} \dl{P}{V \piu Q'}{R} ; (\symmt{P}{V \piu Q'} \piu \symmt{P}{R})
    \end{align*}
\end{proof}

\begin{proof}[Proof of Lemma \ref{lemma:tapesymdis}.\ref{lm:tape:symmt uno}]
We want to prove that $\symmt{P}{\uno} = \id{P}$ by induction on $P$.

    \textbf{Case $P = \zero$:} it holds by definition of $\symmt$.

    \textbf{Case $P = U \piu P'$:} 
    \begin{align*}
        \symmt{U \piu P'}{\uno} &=\tag{Lemma~\ref{lemma:tapesymdis}.\ref{lm:tape:symmtInv}} \Isymmt{\uno}{U \piu P'} \\
        &=\tag{Def. $\symmt$} (\tapesymm{U}{\uno} \piu \symmt{P'}{\uno}) ; \Idl{\uno}{U}{P'} \\
        &=\tag{Lemma~\ref{lemma:tapesymdis}.\ref{lm:tape:dl U}} (\tapesymm{U}{\uno} \piu \symmt{P'}{\uno}) ; \id{UP'} \\
        &=\tag{\ref{eq:symmax2}, ind. hp.} \id{U} \piu \id{P'} \\
        &= \id{U \piu P'}
    \end{align*}
\end{proof}

\subsection{Additional lemmas for monomial whiskerings}\label{appTape2}
\begin{lemma}\label{lm:whisk}
    Let $\t \colon \Piu[i]{U_i} \to \Piu[j]{V_j}$ be a tape diagram, then the following holds
    \[ \RW{W} \t = \Piu[i]{\tapesymm{U_i}{W}} ; \LW{W} \t ; \Piu[j]{\tapesymm{W}{V_j}} \]
\end{lemma}
\begin{proof} 
    By induction on $\t$.
    
    \textbf{Case $\t = \id\zero$:} Notice that $\PiuL[i=1][0]{\tapesymm{U_i}{W}} = \PiuL[j=1][0]{\tapesymm{W}{V_j}} = \id\zero$ and thus $\RW W {\id\zero} = \id\zero = \LW W {\id\zero}$.

    \textbf{Case $\t = \tape{c}$:} Suppose $c \colon U \to V$, then 
    \begin{align*}
        \RW W {\tape{c}} &=\tag{Def. $R$} \tape{c \per \id W} \\
        &=\tag{\ref{eq:symmax1}} \tape{c \per \id W} ; \tapesymm{V}{W} ; \tapesymm{W}{V} \\
        &=\tag{Nat. $\tape{\sigma}$} \tapesymm{U}{W} ; \tape{\id W \per c} ; \tapesymm{W}{V} \\
        &=\tag{Def. $L$} \tapesymm{U}{W} ; \LW W {\tape{c}} ; \tapesymm{W}{V}
    \end{align*}
    
    \textbf{Case $\t = \symmp{U}{V}$:} 
    \begin{align*}
        \RW W {\symmp{U}{V}} &=\tag{Def. $R$} \symmp{UW}{VW} \\
        &=\tag{\ref{eq:symmax1}} \symmp{UW}{VW} ; ((\tapesymm{V}{W} ; \tapesymm{W}{V}) \piu (\tapesymm{U}{W} ; \tapesymm{W}{U})) \\
        &=\tag{\ref{ax:symmpnat}} (\tapesymm{U}{W} \piu \tapesymm{V}{W}) ; \symmp{WU}{WV} ; (\tapesymm{W}{V} \piu \tapesymm{W}{U}) \\
        &=\tag{Def. $L$} (\tapesymm{U}{W} \piu \tapesymm{V}{W}) ; \LW W {\symmp{U}{V}} ; (\tapesymm{W}{V} \piu \tapesymm{W}{U})
    \end{align*}

    \textbf{Case $\t = \diag U$:} (similarly for $\codiag{U}$)
    \begin{align*}
        \RW W {\diag U} &=\tag{Def. $R$} \diag{UW} \\
        &=\tag{\ref{eq:symmax1}} \diag{UW} ; ((\tapesymm{U}{W} ; \tapesymm{W}{U})  \piu (\tapesymm{U}{W} ; \tapesymm{W}{U}))\\
        &=\tag{\ref{ax:diagnat}} \tapesymm{U}{W} ; \diag{WU} ; (\tapesymm{W}{U} \piu \tapesymm{W}{U}) \\
        &=\tag{Def. $L$} \tapesymm{U}{W} ; \LW W {\diag{U}} ; (\tapesymm{W}{U} \piu \tapesymm{W}{U})
    \end{align*}

    \textbf{Case $\t = \bang U$:} (similarly for $\cobang{U}$)
    \begin{align*}
        \RW W {\bang U} &=\tag{Def. $R$} \bang{UW} \\
        &=\tag{\ref{eq:symmax1}} \tapesymm{U}{W} ; \tapesymm{W}{U} ; \bang{UW} \\
        &=\tag{\ref{ax:bangnat}} \tapesymm{U}{W} ; \bang{WU} \\
        &=\tag{Def. $L$} \tapesymm{U}{W} ; \LW W {\bang{U}}
    \end{align*}

    \textbf{Case $\t = \t_1 ; \t_2$:} Suppose $\t_1 \colon \Piu[i]{U_i} \to \Piu[l]{Z_l}, \t_2 \colon \Piu[l]{Z_l} \to \Piu[j]{V_j}$, then 
    \begin{align*}
        \RW W {\t_1 ; \t_2} &=\tag{Def. $R$} \RW W {\t_1} ; \RW W {\t_2} \\
        &=\tag{Ind. hp.} \Piu[i]{\tapesymm{U_i}{W}} ; \LW{W} {\t_1} ; \Piu[l]{\tapesymm{W}{Z_l}} ; \Piu[l]{\tapesymm{Z_l}{W}} ; \LW{W} {\t_2} ; \Piu[j]{\tapesymm{W}{V_j}} \\
        &=\tag{\ref{eq:symmax1}} \Piu[i]{\tapesymm{U_i}{W}} ; \LW{W} {\t_1} ; \LW{W} {\t_2} ; \Piu[j]{\tapesymm{W}{V_j}} \\
        &=\tag{Def. $L$} \Piu[i]{\tapesymm{U_i}{W}} ; \LW{W} {\t_1 ; \t_2} ; \Piu[j]{\tapesymm{W}{V_j}}
    \end{align*}

    \textbf{Case $\t = \t_1 \piu \t_2$:} Suppose $\t_1 \colon \Piu[i]{U_i} \to \Piu[j]{V_j}, \t_2 \colon \Piu[i']{U'_{i'}} \to \Piu[j']{V'_{j'}}$, then 
    \begin{align*}
        \RW W {\t_1 \piu \t_2} &=\tag{Def. $R$} \RW W {\t_1} \piu \RW W {\t_2} \\
        &=\tag{Ind. hp.} (\Piu[i]{\tapesymm{U_i}{W}} ; \LW{W} {\t_1} ; \Piu[j]{\tapesymm{W}{V_j}}) \piu (\Piu[i']{\tapesymm{U'_{i'}}{W}} ; \LW{W} {\t_2} ; \Piu[j']{\tapesymm{W}{V'_{j'}}}) \\
        &=\tag{Funct. $\piu$} (\Piu[i]{\tapesymm{U_i}{W}} \piu \Piu[i']{\tapesymm{U'_{i'}}{W}}) ; (\LW{W} {\t_1} \piu \LW{W}{\t_2} ; (\Piu[j]{\tapesymm{W}{V_{j}}} \piu \Piu[j']{\tapesymm{W}{V'_{j'}}}) \\
        &=\tag{Def. $L$} (\Piu[i]{\tapesymm{U_i}{W}} \piu \Piu[i']{\tapesymm{U'_{i'}}{W}}) ; (\LW{W} {\t_1 \piu \t_2}) ; (\Piu[j]{\tapesymm{W}{V_{j}}} \piu \Piu[j']{\tapesymm{W}{V'_{j'}}})
    \end{align*}
\end{proof}

\begin{lemma}\label{lm:tape:perC} 
    For all $\tape{c} \colon U \to V, \t \colon P \to Q$, $\LW{U}{\t} ; \RW{Q}{\tape{c}} = \RW{P}{\tape{c}} ; \LW{V}{\t}$.
\end{lemma}
\begin{proof} 
We fix $P=\Piu[k]{W_k} $ and $Q= \Piu[l]{Z_l}$. By definition of $\RW{P}{-}$ and $\delta^l$, it is immediate that  $\RW{Q}{\tape{c}}= \Piu[l]{\tape{c \per \id{Z_l}}}$ and $ \RW{P}{\tape{c}} = \Piu[k]{\tape{c \per \id{W_k}}}$.

We proceed by induction on $\t$.

\textbf{Case $\t = \id{\zero}$:} $\LW U {\id{\zero}} ; \Piu[l=1][0]{\tape{c \per \id{Z_l}}} = \id{\zero} = (\Piu[k=1][0]{\tape{c \per \id{W_k}}}) ; \LW V {\id{\zero}}$

\textbf{Case $\t = \tape{d}$:} Suppose $d \colon W \to Z$, then 
\begin{align*}
    \LW U {\tape{d}} ; \tape{c \per \id Z} &=\tag{Def. $L$} \tape{\id U \per d} ; \tape{c \per \id Z} \\
    &=\tag{Funct. $\tape{\per}$} \tape{c \per d} \\
    &=\tag{Funct. $\tape{\per}$} \tape{c \per \id W} ; \tape{\id V \per d} \\
    &=\tag{Def. $L$} \tape{c \per \id W} ; \LW V {\tape{d}}
\end{align*}

\textbf{Case $\t = \symmp{W}{Z}$:} 
\begin{align*}
    \LW U {\symmp{W}{Z}} ; (\tape{c \per \id Z} \piu \tape{c \per \id W}) &=\tag{Def. $L$} \symmp{UW}{UZ} ; (\tape{c \per \id Z} \piu \tape{c \per \id W}) \\
    &=\tag{\ref{ax:symmpnat}} (\tape{c \per \id W} \piu \tape{c \per \id Z}) ; \symmp{VW}{VZ} \\
    &=\tag{Def. $L$} (\tape{c \per \id W} \piu \tape{c \per \id Z}) ; \LW V {\symmp{W}{Z}}
\end{align*}

\textbf{Case $\t = \diag{W}$:} (similarly for $\codiag{W}$)
\begin{align*}
    \LW U {\diag{W}} ; (\tape{c \per \id W} \piu \tape{c \per \id W}) &=\tag{Def. $L$} \diag{UW} ; (\tape{c \per \id W} \piu \tape{c \per \id W}) \\
    &=\tag{\ref{ax:diagnat}} \tape{c \per \id W} ; \diag{VW} \\
    &=\tag{Def. $L$} \tape{c \per \id W} ; \LW V {\diag{W}}
\end{align*}

\textbf{Case $\t = \bang{W}$:} (similarly for $\cobang{W}$)
\begin{align*}
    \LW U {\bang{W}} ; \Piu[l=1][0]{\tape{c \per \id{Z_l}}} &=\tag{Def. $L$} \bang{UW} \\ &=\tag{\ref{ax:bangnat}} \tape{c \per \id W} ; \bang{VW} \\
    &=\tag{Def. $L$} \tape{c \per \id W} ; \LW V {\bang{W}}
\end{align*}

\textbf{Case $\t = \t_1 ; \t_2$:} Suppose $\t_1 \colon \Piu[k]{W_k} \to \Piu[j]{V_j}, \t_2 \colon \Piu[j]{V_j} \to \Piu[l]{Z_l}$, then
\begin{align*}
    \LW U {\t_1 ; \t_2} ; \Piu[l]{\tape{c \per \id{Z_l}}} &=\tag{Def. $L$} \LW U {\t_1} ; \LW U {\t_2} ; \Piu[l]{\tape{c \per \id{Z_l}}} \\
    &=\tag{Ind. hp.} \LW U {\t_1} ; \Piu[j]{\tape{c \per \id{V_j}}} ; \LW V {\t_2}\\
    &=\tag{Ind. hp.} \Piu[k]{\tape{c \per \id{W_k}}} ; \LW V {\t_1} ; \LW V {\t_2}\\
    &=\tag{Def. $L$} \Piu[k]{\tape{c \per \id{W_k}}} ; \LW V {\t_1 ; \t_2}
\end{align*}

\textbf{Case $\t = \t_1 \piu \t_2$:} Suppose $\t_1 \colon \Piu[k]{W_k} \to \Piu[l]{Z_l}, \t_2 \colon \Piu[k']{W'_{k'}} \to \Piu[l']{Z'_{l'}}$, then
\begin{align*}
    &\LW U {\t_1 \piu \t_2} ; (\Piu[l]{\tape{c \per \id{Z_l}}} \piu \Piu[l']{\tape{c \per \id{Z'_{l'}}}}) \\
    &=\tag{Def. $L$} (\LW U {\t_1} \piu \LW U {\t_2}) ; (\Piu[l]{\tape{c \per \id{Z_l}}} \piu \Piu[l']{\tape{c \per \id{Z'_{l'}}}}) \\
    &=\tag{Funct. $\piu$} (\LW U {\t_1} ; \Piu[l]{\tape{c \per \id{Z_l}}}) \piu (\LW U {\t_2} ; \Piu[l']{\tape{c \per \id{Z'_{l'}}}}) \\
    &=\tag{Ind. hp.} (\Piu[k]{\tape{c \per \id{W_k}}} ; \LW V {\t_1}) \piu (\Piu[k']{\tape{c \per \id{W'_{k'}}}} ; \LW V {\t_2}) \\
    &=\tag{Funct. $\piu$} (\Piu[k]{\tape{c \per \id{W_k}}} \piu \Piu[k']{\tape{c \per \id{W'_{k'}}}}) ; (\LW V {\t_1} \piu \LW V {\t_2}) \\
    &=\tag{Def. $L$} (\Piu[k]{\tape{c \per \id{W_k}}} \piu \Piu[k']{\tape{c \per \id{W'_{k'}}}}) ; \LW V {\t_1 \piu \t_2}
\end{align*}
\end{proof}

\subsection{Proof of Lemma~\ref{lm:whisk properties}: the algebra of whiskerings}\label{app:whiskering}

The properties of the left whiskering are all trivial, thus we show only the proofs for their right counterpart.

    Let $P = \Piu[i]{U_i}, Q = \Piu[j]{V_j}, S = \Piu[k]{W_k}, T = \Piu[l]{Z_l}$.
    
    \textsc{Equation~\eqref{eq:whisk:id}:} $\RW S {\id{P}} = \id{PS}$. We prove it by induction on $S$.

\textbf{Case $S=\zero$:} it holds by Definition~\ref{def:tape:whiskG}.
    
    \textbf{Case $S=W \piu S'$:}
    \begin{align*}
        \RW {W \piu S'} {\id{Q}} &=\tag{Def. $R$} \dl{Q}{W}{S'} ; (\RW{W}{\id Q} \piu \RW{S'}{\id Q}) ; \Idl{Q}{W}{S'} \\
        &=\tag{Def. $R$, ind. hp.} \dl{Q}{W}{S'} ; (\id {QW} \piu \id {QS'}) ; \Idl{Q}{W}{S'} \\
        &=\tag{Iso} \id{Q(W \piu S')}
    \end{align*}

    \textsc{Equation~\eqref{eq:whisk:funct}:} $\RW{S}{\t_1 ; \t_2} = \RW{S}{\t_1} ; \RW{S}{\t_2}$. Let $\t_1 \colon P \to Q, \t_2 \colon Q \to T$, we prove it by induction on $S$.
    
\textbf{Case $S=\zero$:} it holds by Definition~\ref{def:tape:whiskG}.
    
    \textbf{Case $S=W \piu S'$:}
    \begin{align*}
        &\RW{W \piu S'}{\t_1 ; \t_2} \\
        &=\tag{Def. $R$} \dl{P}{W}{S'} ; (\RW{W}{\t_1 ; \t_2} \piu \RW{S'}{\t_1;\t_2}) ; \Idl{T}{W}{S'} \\
        &=\tag{Def. $R$, ind. hp.} \dl{P}{W}{S'} ; ((\RW{W}{\t_1} ; \RW{W}{\t_2}) \piu (\RW{S'}{\t_1} ; \RW{S'}{\t_2})) ; \Idl{T}{W}{S'} \\
        &=\tag{Funct. $\piu$} \dl{P}{W}{S'} ; ((\RW{W}{\t_1} \piu \RW{S'}{\t_1});(\RW{W}{\t_2} \piu \RW{S'}{\t_2})) ; \Idl{T}{W}{S'} \\
        &=\tag{Iso} \dl{P}{W}{S'} ; ((\RW{W}{\t_1} \piu \RW{S'}{\t_1}) ; \Idl{Q}{W}{S'} ; \dl{Q}{W}{S'} ; (\RW{W}{\t_2} \piu \RW{S'}{\t_2})) ; \Idl{T}{W}{S'} \\
        &=\tag{Def. $R$} \RW{W \piu S'}{\t_1} ; \RW{W \piu S'}{\t_2} 
    \end{align*}

    \textsc{Equation~\eqref{eq:whisk:uno}:} $\RW{\uno}{\t} = \t$. We prove it by induction on $\t$. All the base cases are trivial and the inductive cases follow by inductive hypothesis and functoriality of $R$ and $\piu$.

    \textsc{Equation~\eqref{eq:whisk:zero}:} $\RW{\zero}{\t} = \id{\zero}$. It holds by Definition~\ref{def:tape:whiskG}.

    \textsc{Equation~\eqref{eq:whisk:sum}:} $\RW{S \piu T}{\t} = \dl{P}{S}{T} ; ( \RW{S}{\t} \piu \RW{T}{\t} ) ; \Idl{Q}{S}{T}$. Let $\t \colon P \to Q$, we prove it by induction on $S$. 

\textbf{Case $S=\zero$:} it holds by Definition~\ref{def:tape:whiskG}.
    
    \textbf{Case $S=W \piu S'$:}
    \begin{align*}
        &\RW {W \piu S' \piu T} {\t} \\
        &=\tag{Def. $R$} \dl{P}{W}{S' \piu T} ; (\RW{W}{\t} \piu \RW{S' \piu T}{\t}) ; \Idl{Q}{W}{S' \piu T} \\
        &=\tag{Ind. hp.} \dl{P}{W}{S' \piu T} ; (\RW{W}{\t} \piu (\dl{P}{S'}{T} ; (\RW{S'}{\t} \piu \RW{T}{\t}) ; \Idl{Q}{S'}{T})) ; \Idl{Q}{W}{S' \piu T} \\
        &=\tag{Funct. $\piu$} \dl{P}{W}{S' \piu T} ; (\id{PW} \piu \dl{P}{S'}{T}) ; (\RW{W}{\t} \piu ((\RW{S'}{\t} \piu \RW{T}{\t}) ; \Idl{Q}{S'}{T})) ; \Idl{Q}{W}{S' \piu T} \\
        &=\tag{Lemma~\ref{lm:tape:dl aux}} \dl{P}{W \piu S'}{T} ; (\dl{P}{W}{S'} \piu \id{PT}) ; (\RW{W}{\t} \piu ((\RW{S'}{\t} \piu \RW{T}{\t}) ; \Idl{Q}{S'}{T})) ; \Idl{Q}{W}{S' \piu T} \\
        &=\tag{Funct. $\piu$} \dl{P}{W \piu S'}{T} ; (\dl{P}{W}{S'} \piu \id{PT}) ; (\RW{W}{\t} \piu \RW{S'}{\t} \piu \RW{T}{\t}) ; (\id{QW} \piu \Idl{Q}{S'}{T}) ; \Idl{Q}{W}{S' \piu T} \\
        &=\tag{Lemma~\ref{lm:tape:dl aux}} \dl{P}{W \piu S'}{T} ; (\dl{P}{W}{S'} \piu \id{PT}) ; (\RW{W}{\t} \piu \RW{S'}{\t} \piu \RW{T}{\t}) ; (\Idl{Q}{W}{S'} \piu \id{QT}) ; \Idl{Q}{W \piu S'}{T} \\
        &=\tag{Funct. $\piu$} \dl{P}{W \piu S'}{T} ; ((\dl{P}{W}{S'} ; (\RW{W}{\t} \piu \RW{S'}{\t}) ; \Idl{Q}{W}{S'}) \piu \RW{T}{\t})  ; \Idl{Q}{W \piu S'}{T} \\
        &=\tag{Def. $R$} \dl{P}{W \piu S'}{T} ; (\RW{W \piu S'}{\t} \piu \RW{T}{\t})  ; \Idl{Q}{W \piu S'}{T}
    \end{align*}

    \textsc{Equation~\eqref{eq:whisk:funct piu}:} $\RW{S}{\t_1 \piu \t_2} = \RW{S}{\t_1} \piu \RW{S}{\t_2}$. Let $\t_1 \colon P \to Q, \t_2 \colon P' \to Q'$, we prove it by induction on $S$.
    
\textbf{Case $S = \zero$:} it holds by Definition~\ref{def:tape:whiskG}.

    \textbf{Case $S = W \piu S'$:}
    \begin{align*}
        &\RW {W \piu S'} {\t_1 \piu \t_2} \\
        &=\tag{\ref{eq:whisk:sum}} \dl{P \piu P'}{W}{S'} ; (\RW W {\t_1 \piu \t_2} \piu \RW {S'} {\t_1 \piu \t_2}) ; \Idl{Q \piu Q'}{W}{S'} \\
        &=\tag{Def. $R$, ind. hp.} \dl{P \piu P'}{W}{S'} ; (\RW W {\t_1} \piu \RW W {\t_2} \piu \RW {S'} {\t_1} \piu \RW {S'} {\t_2}) ; \Idl{Q \piu Q'}{W}{S'} \\
        &=\tag{Lemma~\ref{lemma:tapesymdis}.\ref{lm:tape:dlG}}\begin{multlined}[t]
            (\dl{P}{W}{S'} \piu \dl{P'}{W}{S'}) ; (\id{PW} \piu \symmp{PS'}{P'W} \piu \id{P'S'}) ; \\
            (\RW W {\t_1} \piu \RW W {\t_2} \piu \RW {S'} {\t_1} \piu \RW {S'} {\t_2}) ; \\
            (\id{QW} \piu \symmp{Q'W}{QS'} \piu \id{Q'S'}) ; (\Idl{Q}{W}{S'} \piu \Idl{Q'}{W}{S'})
        \end{multlined} \\
        &=\tag{\ref{ax:symmpnat}} (\dl{P}{W}{S'} \piu \dl{P'}{W}{S'}) ; (\RW W {\t_1} \piu \RW {S'} {\t_1} \piu \RW {W} {\t_2} \piu \RW {S'} {\t_2}) ; (\Idl{Q}{W}{S'} \piu \Idl{Q'}{W}{S'}) \\
        &=\tag{Funct. $\piu$} (\dl{P}{W}{S'} ; (\RW W {\t_1} \piu \RW {S'} {\t_1}) ; \Idl{Q}{W}{S'})  \piu (\dl{P'}{W}{S'} ; (\RW {W} {\t_2} \piu \RW {S'} {\t_2}) ; \Idl{Q'}{W}{S'}) \\
        &=\tag{$\ref{eq:whisk:sum}$} \RW {W \piu S'} {\t_1} \piu  \RW {W \piu S'} {\t_2}
    \end{align*}

    \textsc{Equation~\eqref{eq:whisk:diag}:} $\RW S {\diag{U}} = \diag{US}$. We prove it by induction on $S$.
    
\textbf{Case $S = \zero$:} it holds by Definition~\ref{def:tape:whiskG} and definition of $\diag{}$. 

    \textbf{Case $S = W \piu S'$:}
    \begin{align*}
        &\RW {W \piu S'} {\diag{U}} \\
        &=\tag{\ref{eq:whisk:sum}} \dl{U}{W}{S'} ; (\RW W {\diag{U}} \piu \RW {S'} {\diag{U}}) ; \Idl{U\piu U}{W}{S'} \\
        &=\tag{Lemma~\ref{lemma:tapesymdis}.\ref{lm:tape:dl U}, def. $\delta^l$} (\RW W {\diag{U}} \piu \RW {S'} {\diag{U}}) ; (\id{UW} \piu \symmp{UW}{US'} \piu \id{US'}) ; (\id{U(W\piu S')} \piu \Idl{U}{W}{S'}) \\
        &=\tag{Lemma~\ref{lemma:tapesymdis}.\ref{lm:tape:dl U}} (\RW W {\diag{U}} \piu \RW {S'} {\diag{U}}) ; (\id{UW} \piu \symmp{UW}{US'} \piu \id{US'}) \\
        &=\tag{Def. $R$, ind. hp.} (\diag{UW} \piu \diag{US'}) ; (\id{UW} \piu \symmp{UW}{US'} \piu \id{US'}) \\
        &=\tag{Def. $\diag{}$} \diag{UW \piu US'} \\ 
        &=\tag{Rem.~\ref{rem:per tape is right distributive}} \diag{U(W \piu S')}
    \end{align*}
    The case for $\codiag{U}$ is completely analogous.

    \textsc{Equation~\eqref{eq:whisk:bang}:} $\RW S {\bang{U}} = \bang{US}$. We prove it by induction on $S$.
    
\textbf{Case $S = \zero$:} it holds by Definition~\ref{def:tape:whiskG} and definition of $\bang{}$.
    
    \textbf{Case $S = W \piu S'$:} \begin{align*}
        \RW {W \piu S'} {\bang{U}} &=\tag{\ref{eq:whisk:sum}} \dl{U}{W}{S'} ; (\RW W {\bang{U}} \piu \RW {S'} {\bang{U}}) ; \Idl{\zero}{W}{S'} \\
        &=\tag{Lemma~\ref{lemma:tapesymdis}.\ref{lm:tape:dl U}, def. $\delta^l$} \RW W {\bang{U}} \piu \RW {S'} {\bang{U}} \\
        &=\tag{Def. $R$, ind. hp.} \bang{UW} \piu \bang{US'} \\
        &=\tag{Def. $\bang{}$} \bang{UW \piu US'} \\
        &=\tag{Rem.~\ref{rem:per tape is right distributive}} \bang{U(W \piu S')}
    \end{align*} The case for $\cobang{U}$ is completely analogous.

    \textsc{Equation~\eqref{eq:whisk:symmp}:} $\RW S {\symmp{P}{Q}} = \symmp{PS}{QS}$. We prove it by induction on $P$.

    \textbf{Case $P = \zero$:} \begin{align*}
        \RW{S}{\symmp{\zero}{Q}} &=\tag{\ref{eq:symmax2}} \RW{S}{\id{Q}} \\
        &=\tag{\ref{eq:whisk:id}} \id{QS} \\
        &=\tag{\ref{eq:symmax2}} \symmp{\zero}{QS}
    \end{align*}

    \textbf{Case $P = U \piu P'$:} \begin{align*}
        \RW{S}{\symmp{U \piu P'}{Q}} &=\tag{\ref{eq:symmax3}} \RW{S}{(\id{U} \piu \symmp{P'}{Q}) ; (\symmp{U}{Q} \piu \id{P'})} \\
        &=\tag{\ref{eq:whisk:funct}, \ref{eq:whisk:funct piu} } (\RW{S}{\id{U}} \piu \RW{S}{\symmp{P'}{Q}}) ; (\RW{S}{\symmp{U}{Q}} \piu \RW{S}{\id{P'}}) \\
        &=\tag{\ref{eq:whisk:id}, ind. hp} (\id{US} \piu \symmp{P'S}{QS}) ; (\symmp{US}{QS} \piu \id{P'S}) \\
        &=\tag{\ref{eq:symmax3}} \symmp{US \piu P'S}{QS} \\
        &=\tag{Rem.~\ref{rem:per tape is right distributive}} \symmp{(U \piu P')S}{QS}
    \end{align*}

    \textsc{Equation~\eqref{eq:whisk:dl}:} $\RW S {\dl{P}{Q}{R}} = \dl{P}{QS}{RS}$. We prove it by induction on $P$.
    
    \textbf{Case $P = \zero$:} it follows from Definition~\ref{def:tape:whiskG} and definition of $\delta^l$.

    \textbf{Case $P = U \piu P'$:} 
    \begin{align*}
        &\RW{S}{\dl{U\piu P'}{Q}{R}} \\
        &=\tag{Def. $\delta^l$} \RW{S}{(\id{U(Q\piu R)} \piu \dl{P'}{Q}{R}) ; (\id{UQ} \piu \symmp{UR}{P'Q} \piu \id{P'R})}\\
        &=\tag{\ref{eq:whisk:funct}, \ref{eq:whisk:funct piu}} (\RW{S}{\id{U(Q\piu R)}} \piu \RW{S}{\dl{P'}{Q}{R}}) ; (\RW{S}{\id{UQ}} \piu \RW{S}{\symmp{UR}{P'Q}} \piu \RW{S}{\id{P'R}}) \\
        &=\tag{Ind. hp.} (\RW{S}{\id{U(Q\piu R)}} \piu \dl{P'}{QS}{RS}) ; (\RW{S}{\id{UQ}} \piu \RW{S}{\symmp{UR}{P'Q}} \piu \RW{S}{\id{P'R}})\\
        &=\tag{\ref{eq:whisk:id}, \ref{eq:whisk:symmp}} (\id{U(Q\piu R)S} \piu \dl{P'}{QS}{RS}) ; (\id{UQS} \piu \symmp{URS}{P'QS} \piu \id{P'RS})\\
        &=\tag{Def. $\delta^l$} \dl{U \piu P'}{QS}{RS}
    \end{align*}

    \textsc{Equation~\eqref{eq:symmper}:} $\symmt{PQ}{S} = \LW{P}{ \symmt{Q}{S}} ; \RW{Q}{\symmt{P}{S}}$. First we prove the following equations:

    \begin{align*}
        \LW{U}{\tapesymm{V}{W}};\RW{V}{\tapesymm{U}{W}} &= \tape{\sigma_{UV,W}} \tag{$\ast_1$} \\
        \LW{U}{\Piu[j]{\tapesymm{V_j}{W}}} ; \RW{Q}{\tapesymm{U}{W}} &= \Piu[j]{\tapesymm{UV_j}{W}} \text{\qquad where $Q = \Piu[j]{V_j}$} \tag{$\ast_2$} \\
        \LW{P}{\symmt{Q}{W}} ; \RW{Q}{\symmt{P}{W}} &= \symmt{PQ}{W} \tag{$\ast_3$}
    \end{align*}

    For $(\ast_1)$ observe that the following holds:
    \begin{align*}
        \LW{U}{\tapesymm{V}{W}};\RW{V}{\tapesymm{U}{W}} &=\tag{Def. $L, R$} \tape{\id U \per \sigma_{V,W}};\tape{\sigma_{U,W} \per \id{V}} \\
        &=\tag{\ref{eq:symmax3}} \tape{\sigma_{UV,W}}
    \end{align*}
    
    For $(\ast_2)$ we proceed by induction on $Q$:
    \begin{equation}
        \LW{U}{\Piu[j]{\tapesymm{V_j}{W}}} ; \RW{Q}{\tapesymm{U}{W}} = \Piu[j]{\tapesymm{UV_j}{W}} \text{\qquad where $Q = \Piu[j]{V_j}$}
    \end{equation}

    \textbf{Case $Q = \zero$:} it follows from Definitions~\ref{def:tape:whisk} and~\ref{def:tape:whiskG} and from the fact that $\PiuL[j=1][0]{\tapesymm{V_j}{W}} = \id\zero$.

\textbf{Case $Q = V \piu Q'$:} Let $Q' = \Piu[j']{V'_{j'}}$, then  \begin{align*}
        &\LW{U}{\tapesymm{V}{W} \piu \Piu[j']{\tapesymm{V'_{j'}}{W}}};\RW{V \piu Q'}{\tapesymm{U}{W}} \\
        &=\tag{Def. $L$, \ref{eq:whisk:sum}} (\LW{U}{\tapesymm{V}{W}} \piu \LW{U}{\Piu[j']{\tapesymm{V'_{j'}}{W}}}) ; \dl{UW}{V}{Q'} ; (\RW{V}{\tapesymm{U}{W}} \piu \RW{Q'}{\tapesymm{U}{W}}) ; \Idl{WU}{V}{Q'} \\
        &=\tag{Lemma~\ref{lemma:tapesymdis}.\ref{lm:tape:dl U}} (\LW{U}{\tapesymm{V}{W}} \piu \LW{U}{\Piu[j']{\tapesymm{V'_{j'}}{W}}}) ; (\RW{V}{\tapesymm{U}{W}} \piu \RW{Q'}{\tapesymm{U}{W}}) \\
        &=\tag{Funct. $\piu$} (\LW{U}{\tapesymm{V}{W}};\RW{V}{\tapesymm{U}{W}}) \piu (\LW{U}{\Piu[j']{\tapesymm{V'_{j'}}{W}}} ; \RW{Q'}{\tapesymm{U}{W}}) \\
        &=\tag{$\ast_1$, ind. hp.} \tape{\sigma_{UV,W}} \piu \Piu[j']{\tapesymm{UV'_{j'}}{W}} \\
    \end{align*}

    For $(\ast_3)$ observe that the following holds:
    \begin{align*}
        &\LW{P}{\symmt{Q}{W}} ; \RW{Q}{\symmt{P}{W}} \\
        &=\tag{Lemma~\ref{lemma:tapesymdis}.\ref{lm:tape:symmt V}} \LW{P}{\Piu[j]{\tapesymm{V_j}{W}}} ; \RW{Q}{\Piu[i]{\tapesymm{U_i}{W}}} \\
        &=\tag{Def. $L$,\ref{eq:whisk:funct piu}} \Piu[i]{\LW{U_i}{\Piu[j]{\tapesymm{V_j}{W}}}} ; \Piu[i]{\RW{Q}{\tapesymm{U_i}{W}}} \\
        &=\tag{Funct. $\piu$} \PiuPar[i]{\LW{U_i}{\Piu[j]{\tapesymm{V_j}{W}}} ; \RW{Q}{\tapesymm{U_i}{W}}} \\
        &=\tag{$\ast_2$} \Piu[i]{\Piu[j]{\tapesymm{U_iV_j}{W}}} \\
        &=\tag{Lemma~\ref{lemma:tapesymdis}.\ref{lm:tape:symmt V}} \symmt{PQ}{W} \\
    \end{align*}

    Now we are ready to prove Equation~\eqref{eq:symmper} by induction on $S$:

    \textbf{Case $S=\zero$:} it follows from Definitions~\ref{def:tape:whisk},~\ref{def:tape:whiskG} and definition of $\symmt$.

\textbf{Case $S=W \piu S'$:}
    \begin{align*}
        &\LW{P}{\symmt{Q}{W \piu S'}} ; \RW{Q}{\symmt{P}{W \piu S'}} \\
        &=\tag{Lemma~\ref{lemma:tapesymdis}.\ref{lm:tape:symmt dl}} \LW{P}{\dl{Q}{W}{S'} ; (\symmt{Q}{W} \piu \symmt{Q}{S'})} ; \RW{Q}{\dl{P}{W}{S'} ; (\symmt{P}{W} \piu \symmt{P}{S'})} \\
        &=\tag{\ref{eq:whisk:funct}, \ref{eq:whisk:funct piu}} \begin{multlined}[t]
            \LW{P}{\dl{Q}{W}{S'}} ; \dl{P}{QW}{QS'} ; (\LW{P}{\symmt{Q}{W}} \piu \LW{P}{\symmt{Q}{S'}}) ; \Idl{P}{WQ}{S'Q} ; \\
            \RW{Q}{\dl{P}{W}{S'}} ; (\RW{Q}{\symmt{P}{W}} \piu \RW{Q}{\symmt{P}{S'}})
        \end{multlined} \\
        &=\tag{\ref{eq:whisk:Ldl}, \ref{eq:whisk:dl}} \begin{multlined}[t]
            \dl{PQ}{W}{S'} ; \Idl{P}{QW}{QS'} ; \dl{P}{QW}{QS'} ; (\LW{P}{\symmt{Q}{W}} \piu \LW{P}{\symmt{Q}{S'}}) ; \Idl{P}{WQ}{S'Q} ; \\
            \dl{P}{WQ}{S'Q} ; (\RW{Q}{\symmt{P}{W}} \piu \RW{Q}{\symmt{P}{S'}})
        \end{multlined} \\
        &=\tag{Iso} \dl{PQ}{W}{S'} ; (\LW{P}{\symmt{Q}{W}} \piu \LW{P}{\symmt{Q}{S'}}) ; (\RW{Q}{\symmt{P}{W}} \piu \RW{Q}{\symmt{P}{S'}}) \\
        &=\tag{Funct. $\piu$} \dl{PQ}{W}{S'} ; ((\LW{P}{\symmt{Q}{W}};\RW{Q}{\symmt{P}{W}}) \piu (\LW{P}{\symmt{Q}{S'}};\RW{Q}{\symmt{P}{S'}})) \\
        &=\tag{$\ast_3$, ind. hp.} \dl{PQ}{W}{S'} ; (\symmt{PQ}{W} \piu \symmt{PQ}{S'}) \\
        &=\tag{Lemma~\ref{lemma:tapesymdis}.\ref{lm:tape:symmt dl}} \symmt{PQ}{W \piu S'}
    \end{align*}

    \textsc{Equation~\eqref{eq:LRnatsym}:} $\RW{S}{\t} ; \symmt{Q}{S} = \symmt{P}{S} ; \LW{S}{\t}$. First observe that when $S$ is a monomial $W$, the following holds:
    \begin{equation}\tag{$\ast$}
        \RW{W}{\t} ; \symmt{Q}{W} = \symmt{P}{W} ; \LW{W}{\t}
    \end{equation}
    \begin{align*}
        \RW{W}{\t} ; \symmt{Q}{W} &=\tag{Lemma~\ref{lemma:tapesymdis}.\ref{lm:tape:symmt V}} \RW{W}{\t} ; \Piu[j]{\tapesymm{V_J}{W}} \\
        &=\tag{Lemma~\ref{lm:whisk}} \Piu[i]{\tapesymm{U_i}{W}} ; \LW{W}{\t} \\
        &=\tag{Lemma~\ref{lemma:tapesymdis}.\ref{lm:tape:symmt V}} \symmt{P}{W} ; \LW{W}{\t}
    \end{align*}

    Then we proceed by induction on $S$.
    
    \textbf{Case $S=\zero$} it follows from Definitions~\ref{def:tape:whiskG} and definition of $\symmt$.
    
    \textbf{Case $S=W \piu S'$:} 
    \begin{align*}
        &\RW{W \piu S'}{\t} ; \symmt{Q}{W \piu S'} \\
        &=\tag{\ref{eq:whisk:sum}, Lemma~\ref{lemma:tapesymdis}.\ref{lm:tape:symmt dl}} \dl{P}{W}{S'} ; (\RW{W}{\t} \piu \RW{S'}{\t}) ; \Idl{Q}{W}{S'} ; \dl{Q}{W}{S'} ; (\symmt{Q}{W} \piu \symmt{Q}{S'}) \\
        &=\tag{Iso} \dl{P}{W}{S'} ; (\RW{W}{\t} \piu \RW{S'}{\t}) ; (\symmt{Q}{W} \piu \symmt{Q}{S'}) \\
        &=\tag{Funct. $\piu$} \dl{P}{W}{S'} ; ((\RW{W}{\t};\symmt{Q}{W}) \piu (\RW{S'}{\t};\symmt{Q}{S'})) \\
        &=\tag{$\ast$, ind. hp.} \dl{P}{W}{S'} ; ((\symmt{P}{W};\LW{W}{\t}) \piu (\symmt{P}{S'};\LW{S'}{\t})) \\
        &=\tag{Funct. $\piu$} \dl{P}{W}{S'} ; (\symmt{P}{W} \piu \symmt{P}{S'}) ; (\LW{W}{\t} \piu \LW{S'}{\t}) \\
        &=\tag{Lemma~\ref{lemma:tapesymdis}.\ref{lm:tape:symmt dl}, \ref{eq:whisk:sum}} \symmt{P}{W \piu S'} ; \LW{W \piu S'}{\t}
    \end{align*}

    \textsc{Equations~\eqref{eq:tape:LL}: $\LW{ST}{\t} = \LW{S}{\LW{T}{\t}}$,~\eqref{eq:tape:LR}: $\LW{S}{\RW{T}{\t}} = \RW{T}{\LW{S}{\t}}$,~\eqref{eq:tape:RR}:$\RW{TS}{\t} = \RW{S}{\RW{T}{\t}}$}. First we prove the particular cases in which $S$ and $T$ are monomials:
    \begin{align*}
        \LW{W}{\LW{Z}{\t}} &= \LW{WZ}{\t} &\tag{$\ast_1$}\\
        \LW{W}{\RW{Z}{\t}} &= \RW{Z}{\LW{W}{\t}} &\tag{$\ast_2$}\\
        \RW{W}{\RW{Z}{\t}} &= \RW{ZW}{\t} &\tag{$\ast_3$}
    \end{align*}
    All the three equations are proved by induction on $\t$ as follows:  
    the base cases $\t = \id\zero, \symmp{U}{V}, \diag{U}, \bang{U}, \codiag{U}, \cobang{U}$ are trivial by definition of $L$ and $R$. For $\t = \tape{c}$ we apply the definition of $L$ and $R$ and the fact that $\per$ is associative inside a tape.
    The inductive cases follow from functoriality of $L$ and $R$ and the inductive hypothesis.

    For the general cases, we start with~\eqref{eq:tape:LL}: \begin{align*}
        \LW{S}{\LW{T}{\t}} &=\tag{Def. $L$} \Piu[k]{\Piu[l]{\LW{W_k}{\LW{Z_l}{\t}}}}\\
        &=\tag{$\ast_1$} \Piu[k]{\Piu[l]{\LW{W_kZ_l}{\t}}}\\
        &=\tag{Def. $L$} \LW{ST}{\t}
    \end{align*}

    Regarding {Equation~\eqref{eq:tape:LR}}, first observe that the following holds:
    \begin{equation*}\tag{$\ast$}
        \RW{W}{\LW{T}{\t}} = \LW{T}{\RW{W}{\t}}
    \end{equation*}
	indeed:
    \begin{align*}
        \RW{W}{\LW{T}{\t}} &=\tag{Def. $L$} \RW{W}{\Piu[l]{\LW{Z_l}{\t}}} \\
        &=\tag{Def. $R$} \Piu[l]{\RW{W}{\LW{Z_l}{\t}}} \\
        &=\tag{$\ast_2$} \Piu[l]{\LW{Z_l}{\RW{W}{\t}}} \\
        &=\tag{Def. $L$} \LW{T}{\RW{W}{\t}}
    \end{align*}

    Then we proceed by induction on $S$:

    \textbf{Case $S = \zero$:} if follows from Definition~\ref{def:tape:whiskG} and~\eqref{eq:whisk:zero}.

    \textbf{Case $S = W \piu S'$:} \begin{align*}
        &\RW{W \piu S'}{\LW{T}{\t}} \\
        &=\tag{Def. $R$} \dl{TP}{W}{S'} ; (\RW{W}{\LW{T}{\t}} \piu \RW{S'}{\LW{T}{\t}}) ; \Idl{TQ}{W}{S'} \\
        &=\tag{$\ast$, ind. hp.} \dl{TP}{W}{S'} ; (\LW{T}{\RW{W}{\t}} \piu \LW{T}{\RW{S'}{\t}}) ; \Idl{TQ}{W}{S'} \\
        &=\tag{\ref{eq:whisk:funct piu}} \dl{TP}{W}{S'} ; \Idl{T}{PW}{PS'} ; \LW{T}{\RW{W}{\t} \piu \RW{S'}{\t}} ; \dl{T}{QW}{QS'} ; \Idl{TQ}{W}{S'} \\
        &=\tag{\ref{eq:whisk:Ldl}} \LW{T}{\dl{P}{W}{S'}} ; \LW{T}{\RW{W}{\t} \piu \RW{S'}{\t}} ; \LW{T}{\Idl{Q}{W}{S'}} \\
        &=\tag{\ref{eq:whisk:funct}} \LW{T}{\dl{P}{W}{S'} ; (\RW{W}{\t} \piu \RW{S'}{\t}) ; \Idl{Q}{W}{S'}} \\
        &=\tag{\ref{eq:whisk:sum}} \LW{T}{\RW{W \piu S'}{\t}}
    \end{align*}

    Equation~\eqref{eq:tape:RR} is proved by means of the other two:
    \begin{align*}
        \RW{TS}{\t} & =\tag{\ref{eq:LRnatsym}} \symmt{P}{TS} ; \LW{TS}{\t} ; \symmt{TS}{Q}\\
        & =\tag{\ref{eq:tape:LL}} \symmt{P}{TS} ; \LW{T}{\LW{S}{\t}} ; \symmt{TS}{Q}\\
        & =\tag{\ref{eq:symmper}} \RW{S}{\symmt{P}{T}} ; \LW{T}{\symmt{P}{S}} ; \LW{T}{\LW{S}{\t}} ; \LW{T}{\symmt{S}{Q}}  ;   \RW{S}{\symmt{T}{Q}}  \\
        & =\tag{\ref{eq:whisk:funct}} \RW{S}{\symmt{P}{T}} ; \LW{T}{\symmt{P}{S} ; \LW{S}{\t} ; \symmt{S}{Q}}  ;   \RW{S}{\symmt{T}{Q}}  \\
        & =\tag{\ref{eq:LRnatsym}} \RW{S}{\symmt{P}{T}} ; \LW{T}{\RW{S}{\t} }  ;   \RW{S}{\symmt{T}{Q}}  \\
        & =\tag{\ref{eq:tape:LR}} \RW{S}{\symmt{P}{T}} ; \RW{S}{\LW{T}{\t} }  ;   \RW{S}{\symmt{T}{Q}}  \\
        & =\tag{\ref{eq:whisk:funct}} \RW{S}{\symmt{P}{T} ; \LW{T}{\t} ;\symmt{T}{Q}}  \\
        & =\tag{\ref{eq:LRnatsym}} \RW{S}{\RW{T}{\t} }  \\
    \end{align*}

   \textsc{Equation~\eqref{eq:tape:LexchangeR}:} $\LW{P}{\t_2} ; \RW{S}{\t_1} = \RW{R}{\t_1} ; \LW{Q}{\t_2}$. Let $\t_1 \colon P \to Q, \t_2 \colon R \to S$, then we show by induction on $\t_1$ that 
  \[\LW{P}{\t_2} ; \RW{S}{\t_1} = \RW{R}{\t_1} ; \LW{Q}{\t_2}\]

    \textbf{Case $\t_1 = \id{\zero}$:} by \eqref{eq:whisk:id}, $ \RW{S}{\id{\zero}} = \id{\zero} = \RW{R}{\id{\zero}}$. By definition, $\LW{\zero}{\t_2}=\id{\zero}$.

    \textbf{Case $\t_1 = \tape{c}$:} it holds by Lemma~\ref{lm:tape:perC}.
    
    \textbf{Case $\t_1 = \symmp{U}{V}, \diag{U}, \bang{U}$:} these cases all follow the same pattern, thus we show only the one for $\diag{U}$:
    \begin{align*}
         \LW{U}{\t_2} ; \RW{S}{\diag{U}}&=\tag{\ref{eq:whisk:diag}} \LW{U}{\t_2} ; \diag{US} \\
        &=\tag{\ref{ax:diagnat}} \diag{UR} ; (\LW{U}{\t_2} \piu \LW{U}{\t_2}) \\
        &=\tag{\ref{eq:whisk:diag}} \RW{R}{\diag{U}} ; (\LW{U}{\t_2} \piu \LW{U}{\t_2}) \\
        &=\tag{Def. $L$}\RW{R}{\diag{U}} ; (\LW{U\piu U}{\t_2} )     \end{align*}

    \textbf{Case $\t_1 = \t_1' ; \t_1''$:} Suppose $\t_1' \colon P \to P', \t_1'' \colon P' \to Q, \t_2 \colon R \to S$
    \begin{align*}
\LW{P}{\t_2} ; \RW{S}{\t_1'; \t_1''} & =\tag{\ref{eq:whisk:funct}} \LW{P}{\t_2} ; \RW{S}{\t_1'} ; \RW{S}{ \t_1''}\\
& =\tag{Ind. hp.} \RW{R}{\t_1'} ; \LW{P'}{\t_2} ; \RW{S}{ \t_1''}\\
& =\tag{Ind. hp.} \RW{R}{\t_1'} ; \RW{R}{\t_1''} ; \LW{Q}{\t_2}  \\
        & =\tag{\ref{eq:whisk:funct}} \RW{R}{\t_1' ; \t_1''} ; \LW{Q}{\t_2}  \\
\end{align*}

    \textbf{Case $\t_1 = \t_1' \piu \t_1''$:} Suppose $\t_1' \colon P_1 \to Q_1, \t_1'' \colon P' \to Q', \t_2 \colon R \to S$
    
     \begin{align*}
      \LW{P_1 \piu P'}{\t_2} ; \RW{S}{\t_1' \piu \t_1''}  &=\tag{\ref{eq:whisk:sum}, \ref{eq:whisk:funct piu}} (\LW{P_1}{\t_2} \piu \LW{P'}{\t_2}) ; (\RW{S}{\t_1'} \piu \RW{S}{\t_1''}) \\
        &=\tag{Funct. $\piu$} (\LW{P_1}{\t_2} ; \RW{S}{\t_1'}) \piu ( \LW{P'}{\t_2} ; \RW{S}{\t_1''}) \\
        &=\tag{Ind. hp.} (\RW{R}{\t_1'} ; \LW{Q_1}{\t_2}) \piu (\RW{R}{\t_1''} ; \LW{Q'}{\t_2}) \\
        &=\tag{Funct. $\piu$} (\RW{R}{\t_1'} \piu \RW{R}{\t_1''}) ; (\LW{Q_1}{\t_2} \piu \LW{Q'}{\t_2}) \\
        &=\tag{\ref{eq:whisk:sum}, \ref{eq:whisk:funct piu}}  \RW{R}{\t_1' \piu \t_1''} ; \LW{Q_1 \piu Q'}{\t_2} 
    \end{align*}
This concludes the proof of Lemma~\ref{lm:whisk properties}.

\subsection{Other proofs of Section~\ref{sc:tape}}\label{appTape4}

\begin{proof}[Proof of Lemma~\ref{lemma:F2 left adjoint to U2}]
	Given a strict fb category $\Cat{D}$ and functor $H \colon \Cat{C} \to U_2(\Cat{D})$, one can define the fb-functor $H^{\sharp} \colon  F_2(\Cat{C})  \to \Cat{D}$  inductively on objects of $ F_2(\Cat{C}) $ as
	\[H^{\sharp}(\unoG) = \unoG \qquad H^{\sharp}(AW)=H(A)\perG H^{\sharp}(W) \]
	and on arrows as
	\[H^{\sharp}(\id\unoG) ={\id{\unoG}} \qquad H^{\sharp}(\id{A})= H(\id{A}) \qquad H^{\sharp}(\tape{c})=H(c) \]
	\[H^{\sharp}(f; g)=H^{\sharp}(f); H^{\sharp}(g) \qquad  H^{\sharp}(f \perG g)= H^{\sharp}(f) \perG H^{\sharp}(g)  \qquad  H^{\sharp}(\sigma_{A,B})=\sigma_{H(A),H(B)} \]
	\[ H^{\sharp}(\bang{A})=\bang{H(A)} \qquad  H^{\sharp}(\diag{A})=\diag{H(A)} \qquad  H^{\sharp}(\cobang{A})=\cobang{H(A)} \qquad H^{\sharp}(\codiag{A})=\codiag{H(A)}\]
	
	Observe that $H^{\sharp}$ is well-defined:

	\begin{minipage}{0.45\linewidth}
		\begin{align*}
			H^{\sharp}(\tape{\id{A}}) & =\tag{Def. $H^{\sharp}$}  H(\id{A}) \\
& =\tag{Def. $H^{\sharp}$} H^{\sharp}(\id{A})
		\end{align*}
	\end{minipage}
	\hfill
	\begin{minipage}{0.45\linewidth}
		\begin{align*}
			H^{\sharp}(\tape{c;d}) & =\tag{Def. $H^{\sharp}$}  H(c;d) \\
			& =\tag{Fun. $H$} H(c); H(d)\\
			& =\tag{Def. $H^{\sharp}$} H^{\sharp}(\tape{c}); H^{\sharp}(\tape{d}) \\
			& =\tag{Fun. $H^{\sharp}$} H^{\sharp}(\tape{c}; \tape{d}) \\
		\end{align*}
	\end{minipage}
	
	The axioms in Tables \ref{fig:freestricmmoncatax} and \ref{fig:freestrictfbcat} are preserved by $H^{\sharp}$, since they hold in $\Cat{D}$.
	By definition, $\tapeFunct{\cdot} \, ; H^{\sharp}= H$. Moreover $ H^{\sharp}$ is the unique strict fb functor satisfying this equation. Thus indeed $F_2 \dashv U_2$.
\end{proof}

\begin{proof}[Proof of Theorem~\ref{thm:taperig}]
    Observe that for all objects $S$ and arrows $\t \colon P \to Q$, it holds that
    \begin{equation}\label{eq:mainwhiskering}
        \t \per \id{S} = \RW{S}{\t} \qquad \qquad \id{S} \per \t = \LW{S}{\t}
    \end{equation}
    as shown below:

    \begin{minipage}{0.48\linewidth}
        \begin{align*}
            \t \per \id{S}  &= \LW{P}{\id{S}} ; \RW{S}{\t} \tag{Def. $\per$} \\ 
            &= \RW{S}{\t} \tag{\ref{eq:whisk:id}}
        \end{align*}
    \end{minipage}
    \hfill
    \begin{minipage}{0.48\linewidth}
        \begin{align*}
            \id{S}  \per \t  &= \LW{S}{\t} ; \RW{Q}{\id{S}} \tag{Def. $\per$}\\ 
            &= \LW{S}{\t} \tag{\ref{eq:whisk:id}}
        \end{align*}
    \end{minipage}
    Thus $\id{P} \per \id{Q} \stackrel{\eqref{eq:mainwhiskering}}{=}  \LW{P}{\id{Q}} \stackrel{\eqref{eq:whisk:id}}{=}  \id{P \per Q}$. To conclude functoriality of $\per$, the key is the sliding law \eqref{eq:tape:LexchangeR}:  for all  $\t_1 \colon P \to Q, \t_2 \colon Q \to S, \t_3 \colon P' \to Q', \t_4 \colon Q' \to S'$, 
    \begin{align*}
        (\t_1 ; \t_2) \per (\t_3 ; \t_4) &=\tag{Def. $\per$}   \LW {P} {\t_3 ; \t_4} ; \RW {S'} {\t_1 ; \t_2} \\
        &=\tag{\ref{eq:whisk:funct}} \LW {P} {\t_3} ; \LW {P} {\t_4} ; \RW {S'} {\t_1} ; \RW {S'} {\t_2} \\
        &=\tag{\ref{eq:tape:LexchangeR}} \LW {P} {\t_3} ; \RW {Q'} {\t_1} ; \LW {Q} {\t_4}   ; \RW {S'} {\t_2} \\
        &=\tag{Def. $\per$} (\t_1 \per \t_3) ; (\t_2 \per \t_4) \\
    \end{align*}

    By \eqref{eq:mainwhiskering} and \eqref{eq:whisk:uno} it immediately follows that 
    \[\id{\uno} \per \t \; =  \; \t  \;= \; \t \per \id{\uno}\]
    
    For associativity of $\per$, observe that for all  $\t_1 \colon P_1 \to Q_1, \t_2 \colon P_2 \to Q_2, \t_3 \colon P_3 \to Q_3$, 
    \begin{align*}
            ((\t_1 \per \t_2) \per \t_3 ) &=\tag{Def. $\per$}   \LW {P_1P_2} {\t_3} ; \RW {Q_3} {\LW{P_1}{\t_2}; \RW{Q_2}{\t_1}} \\
            &=\tag{\ref{eq:whisk:funct}} \LW {P_1P_2} {\t_3} ; \RW {Q_3} {\LW{P_1}{\t_2}}; \RW {Q_3} {\RW{Q_2}{\t_1}} \\
            &=\tag{\ref{eq:tape:LL},\ref{eq:tape:LR},\ref{eq:tape:RR}} \LW {P_1} {\LW{P_2}{\t_3}} ;  \LW{P_1}{\RW {Q_3}{\t_2}};  \RW{Q_2Q_3}{\t_1}\\
            &=\tag{\ref{eq:whisk:funct}} \LW {P_1} {\LW{P_2}{\t_3} ; \RW {Q_3}{\t_2}};  \RW{Q_2Q_3}{\t_1} \\
            &=\tag{Def. $\per$}  (\t_1 \per (\t_2 \per \t_3 ))
        \end{align*}

    So far, we proved that $(\CatTape, \per, \uno)$ is a strict monoidal category. For symmetries, observe that naturality follows immediately by \eqref{eq:mainwhiskering} and \eqref{eq:LRnatsym}. Inverses is Lemma~\ref{lemma:tapesymdis}.\ref{lm:tape:symmtInv} and the coherence axioms \eqref{eq:symmax2} and \eqref{eq:symmax3} hold by Lemma~\ref{lemma:tapesymdis}.\ref{lm:tape:symmt uno} and \eqref{eq:symmper}.
    
    By construction $(\CatTape, \piu, \zero, \symmp)$ is a symmetric strict monoidal category. We are left to prove a few laws. First, 
    \[ \id{\zero} \per \t  \; =  \; \id{\zero} = \t \per \id{\zero}\]
    hold by \eqref{eq:mainwhiskering} and \eqref{eq:whisk:zero}. Regarding right distributivity of $\per$ over $\piu$, observe that for all
    $\t_1 \colon P_1 \to Q_1, \, \t_2 \colon P_2 \to Q_2, \, \t_3 \colon P_3 \to Q_3$ it holds that:
        \begin{align*}
            (\t_1 \piu \t_2) \per \t_3 &=\tag{Def. $\per$} \LW{P_1 \piu P_2}{\t_3} ; \RW{Q_3}{\t_1 \piu \t_2} \\
            &=\tag{\ref{eq:whisk:sum}, \ref{eq:whisk:funct piu}} (\LW{P_1}{\t_3} \piu \LW{P_2}{\t_3}) ; (\RW{Q_3}{\t_1} \piu \RW{Q_3}{\t_2}) \\
            &=\tag{Funct. $\piu$} (\LW{P_1}{\t_3} ; \RW{Q_3}{\t_1}) \piu ( \LW{P_2}{\t_3} ; \RW{Q_3}{\t_2}) \\
            &=\tag{Def. $\per$} (\t_1 \per \t_3) \piu (\t_2 \per \t_3)
        \end{align*}
    (This is tantamount to saying that $\CatTape$ has a natural right distributor whose components $\dr{P}{Q}{R}$ are all identity morphisms).
    
    Next we have to check the axioms of coherence for strict rig categories that amount just to \eqref{eq:rigax1},~\eqref{eq:rigax2},~\eqref{eq:rigax5} and~\eqref{eq:rigax9} from Figure~\ref{fig:rigax}. 
    Axiom~\eqref{eq:rigax1} is Lemma~\ref{lemma:tapesymdis}.\ref{lm:tape:symmt dl}.
    Axiom~\eqref{eq:rigax2} follows by \eqref{eq:mainwhiskering} and \eqref{eq:whisk:symmp}.
    Axiom~\eqref{eq:rigax5} is Lemma \ref{lemma:tapesymdis}.\ref{lm:tape:dlG}. Axiom~\eqref{eq:rigax9} holds by definition of $\symmt$ in Table~\ref{table:def symmt}. Finally, from Axiom~\eqref{eq:rigax1} and the naturality of the right distributor and of $\symmt $ we get that also $\delta^l$ is natural.
    This proves that $\CatTape$ is a strict rig category. $(\CatTape, \piu, \zero, \symmp)$  is a finite biproduct category by definition.

        All that is left to do is to prove that the inclusion functor $\sort \to \CatTape$ makes $\CatTape$ a $\sort$-sesquistrict fb rig category according to Definition~\ref{def:sesquistrict rig category}. This means that we have to show that $\dl{A}{Q}{R} = \id{(A \per Q)  \piu (A \per R)}$ in $\CatTape$ for all $A \in \sort$, which is easily inferred from the inductive definition of $\dl{P}{Q}{R}$.
    \end{proof}    

\begin{proof}[Proof of Theorem~\ref{thm:Tapes is free sesquistrict generated by sigma}]

	We have a trivial interpretation of $(\sort,\sign)$ in $\sort \to \CatTape$, given by $(\id\sort,\tapeFunct{\cdot})$. Suppose now that $H \colon \Cat M \to \Cat D$ is a sesquistrict fb rig category with an interpretation $\interpretation=(\alpha_\sort \colon \sort \to {Ob}(\Cat M), \alpha_\sign \colon \sign \to {Ar}(\Cat D))$. We aim to find a sesquistrict fb rig functor $(\alpha \colon \sort \to \Cat M, \beta \colon \CatTape \to \Cat D)$ such that $\id\sort ; \alpha = \alpha_{\sort}$ and $\tapeFunct{\cdot} ; \beta = \alpha_{\sign}$.
	Since $(\sort,\sign)$ is a monoidal signature, $\interpretation$ is in fact a monoidal interpretation of $(\sort,\sign)$ into the ssm category $(\Cat D,\per,\uno)$, hence by freeness of $\CatString$ there exists a unique ssm functor $\dsem{-}_{\interpretation} \colon \CatString \to \Cat D$ that extends $\interpretation$. Now, because $\CatTape = F_2(\CatString)$ is by definition the free fb category generated by $\CatString$ and $\Cat D$ is also a fb category, we have that there is a unique fb functor $\beta \colon \CatTape \to \Cat D$ that extends $\dsem{-}_{\interpretation}$. 
Now we prove that $\beta$ preserves the rest of the rig structure in several steps. 
	For the rest of this proof, to lighten up the calculations, with little abuse of notation we will treat $\GFunct$ as if it were the identity on objects. The context will make clear whether we will be speaking of $P \in Ob(\CatTape)$ or of $\GFunct(P) \in Ob(\Cat D)$.
\begin{enumerate}[label=\roman*.]
		\item\label{lm:tape:Gsymmt.1} We show that $\GFunct(\dl{P}{Q}{R})= \dl{P}{Q}{R}$  by induction on $P$.
	
		\textbf{Case $P=\zero$:} \begin{align*}
			\GFunct(\dl{\zero}{Q}{R}) &=\tag{Def. $\delta^l$} \GFunct(\id\zero) \\
			&=\tag{Def. $\GFunct$} \id\zero \\
			&=\tag{\ref{eq:dl4}} \dl{\zero}{Q}{R}
		\end{align*}
		
		\textbf{Case $P=U \piu P'$:} \begin{align*}
			&\GFunct(\dl{U \piu P'}{Q}{R}) \\
			&=\tag{Def. $\delta^l$} \GFunct((\id{U(Q\piu R)} \piu \dl{P'}{Q}{R});(\id{UQ} \piu \symmp{UR}{P'Q} \piu \id{P'R})) \\
			&=\tag{Funct. $\GFunct$} (\GFunct(\id{U(Q\piu R)}) \piu \GFunct(\dl{P'}{Q}{R})) ; (\GFunct(\id{UQ}) \piu \GFunct(\symmp{UR}{P'Q}) \piu \GFunct(\id{P'R})) \\
			&=\tag{Def. $\GFunct$} (\id{U(Q\piu R)} \piu \dl{P'}{Q}{R}) ; (\id{UQ} \piu \symmp{UR}{P'Q} \piu \id{P'R}) \\
			&=\tag{Def. $\delta^l$} (\dl{U}{Q}{R} \piu \dl{P'}{Q}{R}) ; (\id{UQ} \piu \symmp{UR}{P'Q} \piu \id{P'R}) \\
			&=\tag{\ref{eq:rigax5}} \dl{U \piu P'}{Q}{R}.
		\end{align*}
	\item\label{lm:tape:Gsymmt.2} Next we prove that $\GFunct(\symmt{P}{Q}) = \symmt{P}{Q}$ by induction on $Q$.

\textbf{Case $Q=\zero$:} it follows from definitions of $\symmt$ and $\GFunct$.
	
	\textbf{Case $Q=V \piu Q'$:} 
	\begin{align*}
		\GFunct(\symmt{P}{V \piu Q'}) &=\tag{Def. $\symmt$} \GFunct(\dl{P}{V}{Q'} ; (\Piu[i]{\tapesymm{U_i}{V}} \piu \symmt{P}{Q'})) \\
		&=\tag{Def. $\GFunct$} \GFunct(\dl{P}{V}{Q'}) ; (\Piu[i]{\GFunct(\tapesymm{U_i}{V})} \piu \GFunct(\symmt{P}{Q'})) \\
		&=\tag{\ref{lm:tape:Gsymmt.1}} \dl{P}{V}{Q'} ; (\Piu[i]{\GFunct(\tapesymm{U_i}{V})} \piu \GFunct(\symmt{P}{Q'})) \\
		&=\tag{Def. $\GFunct$, ind. hp.} \dl{P}{V}{Q'} ; (\Piu[i]{\symmt{U_i}{V}} \piu \symmt{P}{Q'}) \\
&=\tag{Lemma~\ref{lm:sesqui:symmt}} \dl{P}{V}{Q'} ; (\symmt{P}{V} \piu \symmt{P}{Q'}) \\
		&=\tag{\ref{eq:rigax1}} \symmt{P}{V \piu Q'}
	\end{align*}
	\item\label{lm:tape:GperW1} Now we show that if $\t \colon P \to Q$ be a tape diagram, then $\GFunct(\RW{W}{\t}) = \GFunct(\t) \per \id{W}$. We reason by induction on $\t$.

\textbf{Case $\t=\id{\zero}$:} $\GFunct(\RW{W}{\id{\zero}}) = \GFunct(\id{\zero}) = \id\zero = \id\zero \per \id{W} = \GFunct(\id{\zero}) \per \id{W}$

\textbf{Case $\tape{c}$:} $\GFunct(\RW{W}{\tape{c}}) = \GFunct(\tape{c \per \id{W}}) = c \per \id{W} = \GFunct(\tape{c}) \per \id{W}$

\textbf{Case $\symmp{U}{V}$:} $\GFunct(\RW{W}{\symmp{U}{V}}) = \GFunct(\symmp{UW}{VW}) = \symmp{UW}{VW} \stackrel{\eqref{eq:rigax2}}{=} \symmp{U}{V} \per \id{W} = \GFunct(\symmp{U}{V}) \per \id{W}$

\textbf{Case $\diag{U}$:} $\GFunct(\RW{W}{\diag{U}}) = \GFunct(\diag{UW}) = \diag{UW} \; \stackrel{Prop.\ref{prop:fbrig}}{=} \; \diag{U} \per \id{W} = \GFunct(\diag{U}) \per \id{W}$  and analogously for $\bang U, \codiag U, \cobang U$.

\textbf{Case $\t = \t_1 ; \t_2$:} $\GFunct(\RW{W}{(\t_1 ; \t_2)}) = \GFunct(\RW{W}{\t_1}) ; \GFunct(\RW{W}{\t_2}) = (\GFunct(\t_1) \per \id{W}) ; (\GFunct(\t_2) \per \id{W}) = (\GFunct(\t_1) ; \GFunct(\t_2)) \per \id{W} = \GFunct(\t_1 ; \t_2) \per \id{W}$

\textbf{Case $\t = \t_1 \piu \t_2$:} $\GFunct(\RW{W}{(\t_1 \piu \t_2)}) = \GFunct(\RW{W}{\t_1}) \piu \GFunct(\RW{W}{\t_2}) = (\GFunct(\t_1) \per \id{W}) \piu (\GFunct(\t_2) \per \id{W}) = (\GFunct(\t_1) \piu \GFunct(\t_2)) \per \id{W} = \GFunct(\t_1 \piu \t_2) \per \id{W}$.
	\item\label{lm:tape:Gper1} Next we show that if $\t \colon P \to Q$ be a tape diagram, then 
	\begin{enumerate}
		\item $\GFunct(\RW{S}{\t}) = \GFunct(\t) \per \id{S}$
		\item $\GFunct(\LW{S}{\t}) = \id{S} \per \GFunct(\t)$
	\end{enumerate}
     We prove the first point by induction on $S$:

\textbf{Case $S = \zero$:} $\GFunct(\RW{\zero}{\t}) = \GFunct(\id\zero) = \id\zero = \GFunct(\t) \per \id\zero$.

\textbf{Case $S = W \piu S'$:}
\begin{align*}
	\GFunct(\RW{W\piu S'}{\t}) &=\tag{\ref{eq:whisk:sum}} \GFunct(\dl{P}{W}{S'} ; (\RW{W}{\t} \piu \RW{S'}{\t}) ; \dl{Q}{W}{S'}) \\
	&=\tag{Def. $\GFunct$} \GFunct(\dl{P}{W}{S'}) ; (\GFunct(\RW{W}{\t}) \piu \GFunct(\RW{S'}{\t})) ; \GFunct(\dl{Q}{W}{S'}) \\
	&=\tag{\ref{lm:tape:Gsymmt.1}} \dl{P}{W}{S'} ; (\GFunct(\RW{W}{\t}) \piu \GFunct(\RW{S'}{\t})) ; \dl{Q}{W}{S'} \\
	&=\tag{\ref{lm:tape:GperW1}, ind. hp.} \dl{P}{W}{S'} ; ((\GFunct(\t) \per \id W) \piu (\GFunct(\t) \per \id {S'})) ; \dl{Q}{W}{S'} \\
	&=\tag{Nat. $\delta^l$} \GFunct(\t) \per (\id W \piu \id {S'}) \\
	&=\tag{Funct. $\piu$} \GFunct(\t) \per \id{W \piu S'}
\end{align*}

For the second point, we use the first point and naturality of $\per$-symmetries.
\begin{align*}
	\GFunct(\LW{S}{\t}) &=\tag{Lemma~\ref{lemma:tapesymdis}.\ref{lm:tape:symmtInv}} \GFunct(\symmt{S}{P} ; \symmt{P}{S} ; \LW{S}{\t}) \\ 
	&=\tag{\ref{eq:LRnatsym}} \GFunct(\symmt{S}{P} ; \RW{S}{\t} ; \symmt{Q}{S}) \\ 
	&=\tag{Funct. $\GFunct$} \GFunct(\symmt{S}{P}) ; \GFunct(\RW{S}{\t}) ; \GFunct(\symmt{Q}{S}) \\ 
	&=\tag{\ref{lm:tape:Gsymmt.2}} \symmt{S}{P} ; \GFunct(\RW{S}{\t}) ; \symmt{Q}{S} \\ 
	&=\tag{\ref{lm:tape:Gper1}.a} \symmt{S}{P} ; (\GFunct(\t) \per \id{S}) ; \symmt{Q}{S} \\ 
	&=\tag{Nat. $\symmt$} \id{S} \per \GFunct(\t)
\end{align*}
\item Finally, we prove that if $\t_1 \colon P \to Q, \t_2 \colon R \to S$ are tape diagrams, then $\GFunct(\t_1 \per \t_2 ) =  \GFunct(\t_1) \per \GFunct(\t_2)$:
\begin{align*}
	\GFunct(\t_1 \per \t_2 ) &=\tag{Def. $\per$} \GFunct(\LW{P}{\t_2};\RW{S}{\t_1}) \\
	&=\tag{Funct. $\GFunct$} \GFunct(\LW{P}{\t_2});\GFunct(\RW{S}{\t_1}) \\
	&=\tag{\ref{lm:tape:Gper1}} (\id{P} \per \GFunct(\t_2)) ; (\GFunct(\t_1) \per \id{S}) \\
	&=\tag{Funct. $\per$} \GFunct(\t_1) \per \GFunct(\t_2)
\end{align*}
	\end{enumerate}
This concludes the proof of Theorem~\ref{thm:Tapes is free sesquistrict generated by sigma}.
\end{proof}

\begin{proof}[Proof of Corollary~\ref{co:rig semantics}]
	Recall our convention following Lemma~\ref{lemma:every functor in strict(D) is well behaved} that in fact we are considering the codomain of $F$ to be $\sCatT D$, and that we will expect the codomain of $F^\sharp$ also to be $\sCatT D$. Let us call $\interpretation_M$ and $\interpretation_R$ the obvious interpretations of $(\sort,\sign)$ in $\CatString$ and in $\CatTape$, respectively (the first is a monoidal, the second a sesquistrict rig interpretation). Notice that $\interpretation_R = \tapeFunct{\cdot} \circ \interpretation_M$. We have that $F$ induces a sesquistrict rig interpretation $F \circ \interpretation_M$ of $(\sort,\sign)$ in $\sCatT D$, given by
	\[
	\begin{tikzcd}[row sep=0em]
		\sort \ar[r,"\alpha_\sort"] & Ob(\Cat D) \\
		A \ar[r,|->] & F(A)
	\end{tikzcd} 
	\qquad
	\begin{tikzcd}[row sep=0em]
		\sign \ar[r,"\alpha_\sign"] & Ar(\sCatT D) \\
		f \ar[r,|->] & F(f)
	\end{tikzcd}
	\]
	(we are assuming that $F(A)$ is a unary word of a unary word as a consequence of Lemma~\ref{lemma:every functor in strict(D) is well behaved}). This makes $(\alpha_\sort,\alpha_\sign)$ an interpretation because $F$ is strict monoidal. Indeed, if $\ar(f) = U$ with $U=A_1\dots A_n$, then the domain of $\alpha_\sign (f) = F(f)$ in $\sCatT D$ is the formal monomial $F(U)=F(A_1) \per \dots \per F(A_n)$, which is the result of applying the inductive extension of $\iota F$ to $U \in \sort^\star$ ($\iota \colon \Cat D \to \sCatT D$ being the inclusion functor). Since $\CatTape$ is the free sesquistrict  fb rig category generated by $(\sort,\sign)$, there exists a unique strict fb rig functor $F^\sharp \colon \CatTape \to \sCat D$ such that
\begin{itemize}
		\item the pair $(\alpha_\sort,F^\sharp)$ is a sesquistrict fb rig functor from $(\sort \to \CatTape)$ to $(Ob(\Cat D) \to \sCatT D)$,
		\item $F^\sharp \circ \interpretation_R = F \circ \interpretation_M$.
	\end{itemize}
	The second condition means that $F^\sharp(\tape f) = F(f)$ for all $f \in \sign$. Now we want to prove that $F^\sharp(\tapeFunct c) = F(\tapeFunct c)$ for every circuit $c \in \CatString$.

	We note that $F^\sharp \circ \interpretation_R$ is in fact a monoidal interpretation of $(\sort,\sign)$ in the ssm category $(\sCat D,\per,\uno)$, given by
	\[
	\alpha_\sort ' = \alpha_\sort \qquad 	
	\begin{tikzcd}[row sep=0em]
		\sign \ar[r,"\alpha_\sign '"] & Ar(\sCat D) \\
		f \ar[r,|->] & F^\sharp(\tape f)
	\end{tikzcd}
	\]
	Now, since $\CatString$ is the free strict symmetric monoidal category generated by $(\sort,\sign)$, there exists a unique strict monoidal functor $G \colon \CatString \to \sCat D$ such that $G \circ \interpretation_M = F^\sharp \circ \interpretation_R$. But both $G=F$ and $G=F^\sharp \circ \tapeFunct{\cdot}$ satisfy the above equation. Hence $F=F^\sharp \circ \tapeFunct{\cdot}$. 
	
	Regarding uniqueness: we have shown that every functor $H \colon \CatTape \to \sCat D$ extending the interpretation $F \circ \interpretation_M$ also satisfies $F = H \circ \tapeFunct{\cdot}$. Vice versa, any functor $H$ extending $F$ via $\tapeFunct{\cdot}$ clearly extends the interpretation $F \circ \interpretation_M$. Hence these two sets are in bijective correspondence, and by freeness of $\CatTape$ we know that there is only one $F^\sharp$ extending the interpretation $F \circ \interpretation_M$. Thus we conclude.
\end{proof}

We conclude by proving Theorem~\ref{thm:contextual}. For this, it is convenient to first prove the following two lemmas.

\begin{proof}[Proof of Theorem \ref{thm:contextual}]
	In the main text, we claimed that $\WprecongBA$ is closed under $\per$, i.e.\ that 
	if $\t_1 \WprecongBA \t_2$ and  $\s_1 \WprecongBA \s_2$, then  $\t_1\per \s_1 \WprecongBA \t_2 \per \s_2$. This is immediate using the definition of $\per$ and Lemma~\ref{lemmaConWi}, which we prove below.
\end{proof}
	
	\begin{lemma}\label{lemmaCont}
		If $\t_1 \WprecongBA \t_2 $, then $\RW{U}{\t_1}\WprecongBA \RW{U}{\t_2}$ for all monomials $U$.
	\end{lemma}
	\begin{proof}
		We proceed by induction on the rule generating $\WprecongBA$. 

		If $\t_1  \WprecongBA \t_2$ by rule $(\wiskbasicR)$, then $\t_1   \WprecongBA \t_2$, i.e.\ there exists $(\t_1',\t_2')\in \basicR$ and $V\in S^*$ such that $\t_i=\RW{V}{\t_i'}$. Thus $\RW{U}{\t_i}= \RW{U}{\RW{V}{\t_i'}} \stackrel{\eqref{eq:tape:RR}}{=} \RW{VU}{\t_i'}$.
		Since $(\t_1',\t_2')\in \basicR$, by definition of $ \wiskbasicR$, it holds that $\t_1 \wiskbasicR \t_2$ and thus, by rule $(\wiskbasicR)$, it holds that $\RW{U}{\t_1}   \WprecongBA \RW{U}{\t_2}$
		
		If $\t_1   \WprecongBA \t_2$ by rule ($t$), then there exists $\t$ such that $\t_1   \WprecongBA \t$ and $\t   \WprecongBA \t_2$. By induction hypothesis, it holds that $\RW{U}{\t_1}   \WprecongBA \RW{U}{\t}$ and $\RW{U}{\t}   \WprecongBA \RW{U}{\t_2}$. Thus by rule ($t$), 
		$\RW{U}{\t_1}   \WprecongBA \RW{U}{\t_2}$.
		
		The cases for $;$ and $\piu$ exploit induction in the same way and the laws \eqref{eq:whisk:funct} and \eqref{eq:whisk:funct piu} expressing the fact that $\RW{U}{-}$ preserves both $;$ and $\piu$.
		The case for ($r$) is trivial.
	\end{proof} 
	
	\begin{lemma}\label{lemmaConWi}
		If $\t_1 \WprecongBA \t_2 $, then $\RW{S}{\t_1}\WprecongBA \RW{S}{\t_2}$ and $\LW{S}{\t_1}\WprecongBA \LW{S}{\t_2}$ for all polynomials $S$.
	\end{lemma}
	\begin{proof}
		We first show the case for $\RW{S}{-}$. The proof proceed by induction on $S$.
		
		\textbf{Case $S = \zero$:} $\RW{\zero}{\t_1}=\id{\zero} \WprecongBA \id{\zero} = \RW{\zero}{\t_2}$.
		
		\textbf{Case $S = W \piu S'$:} 
		\begin{align*}
			\RW{W \piu S'}{\t_1} &= \dl{P}{W}{S'} ; (\RW{W}{\t_1} \piu \RW{S'}{\t_1}) ; \Idl{Q}{W}{S'}  \tag{Def. $R$} \\
			&\WprecongBA  \dl{P}{W}{S'} ; (\RW{W}{\t_2} \piu \RW{S'}{\t_1}) ; \Idl{Q}{W}{S'}  \tag{Lemma \ref{lemmaCont}, Rules ($r$), ($;$), ($\piu$)} \\
			&\WprecongBA \dl{P}{W}{S'} ; (\RW{W}{\t_2} \piu \RW{S'}{\t_2}) ; \Idl{Q}{W}{S'} \tag{Ind. Hyp., Rules ($r$), ($;$), ($\piu$)}\\
			& = \RW{W \piu S'}{\t_2} \tag{Def. $R$}
		\end{align*}
		The proof for $\LW{S}{-}$ is trivial using the result for $\RW{S}{-}$ and \eqref{eq:LRnatsym}.
	\end{proof}


\section{Appendix to Section~\ref{sc:matrix}}\label{appendix:matrix}
In this Appendix we dive into the details of the calculus of matrices for categories with finite biproducts. In Section~\ref{sc:appmatrix1} we present the standard calculus in the general setting of an arbitrary fb category. In Section~\ref{sc:appmatrix2} we prove Theorem~\ref{thm:tensor of matrices is Kronecker}, showing how $F_2(\Cat C)$ and $\Mat{\fCMon{\Cat C}}$ are isomorphic as fb rig categories. In Section~\ref{sec:appmatrix3} we define in detail the Kronecker product on $\Mat{\fCMon{\CatString}}$ and we show that it coincides with the monoidal product inherited from $F_2(\CatString)=\CatTape$. Finally, Section~\ref{sec:appmatrix4} is dedicated to the proof of Theorem~\ref{thm:tapes as matrices poset version}.
\subsection{Matrix calculus for categories with finite biproducts}\label{sc:appmatrix1}
It is well known that morphisms in a fb category have a matrix representation such that composition corresponds to the usual matrix multiplication (see, for example, \cite{mac_lane_categories_1978}). Here we recall how the correspondence is defined and a few useful properties. For the rest of this section we denote the composite of two morphisms $f_1 \colon X \to Y$ and $f_2 \colon Y \to Z$ as $f_2 \circ f_1 \colon X \to Z$.

\begin{definition}\label{def:associated matrix in biproduct categories}
	Let $f \colon \PiuL[k=1][n]{A_k} \to \PiuL[k=1][m]{B_k}$ in $\Cat C$. The \emph{matrix associated to} $f$ is the $m \times n$ matrix $\matr f$ whose $(j,i)$ entry (row $j$, column $i$), for $j=1,\dots,m$ and $i=1,\dots,n$, is
	\[
	f_{ji} \defeq \Bigl( 
	\begin{tikzcd}
	A_i \ar[r,"\inj i"] & \PiuL[k=1][n]{A_k} \ar[r,"f"] & \PiuL[k=1][m]{B_k} \ar[r,"\proj j"] & B_j
	\end{tikzcd}
	\Bigr).
	\]
\end{definition}

For $f \colon \PiuL[k=1][n]{A_k} \to \PiuL[k=1][m]{B_k}$, we have

\[
\matr f =
\begin{pNiceMatrix}[first-row,first-col]
\rotatebox{90}{$\Lsh$} & A_1 & \dots & A_n \\
B_1&f_{11} &  \dots & f_{1n} \\
\vdots   & \vdots & \ddots & \vdots \\
B_m& f_{m1}  &   \dots      & f_{mn}
\end{pNiceMatrix}
\]
Notice how all morphisms in the $i$-th column have domain $A_i$, hence it makes sense to consider their pairing $\pairing{f_{1i},\dots,f_{mi}} \colon A_i \to \Piu[k]{B_k}$. Ranging over $i=1,\dots,n$, we can now consider their copairing $\copairing{\pairing{f_{11},\dots,f_{m1}},\dots,\pairing{f_{1n},\dots,f_{mn}}}$, which is now a morphism of type $\Piu[k]{A_k} \to \Piu[k]{B_k}$, just like $f$. Also, all morphisms in the $j$-th row have codomain $B_j$, hence we can take their copairing $\copairing{f_{j1},\dots,f_{jn}} \colon \Piu[k]{A_k} \to B_j$ and then, ranging over $j=1,\dots,m$, their collective pairing $\pairing{\copairing{f_{11},\dots,f_{1n}},\dots,\copairing{f_{m1},\dots,f_{mn}}}$, which is also of type $\Piu[k]{A_k} \to \Piu[k]{B_k}$. It turns out these morphisms are all equal to each other.
\begin{lemma}\label{lemma:f=copairing of pairings of columns=pairing of copairings of rows}
	Let $f \colon \PiuL[k=1][n]{A_k} \to \PiuL[k=1][m]{B_k}$. Then
	\begin{align*}
	f &= \copairing{\pairing{f_{11},\dots,f_{m1}},\dots,\pairing{f_{1n},\dots,f_{mn}}} \\
	&= \pairing{\copairing{f_{11},\dots,f_{1n}},\dots,\copairing{f_{m1},\dots,f_{mn}}}
	\end{align*}
\end{lemma}
\begin{proof}
	For the first equality we need to prove that $\inj i ; f = \pairing{f_{1i},\dots,f_{mi}}$ for all $i \in \{1,\dots,n\}$. This holds for any $i$ because $(\inj i ; f) ; \proj j = f_{ji}$ for all $j \in \{1,\dots,m\}$.
	
	Similarly, for the second equality we need to prove that for all $j \in \{1,\dots,m\}$ we have $f ; \proj j = \copairing{f_{j1},\dots,j_{jn}}$, which is true because for all $i \in \{1,\dots,n\}$ it holds that $\inj i ; (f ; \proj j) = f_{ji}$.
\end{proof}

\begin{corollary}\label{cor:M(f)=M(g) iff f=g}
	Let $f$ and $g$ have same domain and codomain. Then $\matr f = \matr g$ if and only if $f = g$.
\end{corollary}

We denote with $\diagg X m \colon X \to \PiuL[j=1][m] X$ the generalised diagonal of $X$, inductively defined in the obvious way:
\begin{equation}\label{eq:generalised diagonal}
\diagg X 0 \defeq \bang X \qquad \diagg X {m+1} \defeq \Bigl(
\begin{tikzcd}[column sep=4em]
X \ar[r,"\diag X"] & X \piu X \ar[r,"\id X \piu \diagg X {m}"] & \PiuL[j=1][m+1] X
\end{tikzcd}
\Bigr)	
\end{equation}
Similarly, we denote with $\codiagg X n \colon \PiuL[i=1][n]{X} \to X$ the generalised codiagonal of $X$, inductively defined as:
\begin{equation}\label{eq:generalised codiagonal}
\codiagg X 0 \defeq \cobang X \qquad \codiagg X {n+1} \defeq \Bigl(
\begin{tikzcd}[column sep=4em]
\PiuL[i=1][n+1]{X} \ar[r,"\id X \piu \codiagg X {n}"] & X \piu X \ar[r,"\codiag X"] & X
\end{tikzcd}
\Bigr).
\end{equation}

\begin{proposition}\label{prop:f as composite of diagonals-fji-codiagonals}
	Let $f \colon \PiuL[k=1][n]{A_k} \to \PiuL[k=1][m]{B_k}$. Then
	\begin{align*}
	f &= \Bigl(
	\begin{tikzcd}[ampersand replacement=\&,column sep=3em]
	\PiuL[i=1][n]{A_i} \ar[r,"{\PiuL[i=1][n]{\diagg{A_i} m}}"] \& \PiuL[i=1][n]{\PiuL[j=1][m]{A_i}} \ar[r,"{\PiuL[i=1][n]{\PiuL[j=1][m]{f_{ji}}}}"] \& \PiuL[i=1][n]{\PiuL[j=1][m]{B_j}} \ar[r,"\codiagg{\Piu[j=1][m]{B_j}}{n}"] \& \PiuL[j=1][m]{B_j}
	\end{tikzcd}
	\Bigr) \\
	&= \Bigl(
	\begin{tikzcd}[ampersand replacement=\&,column sep=3em]
	\PiuL[i=1][n]{A_i} \ar[r,"\diagg{\Piu[i=1][n]{A_i}} m"] \& \PPiuL j m i n {A_i} \ar[r,"{\PPiuL j m i n {f_{ji}}}"] \& \PPiuL j m i n {B_j} \ar[r,"{\PiuL[j=1][m]{\codiagg{B_j} n}}"] \& \PiuL[j=1][m]{B_j}
	\end{tikzcd}
	\Bigr)
	\end{align*}
\end{proposition}
\begin{proof}
	In general, if $u_j \colon X \to Y_j$, with $j=1,\dots,m$, we have
	\[
	\pairing{u_1,\dots,u_m} = \Bigl(
	\begin{tikzcd}[column sep=4em] 
	X \ar[r,"\diagg X m"] & \PiuL[j=1][m]{X} \ar[r,"{\Piu[j]{u_j}}"] & \PiuL[j=1][m]{U_j} 
	\end{tikzcd}
	\Bigr)
	\]
	(one can prove it by showing that the above satisfies the same universal property of $\pairing{u_1,\dots,u_m}$). Hence, for any $i \in \{1,\dots,n\}$,
	\[
	\pairing{f_{1i},\dots,f_{mi}} = \Bigl(
	\begin{tikzcd}
	A_i \ar[r,"\diagg {A_i} m"] & \PiuL[j=1][m]{A_j} \ar[r,"{\PiuL[j=1][m]{f_{ji}}}"] & \PiuL[j=1][m]{B_j}
	\end{tikzcd}
	\Bigr)
	\]
	Analogously, if $v_i \colon W_i \to Z$, with $i=1,\dots,n$,
	\[
	\copairing{v_1,\dots,v_n} = \Bigl(
	\begin{tikzcd}
	\PiuL[i=1][n]{W_i} \ar[r,"{\PiuL[i=1][n]{v_i}}"] & \PiuL[i=1][n]{Z} \ar[r,"\codiagg Z n"] & Z
	\end{tikzcd}
	\Bigr)
	\]
	Using these facts and Lemma~\ref{lemma:f=copairing of pairings of columns=pairing of copairings of rows}, we get the two equations of the statement.
\end{proof}

\begin{example}
	Let $A = \PiuL[k=1][n]{A_k}$ and $B=\PiuL[k=1][m]{B_k}$. Then
	\[
	\matr{\id A}
	= \begin{pmatrix}
	\delta_{A_i,A_j}
	\end{pmatrix}
	= 
	\begin{pNiceMatrix}[first-row,first-col]
	& A_1      			& A_2 			          & A_{n-1} 			   & A_n \\
	A_1 	& \id{A_1} 			& \zero_{A_2,A_1} & \Cdots   & \zero_{A_n,A_1}	\\
	A_2 	& \zero_{A_1,A_2}	& 	\Ddots	  &  \Ddots  & \Vdots	\\
	A_{n-1}  & \vdots 			& \Ddots 		  &  & \zero_{A_n,A_{n-1}} \\
	A_n     & \zero_{A_1,A_n}	& \Cdots & \zero_{A_{n-1},A_n}   & \id{A_n} 
	\CodeAfter
	\line{2-0}{3-0}
	\line{0-2}{0-3}
	\end{pNiceMatrix}
	\]
	while
	\[
	\matr{\zero_{A,B}} = \begin{pmatrix}
	\zero_{A_i,B_j}
	\end{pmatrix}
	= 
	\begin{pNiceMatrix}[first-row,first-col] 
	& A_1 & \dots & A_n\\
	B_1&\zero_{A_1,B_1} &  \Cdots & \zero_{A_n,B_1} \\
	\vdots   & \vdots & \ddots & \vdots\\
	B_m& \zero_{A_1,B_m}  &   \Cdots      & \zero_{A_n,B_m} \\
	\end{pNiceMatrix}
	\]
	For this reason we will sometimes simply write $\zero_{A,B}$ for the matrix $\matr{\zero_{A,B}}$ and $\id{A}$ for the matrix $\matr{\id A}$. The context they appear in will make clear whether we are referring to a matrix or to a morphism of $\Cat C$.
\end{example}

The matrix representation of morphisms in fb categories enjoys several nice properties. 

\begin{proposition}\label{prop:proprietà matrici}
For $f$ and $g$ of the appropriate type, we have:
	\begin{enumerate}
	\item $\matr{g \circ f} = \matr g  \matr f$,
	\item $\matr{f \piu g} = \begin{pmatrix}
	\matr f & \zero \\ 
	\zero & \matr g
	\end{pmatrix} $ \label{eq:matrix of sum is the block matrix of addends on diagonal}\\
	\item $\matr{\pairing{f,g}} = \begin{pmatrix}
	\matr f \\ \matr g
	\end{pmatrix}$
	\item $\matr{\copairing{f,g}} = \begin{pmatrix}
	\matr f & \matr g
	\end{pmatrix}$
	\item $\matr{\inj i \colon A_i \to \PiuL[k=1][n]{A_k}} = \begin{pmatrix}
	\delta_{i,1} \\
	\vdots \\
	\delta_{i,n}
	\end{pmatrix}$, $\matr{\proj j \colon \PiuL[k=1][n]{A_k} \to A_j} = \begin{pmatrix}
	\delta_{1,j} & \Cdots & \delta_{m,j}
	\end{pmatrix}$
	\item For $A = \PiuL[k=1][n]{A_k}$ and $B=\PiuL[k=1][m]{B_k}$,
	\[
	\matr{\symm{A}{B}} = \begin{pmatrix}
	\matr{\zero_{B,A}} & \matr{\id B} \\
	\matr{\id A} & \matr{\zero_{A,B}}
	\end{pmatrix}
	\]
	matrix of size $(m+n) \times (n+m)$.
\end{enumerate}
\end{proposition}

\subsection{Additional details of the Proof of Theorem~\ref{thm:F_2(C) isomorphic to biproduct completion}}\label{sc:appmatrix2}

We present here the two functors 
	\begin{equation}\label{eq:iso between CatTape and biproduct completion}
		\begin{tikzcd}[column sep={8em,between origins}]
			F_2(\Cat C) \ar[r,bend left,"\FF"name=F]   & \Mat{\fCMon{\Cat C}} \ar[l,bend left,"\GG"name=G]
			\arrow[from=F,to=G,draw=none,"\cong"description]
			\end{tikzcd}
		\end{equation}
	 of Theorem~\ref{thm:F_2(C) isomorphic to biproduct completion} explicitly.

$\FF$ was defined in the proof of Theorem~\ref{thm:F_2(C) isomorphic to biproduct completion}, so let us define $\GG$. 
On objects it simply is the identity. On morphisms: let 
\[
M = (M_{ji})_{\substack{j=1\dots m \\ i=1 \dots n}} \colon \PiuL[i=1][n]{A_i} \to \PiuL[j=1][m]{B_j} \text{ in } \Mat{\fCMon{\Cat{C}}}.
\]
If $m=0$, then define $\GG(M)=\bang{\Piu[j]{B_j}}$; if $n=0$ define instead $\GG(M) = \cobang{\Piu[i]{A_i}}$. (This is consistent with the case $n=0=m$, because $\bang \zero = \cobang \zero = \id\zero$ by Axioms~\eqref{eq:bang I = id I} and~\eqref{eq:cobang I = id I}.) 

Suppose now that $n \ne 0 \ne m$. 
We have that $M_{ji}$ is a multiset of morphisms of type $A_i \to B_j$ in $\Cat{C}$. $M_{ji}$ determines a morphism in $F_2(\Cat{C})$:

\[
\Gg(M_{ji}) \defeq \sum_{a \in M_{ji}} \tapeFunct{a} \colon A_i \to B_j 
\]
Notice that for $M_{ji}=\emptyset$ we get $\Gg(\emptyset)=\zero_{A_i,B_j}$. Now, $\bigl(\Gg(M_{ji})\bigr)_{j=1\dots m, i=1 \dots n}$ is a matrix of morphisms of $F_2(\Cat{C})$, to which corresponds a unique arrow of the same type of $M$:
\[
\GG(M) \defeq 
\Bigl(
\begin{tikzcd}[ampersand replacement=\&,column sep=4em]
	\PiuL[i=1][n]{A_i} \ar[r,"{\PiuL[i=1][n]{\diagg{A_i} m}}"] \& \PiuL[i=1][n]{\PiuL[j=1][m]{A_i}} \ar[r,"{\PiuL[i=1][n]{\PiuL[j=1][m]{\Gg(M_{ji})}}}"] \& \PiuL[i=1][n]{\PiuL[j=1][m]{B_j}} \ar[r,"\codiagg{\Piu[j=1][m]{B_j}}{n}"] \& \PiuL[j=1][m]{B_j}
\end{tikzcd}
\Bigr)
\]
where $\diagg X m \colon X \to \Piu[][m]{X}$ and $\codiagg Y n \colon Y \to \Piu[][n]{Y}$ are defined as:
\begin{equation}
	\begin{array}{rclcrcl}
		\diagg X 0 &\defeq& \bang X  & \qquad \qquad&\codiagg Y 0 &\defeq& \cobang{Y}  \\
		\diagg X {m+1} &\defeq& \diag X ; \diagg X m & \qquad \qquad &\codiagg Y {n+1} &\defeq& \codiagg Y n ; \codiag Y
	\end{array}	
\end{equation}

Now we start showing that $\FF$ and $\GG$ are mutually inverse. We first observe a few properties of $\FF$: if $U = \PiuL[i=1][n]{A_i}$ then
\[
\FF(\id U) =
\begin{pNiceMatrix}
\multiset{\id{A_1}}  	& \emptyset & \Cdots & \emptyset  \\
\emptyset  &   & \Ddots & \Vdots \\	
\Vdots & \Ddots &   & \emptyset  \\
\emptyset  & \Cdots & \emptyset  & \multiset{\id{A_{n}}} 
\CodeAfter
\line{1-1}{4-4}
\end{pNiceMatrix}
\]

Since $\FF$ sends symmetries, (co)diagonals and (co)bangs into symmetries, (co)diagonals and (co)bangs of $\Mat{\fCMon{\Cat C}}$, if $U = \PiuL[i=1][n]{A_i}$ and $V=\PiuL[j=1][m]{B_j}$ then:

\[
\FF(\symm{U}{V}) =
\begin{pNiceMatrix}
\emptyset_{m \times n} & \FF(\id V)  \\
\FF(\id U) & \emptyset_{n \times m}  \\
\end{pNiceMatrix}
\]

\[
\FF(\diag U) =
\begin{pNiceMatrix}
\FF(\id U) & \FF(\id U)
\end{pNiceMatrix}
\qquad
\FF(\codiag U) = 
\begin{pNiceMatrix}
\FF(\id U) \\
\FF(\id U)
\end{pNiceMatrix}
\]
\[
\FF(\bang U) = \text{empty matrix of size $0 \times n$} \qquad \FF(\cobang U) = \text{empty matrix of size $n \times 0$}
\]

\begin{remark}\label{rem:FF(t1+t2)=FF(t1)+FF(t2)}
	Since $\Cat{C}$ has all finite biproducts, $\Cat{C}[U,V]$ is a commutative monoid where addition is given using diagonals and codiagonals as in~\eqref{eq:+zero}.
	The action of $\FF$ on $f_1 + f_2 \colon U \to V$, with $U = \PiuL[i=1][n]{A_i}$ and $V=\PiuL[j=1][m]{B_j}$, is the usual entry-by-entry addition of matrices:
	\begin{align*}
	\FF(f_1 + f_2) &= \FF(\codiag V) \circ \FF(f_1 \piu f_2) \circ \FF(\diag U) \\
	&=
	\begin{pNiceMatrix}
	\FF(\id V) & \FF(\id V)
	\end{pNiceMatrix}
	\begin{pNiceMatrix}
	\FF(f_1) & \emptyset_{m \times n} \\
	\emptyset_{m \times n} & \FF(f_2)
	\end{pNiceMatrix}
	\begin{pNiceMatrix}
	\FF(\id U) \\
	\FF(\id U)
	\end{pNiceMatrix} \\
	&=
	\begin{pNiceMatrix}
	\FF(f_1) & \FF(f_2)
	\end{pNiceMatrix}
	\begin{pNiceMatrix}
	\FF(\id U) \\
	\FF(\id U)
	\end{pNiceMatrix} \\
	&= \FF(f_1) + \FF(f_2).
	\end{align*}
\end{remark}

\begin{remark}\label{rem: FF of pairing and copairings}
	If $f_1 \colon W \to U$ and $f_2 \colon W \to V$ then
	\begin{align*}
	\FF(\pairing{f_1,f_2}) &= \FF((f_1 \piu f_2)  \circ \pairing{\id W, \id W}) \\
	&= \FF(f_1 \piu f_2) \circ \FF(\diag W) \\
	&= \begin{pmatrix}
	\FF(f_1) & \emptyset \\
	\emptyset & \FF(f_2)
	\end{pmatrix}
	\begin{pNiceMatrix}
	\FF(\id W) \\
	\FF(\id W)
	\end{pNiceMatrix} \\
	&= \begin{pNiceMatrix}
	\FF(f_1) \\
	\FF(f_2)
	\end{pNiceMatrix}
	\end{align*}
	Analogously, if $f_1 \colon U \to W$ and $f_2 \colon V \to W$ then $\FF(\copairing{f_1,f_2}) = \begin{pNiceMatrix}
	\FF(f_1) & \FF(f_2)
	\end{pNiceMatrix}$.
\end{remark}

\begin{proposition}
	$\GG \colon \Mat{\fCMon{\Cat C}} \to F_2(\Cat C)$ is a functor that preserves biproducts.
\end{proposition}
\begin{proof}
	Let
	\[
	\begin{tikzcd}
	\PiuL[i=1][n]{A_i} \ar[r,"M"] & \PiuL[k=1][m]{V_k} \ar[r,"N"] & \PiuL[j=1][l]{W_j}.
	\end{tikzcd}
	\]
	We aim to compute $\GG(N \circ M)$. We have
	\begin{align*}
	(N \circ M)_{ji} = (N \cdot M)_{ji} &= \sum_{k=1}^m N_{jk} \circ M_{ki} & \tag{Sum and composition in $\fCMon{\CatString}$} \\
	&= \bigcup_{k=1}^m {\multiset{ b \circ_{\Cat C} a \mid a \in M_{ki}, \, b \in N_{jk}} } & \tag{Union of multisets}
	\end{align*}
	therefore
	\[
	\Gg\bigl( (N\circ M)_{ji}  \bigr) = \sum_{k=1}^m \, \sum_{\substack{a \in M_{ki} \\ b \in N_{jk}}} \tape{b \circ a}
	\]
	Now, $\GG(N) \circ \GG(M)$ is a morphism in $F_2(\Cat C)$ whose associated matrix, by Proposition~\ref{prop:proprietà matrici},
	is
	\[
	\bigl( \Gg(N_{wv})  \bigr)_{\substack{w=1\dots l \\ v=1\dots m}} \cdot \bigl( \Gg(M_{vu})  \bigr)_{\substack{v=1 \dots m \\ u=1 \dots n}}
	\]
	whose entry $(j,i)$, for $i \in \{1,\dots, n\}$ and $j \in \{1,\dots, m\}$, is the following sum of compositions in $F_2(\Cat C)$:
	\begin{align*}
	\bigl( \matr{\GG(N) \circ \GG(M)}  \bigr)_{ji} &= \sum_{k=1}^m \Gg(N_{jk}) \circ \Gg(M_{ki}) \\
	&= \sum_{k=1}^m \Bigl( \sum_{b \in N_{jk}} \tape b \Bigr) \circ \Bigl( \sum_{a \in M_{ki}} \tape a \Bigr) \\
	&= \sum_{k=1}^m \sum_{a \in M_{ki}} \Bigl( \sum_{b \in N_{jk}} \tape b \Bigr) \circ \tape a \tag{$\CMon$-enrichment: $v \circ (f+g)= v \circ f + v \circ g$} \\ 
	&= \sum_{k=1}^m \sum_{a \in M_{ki}} \sum_{b \in N_{jk}} \tape b \circ \tape a \tag{$\CMon$-enrichment: $(f+g) \circ u = f \circ u + g \circ u$} \\ 
	&= \sum_{k=1}^m \sum_{a \in M_{ki}} \sum_{b \in N_{jk}} \tape{b \circ a} \tag{\refeq{ax:tape}} \\
	&= \Gg\bigl( (N\circ M)_{ji}  \bigr) \\
	&= \matr{\GG(N \circ M)}_{ji}.
	\end{align*}
	Hence $\GG(N) \circ \GG(M) = \GG(N \circ M)$ by Corollary~\ref{cor:M(f)=M(g) iff f=g}.
	
	Regarding preservation of biproducts, if $M \colon \PiuL[k=1][n]{A_k} \to \PiuL[k=1][n']{A_k'}$ and $N \colon \PiuL[k=1][m]{B_k} \to \PiuL[k=1][m']{B_k}$ are any two matrices, we have that 
	\[
	\GG(M \piu N) = \GG \begin{pNiceMatrix}
		M & \emptyset \\
		\emptyset & N
	\end{pNiceMatrix} \qquad \text{(block matrix)}
	\]

	thus $\GG(M \piu N)$ is the morphism in $\CatTape$ whose associated matrix is 
	\[
	\begin{pNiceMatrix}
	\Gg(M_{11}) & \Cdots & \Gg(M_{1n}) & \zero & \Cdots & \zero \\
	\vdots & \Ddots & \vdots & \vdots &  & \vdots \\
	\Gg(M_{n'1}) & \dots & \Gg(M_{n'n}) & \zero & \dots & \zero \\
	\zero & \Cdots & \zero & \Gg(N_{11}) & \Cdots & \Gg(N_{1m}) \\
	\vdots &  & \vdots & \vdots & \ddots & \vdots \\
	\zero & \Cdots & \zero & \Gg(N_{m'1}) & \Cdots & \Gg(N_{m' m})
	\CodeAfter
	\line{1-4}{3-6}
	\line{4-1}{6-3}
	\end{pNiceMatrix}
	\]
	which is exactly the block matrix $\begin{pNiceMatrix}
	\matr{\GG(M)} & \zero \\
	\zero & \matr{\GG(N)}
	\end{pNiceMatrix}$, which in turn is $\matr{\GG(M) \piu \GG(N)}$ by Proposition~\ref{prop:proprietà matrici}~\eqref{eq:matrix of sum is the block matrix of addends on diagonal}. Hence $\GG(M \piu N) = \GG(M) \piu \GG(N)$.
	
	In light of this, in order to check that $\GG$ preserves identities it suffices to make sure that $\GG(\id A) = \id A$ for $A$ object in $\Cat C$ (seen as a unary list). And indeed $\GG(\id A)$ is the morphism whose associated matrix is the $1 \times 1$ matrix 
	$
	\begin{pNiceMatrix}
	\multiset{\id A}
	\end{pNiceMatrix}
	$,
	hence $\GG(\id A) = \id A$.
\end{proof}

\begin{proposition}
	$\FF$ and $\GG$ are mutually inverse.
\end{proposition}
\begin{proof}
	It is not difficult to prove by induction on $f$ that $\GG\FF(f) = f$, using Proposition~\ref{prop:proprietà matrici}. We show here that $\FF\GG = \id{}$: 
	let $M \colon \PiuL[i=1][n]{A_i} \to \PiuL[j=1][m]{B_j}$. First notice that 
	\[
	\GG(M) = \copairing{\pairing{\Gg(M_{11}),\dots,\Gg(M_{n1})}, \dots, \pairing{\Gg(M_{1m}),\dots,\Gg(M_{mn}) }}
	\]
	by Lemma~\ref{lemma:f=copairing of pairings of columns=pairing of copairings of rows} and Proposition~\ref{prop:f as composite of diagonals-fji-codiagonals}.
	Then 
	\begin{align*}
	\FF\GG(M) &= \FF \Bigl( \copairing{\pairing{\sum_{f \in M_{11}} \tape f,\dots,\sum_{f \in M_{n1}} \tape f}, \dots, \pairing{\sum_{f \in M_{1m}} \tape f,\dots,\sum_{f \in M_{mn}} \tape f}} \Bigr) & \tag{Definition of $\GG$} \\
	&= 
	\begin{pNiceMatrix}
	\FF\bigl( \sum_{f \in M_{11}} \tape f \Bigr) & \dots & \FF\bigl( \sum_{f \in M_{1m}} \tape f \Bigr) \\
	\vdots & \ddots & \vdots \\
	\FF\bigl( \sum_{f \in M_{n1}} \tape f \Bigr) & \dots & \FF\bigl( \sum_{f \in M_{mn}} \tape f \Bigr)
	\end{pNiceMatrix} 
	& \tag{Remark~\ref{rem: FF of pairing and copairings}} \\
	&= M
	\end{align*}
	because for every $j \in \{1,\dots,m\}$ and $i \in \{1,\dots,n\}$
	\begin{align*}
	\FF\bigl( \sum_{f \in M_{ji}} \tape f \Bigr) &= \sum_{f \in M_{ji}} \FF(\tape f) & \tag{Remark~\ref{rem:FF(t1+t2)=FF(t1)+FF(t2)}} \\
	&= \sum_{f \in M_{ji}} 
	\begin{pNiceMatrix}
	\multiset{f}
	\end{pNiceMatrix} \\
	&= M_{ji}
	\end{align*}
	where above $\sum_{f \in M_{ji}} 
	\begin{pNiceMatrix}
	\multiset{f}
	\end{pNiceMatrix} $ is an entry-by-entry sum of $1 \times 1$ matrices, which yields the matrix whose only entry is $\sum_{f \in M_{ji}} \multiset f$: a sum of multisets equal to $M_{ji}$ itself. 
\end{proof}

\subsection{Kronecker product. Details of the proof of Theorem~\ref{thm:tensor of matrices is Kronecker}}\label{sec:appmatrix3}
We will work using $\Cat C = \CatString$ here, but everything that follows can be generalised to an arbitrary SMC. Recall the definition of $\kron$. On objects it is set as follows:
\[
\Bigl(\PiuL[i=1][n]{U_i} \Bigr) \kron \Bigl( \PiuL[j=1][m]{V_j} \Bigr) \defeq \PiuL[i=1][n]{\PiuL[j=1][m]{U_iV_j}}.
\]
If $M \colon \PiuL[i=1][n]{U_i} \to \PiuL[i'=1][n']{U_{i'}'}$ and $N \colon \PiuL[j=1][m]{V_j} \to \PiuL[j'=1][m']{V_{j'}'}$, define $M \kron N$ as the block matrix:
\[
M \kron N =
\begin{pNiceMatrix}[first-row,first-col]
& U_1V_1 & \Cdots & U_1V_m & \Cdots & U_nV_1 & \Cdots & U_nV_m \\
U_1'V_1' 	& \Block[borders={right,bottom}]{3-3}{M_{11} \per N} & & &  & \Block[borders={left,bottom}]{3-3}{M_{1n} \per N}\\
\Vdots & & & & \Hdotsfor{1}	& \\
U_1'V_{m'}' & &  & & & &   \\
\Vdots 	& & \Hdotsfor{1} & &  & & \Hdotsfor{1} & \\
U_{n'}'V_1' 	& \Block[borders={right,top}]{3-3}{M_{n'1} \per N} &  & &  & \Block[borders={left,top}]{3-3}{M_{n'n} \per N}\\
\Vdots & & & & \Hdotsfor{1} & \\
U_{n'}'V_{m'}'
\end{pNiceMatrix}
\]
where 
\[
M_{i'i} \per N =
\begin{pNiceMatrix}
M_{i'i} \per N_{11} & \dots & M_{i'i} \per N_{1m} \\
\vdots & \ddots & \vdots \\
M_{i'i} \per N_{m'1} & \dots & M_{i'i} \per N_{m'n'}
\end{pNiceMatrix}
\]
In other words, $M \kron N$ is a matrix of size $n' m' \times n  m$, whose rows are indexed by $(i',j') \in \{1,\dots,n'\} \times \{1,\dots,m'\}$ and whose columns by $(i,j) \in \{1,\dots,n\} \times \{1,\dots,m\}$; then we have
\[
(M \kron N)_{(i',j'),(i,j)} = M_{i'i} \per N_{j'j} = \multiset{a \per_{\CatString} b \mid a \in M_{i'i},\, b \in N_{j'j}}.
\]

We will show that $\kron$ coincides with the tensor product inherited from $\CatTape$. We begin by proving some preliminary results.

\begin{proposition}\label{prop:partial functoriality of kronecker}
	Let $M \colon P \to P'$ and $N \colon Q \to Q'$, with $P=\PiuL[i=1][n]{U_i}$, $P'= \PiuL[i'=1][n']{U_{i'}'}$, $Q=\PiuL[j=1][m]{V_j}$ and $Q'=\PiuL[j'=1][m']{V_{j'}'}$. Then
	\begin{align*}
	M \kron N &= (\id{P'} \kron N) \circ (M \kron \id{Q}) \\
	&= (M \kron \id{Q'}) \circ (\id P \kron N)
	\end{align*}
\end{proposition}
\begin{proof}
	We only show the first equality: the other is analogous. We have:
	\[
	\id{P'} \kron N =
	\begin{pNiceMatrix}
	\multiset{\id{U_1'}} \per N 	& \emptyset \per N & \Cdots & \emptyset \per N \\
	\emptyset \per N &  & \Ddots & \Vdots \\
	
	\Vdots & \Ddots & \Ddots[shorten=0.2em] & \emptyset \per N \\
	\emptyset \per N & \Cdots & \emptyset \per N & \multiset{\id{U_{n'}'}} \per N
	\end{pNiceMatrix}
	\]
	Notice that $\emptyset \per N$ is $\emptyset_{m' \times m}$, the $m' \times m$ matrix whose every entry is the empty multiset. Next,
	\[
	M \kron \id Q =
	\begin{pNiceMatrix}
	M_{11} \per \id Q & \dots & M_{1n} \per \id Q \\
	\vdots & \ddots & \vdots \\
	M_{n'1} \per \id Q & \dots & M_{n'n} \per \id Q
	\end{pNiceMatrix}
	\]
	(in the above $\id{P'}$ and $\id Q$ are identity morphisms in $\Mat{\fCMon{\CatString}}$, hence they are matrices). To save space, we shall drop the multiset brackets in $\multiset{\id{U_{i'}'}}$, since it is a singleton, and simply write $\id{U_{i'}'}$. Therefore $(\id{P'} \kron N) \circ (M \kron \id{Q})$ is equal to:
	\[
	\scalebox{0.9}{$
		\begin{pNiceMatrix}
		(\id{U_1'} \per N) \circ (M_{11} \per \id Q) & (\id{U_1'} \per N) \circ (M_{12} \per \id Q) & \dots & (\id{U_1'} \per N) \circ (M_{1n} \per \id Q) \\
		(\id{U_2'} \per N) \circ (M_{21} \per \id Q) & (\id{U_2'} \per N) \circ (M_{22} \per \id Q) & \dots & (\id{U_2'} \per N) \circ (M_{2n} \per \id Q) \\
		\vdots & \vdots & \ddots & \vdots \\
		(\id{U_1{n'}'} \per N) \circ (M_{n'1} \per \id Q) & (\id{U_{n'}'} \per N) \circ (M_{n'2} \per \id Q) & \dots & (\id{U_{n'}'} \per N) \circ (M_{n'n} \per \id Q) \\
		\end{pNiceMatrix}$}
	\]
	In other words, $(\id{P'} \kron N) \circ (M \kron \id{Q})$ is a matrix consisting of $n'n$ blocks (the entries in the matrix above), each of size $m' \times m$, where the block at row $i'$ and column $i$ is
	\[
	B^{i'i} = \underbrace{(\id{U_{i'}'} \per N)}_{m' \times m} \circ \underbrace{(M_{i'i} \per \id Q)}_{m' \times m}.
	\]
	The entry $(j',j)$ of the above block is the following sum of composite of multisets:
	\begin{align*}
	B_{j'j}^{i'i} &=	\sum_{k=1}^m (\id{U_{i'}'} \per N)_{j'k} \circ (M_{i'i} \per \id Q)_{kj} \\
	&= \sum_{k=1}^m (\id{U_{i'}'} \per N_{j'k}) \circ (M_{i'i} \per (\id Q)_{kj})
	\end{align*}
	Now, if $k \ne j$ then $(\id Q)_{kj} = \emptyset$, hence $M_{i'i} \per \emptyset = \emptyset$ and $(\id{U_{i'}'} \per N_{j'k}) \circ \emptyset =\emptyset$. Therefore,
	\begin{align*}
	B_{j'j}^{i'i} &= (\id{U_{i'}'} \per N_{j'j}) \circ (M_{i'i} \per \id{V_j}) \\
	&= \multiset{\id{U_{i'}'} \per_{\CatString} b \mid b \in N_{j'j}} \circ 
	\multiset{a \per_{\CatString} \id{V_j} \mid a \in M_{i'i}} & \tag{Definition of $\per_{\fCMon{\CatString}}$} \\
	&= \multiset{(\id{U_{i'}'} \per_{\CatString} b) \circ (a \per_{\CatString} \id{V_j}) \mid a \in M_{i'i},\, b \in N_{j'j} } & \tag{Definition of $\circ$ in $\fCMon{\CatString}$} \\
	&= \multiset{ a \per_{\CatString} b \mid a \in M_{i'i},\, b \in N_{j'j} } & \tag{Functoriality of $\per_{\CatString}$} \\
	&= (M \kron N)_{(i',j'),(i,j)}. \tag*{\qedhere}
	\end{align*}
\end{proof}

$\Mat{\fCMon{\CatString}}$ inherits a tensor functor $\perM$ from $\CatTape$, via the isomorphism~\eqref{eq:iso between CatTape and biproduct completion}, defined as
	 \[
	 \text{$M \perM N$} \defeq {\FF(\GG(M) \perT \GG(N))}.
	 \]
	 This makes $\Mat{\fCMon{\CatString}}$ a strict fb rig category, where $\FF$ and $\GG$ strictly preserve the rig structure. Now we show that $\perM$ actually coincides with the Kronecker product.

\begin{proposition}\label{prop:Kronecker preserves identities}
	Let $P=\PiuL[i=1][n]{U_i}$ and $Q=\PiuL[j=1][m]{V_j}$. Then $\id P \kron \id Q = \id{P \kron Q}$.
\end{proposition}
\begin{proof}
	We have:
	\begin{align*}
	\id P \kron \id Q &=
	\begin{pNiceMatrix}
	(\id P)_{11} \per \id Q & \dots & (\id P)_{1n} \per \id Q \\
	\vdots & \ddots & \vdots \\
	(\id P)_{n1} \per \id Q & \dots & (\id P)_{nn} \per \id Q
	\end{pNiceMatrix}
	\\
	&=
	\begin{pNiceMatrix}
	\multiset{\id{U_1}} \per \id Q 	& \emptyset & \Cdots & \emptyset  \\
	\emptyset  &  & \Ddots & \Vdots \\	
	\Vdots & \Ddots &  & \emptyset  \\
	\emptyset  & \Cdots & \emptyset  & \multiset{\id{U_{n}}} \per \id Q
	\CodeAfter
	\line{1-1}{4-4}
	\end{pNiceMatrix}
	\\
	&=
	\begin{pNiceMatrix}
	\multiset{\id{U_1} \per \id{V_1}}  	& \emptyset & \Cdots & \emptyset  \\
	\emptyset  & \multiset{\id{U_1} \per \id{V_m}} & \Ddots & \Vdots \\	
	\Vdots & \Ddots & \multiset{\id{U_n} \per \id{V_1}}  & \emptyset  \\
	\emptyset  & \Cdots & \emptyset  & \multiset{\id{U_{n}} \per \id{V_m}} 
	\CodeAfter
	\line{1-1}{2-2}
	\line{2-2}{3-3}
	\line{3-3}{4-4}
	\end{pNiceMatrix}
	\\
	&= \id{\Piu[i]{\Piu[j]{U_iV_j}}} \\
	&= \id{P \kron Q}. \qedhere
	\end{align*}
\end{proof}

\begin{theorem}
	$\perM = \kron$ in $\Mat{\fCMon{\CatString}}$.
\end{theorem}
\begin{proof}
On objects: if $P=\PiuL[i=1][n]{U_i}$ and $Q = \PiuL[j=1][m]{V_j}$ we have $P \perM Q = \FF(P \perT Q) = \FF(\Piu[i]{\Piu[j]{U_iV_j}} = P \kron Q$.

\textbf{Claim}: $\FF( \t {}\perT$ $\id{})$ $=$ $\FF(\t) \kron \id{}$ and $\FF(\id{}$ $\perT$ $\t)$ $=$ $\id{} \kron \FF(\t)$ for every $\t \in \CatTape$.

The claim will allow us to conclude the proof, because we will have:
\begin{align*}
M \perM N &= \FF(\GG(M) \perT \GG(N)) & \tag{Definition of $\perM$} \\
&= \FF\bigl( (\id{} \perT \GG(N) )\circ (\GG(M) \perT \id{}) \bigr) & \tag{Functoriality of $\perT$} \\
&= \FF \bigl( (\id{} \perT \GG(N) )\bigr) \circ \FF \bigl(\GG(M) \perT \id{}\bigr) & \tag{Functoriality of $\FF$} \\
&= \bigl(\id{} \kron \FF\GG(N)\bigr) \circ \bigl(\FF\GG(M) \kron \id{} \bigr) & \tag{Claim} \\
&= (\id{} \kron N) \circ (M \kron \id{}) & \tag{$\FF\GG=\id{}$} \\
&= M \kron N & \tag{Proposition~\ref{prop:partial functoriality of kronecker}}
\end{align*}

We now proceed to prove the claim. First of all, if $\t=\id P$ say, then
\begin{align*}
\FF(\id P \perT \id R) &= \FF(\id{P \perT R}) & \tag{Functoriality of $\perT$} \\
&= \id{P \perM  R} & \tag{Functoriality of $\FF$} \\
&= \id{P \kron  R} & \tag{$\perM$ and $\kron$ coincide on objects} \\
&= \id P \kron \id R & \tag{Proposition~\ref{prop:Kronecker preserves identities}}.
\end{align*}
therefore the claim holds. 

Next, suppose $\t \ne \id{}$. Then $\t$ can be written as a composite of morphisms of the form $\id P \piu \t' \piu \id Q$ with $\t' = \tapeFunct{c},\,\symmp{U}{V},\,\diag U,\, \bang U,\, \codiag U, \, \cobang U$ with $U,V$ tensors of basic sorts. Now, since $\perT$ is right distributive, 
\[
(\id P \piu \t' \piu \id Q) \perT \id{R} = (\id P \perT \id R) \piu (\t' \perT \id R) \piu (\id Q \perT \id R).
\]
Since $\FF$ preserves biproducts, it suffices to prove that
\[
\FF(\t \per \id R) = \FF(\t) \kron \id R
\]
for $\t = \tapeFunct c, \,\symmp{U}{V},\,\diag U,\, \bang U,\, \codiag U\, \cobang U$ and $R = \PiuL[k=1][l]{Z_k}$ in order to prove the first half of the claim.

\textbf{Case 1}: $\t = \tapeFunct c \colon U \to U'$. We have that
\[
\t \perT \id R = \PiuL[k=1][l]{\tape{c \per_{\CatString} \id{Z_k}}} \colon \PiuL[k=1][l]{UZ_k} \to \PiuL[k=1][l]{U'Z_k}.
\]
Therefore
\begingroup
\allowdisplaybreaks
\begin{align*}
\FF(\t \perT \id R) &=	
\begin{pNiceMatrix}
\FF(\tape{c \per_{\CatString} \id{Z_1}})  	& \emptyset & \Cdots & \emptyset  \\
\emptyset  &  & \Ddots & \Vdots \\	
\Vdots & \Ddots &  & \emptyset  \\
\emptyset  & \Cdots & \emptyset  & \FF(\tape{c \per_{\CatString} \id{Z_l}})
\CodeAfter
\line{1-1}{4-4}
\end{pNiceMatrix} \\
&=
\begin{pNiceMatrix}
\multiset{c \per_{\CatString} \id{Z_1}}  	& \emptyset & \Cdots & \emptyset  \\
\emptyset  &  & \Ddots & \Vdots \\	
\Vdots & \Ddots &  & \emptyset  \\
\emptyset  & \Cdots & \emptyset  & \multiset{c \per_{\CatString} \id{Z_l}}
\CodeAfter
\line{1-1}{4-4}
\end{pNiceMatrix} \\
&= 
\begin{pNiceMatrix}
\multiset c
\end{pNiceMatrix}
\kron
\begin{pNiceMatrix}
\multiset{\id{Z_1}}  	& \emptyset & \Cdots & \emptyset  \\
\emptyset  &  & \Ddots & \Vdots \\	
\Vdots & \Ddots &  & \emptyset  \\
\emptyset  & \Cdots & \emptyset  & \multiset{\id{Z_l}}
\CodeAfter
\line{1-1}{4-4}
\end{pNiceMatrix} \\
&= \FF(\tapeFunct c) \kron \id R.
\end{align*}
\endgroup

\textbf{Case 2}: $\t = \symmp{U}{V}$. We have:
\begingroup
\allowdisplaybreaks
\begin{align*}
\FF(\t \perT \id R) &= \FF(\symmp{\Piu[k]{UZ_k}}{\Piu[k]{VZ_k}}) \\
&= \begin{pNiceMatrix}
\emptyset_{l \times l} & \id{\Piu[k]{VZ_k}} \\
\id{\Piu[k]{UZ_k}} & \emptyset_{l \times l}
\end{pNiceMatrix} \\
&= 	\begin{pNiceMatrix}
\FF(\tape{c \per_{\CatString} \id{Z_1}})  	& \emptyset & \Cdots & \emptyset  \\
\emptyset  &  & \Ddots & \Vdots \\	
\Vdots & \Ddots &  & \emptyset  \\
\emptyset  & \Cdots & \emptyset  & \FF(\tape{c \per_{\CatString} \id{Z_l}})
\CodeAfter
\line{1-1}{4-4}
\end{pNiceMatrix} \\
&=
\begin{pNiceMatrix}
\Block[borders={right,bottom}]{4-4}<\Large>{\emptyset} & & & & \multiset{\id{VZ_1}}  	& \emptyset & \Cdots & \emptyset  \\
& & & & \emptyset  &  & \Ddots & \Vdots \\	
& & & & \Vdots & \Ddots &  & \emptyset  \\
& & & & \emptyset  & \Cdots & \emptyset  & \multiset{\id{VZ_l}} \\
\multiset{\id{UZ_1}}  	& \emptyset & \Cdots & \emptyset & \Block[borders={left,top}]{4-4}<\Large>{\emptyset}  \\
\emptyset  &  & \Ddots & \Vdots \\	
\Vdots & \Ddots &  & \emptyset  \\
\emptyset  & \Cdots & \emptyset  & \multiset{\id{UZ_l}}
\CodeAfter
\line{1-5}{4-8}
\line{5-1}{8-4}
\end{pNiceMatrix} \\
&= \begin{pNiceMatrix}
\emptyset & \multiset{\id V} \\
\multiset{\id U} & \emptyset
\end{pNiceMatrix}
\kron
\begin{pNiceMatrix}
\multiset{\id{Z_1}}  	& \emptyset & \Cdots & \emptyset  \\
\emptyset  &  & \Ddots & \Vdots \\	
\Vdots & \Ddots &  & \emptyset  \\
\emptyset  & \Cdots & \emptyset  & \multiset{\id{Z_l}}
\CodeAfter
\line{1-1}{4-4} 
\end{pNiceMatrix} \\
&= \FF(\symmp{U}{V}) \kron \id R.
\end{align*}
\endgroup

\textbf{Case 3}: $\t = \diag U$. We have:
\begingroup
\allowdisplaybreaks
\begin{align*}
\FF(\diag U \perT \id R) &= \FF(\diag{\Piu[k]{UZ_k}}) \\
&= \begin{pNiceMatrix}
\multiset{\id{UZ_1}}  	& \emptyset & \Cdots & \emptyset  \\
\emptyset  &  & \Ddots & \Vdots \\	
\Vdots & \Ddots &  & \emptyset  \\
\emptyset  & \Cdots & \emptyset  & \multiset{\id{UZ_l}} \\
\multiset{\id{UZ_1}}  	& \emptyset & \Cdots & \emptyset  \\
\emptyset  &  & \Ddots & \Vdots \\	
\Vdots & \Ddots &  & \emptyset  \\
\emptyset  & \Cdots & \emptyset  & \multiset{\id{UZ_l}}
\CodeAfter
\line{1-1}{4-4}
\line{5-1}{8-4}
\end{pNiceMatrix} \\
&= \begin{pNiceMatrix}
\multiset{\id U} \\
\multiset{\id U}
\end{pNiceMatrix}
\kron
\begin{pNiceMatrix}
\multiset{\id{UZ_1}}  	& \emptyset & \Cdots & \emptyset  \\
\emptyset  &  & \Ddots & \Vdots \\	
\Vdots & \Ddots &  & \emptyset  \\
\emptyset  & \Cdots & \emptyset  & \multiset{\id{UZ_l}}
\CodeAfter
\line{1-1}{4-4}
\end{pNiceMatrix} \\
&= \FF(\diag U) \kron \id R.
\end{align*}
\endgroup

\textbf{Case 4}: $\t=\bang U$. We have:
\begingroup
\allowdisplaybreaks
\begin{align*}
\FF(\bang U \perT \id R) &= \FF(\bang{\Piu[k]{UZ_k}}) = \emptyset_{0 \times l} = \emptyset_{0 \times 1} \kron \begin{pNiceMatrix}
\multiset{\id{Z_1}}  	& \emptyset & \Cdots & \emptyset  \\
\emptyset  &  & \Ddots & \Vdots \\	
\Vdots & \Ddots &  & \emptyset  \\
\emptyset  & \Cdots & \emptyset  & \multiset{\id{Z_l}}
\CodeAfter
\line{1-1}{4-4}
\end{pNiceMatrix} \\
&= \FF(\bang U) \kron \id R.
\end{align*}
\endgroup

The other cases concerning $\codiag U$ and $\cobang U$ are analogous. Now we prove that 
\[
\FF(\id R \perT (\id P \piu \t \piu \id Q)) = \id R \kron \FF(\id P \piu \t \piu Q)
\]
with $P= \PiuL[i=1][n]{X_i}$, $Q=\PiuL[j=1][m]{Y_j}$, $R = \PiuL[k=1][l]{Z_k}$. To lighten notation, for the rest of the proof we will often denote identity morphisms with their corresponding objects, that is we will simply write $X$ instead of $\id X$.

\textbf{Case 1}: $\t = \tape c \colon U \to V$.
\begingroup
\allowdisplaybreaks
\begin{align*}
\FF(R \perT (P \piu \t \piu  Q)) &= \FF \Bigl( \PiuL[k=1][l]{\bigl( \PiuL[i=1][n]{Z_kX_i} \piu \tape{Z_k \per_{\CatString} c} \piu \PiuL[j=1][m]{Z_kY_j} \bigr)} \Bigr) \\
&= \PiuL[k=1][l]{\Biggl[ \PiuL[i=1][n]{\begin{pNiceMatrix}
		\multiset{Z_kX_i}
		\end{pNiceMatrix}} 
	\piu \begin{pNiceMatrix}
	\multiset{Z_k \per_{\CatString} c}
	\end{pNiceMatrix}
	\piu \PiuL[j=1][m]{\begin{pNiceMatrix}
		\multiset{Z_kY_j}
		\end{pNiceMatrix}} \Biggr]} \\
&= \PiuL[k=1][l]{
	\begin{pNiceMatrix}
	\multiset{Z_kX_1} & \emptyset & \Cdots & &  \emptyset \\
	\emptyset & \multiset{Z_kX_n} & \Ddots & & \Vdots \\
	\Vdots & \Ddots & \multiset{Z_k \per_{\CatString} c} \\
	& & & \multiset{Z_k Y_1} & \emptyset \\
	\emptyset & \Cdots & & \emptyset & \multiset{Z_kY_m}
	\CodeAfter
	\line{1-1}{2-2}
	\line{4-4}{5-5}
	\end{pNiceMatrix}
} \\
&= \id R \kron
\begin{pNiceMatrix}
\Block[borders={right,bottom}]{3-3}<\Large>{\id P} & & &\emptyset & \Block[borders={left,bottom}]{3-3}<\Large>{\emptyset} \\
& & & \Vdots \\
& & & \emptyset \\
\emptyset & \Cdots & \emptyset & c & \emptyset & \Cdots & \emptyset \\
\Block[borders={right,top}]{3-3}<\Large>{\emptyset} & & & \emptyset & \Block[borders={left,top}]{3-3}<\Large>{\id Q}\\
& & & \Vdots \\
& & & \emptyset
\end{pNiceMatrix} \\
&= \id R \kron \FF(P \piu \t \piu Q).
\end{align*}
\endgroup

\textbf{Case 2}: $\t = \symmp{U}{V}$.
\begingroup
\allowdisplaybreaks
\begin{align*}
&\FF( R \perT (P \piu \t \piu Q)) \\
&= \PiuL[k=1][l]{\Biggl[ \PiuL[i=1][n]{\begin{pNiceMatrix}
		\multiset{Z_kX_i}
		\end{pNiceMatrix}} 
	\piu \begin{pNiceMatrix}
	\emptyset & \multiset{Z_kV} \\
	\multiset{Z_kU} & \emptyset
	\end{pNiceMatrix}
	\piu \PiuL[j=1][m]{\begin{pNiceMatrix}
		\multiset{Z_kY_j}
		\end{pNiceMatrix}} \Biggr]} \\
&= \PiuL[k=1][l]{
	\begin{pNiceMatrix}
	\multiset{Z_kX_1} & \emptyset & \Cdots & \emptyset & \Block[borders={left,bottom,right}]{4-2}<\Large>{\emptyset} & & \Block[borders={bottom}]{4-4}<\Large>{\emptyset} \\
	\emptyset &  & \Ddots & \Vdots \\
	\Vdots & \Ddots & & \emptyset \\
	\emptyset & \Cdots & \emptyset & \multiset{Z_kX_n} \\
	\Block[borders={top,right,bottom}]{2-4}<\Large>{\emptyset} & & & & \emptyset & \multiset{Z_k V} & \Block[borders={top,left,bottom}]{2-4}<\Large>{\emptyset}\\
	& & & & \multiset{Z_k U} & \emptyset \\
	\Block[borders={top,right}]{4-4}<\Large>{\emptyset}& & & & \Block[borders={top,right}]{4-2}<\Large>{\emptyset} & &	\multiset{Z_kY_1} & \emptyset & \Cdots & \emptyset \\
	& & & & & &		\emptyset &  & \Ddots & \Vdots \\
	& & & & & &		\Vdots & \Ddots & & \emptyset \\
	& & & & & &		\emptyset & \Cdots & \emptyset & \multiset{Z_kY_m}
	\CodeAfter
	\line{1-1}{4-4}
	\line{7-7}{10-10}
	\end{pNiceMatrix}
} \\
&= \id R \kron
\begin{pNiceMatrix}
\multiset{X_1} & \emptyset & \Cdots & \emptyset & \Block[borders={left,bottom,right}]{4-2}<\Large>{\emptyset} & & \Block[borders={bottom}]{4-4}<\Large>{\emptyset} \\
\emptyset &  & \Ddots & \Vdots \\
\Vdots & \Ddots & & \emptyset \\
\emptyset & \Cdots & \emptyset & \multiset{X_n} \\
\Block[borders={top,right,bottom}]{2-4}<\Large>{\emptyset} & & & & \emptyset & \multiset{V} & \Block[borders={top,left,bottom}]{2-4}<\Large>{\emptyset}\\
& & & & \multiset{U} & \emptyset \\
\Block[borders={top,right}]{4-4}<\Large>{\emptyset}& & & & \Block[borders={top,right}]{4-2}<\Large>{\emptyset} & &	\multiset{Y_1} & \emptyset & \Cdots & \emptyset \\
& & & & & &		\emptyset &  & \Ddots & \Vdots \\
& & & & & &		\Vdots & \Ddots & & \emptyset \\
& & & & & &		\emptyset & \Cdots & \emptyset & \multiset{Y_m}
\CodeAfter
\line{1-1}{4-4}
\line{7-7}{10-10}
\end{pNiceMatrix} \\
&= \id R \kron \FF(P \piu \t \piu Q).
\end{align*}
\endgroup

\textbf{Case 3}: $\t = \diag U$.
\begingroup
\allowdisplaybreaks
\begin{align*}
&\FF( R \perT (P \piu \t \piu Q)) \\
&= \PiuL[k=1][l]{\Biggl[ \PiuL[i=1][n]{\begin{pNiceMatrix}
		\multiset{Z_kX_i}
		\end{pNiceMatrix}} 
	\piu \begin{pNiceMatrix}
	\multiset{Z_kU} \\
	\multiset{Z_kU}
	\end{pNiceMatrix}
	\piu \PiuL[j=1][m]{\begin{pNiceMatrix}
		\multiset{Z_kY_j}
		\end{pNiceMatrix}} \Biggr]} \\
&=\PiuL[k=1][l]{
	\begin{pNiceMatrix}
	\multiset{Z_kX_1} & \emptyset & \Cdots & \emptyset & \Block[borders={left,bottom,right}]{4-1}<\Large>{\emptyset} &  \Block[borders={bottom}]{4-4}<\Large>{\emptyset} \\
	\emptyset &  & \Ddots & \Vdots \\
	\Vdots & \Ddots & & \emptyset \\
	\emptyset & \Cdots & \emptyset & \multiset{Z_kX_n} \\
	\Block[borders={top,right,bottom}]{2-4}<\Large>{\emptyset} & & & &  \multiset{Z_k U} & \Block[borders={top,left,bottom}]{2-4}<\Large>{\emptyset}\\
	& & & & \multiset{Z_k U} \\
	\Block[borders={top,right}]{4-4}<\Large>{\emptyset}& & & & \Block[borders={top,right}]{4-1}<\Large>{\emptyset}  &	\multiset{Z_kY_1} & \emptyset & \Cdots & \emptyset \\
	& & & & & 		\emptyset &  & \Ddots & \Vdots \\
	& & & & & 		\Vdots & \Ddots & & \emptyset \\
	& & & & & 		\emptyset & \Cdots & \emptyset & \multiset{Z_kY_m}
	\CodeAfter
	\line{1-1}{4-4}
	\line{7-6}{10-9}
	\end{pNiceMatrix}
} \\
&= \id R \kron
\begin{pNiceMatrix}
\multiset{X_1} & \emptyset & \Cdots & \emptyset & \Block[borders={left,bottom,right}]{4-1}<\Large>{\emptyset} &  \Block[borders={bottom}]{4-4}<\Large>{\emptyset} \\
\emptyset &  & \Ddots & \Vdots \\
\Vdots & \Ddots & & \emptyset \\
\emptyset & \Cdots & \emptyset & \multiset{X_n} \\
\Block[borders={top,right,bottom}]{2-4}<\Large>{\emptyset} & & & &  \multiset{ U} & \Block[borders={top,left,bottom}]{2-4}<\Large>{\emptyset}\\
& & & & \multiset{ U} \\
\Block[borders={top,right}]{4-4}<\Large>{\emptyset}& & & & \Block[borders={top,right}]{4-1}<\Large>{\emptyset}  &	\multiset{Y_1} & \emptyset & \Cdots & \emptyset \\
& & & & & 		\emptyset &  & \Ddots & \Vdots \\
& & & & & 		\Vdots & \Ddots & & \emptyset \\
& & & & & 		\emptyset & \Cdots & \emptyset & \multiset{Y_m}
\CodeAfter
\line{1-1}{4-4}
\line{7-6}{10-9}
\end{pNiceMatrix} \\
&= \id R \kron \FF(P \piu \t \piu Q).
\end{align*}
\endgroup

\textbf{Case 4}: $\t = \bang U$.
\begingroup
\allowdisplaybreaks
\begin{align*}
&\FF( R \perT (P \piu \t \piu Q)) \\
&= \PiuL[k=1][l]{\Biggl[ \PiuL[i=1][n]{\begin{pNiceMatrix}
		\multiset{Z_kX_i}
		\end{pNiceMatrix}} 
	\piu \emptyset_{0 \times 1}
	\piu \PiuL[j=1][m]{\begin{pNiceMatrix}
		\multiset{Z_kY_j}
		\end{pNiceMatrix}} \Biggr]} \\
&= \PiuL[k=1][l]{
	\begin{pNiceMatrix}
	\multiset{Z_kX_1}  	& \emptyset & \Cdots & \emptyset & \Block[borders={bottom,left}]{4-4}<\Large>{\emptyset} \\
	\emptyset  &  & \Ddots & \Vdots \\	
	\Vdots & \Ddots &  & \emptyset  \\
	\emptyset  & \Cdots & \emptyset  & \multiset{Z_kX_n} \\
	\Block[borders={top,right}]{4-4}<\Large>{\emptyset}& & & &		\multiset{Z_kY_1}  	& \emptyset & \Cdots & \emptyset  \\
	& & & & \emptyset  &  & \Ddots & \Vdots \\	
	& & & & \Vdots & \Ddots &  & \emptyset  \\
	& & & & \emptyset  & \Cdots & \emptyset  & \multiset{Z_kXY_m}
	\CodeAfter
	\line{1-1}{4-4}
	\line{5-5}{8-8}
	\end{pNiceMatrix}
} \\
&= \id R \kron (P \piu \emptyset_{0 \times 1} \piu Q).
\end{align*}
\endgroup
The other cases concerning $\t=\codiag U$ and $\t=\cobang U$ are analogous. This concludes the proof of the claim, thus of the Theorem.
\end{proof}

\begin{proof}[Proof of Corollary~\ref{cor:monomials in tape are isomorphic to multisets of string diagrams}]
	The functor $\FF \colon \CatTape \to \Mat{\fCMon{\CatString}}$ restricts and corestricts to an isomorphism of symmetric monoidal categories (with respect to $\per$) between $\monomial$ and the subcategory of $\Mat{\fCMon{\CatString}}$ consisting only of $1 \times 1$ matrices, which is monoidally isomorphic to $\fCMon\CatString$.
\end{proof}

\subsection{Proof of Theorem~\ref{thm:tapes as matrices poset version}}\label{sec:appmatrix4}
	
Let $\Cat C$ be any preorder-enriched category. Then we can equip $\fCMon{\Cat C}[X,Y]$ with the Egli-Milner preorder, which is defined as
\[
\multiset{c_1,\dots,c_n} \le^{EM} \multiset{d_1,\dots,d_m} \iff \forall i \ldotp \exists j \ldotp c_i \le_{\Cat C} d_j,
\]
and this makes $\fCMon{\Cat C}$ a preorder-enriched category. Then also $\Mat{\fCMon{\Cat{C}}}$ is preorder-enriched, by simply setting
\[
M \le N \iff \forall j,i \ldotp M_{ji} \le^{EM} N_{ji}.
\]

We can also enrich $F_2(\Cat C)$ in a way that the isomorphism~\eqref{eq:iso between CatTape and biproduct completion} becomes an isomorphism of preorder-enriched fb rig categories. To do so we need to add on the homsets of $F_2(\Cat C)$ the precongruence, with respect to composition, $\piu$ and $\per$, generated by the following axioms:
\[
f \preceq_{\Cat C} g \implies \tape{f} \preceq \tape{g}
\]
\begin{align*}
\diag A ; \codiag A &\preceq \id A & \id{A \piu A} &\preceq {\codiag{A} ; \diag A} \\
\bang A ; \cobang A &\preceq \id A & \id\zero &\preceq {\cobang{A} ; \bang{A}}
\end{align*}
It is straightforward to see that the functor $\FF$ of~\eqref{eq:iso between CatTape and biproduct completion} preserves the inequalities of $F_2(\Cat C)$, while to show that $\GG$ respects the preorder on $\Mat{\fCMon{\Cat C}}$ we need the following result.

\begin{lemma}
	Let $\Cat C$ be a preorder-enriched category and $c_1,\dots,c_n, d_1,\dots,d_m \colon A \to B$ in $\Cat C$. If for all $i=1,\dots,n$ there exists $j=1,\dots,m$ such that $c_i \preceq_{\Cat C} d_j$, then
	\[
	\sum_{i=1}^{n} \tape{c_i} \preceq \sum_{j=1}^{m} \tape{d_j}
	\] 
	where the above sums are induced by the finite biproduct structure of $F_2(\Cat C)$.
\end{lemma}
\begin{proof}
	By induction on $n$. Suppose $n=0$: the premiss of the implication is trivially satisfied, and we have that
	\begin{align*}
	\sum_{i=1}^n \tape{c_i} &= \zero_{A,B} \\\displaybreak[0]
	&= \bang A ; \cobang B  \tag{definition of $\zero_{A,B}$} \\\displaybreak[0]
	&= \bang A ; \cobang A ; \sum_{j=1}^m \tape{d_j}  \tag{naturality of $\cobang{}$} \\\displaybreak[0]
	&\preceq \sum_{j=1}^m \tape{d_j} \tag{$\bang A ; \cobang A \preceq \id A$}
	\end{align*}
	Let now $n \ge 0$, suppose the statement is true for $n$, consider $c_1,\dots,c_{n+1},d_1,\dots,d_m \colon A \to B$ in $\Cat C$ such that for all $i = 1,\dots,n+1$ there is $j=1,\dots,m$ for which $c_i \preceq_{\Cat C} d_j$. Then we have
	\begin{align*}
	\sum_{i=1}^{n+1} \tape{c_i} &= \sum_{i=1}^n \tape{c_i} + \tape{c_{n+1}} \tag{associativity of $+$} \\\displaybreak[0]
	&\preceq \sum_{j=1}^{m} \tape{d_j} + \tape{c_{n+1}} \tag{inductive hypothesis} \\\displaybreak[0]
	&\preceq \sum_{j=1}^{m} \tape{d_j} + \tape{d_k} \tag{$\tape{c_{n+1}} \preceq \tape{d_k}$ for some $k$} \\\displaybreak[0]
	&= \sum_{\substack{1 \le j \le m \\ j \ne k}} \tape{d_j} + \tape{d_k} + \tape{d_k} \tag{commutativity of $+$} \\\displaybreak[0]
	&= \sum_{\substack{1 \le j \le m \\ j \ne k}} \tape{d_j} + (\diag A ; \tape{d_k} \piu \tape{d_k} ; \codiag B ) \tag{definition of $+$} \\\displaybreak[0]
	&= \sum_{\substack{1 \le j \le m \\ j \ne k}} \tape{d_j} + ( \tape{d_k} ; \diag B ; \codiag B ) \tag{naturality of $\diag{}$} \\\displaybreak[0]
	&\preceq \sum_{\substack{1 \le j \le m \\ j \ne k}} \tape{d_j} + \tape{d_k}  \tag{$\diag B ; \codiag B \preceq \id B$} \\\displaybreak[0]
	&= \sum_{j=1}^{m} \tape{d_j} \tag*{\qedhere}
	\end{align*}
\end{proof}

Taking now $\posetification{F_2(\Cat C)}$ and $\posetification{\Mat{\fCMon{\Cat C} } } $, we obtain a \emph{poset}-enriched isomorphism. Observe that $\posetification{\Mat{\fCMon{\Cat C}}} \cong \Mat{\posetification{ (\fCMon{\Cat C}) } }$, because the order on matrices is entry by entry, and that $\posetification{(\fCMon{\Cat C})} \cong \downsetification{\Cat C}$, because it holds that for all $S,T \in \fCMon{\Cat C}[X,Y]$:
\[
S \le^{EM} T \land T \le^{EM} S \iff \downward{S} = \downward{T}.
\]

If we apply this to $\Cat C={\Cat C}_{\sign,\basicR}$, then we have that  $\posetification{F_2(\Cat C)} = \CatTapeI{\tilde{\basicR}}$ and we obtain
\[
\CatTapeI{\tilde{\basicR}} \cong \Mat{\downsetification{{\Cat C}_{\sign,\basicR}}}.
\]
Theorem~\ref{thm:tapes as matrices poset version} now follows from the simple observation that $\downsetification{{\Cat C}_{\sign,\basicR}} \cong \downsetification{\posetification{\Cat{C}_{\sign,\basicR}}}$.

\section{Proofs of Section \ref{sec:CBPOPL}}\label{sc:appendixCB}
In this Appendix we provide the details for proofs in Section~\ref{sec:CBPOPL}. In Section~\ref{sec:appCB1} we prove Theorem~\ref{thm:Tape Cartesian}, which states that $\CatTapeCB$ is a cartesian bicategory. In Section~\ref{sec:appCB2} we prove Proposition~\ref{prop:bijective}, regarding the bijective correspondence between interpretations and morphisms of sesquistrict fb-cb rig categories. In Section~\ref{app:Completeness} we prove our completeness result (Theorem~\ref{thm:completeness}). Finally in Section~\ref{sec:appCB4} we show the proof for the other results stated in Section~\ref{sec:CBPOPL}, that is Proposition~\ref{prop:encodingsound} and Corollary~\ref{crlFinal}.

\subsection{Proof of Theorem~\ref{thm:Tape Cartesian}}\label{sec:appCB1}

\begin{lemma}\label{lm:copier}
    For all $P_1, P_2$ polynomials the following holds:
    \[ \copier{P_1 \piu P_2} = (\copier{P_1} \piu \cobang{P_1 P_2});(\Idl{P_1}{P_1}{P_2} \piu (\cobang{P_2 P_1} \piu \copier{P_2}));\Idl{P_2}{P_1}{P_2} \]
\end{lemma}
\begin{proof}
    First observe that the following holds for all polynomials $P,Q,R$:
    \begin{equation*}\tag{$\ast$}
        \cobang{P(Q \piu R)} \axeq{\refeq{eq:whisk:bang}} \RW{Q \piu R}{\cobang{P}} \axeq{\refeq{eq:whisk:sum}} (\RW{Q}{\cobang{P}} \piu \RW{R}{\cobang{P}}) ; \Idl{P}{Q}{R} \axeq{\refeq{eq:whisk:bang}} (\cobang{PQ} \piu \cobang{PR}) ; \Idl{P}{Q}{R}
    \end{equation*}
    Then we proceed by induction on $P_1$.
    
    \textbf{Case $P = \zero$:} \begin{align*}
        &(\copier{\zero} \piu \cobang{\zero P_2});\Idl{\zero}{\zero}{P_2} \piu (\cobang{P_2 \zero} \piu \copier{P_2});\Idl{P_2}{\zero}{P_2} \\\displaybreak[0]
        &=(\id\zero \piu \cobang{\zero P_2});\Idl{\zero}{\zero}{P_2} \piu (\cobang{P_2 \zero} \piu \copier{P_2});\Idl{P_2}{\zero}{P_2} \tag{Def. $\copier{}$} \\\displaybreak[0]
        &=(\id\zero \piu \id\zero);\Idl{\zero}{\zero}{P_2} \piu (\id\zero \piu \copier{P_2});\Idl{P_2}{\zero}{P_2} \tag{Def. $\cobang{}$} \\\displaybreak[0]
        &=(\id\zero \piu \id\zero);\id\zero \piu (\id\zero \piu \copier{P_2});\Idl{P_2}{\zero}{P_2} \tag{Def. $\delta^l$} \\\displaybreak[0]
        &=(\id\zero \piu \id\zero);\id\zero \piu (\id\zero \piu \copier{P_2});\id{P_2 P_2} \tag{Lemma~\ref{lemma:tapesymdis}.3} \\\displaybreak[0]
        &=\copier{P_2} \\\displaybreak[0]
        &= \copier{\zero \piu P_2}\tag{Funct. $\per$}
    \end{align*}

    \textbf{Case $P = U \piu P'$:}
    \begingroup
    \allowdisplaybreaks
    \begin{align*}
        &\scalebox{0.7}{
} \tag*{\qedhere} \end{align*}
    \endgroup
\end{proof}

\begin{lemma}\label{lm:discharger}
    For all $P_1, P_2$ polynomials the following holds:
    \[ \discharger{P_1 \piu P_2} = (\discharger{P_1} \piu \discharger{P_2}) ; \codiag{\uno} \]
\end{lemma}
\begin{proof} By induction on $P_1$.
    
    \textbf{Case $P = \zero$:} \begin{align*}
        (\discharger{\zero} \piu \discharger{P_2}) ; \codiag{\uno} \axeq{\text{def. }\discharger{}} (\cobang{\uno} \piu \discharger{P_2}) ; \codiag{\uno} \axeq{\text{funct. }\piu} (\id\zero \piu \discharger{P_2});(\cobang{\uno} \piu \id\uno);\codiag{\uno} \axeq{\refeq{ax:codiagun}} \discharger{P_2} \axeq{\text{funct. }\piu} \discharger{\zero \piu P_2}
    \end{align*}

    \textbf{Case $P = U \piu P'$:}
    \begin{align*}
        \discharger{(U \piu P') \piu P_2} &= \discharger{U \piu (P' \piu P_2)} \tag{Assoc. $\piu$}\\
        &= (\Tdischarger{U} \piu \discharger{P' \piu P_2}) ; \codiag{\uno} \tag{Def. $\discharger{}$} \\
        &= (\Tdischarger{U} \piu ((\discharger{P'} \piu \discharger{P_2}) ; \codiag{\uno})) ; \codiag{\uno} \tag{Ind. hp.}\\
        &= (((\Tdischarger{U} \piu \discharger{P'});\codiag{\uno}) \piu \discharger{P_2}) ; \codiag{\uno} \tag{\refeq{ax:codiagas}} \\
        &= (\discharger{U \piu P'} \piu \discharger{P_2}) ; \codiag{\uno} \tag{Def. $\discharger{}$}
    \end{align*}
\end{proof}

\begin{lemma}[Adjointness]
    The following hold for any polynomial $P$
    \[ \codischarger{P} ; \discharger{P} \leq \id\uno \qquad\quad \id P \leq \discharger{P} ; \codischarger{P} \qquad\quad  \cocopier{P} ; \copier{P} \leq \id{P \per P} \qquad\quad  \id P \leq \copier{P} ; \cocopier{P} \]
\end{lemma}
\begin{proof} \vphantom{.}

    \textsc{Equation $\codischarger{P} ; \discharger{P} \leq \id\uno$}: by induction on $P$.
    
    \textbf{Case $P = \zero$:} $\codischarger{\zero} ; \discharger{\zero} \axeq{\text{def. }\codischarger{}, \discharger{}} \bang{\uno} ; \cobang{\uno} \axsubeq{\refeq{ax:relBangCobang}} \id\uno$

    \textbf{Case $P = U \piu P'$:} 
    \begin{align*}

        \axsubeq{\text{ind. hp.}}
        \Tcodiagdiag{\uno}
        \axsubeq{\refeq{ax:relDiagCodiag}}
        \Twire{\uno} 
    \end{align*}

    \textsc{Equation $\id P \leq \discharger{P} ; \codischarger{P}$}: by induction on $P$.
    
    \textbf{Case $P = \zero$:} $\id\zero \axeq{\refeq{ax:bo}} \cobang{\uno} ; \bang{\uno} \axeq{\text{def. }\discharger{},\codischarger{}} \discharger{\zero} ; \codischarger{\zero}$

    \textbf{Case $P = U \piu P'$:}
    \begin{align*}
        %
        
    \end{align*}

    \textsc{Equation $\cocopier{P} ; \copier{P} \leq \id{P \per P}$}: by induction on $P$.

    \textbf{Case $P = \zero$:} $\id\zero ; \id\zero = \id\zero = \id{\zero \per \zero}$

    \textbf{Case $P = U \piu P'$:} \begin{align*}
        &%
                                       
    \end{align*}

    \textsc{Equation $\id P \leq \copier{P} ; \cocopier{P}$}: by induction on $P$.

    \textbf{Case $P = \zero$:} $\id\zero \leq \id\zero = \id{\zero} ; \id\zero$

    \textbf{Case $P = U \piu P'$:} \begin{align*}
        &\begin{aligned}\begin{gathered} \Twire{U} \\ \TPolyWire{P'} \end{gathered}\end{aligned}
        \axsubeq{\refeq{ax:copieradj2}}
        \begin{aligned}\begin{gathered} %
        
        \axeq{\text{def. } \copier{}, \cocopier{}} \copier{U \piu P'} ; \cocopier{U \piu P'} \tag*{\qedhere}
    \end{align*}
\end{proof}

\begin{lemma}[Comonoid]
    The following hold for any polynomial $P$
    \[ \copier{P} ; \symmt{P}{P} = \copier{P} \qquad\quad \copier{P} ; \RW{P}{\discharger{P}} = \id{P} \qquad\quad  \copier{P} ; \RW{P}{\copier{P}} = \copier{P} ; \LW{P}{\copier{P}} \]
\end{lemma}
\begin{proof} \vphantom{.}
    
    \textsc{Equation $\copier{P} ; \symmt{P}{P} = \copier{P}$}: by induction on $P$.

    \textbf{Case $P = \zero$:} $\copier{\zero} ; \symmt{\zero}{\zero} = \id\zero ; \id\zero = \id\zero = \copier{\zero}$
    
    \textbf{Case $P = U \piu P'$:} 
    \begingroup
    \allowdisplaybreaks
    \begin{align*}
        &\copier{U \piu P'} ;  \symmt{U \piu P'}{U \piu P'}
        \axeq{\text{def. }\copier{}, \sigma^\per}
 
        \axeq{\text{def. }\copier{}} 
        \copier{U \piu P'}                  
    \end{align*}
    \endgroup
    
    \textsc{Equation $\copier{P} ; \RW{P}{\discharger{P}} = \id{P}$}: by induction on $P$.

    \textbf{Case $P = \zero$:} $\id\zero ; \RW{\zero}{\cobang{\uno}} = \id\zero ; \id\zero = \id\zero$

    \textbf{Case $P = U \piu P'$:} 
    \begingroup
    \allowdisplaybreaks
    \begin{align*}
        &\copier{U \piu P'} ; \RW{U \piu P'}{\discharger{U \piu P'}}  \\
        &= (\Tcopier{U} \piu \cobang{UP'} \piu ((\cobang{P'U} \piu \copier{P'}) ; \Idl{P'}{U}{P'})) ; \RW{U \piu P'}{(\Tdischarger{U} \piu \discharger{P'}) ; \codiag{\uno}} \tag{Def. $\copier{}, \discharger{}$}\\
        &= \begin{multlined}[t]
            (\Tcopier{U} \piu \cobang{UP'} \piu ((\cobang{P'U} \piu \copier{P'}) ; \Idl{P'}{U}{P'})) ; \\
                \dl{U \piu P'}{U}{P'} ; (\RW{U}{(\Tdischarger{U} \piu \discharger{P'}) ; \codiag{\uno}} \piu \RW{P'}{(\Tdischarger{U} \piu \discharger{P'}) ; \codiag{\uno}})
        \end{multlined} \tag{\refeq{eq:whisk:sum}}\\
        &= \begin{multlined}[t]
            (\Tcopier{U} \piu \cobang{UP'} \piu ((\cobang{P'U} \piu \copier{P'}) ; \Idl{P'}{U}{P'})) ; \\
                (\id{U(U \piu P')} \piu \dl{P'}{U}{P'}) ; (\id{UU} \piu \symmp{UP'}{P'U} \piu \id{P'P'}) \\
                (\RW{U}{(\Tdischarger{U} \piu \discharger{P'}) ; \codiag{\uno}} \piu \RW{P'}{(\Tdischarger{U} \piu \discharger{P'}) ; \codiag{\uno}})
        \end{multlined} \tag{Def. $\delta^l$}\\
        &= \begin{multlined}[t]
            (\Tcopier{U} \piu \cobang{UP'} \piu ((\cobang{P'U} \piu \copier{P'}) ; \Idl{P'}{U}{P'} ; \dl{P'}{U}{P'})) ; \\
                (\id{UU} \piu \symmp{UP'}{P'U} \piu \id{P'P'}) ; (\RW{U}{(\Tdischarger{U} \piu \discharger{P'}) ; \codiag{\uno}} \piu \RW{P'}{(\Tdischarger{U} \piu \discharger{P'}) ; \codiag{\uno}})
        \end{multlined} \tag{Funct. $\piu$}\\
        &= \begin{multlined}[t]
            (\Tcopier{U} \piu \cobang{UP'} \piu \cobang{P'U} \piu \copier{P'}) ; \\
                (\id{UU} \piu \symmp{UP'}{P'U} \piu \id{P'P'}) ; (\RW{U}{(\Tdischarger{U} \piu \discharger{P'}) ; \codiag{\uno}} \piu \RW{P'}{(\Tdischarger{U} \piu \discharger{P'}) ; \codiag{\uno}})
        \end{multlined} \tag{Iso}\\
        &= (\Tcopier{U} \piu \cobang{P'U} \piu \cobang{UP'} \piu \copier{P'}) ; (\RW{U}{(\Tdischarger{U} \piu \discharger{P'}) ; \codiag{\uno}} \piu \RW{P'}{(\Tdischarger{U} \piu \discharger{P'}) ; \codiag{\uno}}) \tag{\refeq{ax:symmpnat}}\\
        &= (\Tcopier{U} \piu \cobang{P'U} \piu \cobang{UP'} \piu \copier{P'}) ; ((\RW{U}{\Tdischarger{U}} \piu \RW{U}{\discharger{P'}}) ; \RW{U}{\codiag{\uno}} \piu \RW{P'}{\Tdischarger{U}} \piu \RW{P'}{\discharger{P'}}) ; \RW{P'}{\codiag{\uno}}) \tag{\refeq{eq:whisk:funct}}\\
        &= (\Tcopier{U} \piu \cobang{P'U} \piu \cobang{UP'} \piu \copier{P'}) ; ((\begin{tikzpicture}
                \begin{pgfonlayer}{nodelayer}
                    \node [style=none] (0) at (-0.25, -0.5) {};
                    \node [style=none] (3) at (-0.25, 0.5) {};
                    \node [style=none] (4) at (1.25, 0.5) {};
                    \node [style=none] (5) at (1.25, -0.5) {};
                    \node [style=none] (6) at (-0.25, 0.225) {};
                    \node [style=black] (7) at (0.5, 0.225) {};
                    \node [style=label] (8) at (-0.75, 0.25) {$U$};
                    \node [style=none] (9) at (-0.25, -0.25) {};
                    \node [style=none] (10) at (1.25, -0.25) {};
                    \node [style=label] (11) at (-0.75, -0.3) {$U$};
                    \node [style=label] (12) at (1.75, -0.25) {$U$};
                \end{pgfonlayer}
                \begin{pgfonlayer}{edgelayer}
                    \draw [style=tape] (5.center)
                         to (4.center)
                         to (3.center)
                         to (0.center)
                         to cycle;
                    \draw (6.center) to (7);
                    \draw (9.center) to (10.center);
                \end{pgfonlayer}
            \end{tikzpicture}
             \piu \RW{U}{\discharger{P'}}) ; \codiag{U} \piu \RW{P'}{\Tdischarger{U}} \piu \RW{P'}{\discharger{P'}}) ; \codiag{P'}) \tag{Def. $R$}\\
        &= (\begin{tikzpicture}
            \begin{pgfonlayer}{nodelayer}
                \node [style=none] (0) at (-0.75, -0.5) {};
                \node [style=none] (3) at (-0.75, 0.5) {};
                \node [style=none] (4) at (1.25, 0.5) {};
                \node [style=none] (5) at (1.25, -0.5) {};
                \node [style=none] (6) at (0.5, 0.25) {};
                \node [style=black] (7) at (0.75, 0.225) {};
                \node [style=label] (8) at (-1.25, 0) {$U$};
                \node [style=none] (9) at (0.5, -0.25) {};
                \node [style=none] (10) at (1.25, -0.25) {};
                \node [style=label] (12) at (1.75, -0.25) {$U$};
                \node [style=black] (13) at (0, 0) {};
                \node [style=none] (14) at (-0.75, 0) {};
            \end{pgfonlayer}
            \begin{pgfonlayer}{edgelayer}
                \draw [style=tape] (5.center)
                     to (4.center)
                     to (3.center)
                     to (0.center)
                     to cycle;
                \draw (6.center) to (7);
                \draw (9.center) to (10.center);
                \draw [bend right] (13) to (9.center);
                \draw [bend right] (6.center) to (13);
                \draw (14.center) to (13);
            \end{pgfonlayer}
        \end{tikzpicture} \piu \cobang{P'U} ; \RW{U}{\discharger{P'}}) ; \codiag{U} \piu (\cobang{P'U} ; \RW{P'}{\Tdischarger{U}} \piu \copier{P'} ; \RW{P'}{\discharger{P'}}) ; \codiag{P'} \tag{Funct. $\piu$}\\
        &= (\id U \piu \cobang{P'U} ; \RW{U}{\discharger{P'}}) ; \codiag{U} \piu (\cobang{P'U} ; \RW{P'}{\Tdischarger{U}} \piu \id{P'}) ; \codiag{P'} \tag{\ref{ax:copierun}, ind. hp.}\\
        &= (\id U \piu \cobang{U}) ; \codiag{U} \piu (\cobang{P'} \piu \id{P'}) ; \codiag{P'} \tag{\refeq{ax:bangnat}} \\
        &= \id U \piu \id{P'} \tag{\refeq{ax:codiagun}}\\
        &= \id{U \piu P'} \tag{Funct. $\piu$}
    \end{align*}
    \endgroup

    \textsc{Equation $\copier{P} ; \RW{P}{\copier{P}} = \copier{P} ; \LW{P}{\copier{P}}$}: by induction on $P$.

    \textbf{Case $P = \zero$:} $\copier{\zero} ; \RW{\zero}{\copier{\zero}} \axeq{\text{def. }\copier{}, R} \id\zero ; \id\zero \axeq{\text{def. }\copier{}, L} \copier{\zero} ; \LW{\zero}{\copier{\zero}}$
    
    \textbf{Case $P = U \piu P'$:}
    We will split the proof in two parts. The first part proves the following statement:
    \begin{equation*}\tag{$\ast_1$}
        (\copier{U} \piu \cobang{UP'});\RW{U \piu P'}{\copier{U} \piu \cobang{UP'}} = (\copier{U} \piu \cobang{UP'}) ; \LW{U}{\copier{U \piu P'}}
    \end{equation*}
    \begin{align*}
        &(\copier{U} \piu \cobang{UP'});\RW{U \piu P'}{\copier{U} \piu \cobang{UP'}} \\
        &= (\copier{U} \piu \cobang{UP'});(\RW{U \piu P'}{\copier{U}} \piu \RW{U \piu P'}{\cobang{UP'}}) \tag{\ref{eq:whisk:funct piu}} \\
        &= (\copier{U} \piu \cobang{UP'});(\RW{U}{\copier{U}} \piu \RW{P'}{\copier{U}} \piu \RW{U \piu P'}{\cobang{UP'}}) \tag{\ref{eq:whisk:sum}} \\
        &= (\copier{U};\RW{U}{\copier{U}}) \piu \cobang{UP'};(\RW{P'}{\copier{U}} \piu \RW{U \piu P'}{\cobang{UP'}}) \tag{Funct. $\piu$} \\
        &= (\copier{U};\RW{U}{\copier{U}}) \piu \cobang{UUP' \piu UP'(U\piu P')} \tag{\ref{ax:cobangnat}} \\
        &= (\copier{U};\LW{U}{\copier{U}}) \piu \cobang{UUP' \piu UP'(U\piu P')} \tag{\ref{ax:copieras}} \\
        &= (\copier{U};\LW{U}{\copier{U}}) \piu (\cobang{UP'};\LW{U}{\cobang{UP'} \piu (\cobang{P'U} \piu \copier{P'});\dl{P'}{U}{P'}}) \tag{\ref{ax:cobangnat}} \\
        &= (\copier{U} \piu \cobang{UP'}) ; (\LW{U}{\copier{U}} \piu \LW{U}{\cobang{UP'} \piu (\cobang{P'U} \piu \copier{P'});\dl{P'}{U}{P'}}) \tag{Funct. $\piu$} \\
        &= (\copier{U} \piu \cobang{UP'}) ; (\LW{U}{\copier{U} \piu \cobang{UP'} \piu (\cobang{P'U} \piu \copier{P'});\dl{P'}{U}{P'}}) \tag{\ref{eq:whisk:funct piu}} \\
        &= (\copier{U} \piu \cobang{UP'}) ; \LW{U}{\copier{U \piu P'}} \tag{Def. $\copier{}$} \\
    \end{align*}

    For the second part we prove the following statement 
    \begin{equation*}\tag{$\ast_2$}
        (\cobang{P'U} \piu \copier{P'});\Idl{P'}{U}{P'};\RW{U \piu P'}{(\cobang{P'U} \piu \copier{P'});\Idl{P'}{U}{P'}} = (\cobang{P'U} \piu \copier{P'});\Idl{P'}{U}{P'};\LW{P'}{\copier{U \piu P'}}
    \end{equation*}
    Starting from the right hand side of the equation
    \begingroup
    \allowdisplaybreaks
    \begin{align*}
        &(\cobang{P'U} \piu \copier{P'});\Idl{P'}{U}{P'};\LW{P'}{\copier{U \piu P'}} \\
        &= (\cobang{P'U} \piu \copier{P'});\Idl{P'}{U}{P'};\LW{P'}{\copier{U} \piu \cobang{UP'} \piu (\cobang{P'U} \piu \copier{P'});\Idl{P'}{U}{P'}} \tag{Def. $\copier{}$} \\
        &= (\cobang{P'U} \piu \copier{P'});(\LW{P'}{\copier{U} \piu \cobang{UP'}} \piu \LW{P'}{(\cobang{P'U} \piu \copier{P'});\Idl{P'}{U}{P'}});\Idl{P'}{U(U \piu P')}{P'(U \piu P')} \tag{\ref{eq:whisk:funct piu}} \\
        &= ((\cobang{P'U};\LW{P'}{\copier{U} \piu \cobang{UP'}}) \piu (\copier{P'};\LW{P'}{(\cobang{P'U} \piu \copier{P'});\Idl{P'}{U}{P'}}));\Idl{P'}{U(U \piu P')}{P'(U \piu P')} \tag{Funct. $\piu$} \\
        &= (\cobang{P'U(U \piu P')} \piu (\copier{P'};\LW{P'}{(\cobang{P'U} \piu \copier{P'});\Idl{P'}{U}{P'}}));\Idl{P'}{U(U \piu P')}{P'(U \piu P')} \tag{\ref{ax:cobangnat}} \\
        &= (\cobang{P'U(U \piu P')} \piu (\copier{P'};\LW{P'}{\copier{P'};(\cobang{P'U} \piu \id{P'P'});\Idl{P'}{U}{P'}}));\Idl{P'}{U(U \piu P')}{P'(U \piu P')} \tag{Funct. $\piu$} \\
        &= (\cobang{P'U(U \piu P')} \piu (\copier{P'};\LW{P'}{\copier{P'}};\LW{P'}{\cobang{P'U} \piu \id{P'P'}};\LW{P'}{\Idl{P'}{U}{P'}}));\Idl{P'}{U(U \piu P')}{P'(U \piu P')} \tag{\ref{eq:whisk:funct}} \\
        &= (\cobang{P'U(U \piu P')} \piu (\copier{P'};\LW{P'}{\copier{P'}};\LW{P'}{\cobang{P'U} \piu \id{P'P'}};\dl{P'}{P'U}{P'P'};\Idl{P'P'}{U}{P'}));\Idl{P'}{U(U \piu P')}{P'(U \piu P')} \tag{\ref{eq:whisk:dl}} \\
        &= (\cobang{P'U(U \piu P')} \piu (\copier{P'};\LW{P'}{\copier{P'}};(\LW{P'}{\cobang{P'U}} \piu \LW{P'}{\id{P'P'}});\Idl{P'P'}{U}{P'}));\Idl{P'}{U(U \piu P')}{P'(U \piu P')} \tag{\ref{eq:whisk:funct piu}} \\
        &= (\cobang{P'U(U \piu P')} \piu (\copier{P'};\LW{P'}{\copier{P'}};(\cobang{P'P'U} \piu \id{P'P'P'});\Idl{P'P'}{U}{P'}));\Idl{P'}{U(U \piu P')}{P'(U \piu P')} \tag{\ref{eq:whisk:bang}, \ref{eq:whisk:id}} \\
        &= (\cobang{P'U(U \piu P')} \piu ((\cobang{P'P'U} \piu \copier{P'};\LW{P'}{\copier{P'}});\Idl{P'P'}{U}{P'}));\Idl{P'}{U(U \piu P')}{P'(U \piu P')} \tag{Funct. $\piu$} \\
        &= (\RW{U \piu P'}{\cobang{P'U}} \piu ((\cobang{P'P'U} \piu \copier{P'};\LW{P'}{\copier{P'}});\Idl{P'P'}{U}{P'}));\Idl{P'}{U(U \piu P')}{P'(U \piu P')} \tag{\ref{eq:whisk:bang}} \\
        &= (((\cobang{P'UU} \piu \cobang{P'UP'});\Idl{P'U}{U}{P'}) \piu ((\cobang{P'P'U} \piu \copier{P'};\LW{P'}{\copier{P'}});\Idl{P'P'}{U}{P'}));\Idl{P'}{U(U \piu P')}{P'(U \piu P')} \tag{\ref{eq:whisk:sum}} \\
        &= (\cobang{P'UU} \piu \cobang{P'UP'} \piu \cobang{P'P'U} \piu \copier{P'};\LW{P'}{\copier{P'}});(\Idl{P'U}{U}{P'} \piu \Idl{P'P'}{U}{P'});\Idl{P'}{U(U \piu P')}{P'(U \piu P')} \tag{Funct. $\piu$} \\
        &=\begin{multlined}[t]
            (\cobang{P'UU} \piu \cobang{P'UP'} \piu \cobang{P'P'U} \piu \copier{P'};\LW{P'}{\copier{P'}});\\
            (\id{P'UU} \piu \symmp{P'UP'}{P'P'U} \id{P'P'P'}); \Idl{P'U \piu P'P'}{U}{P'};\Idl{P'}{U(U \piu P')}{P'(U \piu P')} 
        \end{multlined}\tag{Lemma~\ref{lemma:tapesymdis}.2} \\
        &= (\cobang{P'UU} \piu \cobang{P'P'U} \piu \cobang{P'UP'} \piu \copier{P'};\LW{P'}{\copier{P'}});\Idl{P'U \piu P'P'}{U}{P'};\Idl{P'}{U(U \piu P')}{P'(U \piu P')} \tag{\ref{ax:symmpnat}} \\
        &= (\cobang{P'UU} \piu \cobang{P'P'U} \piu \cobang{P'UP'} \piu \copier{P'};\RW{P'}{\copier{P'}});\Idl{P'U \piu P'P'}{U}{P'};\Idl{P'}{U(U \piu P')}{P'(U \piu P')} \tag{Ind. hp.} \\
        &= (\cobang{P'UU} \piu \cobang{P'P'U} \piu \RW{P'}{\cobang{P'U}} \piu \copier{P'};\RW{P'}{\copier{P'}});\Idl{P'U \piu P'P'}{U}{P'};\RW{U \piu P'}{\Idl{P'}{U}{P'}}  \tag{\ref{eq:whisk:bang}, \ref{eq:whisk:dl}} \\
        &= (\cobang{P'UU} \piu \cobang{P'P'U} \piu  \copier{P'};(\RW{P'}{\cobang{P'U}} \piu \RW{P'}{\copier{P'}}));\Idl{P'U \piu P'P'}{U}{P'};\RW{U \piu P'}{\Idl{P'}{U}{P'}}  \tag{Funct. $\piu$} \\
        &= (\cobang{P'UU} \piu \cobang{P'P'U} \piu  \copier{P'};\RW{P'}{\cobang{P'U} \piu \copier{P'}});\Idl{P'U \piu P'P'}{U}{P'};\RW{U \piu P'}{\Idl{P'}{U}{P'}}  \tag{\ref{eq:whisk:funct piu}} \\
        &= ((\cobang{P'U};\RW{U}{\cobang{P'U} \piu \copier{P'}}) \piu  \copier{P'};\RW{P'}{\cobang{P'U} \piu \copier{P'}});\Idl{P'U \piu P'P'}{U}{P'};\RW{U \piu P'}{\Idl{P'}{U}{P'}}  \tag{\ref{ax:cobangnat}} \\
        &= (\cobang{P'U} \piu  \copier{P'});(\RW{U}{\cobang{P'U} \piu \copier{P'}} \piu \RW{P'}{\cobang{P'U} \piu \copier{P'}});\Idl{P'U \piu P'P'}{U}{P'};\RW{U \piu P'}{\Idl{P'}{U}{P'}}  \tag{Funct. $\piu$} \\
        &= (\cobang{P'U} \piu  \copier{P'});\Idl{P'}{U}{P'};\RW{U \piu P'}{\cobang{P'U} \piu \copier{P'}};\RW{U \piu P'}{\Idl{P'}{U}{P'}}  \tag{\ref{eq:whisk:sum}} \\
        &= (\cobang{P'U} \piu  \copier{P'});\Idl{P'}{U}{P'};\RW{U \piu P'}{(\cobang{P'U} \piu \copier{P'});\Idl{P'}{U}{P'}}  \tag{\ref{eq:whisk:funct}}
    \end{align*}
    \endgroup

    We conclude the proof by showing that
    \begingroup
    \allowdisplaybreaks
    \begin{align*}
        &\copier{U \piu P'};\RW{U \piu P'}{\copier{U \piu P'}} \\
        &= (\copier{U} \piu \cobang{UP'} \piu (\cobang{P'U} \piu \copier{P'});\Idl{P'}{U}{P'});\RW{U \piu P'}{\copier{U} \piu \cobang{UP'} \piu (\cobang{P'U} \piu \copier{P'});\Idl{P'}{U}{P'}} \tag{Def. $\copier{}$} \\
        &= (\copier{U} \piu \cobang{UP'} \piu (\cobang{P'U} \piu \copier{P'});\Idl{P'}{U}{P'});(\RW{U \piu P'}{\copier{U} \piu \cobang{UP'}} \piu \RW{U \piu P'}{(\cobang{P'U} \piu \copier{P'});\Idl{P'}{U}{P'}}) \tag{\ref{eq:whisk:funct piu}} \\
        &= (\copier{U} \piu \cobang{UP'}) ; \RW{U \piu P'}{\copier{U} \piu \cobang{UP'}} \piu  (\cobang{P'U} \piu \copier{P'});\Idl{P'}{U}{P'};\RW{U \piu P'}{(\cobang{P'U} \piu \copier{P'});\Idl{P'}{U}{P'}} \tag{Funct. $\piu$} \\
        &= (\copier{U} \piu \cobang{UP'}) ; \LW{U}{\copier{U \piu P'}} \piu  (\cobang{P'U} \piu \copier{P'});\Idl{P'}{U}{P'};\LW{P'}{\copier{U \piu P'}} \tag{$\ast_1$, $\ast_2$} \\
        &= (\copier{U} \piu \cobang{UP'} \piu (\cobang{P'U} \piu \copier{P'});\Idl{P'}{U}{P'}) ; (\LW{U}{\copier{U \piu P'}} \piu  \LW{P'}{\copier{U \piu P'}}) \tag{Funct. $\piu$} \\
        &= (\copier{U} \piu \cobang{UP'} \piu (\cobang{P'U} \piu \copier{P'});\Idl{P'}{U}{P'}) ; \LW{U \piu P'}{\copier{U \piu P'}} \tag{\ref{eq:whisk:sum}} \\
        &= \copier{U \piu P'} ; \LW{U \piu P'}{\copier{U \piu P'}} \tag{Def. $\copier{}$} 
    \end{align*}
    \endgroup
    as required.
\end{proof}

\begin{lemma}[Frobenius]
    The following hold for any polynomial $P$
    \[ \copier{P} ; \cocopier{P} = \id{P} \qquad\quad \RW{P}{\copier{P}} ; \LW{P}{\cocopier{P}} = \cocopier{P} ; \copier{P} \]
\end{lemma}
\begin{proof} \vphantom{.}

    \textsc{Equation $\copier{P} ; \cocopier{P} = \id{P}$}: analogous to the proof of $\id{P} \leq \copier{P} ; \cocopier{P}$.

    \textsc{Equation $\RW{P}{\copier{P}} ; \LW{P}{\cocopier{P}} = \cocopier{P} ; \copier{P}$}: by induction on $P$.

    \textbf{Case $P = \zero$:} $\cocopier{\zero} ; \copier{\zero} = \id\zero ; \id\zero = \RW{\zero}{\copier{\zero}} ; \LW{P}{\cocopier{\zero}}$

    \textbf{Case $P = U \piu P'$:}
    We will split the proof in two parts. The first part proves the following statement:
    \begin{equation*}\tag{$\ast_1$}
        \RW{U \piu P'}{\Tcopier{U} \piu \cobang{UP'}} ; \LW{U}{\cocopier{U \piu P'}} = (\Tcocopier{U};\Tcopier{U}) \piu (\bang{UP'};\cobang{UP'})
    \end{equation*}
        \begin{align*}
            &\RW{U \piu P'}{\Tcopier{U} \piu \cobang{UP'}} ; \LW{U}{\cocopier{U \piu P'}} \\
            &= (\RW{U \piu P'}{\Tcopier{U}} \piu \RW{U \piu P'}{\cobang{UP'}}) ; \LW{U}{\cocopier{U \piu P'}} \tag{\refeq{eq:whisk:funct piu}}\\
            &= (\RW{U \piu P'}{\Tcopier{U}} \piu \RW{U \piu P'}{\cobang{UP'}}) ; \LW{U}{\Tcocopier{U} \piu \bang{UP'} \piu \dl{P'}{U}{P'};(\bang{P'U} \piu \cocopier{P'})} \tag{Def. $\cocopier{}$}\\
            &= (\RW{U \piu P'}{\Tcopier{U}} \piu \RW{U \piu P'}{\cobang{UP'}}) ; (\LW{U}{\Tcocopier{U} \piu \bang{UP'}} \piu \LW{U}{\dl{P'}{U}{P'};(\bang{P'U} \piu \cocopier{P'})}) \tag{\refeq{eq:whisk:funct piu}}\\
            &= (\RW{U \piu P'}{\Tcopier{U}};\LW{U}{\Tcocopier{U} \piu \bang{UP'}}) \piu (\RW{U \piu P'}{\cobang{UP'}};\LW{U}{\dl{P'}{U}{P'};(\bang{P'U} \piu \cocopier{P'})}) \tag{Funct. $\piu$}\\
            &= (\RW{U \piu P'}{\Tcopier{U}};\LW{U}{\Tcocopier{U} \piu \bang{UP'}}) \piu (\cobang{UP'(U \piu P')};\LW{U}{\dl{P'}{U}{P'};(\bang{P'U} \piu \cocopier{P'})}) \tag{\refeq{eq:whisk:bang}}\\
            &= (\RW{U \piu P'}{\Tcopier{U}};\LW{U}{\Tcocopier{U} \piu \bang{UP'}}) \piu \cobang{UP'} \tag{\refeq{ax:cobangnat}}\\
            &= ((\RW{U}{\Tcopier{U}} \piu \RW{P'}{\Tcopier{U}});(\LW{U}{\Tcocopier{U}} \piu \LW{U}{\bang{UP'}})) \piu \cobang{UP'} \tag{\refeq{eq:whisk:sum}, \refeq{eq:whisk:funct piu}}\\
            &= (\RW{U}{\Tcopier{U}};\LW{U}{\Tcocopier{U}}) \piu (\RW{P'}{\Tcopier{U}};\LW{U}{\bang{UP'}}) \piu \cobang{UP'} \tag{Funct. $\piu$}\\
            &= (\RW{U}{\Tcopier{U}};\LW{U}{\Tcocopier{U}}) \piu (\RW{P'}{\Tcopier{U}};\bang{UUP'}) \piu \cobang{UP'} \tag{\refeq{eq:whisk:bang}}\\
            &= (\RW{U}{\Tcopier{U}};\LW{U}{\Tcocopier{U}}) \piu \bang{UP'} \piu \cobang{UP'} \tag{\refeq{ax:bangnat}}\\
            &= (\Tcocopier{U};\Tcopier{U}) \piu \bang{UP'} \piu \cobang{UP'} \tag{\ref{ax:frob}}
        \end{align*}
    For the second part we prove the following statement 
    \begin{equation*}\tag{$\ast_2$}
        \RW{U \piu P'}{(\cobang{P'U} \piu \copier{P'});\Idl{P'}{U}{P'}} ; \LW{P'}{\cocopier{U \piu P'}} = \dl{P'}{U}{P'} ; (\bang{P'U} \piu \cocopier{P'}) ; (\cobang{P'U} \piu \copier{P'}) ; \Idl{P'}{U}{P'}
    \end{equation*}
    \begingroup
    \allowdisplaybreaks
    \begin{align*}
        &\RW{U \piu P'}{(\cobang{P'U} \piu \copier{P'});\Idl{P'}{U}{P'}} ; \LW{P'}{\cocopier{U \piu P'}} \\
        &=\RW{U \piu P'}{\cobang{P'U} \piu \copier{P'}};\RW{U \piu P'}{\Idl{P'}{U}{P'}} ; \LW{P'}{\Tcocopier{U} \piu \bang{UP'} \piu \dl{P'}{U}{P'};(\bang{P'U} \piu \cocopier{P'})} \tag{\refeq{eq:whisk:funct}, Def. $\cocopier{}$}\\
        &=\begin{multlined}[t]
            \RW{U \piu P'}{\cobang{P'U} \piu \copier{P'}};\RW{U \piu P'}{\Idl{P'}{U}{P'}} ; \\
            \dl{P'}{U(U \piu P')}{P'(U \piu P')} ; (\LW{P'}{\Tcocopier{U} \piu \bang{UP'}} \piu \LW{P'}{\dl{P'}{U}{P'};(\bang{P'U} \piu \cocopier{P'})}) ; \Idl{P'}{U}{P'}
        \end{multlined} \tag{\refeq{eq:whisk:funct piu}}\\
        &=\begin{multlined}[t]
            \RW{U \piu P'}{\cobang{P'U} \piu \copier{P'}};\RW{U \piu P'}{\Idl{P'}{U}{P'}} ; \\
            \RW{U \piu P'}{\dl{P'}{U}{P'}} ; (\LW{P'}{\Tcocopier{U} \piu \bang{UP'}} \piu \LW{P'}{\dl{P'}{U}{P'};(\bang{P'U} \piu \cocopier{P'})}) ; \Idl{P'}{U}{P'}
        \end{multlined} \tag{\refeq{eq:whisk:dl}}\\
        &=\RW{U \piu P'}{\cobang{P'U} \piu \copier{P'}};(\LW{P'}{\Tcocopier{U} \piu \bang{UP'}} \piu \LW{P'}{\dl{P'}{U}{P'};(\bang{P'U} \piu \cocopier{P'})}) ; \Idl{P'}{U}{P'} \tag{Iso}\\
        &=(\RW{U \piu P'}{\cobang{P'U}} \piu \RW{U \piu P'}{\copier{P'}});(\LW{P'}{\Tcocopier{U} \piu \bang{UP'}} \piu \LW{P'}{\dl{P'}{U}{P'};(\bang{P'U} \piu \cocopier{P'})}) ; \Idl{P'}{U}{P'} \tag{\refeq{eq:whisk:funct piu}}\\
        &=(\cobang{P'U(U \piu P')} \piu \RW{U \piu P'}{\copier{P'}});(\LW{P'}{\Tcocopier{U} \piu \bang{UP'}} \piu \LW{P'}{\dl{P'}{U}{P'};(\bang{P'U} \piu \cocopier{P'})}) ; \Idl{P'}{U}{P'} \tag{\refeq{eq:whisk:bang}}\\
        &=((\cobang{P'U(U \piu P')};\LW{P'}{\Tcocopier{U} \piu \bang{UP'}}) \piu (\RW{U \piu P'}{\copier{P'}};\LW{P'}{\dl{P'}{U}{P'};(\bang{P'U} \piu \cocopier{P'})})) ; \Idl{P'}{U}{P'} \tag{Funct. $\piu$}\\
        &=(\cobang{P'U} \piu (\RW{U \piu P'}{\copier{P'}};\LW{P'}{\dl{P'}{U}{P'};(\bang{P'U} \piu \cocopier{P'})})) ; \Idl{P'}{U}{P'} \tag{\refeq{ax:cobangnat}}\\
        &=(\cobang{P'U} \piu (\dl{P'}{U}{P'} ; (\RW{U}{\copier{P'}} \piu \RW{P'}{\copier{P'}}) ; \Idl{P'P'}{U}{P'} ;\LW{P'}{\dl{P'}{U}{P'}};\LW{P'}{(\bang{P'U} \piu \cocopier{P'})})) ; \Idl{P'}{U}{P'} \tag{\refeq{eq:whisk:sum}}\\
        &=\begin{multlined}[t]
            (\cobang{P'U} \piu (\dl{P'}{U}{P'} ; (\RW{U}{\copier{P'}} \piu \RW{P'}{\copier{P'}}) ; \Idl{P'P'}{U}{P'} ; \\
            \dl{P'P'}{U}{P'};\Idl{P'}{UP'}{P'P'};\LW{P'}{(\bang{P'U} \piu \cocopier{P'})})) ; \Idl{P'}{U}{P'}
        \end{multlined} \tag{\refeq{eq:whisk:Ldl}} \\
        &= (\cobang{P'U} \piu (\dl{P'}{U}{P'} ; (\RW{U}{\copier{P'}} \piu \RW{P'}{\copier{P'}}) ; \Idl{P'}{UP'}{P'P'};\LW{P'}{(\bang{P'U} \piu \cocopier{P'})})) ; \Idl{P'}{U}{P'} \tag{Iso}\\
        &= (\cobang{P'U} \piu (\dl{P'}{U}{P'} ; (\RW{U}{\copier{P'}} \piu \RW{P'}{\copier{P'}}) ; (\LW{P'}{\bang{P'U}} \piu \LW{P'}{\cocopier{P'}}))) ; \Idl{P'}{U}{P'} \tag{\refeq{eq:whisk:funct piu}}\\
        &= (\cobang{P'U} \piu (\dl{P'}{U}{P'} ; ((\RW{U}{\copier{P'}};\LW{P'}{\bang{P'U}}) \piu (\RW{P'}{\copier{P'}};\LW{P'}{\cocopier{P'}})) )) ; \Idl{P'}{U}{P'} \tag{Funct. $\piu$}\\
        &= (\cobang{P'U} \piu (\dl{P'}{U}{P'} ; ((\RW{U}{\copier{P'}};\bang{P'P'U}) \piu (\RW{P'}{\copier{P'}};\LW{P'}{\cocopier{P'}})) )) ; \Idl{P'}{U}{P'}\tag{\refeq{eq:whisk:bang}} \\
        &= (\cobang{P'U} \piu (\dl{P'}{U}{P'} ; (\bang{P'U} \piu (\RW{P'}{\copier{P'}};\LW{P'}{\cocopier{P'}})) )) ; \Idl{P'}{U}{P'} \tag{\refeq{ax:bangnat}}\\
        &= (\cobang{P'U} \piu (\dl{P'}{U}{P'} ; (\bang{P'U} \piu (\cocopier{P'};\copier{P'})) )) ; \Idl{P'}{U}{P'} \tag{Ind. hp.}\\
        &= \dl{P'}{U}{P'} ; (\bang{P'U} \piu \cocopier{P'}) ; (\cobang{P'U} \piu \copier{P'}) ; \Idl{P'}{U}{P'} \tag{Funct. $\piu$}
    \end{align*}
    \endgroup
    We conclude the proof by showing that
    \begin{align*}
        &\RW{U \piu P'}{\copier{U \piu P'}} ; \LW{U \piu P'}{\cocopier{U \piu P'}} \\
        &=\RW{U \piu P'}{\Tcopier{U} \piu \cobang{UP'} \piu (\cobang{P'U} \piu \copier{P'});\Idl{P'}{U}{P'}} ; \LW{U \piu P'}{\cocopier{U \piu P'}} \tag{Def. $\copier{}$} \\
        &=(\RW{U \piu P'}{\Tcopier{U} \piu \cobang{UP'}} \piu \RW{U \piu P'}{(\cobang{P'U} \piu \copier{P'});\Idl{P'}{U}{P'}}); (\LW{U}{\cocopier{U \piu P'}} \piu \LW{P'}{\cocopier{U \piu P'}}) \tag{\refeq{eq:whisk:funct piu}, \refeq{eq:whisk:sum}}\\
        &=(\RW{U \piu P'}{\Tcopier{U} \piu \cobang{UP'}} ; \LW{U}{\cocopier{U \piu P'}}) \piu (\RW{U \piu P'}{(\cobang{P'U} \piu \copier{P'});\Idl{P'}{U}{P'}} ; \LW{P'}{\cocopier{U \piu P'}}) \tag{Funct. $\piu$}\\
        &=(\Tcocopier{U};\Tcopier{U}) \piu (\bang{UP'};\cobang{UP'}) \piu (\dl{P'}{U}{P'} ; (\bang{P'U} \piu \cocopier{P'}) ; (\cobang{P'U} \piu \copier{P'}) ; \Idl{P'}{U}{P'}) \tag{$\ast_1, \ast_2$}\\
        &=(\Tcocopier{U} \piu \bang{UP'} \piu \dl{P'}{U}{P'};(\bang{P'U} \piu \copier{P'})) ; (\Tcopier{U} \piu \cobang{UP'} \piu (\cobang{P'U} \piu \copier{P'});\Idl{P'}{U}{P'}) \tag{Funct. $\piu$}\\
        &=\cocopier{U \piu P'};\copier{U \piu P'} \tag{Def. $\cocopier{}, \copier{}$}
    \end{align*}
as required.
\end{proof}

\begin{lemma}[Lax comonoid homomorphisms]
    The following hold for any $\t \colon P \to Q$
    \[ \t ; \discharger{Q} \leq \discharger{P} \qquad\quad  \t ; \copier{Q} \leq \copier{P} ; (\t \per \t) \]
\end{lemma}
\begin{proof} \vphantom{.}

    \textsc{Equation $\t ; \discharger{Q} \leq \discharger{P}$}: by induction on $\t$.

    \textbf{Case $\t = \id\zero$:} $\id\zero ; \discharger{\zero} \axeq{\text{def. }\discharger{}} \id\zero ; \cobang{\uno} = \cobang{\uno} \axeq{\text{def. }\discharger{}} \discharger{\zero}$

    \textbf{Case $\t = \tape{c}$:} Suppose $c \colon U \to V$, then $\tape{c} ; \discharger{V} \axeq{\text{def. }\discharger{}} \begin{tikzpicture}
        \begin{pgfonlayer}{nodelayer}
            \node [style=none] (73) at (1.75, 0.5) {};
            \node [style=none] (74) at (1.75, -0.5) {};
            \node [style=label] (76) at (-1.5, 0) {$U$};
            \node [style=none] (77) at (-1, -0.5) {};
            \node [style=none] (78) at (-1, 0.5) {};
            \node [style=black] (80) at (1, 0) {};
            \node [style=none] (84) at (0.35, 0) {};
            \node [style=none] (85) at (-0.35, 0) {};
            \node [style=none] (86) at (-1, 0) {};
            \node [style=none] (87) at (-0.35, 0.35) {};
            \node [style=none] (88) at (-0.35, -0.35) {};
            \node [style=none] (89) at (0.35, -0.35) {};
            \node [style=none] (90) at (0.35, 0.35) {};
            \node [style=none] (91) at (0, 0) {$c$};
        \end{pgfonlayer}
        \begin{pgfonlayer}{edgelayer}
            \draw [style=tape] (74.center)
                 to (73.center)
                 to (78.center)
                 to (77.center)
                 to cycle;
            \draw (84.center) to (80);
            \draw (85.center) to (86.center);
            \draw [fill=white] (90.center)
                 to (87.center)
                 to (88.center)
                 to (89.center)
                 to cycle;
        \end{pgfonlayer}
    \end{tikzpicture} \stackrel{\eqref{ax:dischargernat}}{\leq} \Tdischarger{U} \axeq{\text{def. }\discharger{}} \discharger{U}$

    \textbf{Case $\t = \symmp{U}{V}$:}
    \[ 
    \symmp{U}{V} ; \discharger{V \piu U} 
    \axeq{\text{def. }\discharger{}} 
    \begin{tikzpicture}
        \begin{pgfonlayer}{nodelayer}
            \node [style=black] (1) at (1.5, 1) {};
            \node [style=black] (2) at (1.5, -1) {};
            \node [style=label] (5) at (-2.1, 1) {$U$};
            \node [style=none] (6) at (4, 0.5) {};
            \node [style=none] (7) at (4, -0.5) {};
            \node [style=none] (8) at (1.5, -1.5) {};
            \node [style=none] (9) at (1.5, 1.5) {};
            \node [style=none] (10) at (1.5, 0.5) {};
            \node [style=none] (11) at (2, 0) {};
            \node [style=none] (12) at (1.5, -0.5) {};
            \node [style=none] (13) at (3, -0.5) {};
            \node [style=none] (14) at (3, 0.5) {};
            \node [style=none] (17) at (-1.5, 1) {};
            \node [style=none] (18) at (-1.5, 1.5) {};
            \node [style=none] (19) at (-1.5, 0.5) {};
            \node [style=none] (20) at (-1.5, -1) {};
            \node [style=none] (21) at (-1.5, -0.5) {};
            \node [style=none] (22) at (-1.5, -1.5) {};
            \node [style=label] (23) at (-2.1, -1) {$V$};
        \end{pgfonlayer}
        \begin{pgfonlayer}{edgelayer}
            \draw [style=tape] (22.center)
                 to (21.center)
                 to [in=180, out=0, looseness=0.75] (9.center)
                 to [bend left] (14.center)
                 to (6.center)
                 to (7.center)
                 to (13.center)
                 to [bend left] (8.center)
                 to [in=0, out=180, looseness=0.75] (19.center)
                 to (18.center)
                 to [in=180, out=0, looseness=0.75] (12.center)
                 to [bend right=45] (11.center)
                 to [bend right=45] (10.center)
                 to [in=0, out=180, looseness=0.75] cycle;
            \draw [in=0, out=180, looseness=0.75] (2) to (17.center);
            \draw [in=0, out=180, looseness=0.75] (1) to (20.center);
        \end{pgfonlayer}
    \end{tikzpicture}
    \axeq{\refeq{ax:symmpnat}}
    \begin{tikzpicture}
        \begin{pgfonlayer}{nodelayer}
            \node [style=black] (1) at (-1, -1) {};
            \node [style=black] (2) at (-1, 1) {};
            \node [style=label] (5) at (-2.1, 1) {$U$};
            \node [style=none] (6) at (4, 0.5) {};
            \node [style=none] (7) at (4, -0.5) {};
            \node [style=none] (8) at (1.5, -1.5) {};
            \node [style=none] (9) at (1.5, 1.5) {};
            \node [style=none] (10) at (1.5, 0.5) {};
            \node [style=none] (11) at (2, 0) {};
            \node [style=none] (12) at (1.5, -0.5) {};
            \node [style=none] (13) at (3, -0.5) {};
            \node [style=none] (14) at (3, 0.5) {};
            \node [style=none] (17) at (-1.5, 1) {};
            \node [style=none] (18) at (-1.5, 1.5) {};
            \node [style=none] (19) at (-1.5, 0.5) {};
            \node [style=none] (20) at (-1.5, -1) {};
            \node [style=none] (21) at (-1.5, -0.5) {};
            \node [style=none] (22) at (-1.5, -1.5) {};
            \node [style=label] (23) at (-2.1, -1) {$V$};
        \end{pgfonlayer}
        \begin{pgfonlayer}{edgelayer}
            \draw [style=tape] (22.center)
                 to (21.center)
                 to [in=180, out=0, looseness=0.75] (9.center)
                 to [bend left] (14.center)
                 to (6.center)
                 to (7.center)
                 to (13.center)
                 to [bend left] (8.center)
                 to [in=0, out=180, looseness=0.75] (19.center)
                 to (18.center)
                 to [in=180, out=0, looseness=0.75] (12.center)
                 to [bend right=45] (11.center)
                 to [bend right=45] (10.center)
                 to [in=0, out=180, looseness=0.75] cycle;
            \draw [in=0, out=180, looseness=0.75] (2) to (17.center);
            \draw [in=0, out=180, looseness=0.75] (1) to (20.center);
        \end{pgfonlayer}
    \end{tikzpicture}
    \axeq{\refeq{ax:codiagco}}
    \begin{tikzpicture}
        \begin{pgfonlayer}{nodelayer}
            \node [style=black] (1) at (-0.5, -1) {};
            \node [style=black] (2) at (-0.5, 1) {};
            \node [style=label] (5) at (-1.6, 1) {$U$};
            \node [style=none] (6) at (1.5, 0.5) {};
            \node [style=none] (7) at (1.5, -0.5) {};
            \node [style=none] (11) at (-0.5, 0) {};
            \node [style=none] (13) at (0.5, -0.5) {};
            \node [style=none] (14) at (0.5, 0.5) {};
            \node [style=none] (17) at (-1, 1) {};
            \node [style=none] (18) at (-1, 1.5) {};
            \node [style=none] (19) at (-1, 0.5) {};
            \node [style=none] (20) at (-1, -1) {};
            \node [style=none] (21) at (-1, -0.5) {};
            \node [style=none] (22) at (-1, -1.5) {};
            \node [style=label] (23) at (-1.6, -1) {$V$};
        \end{pgfonlayer}
        \begin{pgfonlayer}{edgelayer}
            \draw [style=tape] (13.center)
                 to [bend left] (22.center)
                 to (21.center)
                 to [bend right=45] (11.center)
                 to [bend right=45] (19.center)
                 to (18.center)
                 to [bend left] (14.center)
                 to (6.center)
                 to (7.center)
                 to cycle;
            \draw [in=0, out=180, looseness=0.75] (2) to (17.center);
            \draw [in=0, out=180, looseness=0.75] (1) to (20.center);
        \end{pgfonlayer}
    \end{tikzpicture}
    \axeq{\text{def. }\discharger{}} 
    \discharger{U \piu V}        
    \]

    \textbf{Case $\t = \diag{U}$:} (similarly for $\codiag{U}$)
    \[ \diag{U} ; \discharger{U \piu U} 
    \axeq{\text{def. }\discharger{}} 
    \begin{tikzpicture}
        \begin{pgfonlayer}{nodelayer}
            \node [style=none] (6) at (2.75, 0.5) {};
            \node [style=none] (7) at (2.75, -0.5) {};
            \node [style=none] (11) at (0.75, 0) {};
            \node [style=none] (13) at (1.75, -0.5) {};
            \node [style=none] (14) at (1.75, 0.5) {};
            \node [style=none] (18) at (0.25, 1.5) {};
            \node [style=none] (19) at (0.25, 0.5) {};
            \node [style=none] (21) at (0.25, -0.5) {};
            \node [style=none] (22) at (0.25, -1.5) {};
            \node [style=label] (23) at (-3.35, 0) {$U$};
            \node [style=none] (25) at (-1.25, 0) {};
            \node [style=none] (26) at (-2.75, 0.5) {};
            \node [style=none] (27) at (-2.75, -0.5) {};
            \node [style=none] (28) at (-0.75, 0) {};
            \node [style=none] (29) at (-1.75, -0.5) {};
            \node [style=none] (30) at (-1.75, 0.5) {};
            \node [style=black] (31) at (-0.25, 1) {};
            \node [style=none] (32) at (-0.25, 1.5) {};
            \node [style=none] (33) at (-0.25, 0.5) {};
            \node [style=black] (34) at (-0.25, -1) {};
            \node [style=none] (35) at (-0.25, -0.5) {};
            \node [style=none] (36) at (-0.25, -1.5) {};
            \node [style=none] (37) at (-2.75, 0) {};
        \end{pgfonlayer}
        \begin{pgfonlayer}{edgelayer}
            \draw [style=tape] (6.center)
                 to (7.center)
                 to (13.center)
                 to [bend left] (22.center)
                 to (36.center)
                 to [bend left] (29.center)
                 to (27.center)
                 to (26.center)
                 to (30.center)
                 to [bend left] (32.center)
                 to (18.center)
                 to [bend left] (14.center)
                 to cycle;
            \draw [style=tapeNoFill, fill=white] (19.center)
                 to (33.center)
                 to [bend right=45] (28.center)
                 to [bend right=45] (35.center)
                 to (21.center)
                 to [bend right=45] (11.center)
                 to [bend right=45] cycle;
            \draw [bend left=45] (25.center) to (31);
            \draw (37.center) to (25.center);
            \draw [bend right=45] (25.center) to (34);
        \end{pgfonlayer}
    \end{tikzpicture}
    \axeq{\refeq{ax:diagnat}}
    \begin{tikzpicture}
        \begin{pgfonlayer}{nodelayer}
            \node [style=none] (6) at (2.75, 0.5) {};
            \node [style=none] (7) at (2.75, -0.5) {};
            \node [style=none] (11) at (0.75, 0) {};
            \node [style=none] (13) at (1.75, -0.5) {};
            \node [style=none] (14) at (1.75, 0.5) {};
            \node [style=none] (18) at (0.25, 1.5) {};
            \node [style=none] (19) at (0.25, 0.5) {};
            \node [style=none] (21) at (0.25, -0.5) {};
            \node [style=none] (22) at (0.25, -1.5) {};
            \node [style=label] (23) at (-3.35, 0) {$U$};
            \node [style=black] (25) at (-2, 0) {};
            \node [style=none] (26) at (-2.75, 0.5) {};
            \node [style=none] (27) at (-2.75, -0.5) {};
            \node [style=none] (28) at (-0.75, 0) {};
            \node [style=none] (29) at (-1.75, -0.5) {};
            \node [style=none] (30) at (-1.75, 0.5) {};
            \node [style=none] (32) at (-0.25, 1.5) {};
            \node [style=none] (33) at (-0.25, 0.5) {};
            \node [style=none] (35) at (-0.25, -0.5) {};
            \node [style=none] (36) at (-0.25, -1.5) {};
            \node [style=none] (37) at (-2.75, 0) {};
        \end{pgfonlayer}
        \begin{pgfonlayer}{edgelayer}
            \draw [style=tape] (6.center)
                 to (7.center)
                 to (13.center)
                 to [bend left] (22.center)
                 to (36.center)
                 to [bend left] (29.center)
                 to (27.center)
                 to (26.center)
                 to (30.center)
                 to [bend left] (32.center)
                 to (18.center)
                 to [bend left] (14.center)
                 to cycle;
            \draw [style=tapeNoFill, fill=white] (19.center)
                 to (33.center)
                 to [bend right=45] (28.center)
                 to [bend right=45] (35.center)
                 to (21.center)
                 to [bend right=45] (11.center)
                 to [bend right=45] cycle;
            \draw (37.center) to (25);
        \end{pgfonlayer}
    \end{tikzpicture}    
    \axsubeq{\refeq{ax:relDiagCodiag}}
    \Tdischarger{U}    
     \]

     \textbf{Case $\t = \bang{U}$:} (similarly for $\cobang{U}$) $\begin{tikzpicture}
        \begin{pgfonlayer}{nodelayer}
            \node [style=none] (6) at (2, 0.5) {};
            \node [style=none] (7) at (2, -0.5) {};
            \node [style=none] (13) at (0.75, -0.5) {};
            \node [style=none] (14) at (0.75, 0.5) {};
            \node [style=label] (23) at (-2.6, 0) {$U$};
            \node [style=none] (25) at (-0.25, 0) {};
            \node [style=none] (26) at (-2, 0.5) {};
            \node [style=none] (27) at (-2, -0.5) {};
            \node [style=none] (29) at (-0.75, -0.5) {};
            \node [style=none] (30) at (-0.75, 0.5) {};
            \node [style=none] (37) at (-2, 0) {};
        \end{pgfonlayer}
        \begin{pgfonlayer}{edgelayer}
            \draw [style=tape] (6.center)
                 to (14.center)
                 to [bend right=90, looseness=1.75] (13.center)
                 to (7.center)
                 to cycle;
            \draw [style=tape] (26.center)
                 to (30.center)
                 to [bend left=90, looseness=1.75] (29.center)
                 to (27.center)
                 to cycle;
            \draw (37.center) to (25.center);
        \end{pgfonlayer}
    \end{tikzpicture}
    \axeq{\refeq{ax:cobangnat}}
    \begin{tikzpicture}
        \begin{pgfonlayer}{nodelayer}
            \node [style=none] (6) at (2, 0.5) {};
            \node [style=none] (7) at (2, -0.5) {};
            \node [style=none] (13) at (0.75, -0.5) {};
            \node [style=none] (14) at (0.75, 0.5) {};
            \node [style=label] (23) at (-2.6, 0) {$U$};
            \node [style=none] (25) at (-0.25, 0) {};
            \node [style=none] (26) at (-2, 0.5) {};
            \node [style=none] (27) at (-2, -0.5) {};
            \node [style=none] (29) at (-0.75, -0.5) {};
            \node [style=none] (30) at (-0.75, 0.5) {};
            \node [style=none] (37) at (-2, 0) {};
            \node [style=none] (38) at (0.25, 0) {};
            \node [style=black] (39) at (1.25, 0) {};
        \end{pgfonlayer}
        \begin{pgfonlayer}{edgelayer}
            \draw [style=tape] (6.center)
                 to (14.center)
                 to [bend right=90, looseness=1.75] (13.center)
                 to (7.center)
                 to cycle;
            \draw [style=tape] (26.center)
                 to (30.center)
                 to [bend left=90, looseness=1.75] (29.center)
                 to (27.center)
                 to cycle;
            \draw (37.center) to (25.center);
            \draw (38.center) to (39);
        \end{pgfonlayer}
    \end{tikzpicture}
    \axsubeq{\refeq{ax:relBangCobang}}
    \Tdischarger{U}$

    \textbf{Case $\t = \t_1 ; \t_2$:} Suppose $\t_1 \colon P \to Q, t_2 \colon Q \to R$, then
    \[ \t_1 ; \t_2 ; \discharger{R} \axsubeq{\text{ind. hp.}} \t_1 ; \discharger{Q} \axsubeq{\text{ind. hp.}} \discharger{P} \]

    \textbf{Case $\t = \t_1 \piu \t_2$:} Suppose $\t_1 \colon P_1 \to Q_1, t_2 \colon P_2 \to Q_2$, then
    \begin{align*}
        (\t_1 \piu \t_2) ; \discharger{Q_1 \piu Q_2} &\axeq{\text{Lemma}~\refeq{lm:discharger}} (\t_1 \piu \t_2) ; (\discharger{Q_1} \piu \discharger{Q_2}) ; \codiag{\uno} \axeq{\text{funct. }\piu} ((\t_1 ; \discharger{Q_1}) \piu (\t_2 ; \discharger{Q_2})) ; \codiag\uno \\
        &\axsubeq{\text{ind. hp.}} (\discharger{P_1} \piu \discharger{P_2}) ; \codiag{\uno} \axeq{\text{Lemma}~\refeq{lm:discharger}} \discharger{P_1 \piu P_2}
    \end{align*}
    
    \textsc{Equation $\t ; \copier{Q} \leq \copier{P} ; (\t \per \t)$}: by induction on $\t$.

    \textbf{Case $\t = \id\zero$:} $\id\zero ; \copier{\zero} \axeq{\text{def. }\copier{}} \id\zero ; \id\zero \axeq{\text{def. }\copier{}} \copier{\zero} ; \id\zero = \copier{\zero} ; (\id\zero \per \id \zero) $

    \textbf{Case $\t = \tape{c}$:} Suppose $c \colon U \to V$, then
    \[ \tape{c} ; \copier{V} 
    \axeq{\text{def. }\copier{}} 

        \axeq{\text{def. }\copier{}} 
        \copier{U} ; (\diag{U} \per \diag{U})      
    \end{align*}

     \textbf{Case $\t = \bang{U}$:} (similarly for $\cobang{U}$)
     \[ \bang{U} ; \copier{\zero} \axeq{\text{def. }\copier{}}  \bang{U} ; \id\zero = \bang{U} \axeq{\refeq{ax:bangnat}} \copier{U} ; \bang{UU} \]

    \textbf{Case $\t = \t_1 ; \t_2$:} Suppose $\t_1 \colon P \to Q, t_2 \colon Q \to R$, then
    \[ t_1 ; \t_2 ; \copier{R} \axsubeq{\text{ind. hp.}} \t_1 ; \copier{Q} ; (\t_2 \per \t_2) \axsubeq{\text{ind. hp.}} \copier{P} ; (\t_1 \per \t_1) ; (\t_2 \per \t_2) \axeq{\text{funct. }\per}  \copier{P} ; ((\t_1 ; \t_2) \per (\t_1 ; \t_2)) \]

    \textbf{Case $\t = \t_1 \piu \t_2$:} Suppose $\t_1 \colon P_1 \to Q_1, t_2 \colon P_2 \to Q_2$, then
    \begingroup
    \allowdisplaybreaks
    \begin{align*}
        &\copier{P_1 \piu P_2} ; ((\t_1 \piu \t_2) \per (\t_1 \piu \t_2)) \\
        &=\copier{P_1 \piu P_2} ; \dl{P_1 \piu P_2}{P_1}{P_2} ; (((\t_1 \piu \t_2) \per \t_1) \piu ((\t_1 \piu \t_2) \per \t_2)) ; \Idl{Q_1 \piu Q_2}{Q_1}{Q_2} \tag{Nat. $\delta^l$}\\
        &=\copier{P_1 \piu P_2} ; \dl{P_1 \piu P_2}{P_1}{P_2} ; (\t_1\t_1 \piu \t_2\t_1 \piu \t_1\t_2 \piu \t_2\t_2) ; \Idl{Q_1 \piu Q_2}{Q_1}{Q_2} \tag{Theorem~\refeq{thm:taperig}}\\
        &=((\copier{P_1} \piu \cobang{P_1P_2}) ; \Idl{P_1}{P_1}{P_2} \piu (\cobang{P_2P_1} \piu \copier{P_2}) ; \Idl{P_2}{P_1}{P_2}) ; \dl{P_1 \piu P_2}{P_1}{P_2} ; (\t_1\t_1 \piu \t_2\t_1 \piu \t_1\t_2 \piu \t_2\t_2) ; \Idl{Q_1 \piu Q_2}{Q_1}{Q_2} \tag{Lemma~\refeq{lm:copier}}\\
        &=(\copier{P_1} \piu \cobang{P_1P_2} \piu \cobang{P_2P_1} \piu \copier{P_2}) ; (\Idl{P_1}{P_1}{P_2} \piu \Idl{P_2}{P_1}{P_2}) ; \dl{P_1 \piu P_2}{P_1}{P_2} ; (\t_1\t_1 \piu \t_2\t_1 \piu \t_1\t_2 \piu \t_2\t_2) ; \Idl{Q_1 \piu Q_2}{Q_1}{Q_2} \tag{Funct. $\piu$}\\
        &=(\copier{P_1} \piu \cobang{P_1P_2} \piu \cobang{P_2P_1} \piu \copier{P_2}) ; (\id{P_1P_1} \piu \symmp{P_1P_2}{P_2P_1} \piu \id{P_2P_2}) ; (\t_1\t_1 \piu \t_2\t_1 \piu \t_1\t_2 \piu \t_2\t_2) ; \Idl{Q_1 \piu Q_2}{Q_1}{Q_2} \tag{Lemma~\refeq{lemma:tapesymdis}.2}\\
        &=(\copier{P_1} \piu \cobang{P_1P_2} \piu \cobang{P_2P_1} \piu \copier{P_2}) ; (\t_1\t_1 \piu \t_1\t_2 \piu \t_2\t_1 \piu \t_2\t_2) ; (\id{Q_1Q_1} \piu \symmp{Q_1Q_2}{Q_2Q_1} \piu \id{Q_2Q_2}) ; \Idl{Q_1 \piu Q_2}{Q_1}{Q_2} \tag{\refeq{ax:symmpnat}}\\
        &=(\copier{P_1};\t_1\t_1 \piu \cobang{P_1P_2};\t_1\t_2 \piu \cobang{P_2P_1};\t_2\t_1 \piu \copier{P_2};\t_2\t_2) ; (\id{Q_1Q_1} \piu \symmp{Q_1Q_2}{Q_2Q_1} \piu \id{Q_2Q_2}) ; \Idl{Q_1 \piu Q_2}{Q_1}{Q_2} \tag{Funct. $\piu$}\\
        &=(\copier{P_1};\t_1\t_1 \piu \cobang{Q_1Q_2} \piu \cobang{Q_2Q_1} \piu \copier{P_2};\t_2\t_2) ; (\id{Q_1Q_1} \piu \symmp{Q_1Q_2}{Q_2Q_1} \piu \id{Q_2Q_2}) ; \Idl{Q_1 \piu Q_2}{Q_1}{Q_2} \tag{\refeq{ax:cobangnat}}\\
        &\geq (\t_1;\copier{Q_1} \piu \cobang{Q_1Q_2} \piu \cobang{Q_2Q_1} \piu \t_2;\copier{Q_2}) ; (\id{Q_1Q_1} \piu \symmp{Q_1Q_2}{Q_2Q_1} \piu \id{Q_2Q_2}) ; \Idl{Q_1 \piu Q_2}{Q_1}{Q_2} \tag{Ind. hp.}\\
        &= (\t_1;\copier{Q_1} \piu \cobang{Q_1Q_2} \piu \cobang{Q_2Q_1} \piu \t_2;\copier{Q_2}) ; (\Idl{Q_1}{Q_1}{Q_2} \piu \Idl{Q_2}{Q_1}{Q_2}) \tag{Lemma~\refeq{lemma:tapesymdis}.2} \\
        &= (\t_1 \piu \t_2);(\copier{Q_1} \piu \cobang{Q_1Q_2} \piu \cobang{Q_2Q_1} \piu \copier{Q_2}) ; (\Idl{Q_1}{Q_1}{Q_2} \piu \Idl{Q_2}{Q_1}{Q_2}) \tag{Funct. $\piu$} \\
        &= (\t_1 \piu \t_2);((\copier{Q_1} \piu \cobang{Q_1Q_2});\Idl{Q_1}{Q_1}{Q_2} \piu (\cobang{Q_2Q_1} \piu \copier{Q_2});\Idl{Q_2}{Q_1}{Q_2}\tag{Funct. $\piu$})\\
        &= (\t_1 \piu \t_2);\copier{Q_1 \piu Q_2} \tag{Def. $\copier{}$}\\
    \end{align*}
    \endgroup
    as required.
\end{proof}

\subsection{Proofs of Proposition \ref{prop:bijective}}\label{sec:appCB2}

We start by showing that an arbitrary morphism of strict cartesian bicategories $\mathcal{M} \colon \CB \to \sRel$ can be extended to a morphism of sesquistrict fb-cb rig categories $\mathcal{M}^\sharp \colon \CatTapeCB \to \sRel$.

\begin{lemma}\label{lemma:extending functors from CB to TCB}
	Let $\mathcal{M}\colon \CB \to \sRel$ be a morphism of cartesian bicategories. There exists a morphism of sesquistrict fb-cb rig categories $\mathcal{M}^\sharp \colon \CatTapeCB \to \sRel$ such that $\mathcal{M}^\sharp{(\tape{c})}=\mathcal{M}(c)$ for all $c\in \CB$.
\end{lemma}
\begin{proof}
	Recall our convention following Lemma~\ref{lemma:every functor in strict(D) is well behaved}: we assume that $\mathcal{M}\colon \CB \to \sRelT$. We define $\mathcal{M}^\sharp \colon \CatTapeCB \to \sRelT$ as follows. On objects, $\mathcal {M}^\sharp (\zero) = \zero$, $\mathcal {M}^\sharp (U \piu P) = \mathcal M (U) \piu \mathcal {M}^\sharp (P)$. On morphisms:
	\begin{align*}
		\mathcal{M}^\sharp(\id{U}) &= \id{\mathcal M (U)} & \mathcal{M}^\sharp(\id{\zero}) &= \id{\zero} & \mathcal{M}^\sharp(\tape{c}) &= \mathcal{M}(c) & \mathcal{M}^\sharp(\symmp{U}{U'}) &= \symmp{\mathcal M (U)}{\mathcal M (U')} \\
		\mathcal{M}^\sharp(\codiag{U})&=  \codiag{\mathcal M (U)} & \mathcal{M}^\sharp(\cobang{U}) &= \cobang{\mathcal M (U)}  & \mathcal{M}^\sharp(\diag{U})&= \diag{\mathcal M (U)} & \mathcal{M}^\sharp(\bang{U}) &= \bang{\mathcal M (U)} 
	\end{align*}
	\[
	\mathcal{M}^\sharp(\s;\t) = \mathcal{M}^\sharp(\s) ; \mathcal{M}^\sharp(\t) \qquad \mathcal{M}^\sharp(\s \piu \t) = \mathcal{M}^\sharp(\s) \piu \mathcal{M}^\sharp(\t)
	\]

By definition $\mathcal{M}^\sharp$ is an fb-morphism. Preservation of the rest of the rig structure follows from the inductive definitions of $\dl{P}{Q}{R}$, $\symmp$ and $\per$, which all use the $\piu$-structure and the internal tensor of $\CatString$, that are indeed preserved by $\mathcal{M}^\sharp$. The proof is similar to that of Theorem~\ref{thm:Tapes is free sesquistrict generated by sigma}, in Appendix~\ref{sc:appendixTapes}.
Hence $\mathcal{M}^\sharp$ is a sesquistrict fb rig functor. 
	
	One can check that $\mathcal{M}^\sharp$ preserves the cartesian bicategory structure of $\CatTapeCB$ by induction, knowing that it preserves $\bang{}$, $\cobang{}$, $\delta^l$ and $\piu$ and using the coherence axioms of Definition~\ref{def:fbcbrig}.\ref{fbcb coherence conditions}.
\end{proof}

\begin{proof}[Proof of Proposition \ref{prop:bijective}]
	Recall our convention following Lemma~\ref{lemma:every functor in strict(D) is well behaved}: in the statement we are more precisely interested in morphisms of cartesian bicategories $\CB \to \sRelT$ and of sesquistrict fb-cb rig categories $\CatTapeCB \to \sRelT$. Table~\ref{tabSemantic} provides the recipe for the direction $\eqref{item1} \to \eqref{item2}$. Lemma~\ref{lemma:extending functors from CB to TCB} takes care of $\eqref{item2} \to \eqref{item3}$. For $\eqref{item3} \to \eqref{item1}$, consider $F \colon \CatTapeCB \to \sRelT$ a sesquistrict fb-cb rig functor. Then $F$ induces an interpretation $\interpretation_F = (\alpha_\sort,\alpha_\sign)$ given by
	\[
	\begin{tikzcd}[row sep=0em]
		\sort \ar[r,"\alpha_\sort"] & Ob(\Rel) \\
		A \ar[r,|->] & F(A)
	\end{tikzcd} 
	\qquad
	\begin{tikzcd}[row sep=0em]
		\sign \ar[r,"\alpha_\sign"] & Ar(\Rel) \\
		f \ar[r,|->] & F(\tape f)
	\end{tikzcd}
	\]
	We show that $\interpretation_F$ is actually an interpretation. First of all, notice that since $F$ is sesquistrict it sends elements of $\sort$ into objects of $\Rel$, seen as unary words of unary words in $\sRelT$, so $\alpha_\sort$ type-checks. Next, suppose $\ar(f) = U$ with $U=A_1\dots A_n$, then the domain of $\alpha_\sign (f) = F(\tape f)$ in $\sRel$ is the unary word consisting of the word $F(U)$, where $F(U)=F(A_1) \dots F(A_n)$ because $F$ is strict. By definition of $\sRelT$, we have that $F(\tape f)$ is an actual arrow of $\Rel$ whose domain (in $\Rel$) is the right-bracketing of the set $F(A_1) \per \dots F(A_n)$, which is indeed $\alpha^\sharp_\sort(A_1\dots A_n)$.
	
	It is very easy to see that the functions $\eqref{item1} \to \eqref{item2} \to \eqref{item3}$ are injective, while $\eqref{item3} \to \eqref{item1}$ is also injective because $F$ is completely determined by its action on $\sort$ and $\sign$, hence we conclude.
\end{proof}

\subsection{Proofs of Theorem \ref{thm:completeness} (Completeness)}\label{app:Completeness} 

We prove in detail the five points listed in the proof of Theorem~\ref{thm:completeness} in the main text. We recall first the following result from \cite{GCQ}.

\begin{lemma}[From \cite{GCQ}]\label{lemma:CBprevious}
	For all $c\in \CB[U,V]$, there exist a morphism $\mathcal{U}_c\colon \CB \to \Rel$ and an element $(\iota,\omega)\in \mathcal{U}_c(c)$ such that
	for all $d\in \CB[U,V]$, if $(\iota,\omega)\in \mathcal{U}_c(d)$ then $c \minorinCB d$.
\end{lemma}

The following five results are the points listed in the proof in the main text.

\begin{lemma}\label{Lemma:Unions}
	For all morphisms $\mathcal{M}\colon \CatTapeCB \to \Rel$ and $c,d\in \CB[U,V]$, $\mathcal{M}(\tape{c} + \tape{d}) =  \mathcal{M}(\tape{c}) \cup \mathcal{M}(\tape{d})$.
\end{lemma}
\begin{proof} \begin{align*}
		\mathcal{M}(\tape{c} + \tape{d}) &= \mathcal{M}(\diag{U} ; (\tape{c} \piu \tape{d}) ; \codiag{V}) \tag{def. +}\\
		& = \mathcal{M}(\diag{U}) ; (\mathcal{M}(\tape{c}) \piu \mathcal{M}(\tape{d})) ; \mathcal{M}(\codiag{V}) \tag{funct $\mathcal{M}$}\\
		& = \diag{\mathcal{M}{U}} ; (\mathcal{M}(\tape{c}) \piu \mathcal{M}(\tape{d})) ; \codiag{\mathcal{M}{V}} \tag{$\mathcal{M}$ fb-morphism}\\
		& =\mathcal{M}(\tape{c}) \cup \mathcal{M}(\tape{d}) \tag{Def. $\diag{U}$ $\codiag{V}$ in $\Rel$}
	\end{align*}
\end{proof}

\begin{lemma}\label{lemma:key}
	Let $c,d_1, \dots, d_m\in \CB[U,V]$.
	If for all $\mathcal{M}\colon \CatTapeCB \to \Rel$, $\mathcal{M}(\tape{c}) \leq \mathcal{M}(\sum_{j=1}^{m} \tape{d_j})$, then there exists a $j$ such that $c \minorinCB d_j$. 
\end{lemma}
\begin{proof}
	Let $\mathcal{U}_c\colon \CB \to \Rel$ be the morphism from Lemma \ref{lemma:CBprevious} and  $\mathcal{U}^{\sharp}_c\colon \CatTapeCB \to \Rel$ its extension to $\CatTapeCB$ (Lemma \ref{lemma:extending functors from CB to TCB}).
	By assumption, we have that $\mathcal{U}^{\sharp}_c(\tape{c}) \leq \mathcal{U}^{\sharp}_c((\sum_{j=1}^{m} \tape{d_j})$. By Lemma \ref{Lemma:Unions} this holds iff 
	$\mathcal{U}_c(c) \leq\bigcup_{j=1}^{m}  \mathcal{U}_c(d_j)$. Therefore, the element $(\iota,\omega)\in\mathcal{U}_c(c)$ from Lemma \ref{lemma:CBprevious} should belong to at least one $\mathcal{U}_c(d_j)$.
	By Lemma \ref{lemma:CBprevious}, $c \minorinCB d_j$ follows immediately. 
\end{proof}

\begin{lemma}
	Let $c_1,\dots c_n,d_1, \dots, d_m\in \CB[U,V]$.
	If for all $\mathcal{M}\colon \CatTapeCB \to \Rel$, $\mathcal{M}(\sum_{i=1}^{n} \tape{c_i}) \leq \mathcal{M}(\sum_{j=1}^{m} \tape{d_j})$, then for all $i$, there exists a $j$ such that $c_i \minorinCB d_j$. 
\end{lemma}
\begin{proof}
	By the hypothesis and Lemma \ref{Lemma:Unions}, one has that $\bigcup_{i=1}^{n}  \mathcal{M}(\tape{c_i}) \leq  \mathcal{M}(\sum_{j=1}^{m} \tape{d_j})$. Thus for all $i$,  $\mathcal{M}(\tape{c_i}) \leq \mathcal{M}(\sum_{j=1}^{m} \tape{d_j})$ that, by Lemma \ref{lemma:key}, entails that there exists a $j$ such that $c_i \minorinCB d_j$. 
\end{proof}

From the above result and Corollary \ref{cor:poset}, it follows immediately that:
\begin{corollary}\label{cor:found}
	Let $\s, \t$ be arrows in $\monomial$. If, for all morphisms of fb-cb categories $\mathcal{M} \colon \CatTapeCB \to \Rel$, $\mathcal{M}(\s)\leq \mathcal{M}(\t)$, then
	$\s \minorinTCB \t$.
\end{corollary}

To conclude we only need the following lemma which holds for arbitrary fb-categories.

\begin{lemma}\label{lemma:fbPreordered}
	Let $\Cat{C}$ be a fb category monoidally enriched over a preorder $\leq$.
	Let $f,g\colon \Piu[i=1][n] {A_i} \to \Piu[j=1][m] {B_j}$ be arrows in $\Cat{C}$. Let $f_{ji}$ and $g_{ji}$ be, respectively, $\mu_i ; f; \pi_j$ and $\mu_i ; g; \pi_j$. Then $f\leq g$ iff $f_{ji} \leq g_{ji}$ for all $i,j$.
\end{lemma}
\begin{proof}Since composition is monotone, one has immediately that $f \leq g$ implies $f_{ji} \leq g_{ji}$. Vice versa, suppose that $f_{ji} \leq g_{ji}$ for all $j,i$. It is not difficult to verify that 
	\[
	f = \Bigl(
	\begin{tikzcd}[ampersand replacement=\&,column sep=3em]
		\PiuL[i=1][n]{A_i} \ar[r,"{\PiuL[i=1][n]{\diagg{A_i} m}}"] \& \PiuL[i=1][n]{\PiuL[j=1][m]{A_i}} \ar[r,"{\PiuL[i=1][n]{\PiuL[j=1][m]{f_{ji}}}}"] \& \PiuL[i=1][n]{\PiuL[j=1][m]{B_j}} \ar[r,"\codiagg{\Piu[j=1][m]{B_j}}{n}"] \& \PiuL[j=1][m]{B_j}
	\end{tikzcd}
	\Bigr)
	\]
	and similarly for $g$. Then again because of the monotonicity of composition, we have $f \leq g$.
\end{proof}

\subsection{Other proofs of Section~\ref{sec:CBPOPL}}\label{sec:appCB4}

\begin{proof}[Proof of Proposition \ref{prop:encodingsound}] Recall from Example \ref{ex:signature} that for an interpretation $\mathcal{I}= (\alpha_{\sort},\alpha_{\sign})$, we set $\alpha_{\sort}(A)=X$ and, for all $R\in \sign$, $\alpha_{\sign}(R)=\rho(R)$.
	The proof is by induction on $E$. The base cases are trivial. 
	For the inductive cases, we only illustrate the one for $E=E_1 \cap E_2$, as the remaining ones are simpler.
	\begin{align*}
		\TCBdsem{\encoding{E_1 \cap E_2}}_{\interpretation} &= \TCBdsem{\tape{\copier{A}}; (\encoding{E_1}\times \encoding{E_2}); \tape{\cocopier{A}}}_{\interpretation} \tag{def. $\encoding{-}$}\\
		& = \TCBdsem{\tape{\copier{A}}}_{\interpretation}; (\TCBdsem{\encoding{E_1}}_{\interpretation}\per \TCBdsem{\encoding{E_2}}_{\interpretation}); \TCBdsem{\tape{\cocopier{A}}}_{\interpretation}  \tag{funct. $\TCBdsem{-}$}\\
		& = \copier{X} ; (\TCBdsem{\encoding{E_1}}_{\interpretation}\per \TCBdsem{\encoding{E_2}}_{\interpretation}) ; \cocopier{X} \tag{$\mathcal{M}$ cb-morphism}\\
		& = \copier{X} ; (\dsemRel{E_1}\per \dsemRel{E_2}) ; \cocopier{X} \tag{ind. hyp.}\\
		& =\dsemRel{E_1} \cap \dsemRel{E_2} \tag{Def. $\copier{X}$, $\cocopier{X}$  in $\Rel$}\\
		& =\dsemRel{E_1 \cap E_2} \tag{Def. $\dsemRel{-}$}\\
	\end{align*}
\vphantom{.}
\end{proof}

\begin{proof}[Proof of Corollary \ref{crlFinal}]
	\begin{align*}
		E_1 \minorExpression E_2 &\iff \; \forall \interpretation \, . \; \dsemRel{E_1} \leq \dsemRel{E_2}   \tag{def. $\minorExpression$}\\
		& \iff \; \forall \interpretation \, . \; \TCBdsem{\encoding{E_1}}_{\interpretation} \leq \TCBdsem{\encoding{E_2}}_{\interpretation}   \tag{Proposition \ref{prop:encodingsound}}\\
		& \iff \; \forall \mathcal{M} \, . \; \mathcal{M}{(\encoding{E_1})} \leq \mathcal{M}{(\encoding{E_2})} \tag{Proposition \ref{prop:bijective}}\\
		& \iff \; \encoding{E_1} \minorinTCB \encoding{E_2} \tag{Theorem \ref{thm:completeness}}\\
	\end{align*}
\vphantom{.}
\end{proof}
 
\end{document}

%% file: tikz/stringdiag_ax1_right.tikz
\begin{tikzpicture}
	\begin{pgfonlayer}{nodelayer}
		\node [style=label] (8) at (1.5, 0.5) {$A$};
		\node [style=label] (11) at (-1.5, 0.5) {$A$};
		\node [style=label] (12) at (1.5, -0.5) {$B$};
		\node [style=label] (13) at (-1.5, -0.5) {$B$};
	\end{pgfonlayer}
	\begin{pgfonlayer}{edgelayer}
		\draw (11) to (8);
		\draw (13) to (12);
	\end{pgfonlayer}
\end{tikzpicture}

%% file: tikz/zx/unitary.tikz
\begin{tikzpicture}
	\begin{pgfonlayer}{nodelayer}
		\node [style=none] (129) at (1, 0) {};
		\node [style=box] (134) at (0, 0) {$U$};
		\node [style=none] (160) at (-1, 0) {};
	\end{pgfonlayer}
	\begin{pgfonlayer}{edgelayer}
		\draw (160.center) to (134);
		\draw (134) to (129.center);
	\end{pgfonlayer}
\end{tikzpicture}

%% file: tikz/cb/examples/copierApB.tikz
\begin{tikzpicture}
	\begin{pgfonlayer}{nodelayer}
		\node [style=label] (76) at (-2.25, 2.25) {$A$};
		\node [style=none] (77) at (-1.75, 1.75) {};
		\node [style=none] (78) at (-1.75, 2.75) {};
		\node [style=black] (80) at (0, 2.25) {};
		\node [style=none] (82) at (0.475, 2.55) {};
		\node [style=none] (83) at (0.475, 1.95) {};
		\node [style=none] (84) at (-1.75, 2.25) {};
		\node [style=none] (87) at (0.75, 1.75) {};
		\node [style=none] (88) at (0.75, 2.75) {};
		\node [style=label] (146) at (1.25, 2.55) {$A$};
		\node [style=label] (147) at (1.25, 1.95) {$A$};
		\node [style=label] (153) at (-2.25, -2.25) {$B$};
		\node [style=none] (154) at (-1.75, -2.75) {};
		\node [style=none] (155) at (-1.75, -1.75) {};
		\node [style=black] (156) at (0, -2.25) {};
		\node [style=none] (157) at (0.475, -1.95) {};
		\node [style=none] (158) at (0.475, -2.55) {};
		\node [style=none] (159) at (-1.75, -2.25) {};
		\node [style=none] (160) at (0.75, -2.75) {};
		\node [style=none] (161) at (0.75, -1.75) {};
		\node [style=label] (162) at (1.25, -1.95) {$B$};
		\node [style=label] (163) at (1.25, -2.55) {$B$};
		\node [style=none] (164) at (0.75, 0.975) {};
		\node [style=none] (165) at (0.75, 0.525) {};
		\node [style=label] (166) at (1.25, 1.05) {$A$};
		\node [style=label] (167) at (1.25, 0.45) {$B$};
		\node [style=none] (168) at (-1.05, 0.975) {};
		\node [style=none] (169) at (-1.05, 0.525) {};
		\node [style=none] (175) at (0.8, -0.525) {};
		\node [style=none] (176) at (0.8, -0.975) {};
		\node [style=label] (177) at (1.3, -0.45) {$B$};
		\node [style=label] (178) at (1.3, -1.05) {$A$};
		\node [style=none] (179) at (-1, -0.525) {};
		\node [style=none] (180) at (-1, -0.975) {};
		\node [style=none] (181) at (0.75, 1.25) {};
		\node [style=none] (182) at (-0.5, 1.25) {};
		\node [style=none] (183) at (-0.5, 0.25) {};
		\node [style=none] (184) at (0.75, 0.25) {};
		\node [style=none] (185) at (0.8, -0.25) {};
		\node [style=none] (186) at (-0.45, -0.25) {};
		\node [style=none] (187) at (-0.45, -1.25) {};
		\node [style=none] (188) at (0.8, -1.25) {};
		\node [style=none] (189) at (0.75, 2.55) {};
		\node [style=none] (190) at (0.75, 1.95) {};
		\node [style=none] (191) at (0.75, -1.95) {};
		\node [style=none] (192) at (0.75, -2.55) {};
	\end{pgfonlayer}
	\begin{pgfonlayer}{edgelayer}
		\draw [style=tape] (88.center)
			 to (87.center)
			 to (77.center)
			 to (78.center)
			 to cycle;
		\draw [style=tape] (161.center)
			 to (160.center)
			 to (154.center)
			 to (155.center)
			 to cycle;
		\draw [style=tape] (184.center)
			 to (181.center)
			 to (182.center)
			 to [bend right=90, looseness=2.00] (183.center)
			 to cycle;
		\draw [style=tape] (188.center)
			 to (185.center)
			 to (186.center)
			 to [bend right=90, looseness=2.00] (187.center)
			 to cycle;
		\draw (84.center) to (80);
		\draw [bend left] (80) to (82.center);
		\draw [bend right=330] (83.center) to (80);
		\draw (159.center) to (156);
		\draw [bend left] (156) to (157.center);
		\draw [bend left] (158.center) to (156);
		\draw (168.center) to (164.center);
		\draw (165.center) to (169.center);
		\draw (179.center) to (175.center);
		\draw (176.center) to (180.center);
		\draw (82.center) to (189.center);
		\draw (190.center) to (83.center);
		\draw (191.center) to (157.center);
		\draw (158.center) to (192.center);
	\end{pgfonlayer}
\end{tikzpicture}

%% file: tikz/cb/examples/dischargerApB.tikz
\begin{tikzpicture}
	\begin{pgfonlayer}{nodelayer}
		\node [style=label] (4) at (-1.5, 1) {$A$};
		\node [style=label] (5) at (-1.5, -1) {$B$};
		\node [style=none] (6) at (2.5, 0.5) {};
		\node [style=none] (7) at (2.5, -0.5) {};
		\node [style=none] (8) at (0, -1.5) {};
		\node [style=none] (9) at (0, 1.5) {};
		\node [style=none] (10) at (0, 0.5) {};
		\node [style=none] (11) at (0.5, 0) {};
		\node [style=none] (12) at (0, -0.5) {};
		\node [style=none] (13) at (1.5, -0.5) {};
		\node [style=none] (14) at (1.5, 0.5) {};
		\node [style=none] (16) at (-1, 0.5) {};
		\node [style=none] (17) at (-1, 1.5) {};
		\node [style=none] (18) at (-1, -1.5) {};
		\node [style=none] (19) at (-1, -0.5) {};
		\node [style=black] (20) at (-0.25, 1) {};
		\node [style=black] (21) at (-0.25, -1) {};
		\node [style=none] (22) at (-1, -1) {};
		\node [style=none] (23) at (-1, 1) {};
	\end{pgfonlayer}
	\begin{pgfonlayer}{edgelayer}
		\draw [style=tape] (11.center)
			 to [bend right=45] (10.center)
			 to (16.center)
			 to (17.center)
			 to (9.center)
			 to [bend left] (14.center)
			 to (6.center)
			 to (7.center)
			 to (13.center)
			 to [bend left] (8.center)
			 to (18.center)
			 to (19.center)
			 to (12.center)
			 to [bend right=45] cycle;
		\draw (20) to (23.center);
		\draw (22.center) to (21);
	\end{pgfonlayer}
\end{tikzpicture}

%% file: tikz/lemmaDlG/step1.tikz
\begin{tikzpicture}
	\begin{pgfonlayer}{nodelayer}
		\node [style=none] (8) at (-10.25, 3) {};
		\node [style=none] (15) at (-10.25, 0.5) {};
		\node [style=none] (24) at (-10.25, -5.5) {};
		\node [style=none] (30) at (2, 0.5) {};
		\node [style=none] (64) at (2, -2) {};
		\node [style=label] (65) at (-9.25, -4.5) {$(P'' \piu P')(Q \piu R)$};
		\node [style=label] (66) at (10.5, 1.5) {$P''Q$};
		\node [style=label] (67) at (10.5, -1) {$P'Q$};
		\node [style=label] (69) at (10.5, 4) {$UQ$};
		\node [style=label] (70) at (10.5, -3.5) {$UR$};
		\node [style=label] (71) at (10.5, -6) {$P''R$};
		\node [style=label] (72) at (-9.25, 4) {$UQ$};
		\node [style=label] (74) at (-9.25, 1.5) {$UR$};
		\node [style=dots] (77) at (-9.25, 3.1) {$\vdots$};
		\node [style=dots] (78) at (10.5, 3.1) {$\vdots$};
		\node [style=none] (79) at (11.25, 0.5) {};
		\node [style=none] (80) at (11.25, -2) {};
		\node [style=none] (83) at (11.25, -4.5) {};
		\node [style=none] (84) at (11.25, -7) {};
		\node [style=dots] (85) at (10.5, 0.6) {$\vdots$};
		\node [style=dots] (86) at (10.5, -1.9) {$\vdots$};
		\node [style=dots] (87) at (10.5, -4.4) {$\vdots$};
		\node [style=dots] (88) at (10.5, -6.9) {$\vdots$};
		\node [style=dots] (91) at (-9.25, 0.6) {$\vdots$};
		\node [style=dots] (92) at (-9.25, -5.4) {$\vdots$};
		\node [style=none] (93) at (11.25, 3) {};
		\node [style=none] (94) at (-5.75, -10.5) {};
		\node [style=none] (95) at (-5.75, -1) {};
		\node [style=none] (96) at (1.5, -1) {};
		\node [style=none] (97) at (1.5, -10.5) {};
		\node [style=none] (98) at (-2, -5.5) {$\delta^l$};
		\node [style=none] (99) at (-5.75, -5.5) {};
		\node [style=none] (100) at (1.5, -9.5) {};
		\node [style=none] (102) at (1.5, -2) {};
		\node [style=none] (104) at (1.75, -9.5) {};
		\node [style=none] (105) at (2, -9.5) {};
		\node [style=none] (106) at (1.75, -2) {};
		\node [style=none] (120) at (1.5, -4.5) {};
		\node [style=none] (125) at (1.5, -7) {};
		\node [style=none] (126) at (1.75, -7) {};
		\node [style=none] (127) at (2, -7) {};
		\node [style=none] (130) at (2, -4.5) {};
		\node [style=none] (131) at (9, 0.5) {};
		\node [style=none] (132) at (9, -2) {};
		\node [style=none] (133) at (9, -4.5) {};
		\node [style=none] (134) at (9, -7) {};
		\node [style=label] (135) at (10.5, -8.5) {$P'R$};
		\node [style=none] (136) at (11.25, -9.5) {};
		\node [style=dots] (137) at (10.5, -9.4) {$\vdots$};
		\node [style=none] (138) at (9, -9.5) {};
		\node [style=none] (139) at (1.75, -4.5) {};
	\end{pgfonlayer}
	\begin{pgfonlayer}{edgelayer}
		\draw [line width=3mm, color=tapeBg] (83.center)
			 to (133.center)
			 to [in=0, out=180] (30.center)
			 to (15.center);
		\draw [line width=3mm, color=tapeBg] (79.center)
			 to (131.center)
			 to [in=0, out=-180] (64.center)
			 to (106.center)
			 to (102.center);
		\path [fill=tapeBg] (102.center)
			 to (96.center)
			 to (95.center)
			 to (99.center)
			 to (94.center)
			 to (97.center)
			 to (100.center)
			 to cycle;
		\draw [line width=3mm, color=tapeBg] (100.center)
			 to (104.center)
			 to [in=-180, out=0] (105.center)
			 to (138.center)
			 to (136.center);
		\draw [line width=3mm, color=tapeBg] (24.center) to (99.center);
		\draw [line width=3mm, color=tapeBg] (125.center)
			 to (126.center)
			 to [in=-180, out=0] (127.center)
			 to (134.center)
			 to (84.center);
		\draw [line width=3mm, color=tapeBg] (8.center) to (93.center);
		\draw [line width=3mm, color=tapeBg] (120.center)
			 to (139.center)
			 to (130.center)
			 to [in=-180, out=0] (132.center)
			 to (80.center);
	\end{pgfonlayer}
\end{tikzpicture}

%% file: tikz/lemmaDlG/step2.tikz
\begin{tikzpicture}
	\begin{pgfonlayer}{nodelayer}
		\node [style=none] (8) at (-10.25, 3) {};
		\node [style=none] (15) at (-10.25, 0.5) {};
		\node [style=none] (24) at (-10, -3.25) {};
		\node [style=none] (30) at (2.5, 0.5) {};
		\node [style=none] (64) at (2.5, -2) {};
		\node [style=label] (65) at (-9, -2.25) {$P''(Q \piu R)$};
		\node [style=label] (66) at (10.5, 1.5) {$P''Q$};
		\node [style=label] (67) at (10.5, -1) {$P'Q$};
		\node [style=label] (69) at (10.5, 4) {$UQ$};
		\node [style=label] (70) at (10.5, -3.5) {$UR$};
		\node [style=label] (71) at (10.5, -6) {$P''R$};
		\node [style=label] (72) at (-9.25, 4) {$UQ$};
		\node [style=label] (74) at (-9.25, 1.5) {$UR$};
		\node [style=dots] (77) at (-9.25, 3.1) {$\vdots$};
		\node [style=dots] (78) at (10.5, 3.1) {$\vdots$};
		\node [style=none] (79) at (11.25, 0.5) {};
		\node [style=none] (80) at (11.25, -2) {};
		\node [style=none] (83) at (11.25, -4.5) {};
		\node [style=none] (84) at (11.25, -7) {};
		\node [style=dots] (85) at (10.5, 0.6) {$\vdots$};
		\node [style=dots] (86) at (10.5, -1.9) {$\vdots$};
		\node [style=dots] (87) at (10.5, -4.4) {$\vdots$};
		\node [style=dots] (88) at (10.5, -6.9) {$\vdots$};
		\node [style=dots] (91) at (-9.25, 0.6) {$\vdots$};
		\node [style=dots] (92) at (-9, -3.15) {$\vdots$};
		\node [style=none] (93) at (11.25, 3) {};
		\node [style=none] (104) at (0, -9.5) {};
		\node [style=none] (105) at (2, -9.5) {};
		\node [style=none] (106) at (2.25, -2) {};
		\node [style=none] (120) at (0, -4.5) {};
		\node [style=none] (125) at (0, -7) {};
		\node [style=none] (127) at (2.75, -7) {};
		\node [style=none] (130) at (2.75, -4.5) {};
		\node [style=none] (131) at (9, 0.5) {};
		\node [style=none] (132) at (9, -2) {};
		\node [style=none] (133) at (9, -4.5) {};
		\node [style=none] (134) at (9, -7) {};
		\node [style=label] (135) at (10.5, -8.5) {$P'R$};
		\node [style=none] (136) at (11.25, -9.5) {};
		\node [style=dots] (137) at (10.5, -9.4) {$\vdots$};
		\node [style=none] (138) at (9, -9.5) {};
		\node [style=none] (151) at (-5.75, -5.5) {};
		\node [style=none] (152) at (-5.75, -1) {};
		\node [style=none] (153) at (-1.75, -1) {};
		\node [style=none] (154) at (-1.75, -5.5) {};
		\node [style=none] (155) at (-3.75, -3.25) {$\delta^l$};
		\node [style=none] (156) at (-5.75, -3.25) {};
		\node [style=none] (157) at (-1.75, -4.5) {};
		\node [style=none] (158) at (-1.75, -2) {};
		\node [style=none] (159) at (-5.75, -10.5) {};
		\node [style=none] (160) at (-5.75, -6) {};
		\node [style=none] (161) at (-1.75, -6) {};
		\node [style=none] (162) at (-1.75, -10.5) {};
		\node [style=none] (163) at (-3.75, -8.25) {$\delta^l$};
		\node [style=none] (164) at (-5.75, -8.25) {};
		\node [style=none] (165) at (-1.75, -9.5) {};
		\node [style=none] (166) at (-1.75, -7) {};
		\node [style=none] (167) at (-10, -8.25) {};
		\node [style=label] (168) at (-9, -7.25) {$P'(Q \piu R)$};
		\node [style=dots] (169) at (-9, -8.15) {$\vdots$};
	\end{pgfonlayer}
	\begin{pgfonlayer}{edgelayer}
		\draw [line width=3mm, color=tapeBg] (83.center)
			 to (133.center)
			 to [in=0, out=180] (30.center)
			 to (15.center);
		\draw [line width=3mm, color=tapeBg](158.center)
			 to (106.center)
			 to (64.center)
			 to [in=-180, out=0] (131.center)
			 to (79.center);
		\draw [line width=3mm, color=tapeBg](165.center)
			 to (104.center)
			 to [in=-180, out=0] (105.center)
			 to (138.center)
			 to (136.center);
		\draw [line width=3mm, color=tapeBg] (84.center)
			 to (134.center)
			 to (127.center)
			 to [in=0, out=180, looseness=0.75] (120.center)
			 to (157.center);
		\draw [line width=3mm, color=tapeBg] (8.center) to (93.center);
		\draw [line width=3mm, color=tapeBg] (166.center)
			 to (125.center)
			 to [in=-180, out=0, looseness=0.75] (130.center)
			 to [in=-180, out=0] (132.center)
			 to (80.center);
		\path [fill=tapeBg] (152.center)
			 to (156.center)
			 to (151.center)
			 to (154.center)
			 to (157.center)
			 to (158.center)
			 to (153.center);
		\path [fill=tapeBg] (164.center)
			 to (159.center)
			 to (162.center)
			 to (165.center)
			 to (166.center)
			 to (161.center)
			 to (160.center)
			 to cycle;
		\draw [line width=3mm, color=tapeBg](24.center) to (156.center);
		\draw [line width=3mm, color=tapeBg](167.center) to (164.center);
	\end{pgfonlayer}
\end{tikzpicture}

%% file: tikz/lemmaDlG/step4.tikz
\begin{tikzpicture}
	\begin{pgfonlayer}{nodelayer}
		\node [style=none] (8) at (-10.25, 3) {};
		\node [style=none] (15) at (-10.25, 0.5) {};
		\node [style=none] (24) at (-10, -3.25) {};
		\node [style=none] (30) at (-0.25, 0.5) {};
		\node [style=none] (64) at (-0.25, -2) {};
		\node [style=label] (65) at (-9, -2.25) {$P''(Q \piu R)$};
		\node [style=label] (66) at (10.5, 1.5) {$P''Q$};
		\node [style=label] (67) at (10.5, -1) {$P'Q$};
		\node [style=label] (69) at (10.5, 4) {$UQ$};
		\node [style=label] (70) at (10.5, -3.5) {$UR$};
		\node [style=label] (71) at (10.5, -6) {$P''R$};
		\node [style=label] (72) at (-9.25, 4) {$UQ$};
		\node [style=label] (74) at (-9.25, 1.5) {$UR$};
		\node [style=dots] (77) at (-9.25, 3.1) {$\vdots$};
		\node [style=dots] (78) at (10.5, 3.1) {$\vdots$};
		\node [style=none] (79) at (11.25, 0.5) {};
		\node [style=none] (80) at (11.25, -2) {};
		\node [style=none] (83) at (11.25, -4.5) {};
		\node [style=none] (84) at (11.25, -7) {};
		\node [style=dots] (85) at (10.5, 0.6) {$\vdots$};
		\node [style=dots] (86) at (10.5, -1.9) {$\vdots$};
		\node [style=dots] (87) at (10.5, -4.4) {$\vdots$};
		\node [style=dots] (88) at (10.5, -6.9) {$\vdots$};
		\node [style=dots] (91) at (-9.25, 0.6) {$\vdots$};
		\node [style=dots] (92) at (-9, -3.15) {$\vdots$};
		\node [style=none] (93) at (11.25, 3) {};
		\node [style=none] (104) at (4, -9.5) {};
		\node [style=none] (105) at (6, -9.5) {};
		\node [style=none] (106) at (-0.75, -2) {};
		\node [style=none] (120) at (4, -4.5) {};
		\node [style=none] (125) at (4, -7) {};
		\node [style=none] (127) at (7.25, -7) {};
		\node [style=none] (130) at (7.25, -4.5) {};
		\node [style=none] (131) at (4, 0.5) {};
		\node [style=none] (132) at (10, -2) {};
		\node [style=none] (133) at (10, -4.5) {};
		\node [style=none] (134) at (9, -7) {};
		\node [style=label] (135) at (10.5, -8.5) {$P'R$};
		\node [style=none] (136) at (11.25, -9.5) {};
		\node [style=dots] (137) at (10.5, -9.4) {$\vdots$};
		\node [style=none] (138) at (9, -9.5) {};
		\node [style=none] (151) at (-5.75, -5.5) {};
		\node [style=none] (152) at (-5.75, -1) {};
		\node [style=none] (153) at (-1.75, -1) {};
		\node [style=none] (154) at (-1.75, -5.5) {};
		\node [style=none] (155) at (-3.75, -3.25) {$\delta^l$};
		\node [style=none] (156) at (-5.75, -3.25) {};
		\node [style=none] (157) at (-1.75, -4.5) {};
		\node [style=none] (158) at (-1.75, -2) {};
		\node [style=none] (159) at (-0.25, -10.5) {};
		\node [style=none] (160) at (-0.25, -6) {};
		\node [style=none] (161) at (3.75, -6) {};
		\node [style=none] (162) at (3.75, -10.5) {};
		\node [style=none] (163) at (1.75, -8.25) {$\delta^l$};
		\node [style=none] (164) at (-0.25, -8.25) {};
		\node [style=none] (165) at (3.75, -9.5) {};
		\node [style=none] (166) at (3.75, -7) {};
		\node [style=none] (167) at (-10, -8.25) {};
		\node [style=label] (168) at (-9, -7.25) {$P'(Q \piu R)$};
		\node [style=dots] (169) at (-9, -8.15) {$\vdots$};
		\node [style=none] (170) at (7.25, -2) {};
		\node [style=none] (171) at (4, -2) {};
	\end{pgfonlayer}
	\begin{pgfonlayer}{edgelayer}
		\draw [line width=3mm, color=tapeBg] (15.center)
			 to (30.center)
			 to [in=-180, out=0] (171.center)
			 to (170.center)
			 to [in=180, out=0, looseness=0.75] (133.center)
			 to (83.center);
		\draw [line width=3mm, color=tapeBg](158.center)
			 to (106.center)
			 to (64.center)
			 to [in=-180, out=0] (131.center)
			 to (79.center);
		\draw [line width=3mm, color=tapeBg](165.center)
			 to (104.center)
			 to [in=-180, out=0] (105.center)
			 to (138.center)
			 to (136.center);
		\draw [line width=3mm, color=tapeBg] (84.center)
			 to (134.center)
			 to (127.center)
			 to [in=0, out=180, looseness=0.75] (120.center)
			 to (157.center);
		\draw [line width=3mm, color=tapeBg] (8.center) to (93.center);
		\draw [line width=3mm, color=tapeBg](166.center)
			 to (125.center)
			 to [in=-180, out=0, looseness=0.75] (130.center)
			 to [in=-180, out=0] (132.center)
			 to (80.center);
		\path [fill=tapeBg] (152.center)
			 to (156.center)
			 to (151.center)
			 to (154.center)
			 to (157.center)
			 to (158.center)
			 to (153.center);
		\path [fill=tapeBg] (164.center)
			 to (159.center)
			 to (162.center)
			 to (165.center)
			 to (166.center)
			 to (161.center)
			 to (160.center)
			 to cycle;
		\draw [line width=3mm, color=tapeBg](24.center) to (156.center);
		\draw [line width=3mm, color=tapeBg](167.center) to (164.center);
	\end{pgfonlayer}
\end{tikzpicture}